\pdfoutput=1

\documentclass[ aip,  jmp, amsmath,amssymb,  reprint,]{revtex4-1}%
\usepackage{graphicx}
\usepackage{dcolumn}
\usepackage{bm}
\usepackage{amsmath}
\usepackage{natmove}
\usepackage{longtable}%
\usepackage{amsfonts}%
\usepackage{amssymb}
\providecommand{\U}[1]{\protect\rule{.1in}{.1in}}
\providecommand{\U}[1]{\protect\rule{.1in}{.1in}}
\providecommand{\U}[1]{\protect\rule{.1in}{.1in}}
\providecommand{\U}[1]{\protect\rule{.1in}{.1in}}
\providecommand{\U}[1]{\protect\rule{.1in}{.1in}}

\begin{document}
\title{Modeling techniques for quantum cascade lasers}
\author{Christian Jirauschek}
\affiliation{Institute for Nanoelectronics, Technische Universit\"{a}%
t M\"{u}nchen, D-80333 Munich, Germany}
\author{Tillmann Kubis}
\affiliation
{Network for Computational Nanotechnology, Purdue University, 207 S Martin Jischke Drive, West Lafayette, Indiana 47907, USA}
\date{10 December 2014, published as Appl. Phys. Rev. 1, 011307 (2014)}
\begin{abstract}
Quantum cascade lasers are unipolar semiconductor lasers covering a wide range of the infrared and terahertz spectrum. Lasing action is achieved by using optical intersubband transitions between quantized states in specifically designed multiple-quantum-well heterostructures. A systematic improvement of quantum cascade lasers with respect to operating temperature, efficiency and spectral range requires detailed modeling of the underlying physical processes in these structures. Moreover, the quantum cascade laser constitutes a versatile model device for the development and improvement of simulation techniques in nano- and optoelectronics. This review provides a comprehensive survey and discussion of the modeling techniques used for the simulation of quantum cascade lasers. The main focus is on the modeling of carrier transport in the nanostructured gain medium, while the simulation of the optical cavity is covered at a more basic level. Specifically, the transfer matrix and finite difference methods for solving the one-dimensional Schr{\"o}%
dinger equation and Schr{\"o}%
dinger-Poisson system are discussed, providing the quantized states in the multiple-quantum-well active region. The modeling of the optical cavity is covered with a focus on basic waveguide resonator structures. Furthermore, various carrier transport simulation methods are discussed, ranging from basic empirical approaches to advanced self-consistent techniques. The methods include empirical rate equation and related Maxwell-Bloch equation approaches, self-consistent rate equation and ensemble Monte Carlo methods, as well as quantum transport approaches, in particular the density matrix and non-equilibrium Green's function (NEGF) formalism. The derived scattering rates and self-energies are generally valid for n-type devices based on one-dimensional quantum confinement, such as quantum well structures.\end{abstract}
\maketitle

\tableofcontents

\section*{Glossary of symbols}%

\begin{longtable}
[c]{ll} $\hat{1}$ & Unity operator\\ 1D & One-dimensional\\ 3D &
Three-dimensional\\ $A$ & Spectral function\\ $a$ & Power loss coefficient\\
$a_{\mathrm{m}}$ & Mirror or outcoupling power loss coefficient\\
$a_{\mathrm{w}}$ & Waveguide power loss coefficient\\ CPU & Central processing
unit\\ $C$ & Ideal midpoint of a rough interface of two\\ & \ \ adjacent
materials\\ $c$ & Vacuum speed of light\\ $D_{\mathrm{TO}}$ & Optical
deformation potential field component $\parallel\mathbf{u}$\\ $D^{R}$ &
Retarded environmental Green's function\\ $D^{<}$ & Lesser environmental
Green's function\\ $d_{ij}$ & Optical dipole matrix element for transition\\ &
\ \ from level $i$ to $j$\\ $\mathbf{E}$ & Electric field vector\\ $E_{\xi}$ &
Electric field component ($\xi=x,y,z$)\\ $\mathbf{\hat{E}}$ & Electric field
amplitude vector\\ $\hat{E}_{\xi}$ & Component of $\mathbf{\hat{E}}$
($\xi=x,y,z$)\\ $E$ & Energy\\ $E^{\prime}$ & Energy above conduction band
edge, $E^{\prime}=E-V_{\mathrm{c} }$\\ $E_{i}$/$E_{i\mathbf{k}}$ & Eigenenergy
of state $\left|                i\right\rangle $/$\left|   i\mathbf
{k}\right\rangle $\\ $\tilde{E}_{i}$ & Energy of state $i$ relative to
conduction band edge\\ $E_{i}^{\mathrm{F}}$ & Quasi Fermi level of subband
$i$\\ $E_{\mathrm{p}}$ & Bias drop over a single QCL period\\ $\mathrm
{E}\left(                x\right)                $ & Complete elliptical
integral of the second kind\\ EMC & Ensemble Monte Carlo\\ $e$ & Elementary
charge\\ $f$ & Distribution function\\ $f_{i}$ & Carrier distribution function
in subband $i$\\ $f_{i}^{\mathrm{FD}}$ & Fermi-Dirac distribution function\\
$f_{i}^{\mathrm{MB}}$ & Maxwell-Boltzmann distribution function\\ $G^{R}$ &
Retarded electron Green's function\\ $G^{<}$ & Lesser electron Green's
function\\ $g$ & Power gain coefficient\\ $g_{\mathrm{th}}$ & Threshold power
gain coefficient\\ $\mathbf{H}$ & Magnetic field vector\\ $\mathbf{\hat{H}}$ &
Magnetic field amplitude vector\\ $\hat{H}$ & Total Hamilton operator\\
$\hat{H}_{0}$ & Free electron Hamilton operator\\ $H_{\xi}$ & Magnetic field
component ($\xi=x,y,z$)\\ $\hat{H}_{\xi}$ & Component of $\mathbf{\hat{H}}$
($\xi=x,y,z$)\\ $\hbar$ & Reduced Planck constant\\ $I$ & Optical intensity\\
$\Im$ & Imaginary part\\ $\mathrm{I}_{0}$ & Modified Bessel function\\ $I_{m}$
& Optical intensity in mode $m$\\ $J$ & Current density\\ $\hat{J}$ & Current
density operator\\ $\mathbf{k}$ & In-plane wave vector, $\mathbf{k}=\left[
k_{x},k_{y}\right]              ^{\mathrm{T} }$\\ $\mathbf{k}^{\prime}$ &
Final state in-plane wave vector, $\mathbf{k}^{\prime}=\left[
k_{x} ^{\prime} ,k_{y}^{\prime}\right]          ^{\mathrm{T}}$\\ $k_{0}$ &
Vacuum wave number $\omega/c$\\ $k_{21}$ & Wave number $\omega_{21}n_{0}/c$\\
$k_{n}$ & Wave number (of $\psi$ or $\hat{H} _{y}$) in segment $n$\\
$\tilde{k}_{n}$ & $k_{n}/m_{n}^{\ast}$ (for $\psi$); $k_{n}/\epsilon
_{\mathrm{r}}^{\left(  n\right)                }$ (for $\hat{H}_{y} $)\\
$k_{\mathrm{B}}$ & Boltzmann constant\\ $L$ & Resonator length\\
$L_{\mathrm{p}}$ & Length of a QCL period\\ $\mathcal{L}$ & Lorentzian
lineshape function\\ $\mathcal{L}_{ij}$ & Lorentzian lineshape function for
the transition\\ & \ \ from subband $i$ to $j$\\ LA & Longitudinal acoustic\\
LO & Longitudinal optical\\ $m^{\parallel}$ & In-plane effective mass\\
$m^{\ast}$ & $\Gamma$ valley effective mass\\ $m^{\ast}$ & Effective mass in
the growth direction\\ $N_{\mathrm{e}}\ $ & Number of electrons\\
$N_{\mathrm{p}}\ $ & Number of periods\\ $N_{\mathrm{Ph}}$ & Mode independent
phonon occupation number\\ $N_{\mathbf{Q}}$ & Phonon occupation number of a
mode with\\ & \ \  wave vector $\mathbf{Q}$\\ NEGF & Nonequilibrium Green's
function\\ $n$ & Electron density\\ $n_{0}$ & Refractive index\\
$n_{\mathrm{D}}$ & Donor concentration\\ $n_{\mathrm{eff}}$ & Effective
refractive index $\beta/k_{0} $\\ $n^{\mathrm{s}}$ & Total sheet density per
QCL period\\ $n_{i}^{\mathrm{s}}$ & Electron sheet density of level $i$\\
$n_{i}^{E}$ & Density of states per unit area and energy in\\ & \ \ subband
$i$, $n_{i}^{E}=m_{i}^{\parallel}/\left(                \pi\hbar
^{2}\right)        $\\ $P$ & Probability to find a single impurity\\
$\mathrm{\Pr}$ & Principal value integral\\ $\mathbf{Q}$ & Phonon wave vector
$\left[                \mathbf{q} ,q_{z}\right]               ^{\mathrm{T}}$\\
$\mathbf{q}$ & In-plane component of $\mathbf{Q}$, exchanged wave vector\\
$q_{\mathrm{s}}$ & Inverse screening length\\ QCL & Quantum cascade laser\\
$R$ & Facet reflectance\\ $\Re$ & Real part\\ RPA & Random phase
approximation\\ $\mathbf{r}$ & In-plane position vector $\left(
x,y\right)                ^{\mathrm{T}}$\\ $s$ & Numerical grid spacing\\ $S$
& In-plane cross section area\\ $S_{\mathrm{g}}$ & Gain medium cross section
area\\ $T$ & Facet transmittance\\ $T_{\mathrm{e}}$ & Electron temperature\\
$T_{i}$ & Electron temperature in subband $i$\\ $T_{\mathrm{L}}$ & Lattice
temperature\\ TA & Transverse acoustic\\ TO & Transverse optical\\ $t$ &
Time\\ $\mathbf{u}$ & Displacement vector for lattice vibrations\\ $V$ &
Potential\\ $\tilde{V}$ & Electrostatic potential energy\\ $V_{\mathrm{c}}$ &
Conduction band edge potential\\ $V_{ii_{0}jj_{0} }^{\mathrm{b}}$ & Bare
Coulomb matrix elements\\ $V_{ii_{0}jj_{0}}^{\mathrm{s}}$ & Screened Coulomb
matrix element\\ $V_{\text{imp}}$ & Impurity scattering potential\\
$V_{\mathrm{IR}}$ & Interface roughness potential\\ $V_{\mathrm{o}}$ &
Conduction band offset\\ $\hat{v}$ & Velocity operator\\ $v_{\mathrm{s}}$ &
Longitudinal sound velocity\\ $V_{j\mathbf{k}^{\prime},i\mathbf{k}}$ & Matrix
elements for elastic scattering processes\\ $V_{j\mathbf{k}^{\prime}
,i\mathbf{k}}^{\pm}$ & Matrix elements for emission/absorption processes\\
$W_{i\mathbf{k},j\mathbf{k}^{\prime}}$ & Transition rate from an initial state
$\left|                i\mathbf{k}\right\rangle $\\ & \ \ to a final state
$\left|      j\mathbf{k}^{\prime}\right\rangle $\\ $W_{i\mathbf{k},j\mathbf
{k}^{\prime} }^{\pm}$ & Emission/absorption rate from an initial state
$\left|       i\mathbf{k}\right\rangle $\\ & \ \ to a final state $\left|
j\mathbf{k} ^{\prime}\right\rangle $\\ $W_{ij}^{\mathrm{opt}}$ & Stimulated
transition rate from subband $i$ to $j$\\ $\mathbf{x}$ & Position vector
$\left[      x,y,z\right]             ^{\mathrm{T}}$\\ $\hat{x}$ & Position
operator\\ $x$ & In-plane coordinate in light propagation direction\\ $x$ &
Alloy mole fraction\\ $y$ & In-plane coordinate\\ $z$ & Coordinate in growth
direction\\ $\alpha$ & Optical field amplitude absorption coefficient\\
$\alpha^{\prime}$ & Nonparabolicity parameter\\ $\beta$ & Complex propagation
constant\\ $\beta^{\prime}$ & Nonparabolicity parameter\\ $\Gamma$ & Overlap
factor\\ $\tilde{\Gamma}$ & Scattering rate\\ $\gamma_{E}$ & Energy relaxation
rate\\ $\gamma_{ij}$ & Optical linewidth of transition from level $i$ to $j$\\
$\delta\hat{V}$ & Photonic perturbation potential\\ $\Delta$ & Standard
deviation of interface roughness\\ $\Delta_{21}$ & Inversion, $\Delta
_{21}=\rho_{22}-\rho_{11}$\\ $\Delta_{21}^{\mathrm{eq}}$ & Equilibrium
inversion\\ $\Delta_{n}$ & Length of $n$th segment; thickness of $n$th layer\\
$\Delta_{z}$ & Uniform grid spacing\\ $\epsilon$ & Permittivity,
$\epsilon=\epsilon_{0}\epsilon_{\mathrm{r}}$\\ $\epsilon_{0}$ & Vacuum
permittivity\\ $\epsilon_{\mathrm{r} }$ & Dielectric constant\\ $\epsilon
_{\mathrm{r},0}$ & Static dielectric constant\\ $\epsilon_{\mathrm{r},\infty}$
& Dielectric constant at very high frequencies\\ $\phi$ & Potential drop per
period\\ $\Phi$ & Electrostatic potential\\ $\eta_{21}$ & Slowly varying
envelope function of $\rho_{21}$\\ $\Lambda$ & Interface roughness correlation
length\\ $\mu$ & Chemical potential\\ $\mu_{0}$ & Vacuum permeability\\ $\Xi$
& Deformation potential\\ $\rho$ & Space charge\\ $\rho_{\mathrm{c}}$ &
Density of crystal\\ $\rho_{ij}$ & Density matrix element\\ $\rho_{N}$ &
Number density of crystal\\ $\Sigma^{R}$ & Retarded self-energy\\ $\Sigma^{<}$
& Lesser self-energy\\ $\sigma$ & Complex conductance\\ $\tau_{i}$ & Lifetime
of level $i$\\ $\tau_{ji}$ & Inverse transition rate from subband $j$ to $i$\\
$\psi_{i}\ $ & Wave function of subband $i$\\ $\Omega_{ij}$ & Rabi frequency
for transition from level $i$ to $j$\\ $\Omega_{\mathrm{c}}$ & Crystal
volume\\ $\omega$ & Angular frequency\\ $\omega_{D}$ & Debye frequency\\
$\omega_{ij}$ & Resonance frequency for transition from subband\\ & \ \ $i$ to
$j$, $\left(              E_{i}-E_{j}\right)                /\hbar$\\
$\omega_{\mathrm{LO}}$ & LO\ phonon frequency at $Q=0$\\ $\omega_{\mathrm{TO}
}$ & TO\ phonon frequency at $Q=0$\\ $\left[              \hat{H},\hat{x}
\right]                _{-}$ & Commutator of Hamilton operator $\hat{H}$ and\\
& \ \ position operator $\hat{x}$, $\hat{H}\hat{x}-\hat{x}\hat{H}$
\end{longtable}
\setcounter{table}{0}

\section{\label{sec:Intro}Introduction}

The quantum cascade laser (QCL) is a fascinating device, combining aspects
from different fields such as nanoelectronics and quantum engineering,
plasmonics, as well as subwavelength and nonlinear photonics. Since its first
experimental realization in 1994 \cite{1994Sci...264..553F}, rapid progress
has been achieved with respect to its performance, in particular the output
power and the frequency range covered. The development of innovative types of
QCLs and subsequent design optimization has gone hand in hand with detailed
modeling, involving more and more sophisticated simulation tools. The
advancement of simulation methods is driven by an intrinsic motivation to
describe the underlying physical processes as exactly and realistically as
possible. In addition to scientific motivations improved agreement with
experimental data and enhanced predictive power is desired. Furthermore, the
extension and improvement of QCL simulation methods is largely driven by
experimental progress, such as innovative designs.

The QCL is a special type of semiconductor laser where the optical gain medium
consists of a multiple quantum well heterostructure, giving rise to the
formation of quantized electron states. These so-called subbands assume the
role of the laser levels in the QCL. To date, lasing has only been obtained
for n-type QCLs, i.e., designs using intersubband transitions in the
conduction band. Well established material systems are InGaAs/InAlAs on InP
substrate, and GaAs/AlGaAs on GaAs
\cite{2007NaPho...1..517W,2012NaPho...6..432Y}. Furthermore, antimonide QCLs
have been demonstrated, e.g., near- and mid-infrared QCLs using InAs/AlSb
\cite{2003ApPhL..82.1003O,2010ApPhL..96n1110C} and aluminum-free InGaAs/GaAsSb
QCLs operating in the mid-infrared \cite{5270385} and terahertz
\cite{2010ApPhL..97z1110D} regime. Since all these materials (apart from AlSb)
have direct bandgaps, the development of simulation methods and tools for QCLs
has focused on the conduction band $\Gamma$ valley where the lasing
transitions take place.

\begin{figure}[ptb]
\includegraphics{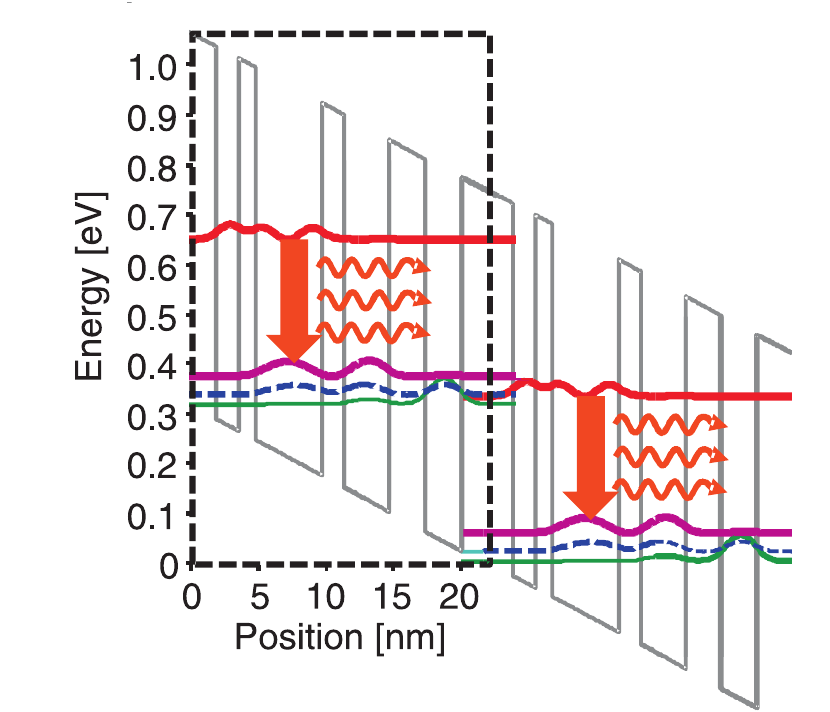}
\caption{{}(Color online) Conduction band profile and probability densities
for a mid-infrared QCL \cite{2010NaPho...4...99B} lasing at $5~\mathrm{\mu m}%
$. Only the relevant energy states are displayed. The upper and lower laser
level are indicated by bold solid lines. Furthermore, the depopulation (dashed
line) and injector (thin solid line) state probability densities are shown.
The rectangle denotes a single QCL\ period.}%
\label{fig:qcl}%
\end{figure}

In contrast to conventional semiconductor lasers where lasing occurs due to
electron-hole recombination between conduction and valence band, the emission
wavelength of QCLs is not determined by the bandgap of the material. Rather,
by adequately designing the nanostructured gain medium, the device can be
engineered to lase at a certain frequency. In this way, the mid/far infrared
and terahertz regimes become also accessible, important for applications such
as chemical and biological sensing, imaging, and special communication
applications. In addition, the QCL offers the typical advantages of
semiconductor devices, such as compactness, reliability, and potentially low
production costs\ \cite{2010CPL...487....1C}. The basic working principle of
the QCL is illustrated in Fig.\thinspace\ref{fig:qcl}. Lasing is obtained due
to stimulated emission between the upper and the lower laser level, here
resulting in mid-infrared radiation at a wavelength of $5~\mathrm{\mu m}$. The
depopulation level is separated from the lower laser level by the longitudinal
optical (LO) phonon energy for InGaAs of $\thicksim30~\mathrm{meV}$, enabling
efficient depopulation of the lower laser level. The injector level, located
between two adjacent periods, ensures electron transport and injection into
the upper laser level of the next period. The optical field confinement and
beam shaping is provided by a specifically designed resonator, which is
frequently based on plasmonic effects and can exhibit subwavelength
structuring for a further performance improvement.

The QCL operating principle was devised in 1971 by Kazarinov and Suris
\cite{Kaz}. However, due to the experimental challenges involved, it took more
than two decades until QCL operation was experimentally demonstrated at Bell
Labs by Capasso and his co-workers, Faist et al. \cite{1994Sci...264..553F}.
Since then, considerable progress has been achieved, e.g., with respect to
operating temperature \cite{a7}, output power \cite{2011ApPhL..98r1102B}, and
the available frequency range \cite{2010ApPhL..97h1110S,2010ApPhL..96n1110C}.
Meanwhile, QCLs cover a spectral range from 2.6
$\mu$%
m to above 400
$\mu$%
m (obtained by applying an external magnetic field)
\cite{2010ApPhL..97h1110S,2010ApPhL..96n1110C}. At room temperature,
continuous wave output powers of several W can now routinely be obtained in
the mid-infrared region \cite{2011ApPhL..98r1102B}, and widely tunable designs
are available \cite{2009ApPhL..95f1103H}. Furthermore, at cryogenic
temperatures wall-plug efficiencies of 50\% and above have been demonstrated
\cite{2010NaPho...4...95L,2010NaPho...4...99B}, and up to 27\% at room
temperature \cite{2011ApPhL..98r1102B}. Commercial QCLs and QCL-based systems
are already available from various companies \cite{2010CPL...487....1C}.

On the other hand, terahertz QCLs, first realized in 2002
\cite{2002Natur.417..156K}, still require cryogenic operating temperatures.
Lasing up to $\thicksim200~\mathrm{K}$ has been obtained
\cite{fathololoumi2012terahertz}, and by applying a strong external magnetic
field, an operating temperature of $225~\mathrm{K}$ has been demonstrated
\cite{2009NaPho...3...41W}. A long standing goal is to achieve operation at
room temperature or at least in the commercially available thermoelectric
cooler range ($\thicksim240~\mathrm{K}$). For a further enhancement of the
operating temperature, it is crucial to improve the gain medium design and
reduce the resonator loss \cite{fathololoumi2012terahertz,2008OExpr..16.3242B}%
. For systematic QCL optimization and exploration of innovative designs,
detailed modeling is essential, requiring advanced simulation approaches which
enable realistic QCL simulations.

Besides its function as the optical gain medium, the QCL quantum well
heterostructure can also exhibit a strong nonlinear optical response. Like the
laser transition itself, such nonlinearities are based on quantized electron
states, and thus can be custom-tailored for specific applications and
optimized to exhibit extremely high nonlinear susceptibilities. This has led
to a new paradigm for QCL-based THz generation at room temperature: Here,
difference frequency mixing is used, where the THz radiation corresponds to
the difference in frequency of two detuned mid-infrared beams, generated by
highly efficient mid-infrared QCLs \cite{belkin2008room}. The optical
nonlinearity is either implemented into the QCL separately in a specifically
designed heterostructure \cite{2011ApPhL..98o1114A}, or, more commonly,
directly integrated into the gain medium \cite{belkin2008room}. With such an
approach, THz generation at room temperature was obtained at power levels of
up to $120~\mathrm{\mu W}$ \cite{2013OExpr..21..968L,a15}. Furthermore,
broadband frequency tunability has been demonstrated, which is important for
various applications such as spectroscopy \cite{a15,2012ApPhL.101y1121L}.
Here, the main goal is to push the available room temperature output power to
a few mW, as required for most technical applications. The modeling of such
terahertz QCL sources is particularly demanding, requiring the coupled
simulation of two mid-infrared QCLs and a careful modeling of the nonlinear
susceptibility and the associated frequency conversion process in the
heterostructure \cite{2013OExpr..21.6180J}. Artificial optical nonlinearities
have also been used to extend QCL operation towards shorter wavelengths, which
cannot be reached directly since the energy spacing of the quantized laser
levels is limited by the quantum well depth of the gain medium. Based on a
frequency doubling structure \cite{2003PhRvL..90d3902O}, a room temperature
QCL source operating at $2.7~\mathrm{\mu m}$ could be demonstrated
\cite{2011IJQE...47..691V}.

The ongoing development and optimization of the QCL heterostructure is
accompanied by improvements of the optical cavity. For example, changes in the
design and material of the plasmonic resonator structure have played a crucial
role for increasing the operating temperature of THz QCLs
\cite{fathololoumi2012terahertz,2008OExpr..16.3242B}. Also the performance of
difference frequency THz sources has largely benefited from special cavity
designs, such as the surface emitting Cherenkov waveguide scheme to obtain
increased output power and efficiency \cite{2012ApPhL.100y1104V,a15}. Thus, a
quantitative modeling and reliable numerical design optimization and
exploration must also include the optical cavity. Moreover, various types of
resonators with periodic subwavelength structuring have been developed in the
THz and mid-infrared regime. This includes distributed feedback structures to
enforce single mode lasing and distributed Bragg reflectors for enhanced facet
reflectivity
\cite{2006OExpr..14.5335D,2009ApPhL..95i1105B,2011ApPhL..99m1106L}. A further
example are surface emission schemes based on one- and two-dimensional
photonic crystal structures
\cite{2006OExpr..14.5335D,2009Natur.457..174C,2003Sci...302.1374C}, offering
tailorable emission properties and improved beam quality. The simulation of
such subwavelength-structured cavities requires advanced electromagnetic
modeling, e.g., based on coupled mode theory \cite{2006IJQE...42..257S} or
even full finite difference time domain simulations of Maxwell's equations
\cite{FDTD,2003Sci...302.1374C}.

The goal of this review is to give a detailed survey and discussion of the
modeling techniques used for QCL simulation, ranging from basic empirical
approaches to advanced self-consistent simulation methods. The focus is here
on the modeling of the heterostructure gain medium. Also the simulation of the
optical cavity will be covered for simple resonator waveguide structures. As
mentioned above, the modeling of complex cavities such as photonic crystal
structures constitutes an advanced electromagnetic modeling task, which is
beyond the scope of this paper. The review is organized as follows: In Section
\ref{sec:SP}, the numerical solution of the one-dimensional Schr\"{o}dinger
equation is discussed, providing the eigenenergies and wave functions of the
energy states in the QCL heterostructure. Furthermore, the inclusion of space
charge effects by solving the Schr\"{o}dinger-Poisson equation system is
treated. Section \ref{sec:Reso} covers the modeling of the optical resonator,
where we focus on a basic waveguide resonator structure. In Section
\ref{sec:Over}, an overview and classification of the different carrier
transport models is given which are commonly used for the theoretical
description of the QCL gain medium. Section \ref{sec:Phen} contains a
discussion of empirical modeling approaches for the gain medium, relying on
experimental or empirical input parameters, namely empirical rate equation and
related Maxwell-Bloch equation approaches, which are discussed in Sections
\ref{sec:PRE} and \ref{sec:PDM}, respectively. Advanced self-consistent
methods, which only require well known material parameters as input, are
covered in Sections \ref{sec:Self}--\ref{sec:NEGF}: Semiclassical approaches
such as the self-consistent rate equation model (Section \ref{sec:SCRE}) and
the ensemble Monte Carlo method (Section \ref{sec:EMC}), as well as quantum
transport approaches such as the density matrix method (Section \ref{sec:DM})
and the non-equilibrium Green's function formalism (Section \ref{sec:NEGF}).
In this context, also the transition rates (Sections \ref{sec:scat}%
--\ref{sec:elast}) and self-energies (Section \ref{sec:NEGF_scattering}) are
derived for the relevant scattering processes in QCLs. The paper is concluded
in Section \ref{sec:Concl} with an outlook on future trends and challenges.

\section{\label{sec:SP}Schr\"{o}dinger-Poisson solver}

A careful design of the quantized states in the QCL heterostructure is crucial
for the development and optimization of experimental QCL devices. In
particular, the lasing frequency is determined by the energy difference
between the upper and lower laser level. Furthermore, a careful energy
alignment of the levels is necessary to obtain an efficient injection into the
upper and depopulation of the lower laser level. For example, for the
structure shown in Fig.\,\ref{fig:qcl}, the depopulation level is separated
from the lower laser level by the InGaAs LO phonon energy of $\thicksim
30~\mathrm{meV}$ to obtain efficient depopulation, while the injector level is
aligned with the upper laser level of the next period
\cite{2010NaPho...4...99B}. Also careful engineering of the wave functions is
important, determining the strength of both the optical and nonradiative
transitions. Moroever, various carrier transport simulation approaches, such
as EMC, require the quantized energy levels as input. The eigenenergies and
wave functions are determined by solving the stationary Schr\"{o}dinger
equation or, if space charge effects are taken into account, the
Schr\"{o}dinger-Poisson equation system.

\subsection{\label{sec:SE}Schr\"{o}dinger equation}

The QCL heterostructure consists of alternately grown thin layers of different
semiconductor materials, resulting in the formation of quantum wells and
barriers in the conduction band. Thus, the potential $V$ and effective masses
vary in the growth direction, here denoted by $z$, whereas $x$ and $y$ refer
to the in-plane directions. The QCL heterostructure is usually treated in the
framework of the Ben Daniel-Duke model which only considers the conduction
band \cite{Bastard:88}, where the lasing transitions and carrier transport
take place. Within this approximation, the stationary Schr\"{o}dinger equation
is given by
\begin{align}
0  &  =\bigg\{-\frac{\hbar^{2}}{2}\left[  \frac{1}{m^{\parallel}\left(
z\right)  }\left(  \partial_{x}^{2}+\partial_{y}^{2}\right)  +\partial
_{z}\frac{1}{m^{\ast}\left(  z\right)  }\partial_{z}\right] \nonumber\\
&  +V\left(  z\right)  -E\bigg\}\psi_{\mathrm{3D}}\left(  x,y,z\right)  ,
\label{eq:s3D}%
\end{align}
where $\psi_{\mathrm{3D}}\ $and $E$ denote the wave function and eigenenergy,
respectively. Furthermore, $m^{\parallel}$ refers to the in-plane effective
mass, and $m^{\ast}$ is the effective mass in the growth direction, i.e.,
perpendicular to epitaxial layers. For bound states, the wave function is
commonly normalized, i.e., $\iiint_{-\infty}^{\infty}\left|  \psi
_{\mathrm{3D}}\right|  ^{2}\mathrm{d}x\mathrm{d}y\mathrm{d}z=1$. Since $V$ and
the effective masses only depend on the $z$ coordinate, we can make the
ansatz
\begin{equation}
\psi_{\mathrm{3D}}\left(  x,y,z\right)  =S^{-1/2}\psi_{\mathbf{k}}\left(
z\right)  \exp\left(  \mathrm{i}k_{x}x+\mathrm{i}k_{y}y\right)  .
\label{eq:psi3D}%
\end{equation}
Here, $S$ is the in-plane cross section area and $\mathbf{k}=\left[
k_{x},k_{y}\right]  ^{\mathrm{T}}$ denotes the in-plane wave vector, where
$\mathrm{T}$ indicates the transpose. The factor $S^{-1/2}$\ is added in
Eq.\thinspace(\ref{eq:psi3D}) to obtain the normalization condition
$\int\left|  \psi_{\mathbf{k}}\right|  ^{2}\mathrm{d}z=1$. Insertion of
Eq.\thinspace(\ref{eq:psi3D}) in Eq.\thinspace(\ref{eq:s3D})\ yields the Ben
Daniel-Duke model%
\begin{equation}
\left\{  \frac{\hbar^{2}}{2}\frac{k_{x}^{2}+k_{y}^{2}}{m^{\parallel}\left(
z\right)  }-\frac{\hbar^{2}}{2}\partial_{z}\frac{1}{m^{\ast}\left(  z\right)
}\partial_{z}+V\left(  z\right)  -E_{\mathbf{k}}\right\}  \psi_{\mathbf{k}%
}\left(  z\right)  =0,
\end{equation}
where the wave function $\psi_{\mathbf{k}}\left(  z\right)  $ and energy
$E_{\mathbf{k}}$ depend on the in-plane electron motion, i.e., on $\mathbf{k}%
$. Decoupling can be obtained by neglecting the $z$ dependence of the in-plane
effective mass $m^{\parallel}$,\ yielding the one-dimensional (1D)
Schr\"{o}dinger equation in its usual form
\begin{equation}
\left[  -\frac{\hbar^{2}}{2}\partial_{z}\frac{1}{m^{\ast}\left(  z\right)
}\partial_{z}+V\left(  z\right)  -E\right]  \psi\left(  z\right)  =0.
\label{eq:s1D}%
\end{equation}
The eigenenergy $E$ due to 1D electron confinement in $z$ direction is related
to the total energy $E_{\mathbf{k}}$ by $E_{\mathbf{k}}=E+E_{\mathrm{kin}}$,
where%
\begin{equation}
E_{\mathrm{kin}}=\hbar^{2}\left(  k_{x}^{2}+k_{y}^{2}\right)  /\left(
2m^{\parallel}\right)  =\hbar^{2}\mathbf{k}^{2}/\left(  2m^{\parallel}\right)
\label{eq:Ekin}%
\end{equation}
is the kinetic electron energy due to the free in-plane motion of the
electrons. The effective mass values not only depend on the material
composition \cite{2001JAP....89.5815V}, but also on the lattice temperature
and doping level \cite{1961PhRv..121..752C}. The latter effects tend to play a
secondary role in QCLs and are thus usually neglected. For strained QCL
structures, the effective masses are additionally affected by the lattice
mismatch between the different semiconductor materials, resulting in modified
values $m^{\parallel}\neq m^{\ast}$ \cite{1993PhRvB..48.8102S}.

\subsection{\label{sec:BC}Boundary conditions}

\begin{figure}[ptb]
\includegraphics{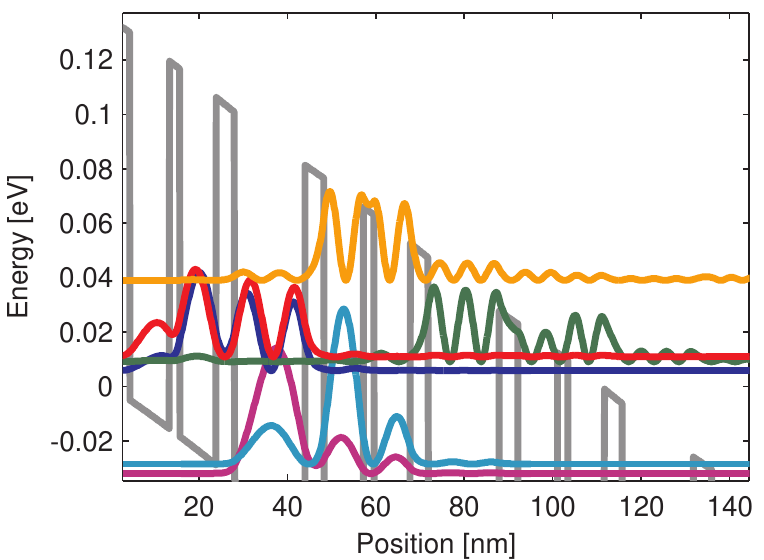}
\caption{{}(Color online) Quasi-bound states in the conduction band of a
QCL\ heterostructure.}%
\label{fig:quasibound}%
\end{figure}

Strictly speaking, the quantum states in the QCL\ heterostructure are not
bound, since the electron energy $E\ $exceeds the barrier potential for large
values of $z$. Thus, the electrons will tunnel out of the multiple quantum
well system after a limited time. This situation is illustrated in Fig.
\ref{fig:quasibound}. If the electron remains in the quantum well system for a
considerable amount of time, the concept of quasi-bound or quasi-stationary
states can be used \cite{1974qmec.book.....L}. Such a state can again be
described by Eq.\,(\ref{eq:s1D}), however the eigenenergy is now a complex
quantity, $E=E_{0}-\mathrm{i}\hbar\gamma_{\mathrm{q}}/2$. The meaning of the
imaginary term $-\mathrm{i}\hbar\gamma_{\mathrm{q}}/2$ for quasi-stationary
states can be understood by considering that the time dependence of a
stationary wave function $\psi$ is given by $\exp\left(  -\mathrm{i}%
Et/\hbar\right)  $. Consequently, for a complex eigenenergy $E$ the density
probability $\left|  \psi\right|  ^{2}$ of the electron in the quantum well
system decays as $\exp\left(  -\gamma_{\mathrm{q}}t\right)  $, thus
$\gamma_{\mathrm{q}}$ is the probability density decay rate of the quasi-bound
state, and $\tau_{\mathrm{q}}=\gamma_{\mathrm{q}}^{-1}$ can be associated with
the lifetime of the particle in the quantum well \cite{1974qmec.book.....L}.

\begin{figure}[ptb]
\includegraphics{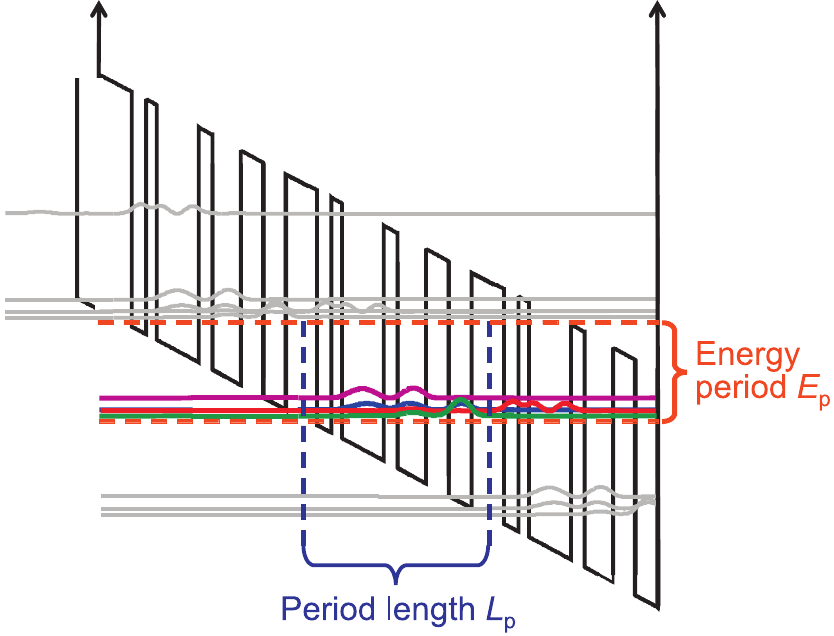}
\caption{{}(Color online) Finite simulation window with artificial boundary
conditions $\psi=0$. The central QCL period of the simulated structure is
indicated, as well as the energy interval used for determining the bound
states, corresponding to an energy period. The states in the other periods are
found by adequate shifts in position and energy.}%
\label{fig:window}%
\end{figure}

Some numerical approaches have been developed to solve Eq.\,(\ref{eq:s1D}) for
quasi-bound states \cite{1990IJQE...26.2025J,1993IJQE...29.2731A}. However,
the relevant states which are considered for the design of a QCL structure,
such as the upper and lower laser level and injector states, are typically
strongly bound to obtain optimum performance. Since the electron lifetime in
these states is governed by scattering processes and resonant tunneling rather
than $\tau_{\mathrm{q}}$, they are usually treated as bound states. This can
be done by restricting the simulation to a finite simulation window containing
a limited number of periods, and imposing artificial boundary conditions
$\psi=0$ at the borders, as illustrated in Fig.\,\ref{fig:window}. The
simulation window has to be chosen sufficiently large so that the portion of
the wave function which lies inside the quantum wells, i.e., which
significantly deviates from $0$, is not markedly affected. For the simulation,
we assume that the QCL heterostructure is periodic, where a period is defined
by a sequence of multiple barriers and wells, as illustrated in
Fig.\,\ref{fig:qcl}. Thus, for a biased QCL structure, it is sufficient to
compute the eigenenergies and corresponding wave functions for a single energy
period of width $E_{\mathrm{p}}$, corresponding to the bias drop over a QCL
period (see Fig. \ref{fig:window}). The solutions of the other periods can
then simply be obtained by shifts in position by multiples of the period
length $L_{\mathrm{p}}$, and corresponding shifts in energy. For example, the
solutions in the right-neighboring period are given by $\psi_{i^{\prime}%
}\left(  z\right)  =\psi_{i}\left(  z-L_{\mathrm{p}}\right)  $, $E_{i^{\prime
}}=E_{i}-E_{\mathrm{p}}$, where $\psi_{i}$ and $E_{i}$ are the wave function
and eigenenergy of the $i$th solution of Eq.\,(\ref{eq:s1D}) obtained in the
central period.

For the laser design and simulation, usually only the relevant laser subbands
such as the upper and lower laser level and injector states are considered.
These typically correspond to the strongly bound states of the QCL
heterostructure. Furthermore, the artificial boundary conditions illustrated
in Fig.\,\ref{fig:window} give rise to spurious solutions of
Eq.\,(\ref{eq:s1D}), which are not localized in the quantum wells. A
systematic selection of the relevant subbands can be achieved by considering
only the most strongly bound levels in each period. In this context, the
energy of state $i$ relative to the conduction band edge%
\begin{equation}
\tilde{E}_{i}=E_{i}-\int V\left|  \psi_{i}\right|  ^{2}\mathrm{d}z
\label{eq:Ei}%
\end{equation}
is a useful quantity \cite{2009IJQE...45..1059J}. An automated selection of
the relevant subbands can then be implemented by choosing the states with the
lowest energies $\tilde{E}_{i}$ in each period, considering all subbands which
contribute significantly to the carrier transport.

\subsection{\label{sec:nonparabolicity}Nonparabolicity}

\begin{figure}[ptb]
\includegraphics{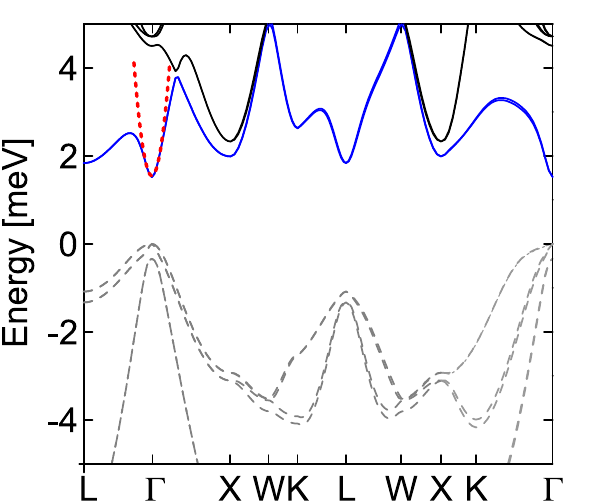}
\caption{{}(Color online) Band structure of gallium arsenide (GaAs), obtained
with NEMO5 in 20 band empirical tight
binding.\cite{2011IEEETransonNano..10.1464S} Shown are the valence bands
(dashed lines), the conduction band (solid line) and the parabolic dispersion
relation used for the $\Gamma$ valley (dotted line).}%
\label{fig:banddiagram}%
\end{figure}

In Fig.\,\ref{fig:banddiagram}, the band structure of GaAs as a typical III-V
semiconductor material is displayed. Shown is the valence band (dashed lines),
consisting of heavy hole, light hole and split-off bands, and the conduction
band (solid line). As pointed out above, only the conduction band is
considered in the Ben Daniel-Duke model, introduced in Section \ref{sec:SE}.
The conduction band has three minima, referred to as the $\Gamma$, L and X (or
$\Delta$) valley according to their position in \textbf{K} space. For direct
bandgap semiconductors typically used in QCLs, the $\Gamma$ valley is the
lowest minimum. Thus, the theoretical treatment is usually restricted to that
valley, although under certain conditions transitions to the X or L valleys
can also affect QCL\ operation
\cite{2006ApPhL..89s1119G,2007JAP...101f3101G,2007JAP...102k3107G,2008JAP...103g3101G}%
. Furthermore, Eqs.\,(\ref{eq:s3D})-(\ref{eq:Ekin}) assume a parabolic
dispersion relation between energy and wave number (dotted line in Fig.
\ref{fig:banddiagram}). This approximation only holds close to the $\Gamma$
valley minimum, i.e., it breaks down for high-lying energy levels. The
deviation from the parabolic dispersion, i.e., the nonparabolicity, scales
roughly inversely with the semiconductor bandgap \cite{1987PhRvB..35.7770N}.
This effect thus plays a role especially for mid-infrared QCLs with lasing
transition energies of up to a few $100~\mathrm{meV}$, which are frequently
based on low-bandgap semiconductors such as InGaAs or InGaSb. A detailed
treatment of nonparabolicity effects in QCLs has been performed by using
\textbf{k.p} theory
\cite{2002PhRvB..65p5220L,2006ApPhL..89s1119G,2007JAP...101f3101G,2007JAP...102k3107G}%
. A related strategy, which we will discuss in the following, is to consider
nonparabolicity in Eq.\,(\ref{eq:s1D}) by implementing an energy dependent
mass, derived from \textbf{k.p} theory. This method involves additional
approximations, but is less complex than full \textbf{k.p} calculations and
can easily be incorporated into the transfer matrix approach, which is widely
used to solve Eq.\,(\ref{eq:s1D}). According to Ekenberg
\cite{1989PhRvB..40.7714E}, we obtain the energy dependent effective masses
\begin{subequations}
\label{eq:meff}%
\begin{align}
m^{\ast}\left(  E^{\prime}\right)   &  =\frac{m^{\ast}}{2\alpha^{\prime
}E^{\prime}}\left[  1-\left(  1-4\alpha^{\prime}E^{\prime}\right)
^{1/2}\right]  ,\label{eq:meff1}\\
m^{\parallel}\left(  E^{\prime}\right)   &  =m^{\ast}\left[  1+\left(
2\alpha^{\prime}+\beta^{\prime}\right)  E^{\prime}\right]  . \label{eq:meff2}%
\end{align}
\end{subequations}
Here, $\alpha^{\prime}$ and $\beta^{\prime}$ are nonparabolicity parameters,
which are defined in the framework of a 14-band \textbf{k.p} calculation
\cite{1989PhRvB..40.7714E,1985JPhC...18.3365B}. Approximately, we have
$\alpha^{\prime}=\left(  E_{\mathrm{g}}+\Delta_{\mathrm{so}}/3\right)  ^{-1}$
\cite{1987PhRvB..35.7770N}, where $E_{\mathrm{g}}$ and $\Delta_{\mathrm{so}}$
are the bandgap energy and the energy difference between the light-hole and
split-of valence bands, respectively. Furthermore, $m^{\ast}$ is the $\Gamma$
valley effective mass. In the QCL heterostructure, the electron kinetic energy
in the conduction band is given by $E^{\prime}\left(  z\right)  =E-V\left(
z\right)  $. Since the semiconductor material changes along the growth
direction, also $m^{\ast}$, $\alpha^{\prime}$ and $\beta^{\prime}$ are $z$
dependent. This approach is valid for moderate energies\ $E^{\prime}%
\ll1/\alpha^{\prime}$. The fact that Eq.\,(\ref{eq:meff1}) even breaks down
for $\alpha^{\prime}E^{\prime}>1/4$ can cause numerical problems for large
values of $z$ close to the right simulation boundary, where $\psi$ is already
close to $0$ but $E^{\prime}$ can assume relatively large values (see
Fig.\,\ref{fig:window}). This issue can be avoided by using
Eq.\,(\ref{eq:meff1}) for $E^{\prime}<0$, and its second order Taylor
expansion,
\begin{equation}
m^{\ast}\left(  E^{\prime}\right)  =m^{\ast}\left(  1+\alpha^{\prime}%
E^{\prime}\right)  ,
\end{equation}
for $E^{\prime}>0$, which is the widely used lowest order implementation of
nonparabolicity \cite{1989PhRvB..40.7714E,1987PhRvB..35.7770N}. For an energy
dependent effective mass, the Hamiltonian in Eq.\,(\ref{eq:s1D}) is not
Hermitian; thus, the obtained wave functions are in general not orthogonal
\cite{1994PhRvB..50.8663S}. Also other approaches are available for
implementing nonparabolicity effects, for example based on the Kane model
\cite{1994PhRvB..50.8663S}.

In self-consistent simulations, each scattering process is evaluated based on
the corresponding Hamiltonian, which depends on the in-plane effective masses
of the subbands involved. Thus, if we want to include corrections due to
nonparabolicity also for the scattering processes
\cite{1988JaJAP..27..563T,2008JAP...104e3719B}, then Eq.\,(\ref{eq:meff2})
should be used to determine $m^{\parallel}\left(  E^{\prime}\right)  $, which
is however $z$ dependent. This problem can be avoided by defining an average
effective mass for the $i$th subband,%
\begin{align}
m_{i}^{\parallel}  &  =\int m^{\ast}\left(  z\right)  \left\{  1+\left[
2\alpha^{\prime}\left(  z\right)  +\beta^{\prime}\left(  z\right)  \right]
\right. \nonumber\\
&  \times\left.  \left[  E_{i}-V\left(  z\right)  \right]  \right\}  \left|
\psi_{i}\left(  z\right)  \right|  ^{2}\mathrm{d}z, \label{eq:mi}%
\end{align}
where the wave function is assumed to be normalized, $\int\left|  \psi
_{i}\right|  ^{2}\mathrm{d}z=1$, and $E_{i}$ denotes the corresponding subband
eigenenergy. This lowest order implementation of nonparabolicity thus yields
different, albeit constant in-plane masses for each subband.

\subsection{Numerical solution}

\begin{figure}[ptb]
\includegraphics{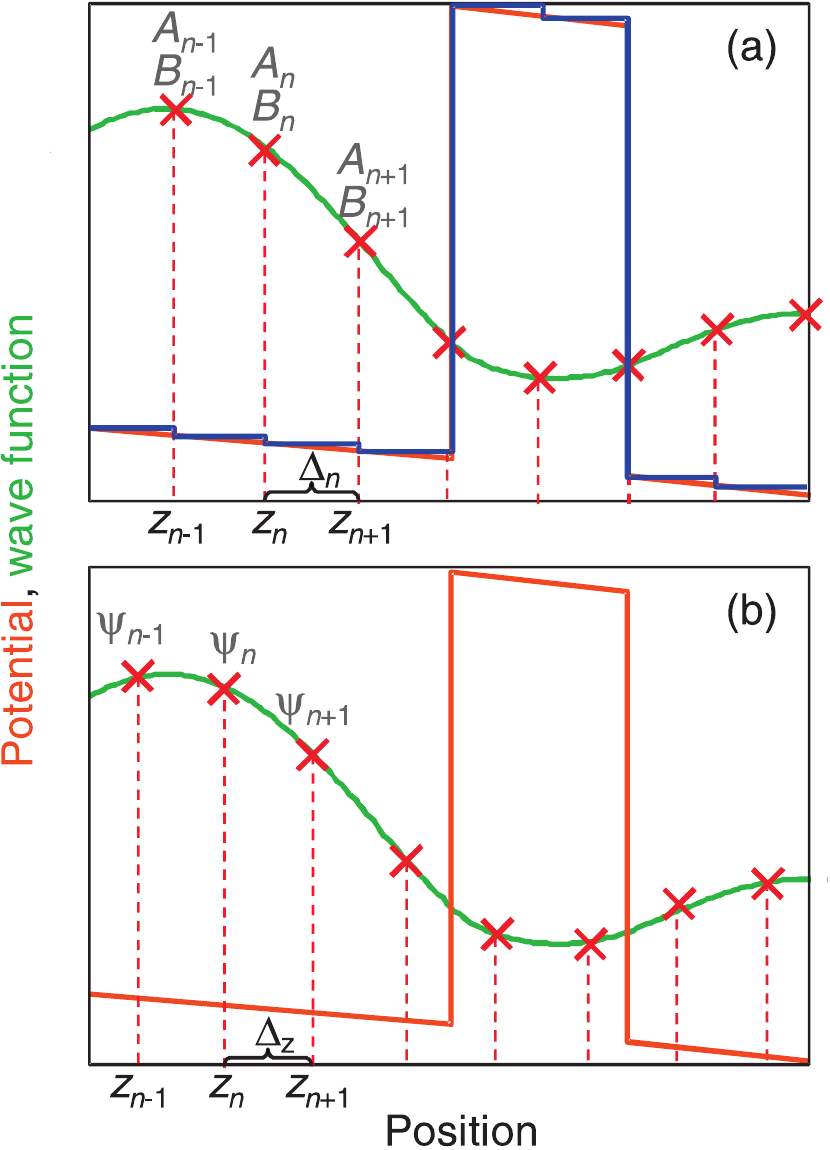}
\caption{{}(Color online) Numerical solution of the one-dimensional effective
mass Schr\"{o}dinger equation: (a) Transfer matrix approach; (b) finite
difference method.}%
\label{fig:transfermat}%
\end{figure}

Numerical approaches for solving the one-dimensional effective mass
Schr\"{o}dinger equation Eq.\,(\ref{eq:s1D}) are required to be robust. Also
computational efficiency is crucial, especially for QCL design and
optimization tasks where many simulations have to be performed. Furthermore, a
straightforward implementation is desirable. Widely used numerical approaches
include the transfer matrix method
\cite{1987JAP....61.1497A,1990IJQE...26.2025J,2000JAP....87.7931C,2009IJQE...45..1059J}
and finite difference scheme \cite{1998hqd.book.....F,1990PhRvB..4112047J}.
\ Both methods have their strengths and shortcomings. In particular, effects
such as nonparabolicity can be included more easily into the transfer matrix
approach. A further advantage is the exact treatment of the potential steps
between barriers and wells in the QCL\ heterostructure. On the other hand,
this method can exhibit numerical instabilities for multiple or extended
barriers due to an exponential blowup caused by roundoff errors
\cite{1998hqd.book.....F}. This issue can however be overcome, for example by
using a somewhat modified approach, the scattering matrix method
\cite{1988PhRvB..38.9945K}. In Fig.\,\ref{fig:transfermat}, the transfer
matrix and finite difference schemes are illustrated.

\subsubsection{\label{sec:transfermatrix}Transfer matrix approach}

The transfer matrix approach uses the fact that analytical solutions of
Eq.\,(\ref{eq:s1D}) are available for constant or linear potential sections
and for potential steps \cite{1997plds.book.....D}. An arbitrary potential can
then be treated by approximating it in terms of piecewise constant or linear
segments, respectively. In the first case, the solution is given by complex
exponentials \cite{1987JAP....61.1497A,1990IJQE...26.2025J}, while the
approximation by linear segments gives rise to Airy function solutions
\cite{1990IJQE...26.2025J}. The Airy function approach provides an exact
solution for structures which consist of segments with constant effective
masses and piecewise linear potentials, such as a biased QCL\ heterostructure
if we neglect nonparabolicity and space charge effects. On the other hand,
Airy functions are much more computationally expensive than exponentials, and
also prone to numerical overflow for segments with nearly flat potential
\cite{1996IJQE...32.1093V}. Thus, great care has to be taken to avoid
numerical problems and to evaluate the Airy functions efficiently
\cite{2001JAP....90.6120D}.

In the following, we focus on the exponential transfer matrix scheme,
illustrated in Fig.\thinspace\ref{fig:transfermat}(a). We start by dividing
the structure into segments which can vary in length, but should be chosen so
that all band edge discontinuities occuring between wells and barriers are
located at the border of a segment, i.e., do not lie within a segment. The
potential and effective mass in each segment $n$ with $z_{n}\leq
z<z_{n}+\Delta_{n}=z_{n+1}$ are approximated by constant values, e.g.,
$V_{n}=V\left(  z_{n}\right)  $, $m_{n}^{\ast}=m^{\ast}\left(  z_{n}\right)
$, resulting in a jump $V_{n}\rightarrow V_{n+1}$, $m_{n}^{\ast}\rightarrow
m_{n+1}^{\ast}$ at the border between the segments $n$ and $n+1$
\cite{1990IJQE...26.2025J}. Nonparabolicity can straightforwardly be
implemented by using Eq.\thinspace(\ref{eq:meff1}) for $m^{\ast}$; then
$m_{n}^{\ast}=m_{n}^{\ast}\left(  E\right)  $ depends on the eigenenergy $E$.
The solution of Eq.\thinspace(\ref{eq:s1D}) in segment $n$ is given by%
\begin{equation}
\psi\left(  z\right)  =A_{n}\exp\left[  \mathrm{i}k_{n}\left(  z-z_{n}\right)
\right]  +B_{n}\exp\left[  -\mathrm{i}k_{n}\left(  z-z_{n}\right)  \right]  .
\label{eq:psi}%
\end{equation}
Here $k_{n}=\sqrt{2m_{n}^{\ast}\left(  E-V_{n}\right)  }/\hbar$ is the wave
number (for $E<V_{n}$, we obtain $k_{n}=\mathrm{i}\kappa_{n}=\mathrm{i}%
\sqrt{2m_{n}^{\ast}\left(  V_{n}-E\right)  }/\hbar$)
\cite{1990IJQE...26.2025J}. The matching conditions for the wave function at a
potential or effective mass discontinuity read%
\begin{align}
\psi\left(  z_{0}+\right)   &  =\psi\left(  z_{0}-\right)  ,\nonumber\\
\left[  \partial_{z}\psi\left(  z_{0}+\right)  \right]  /m^{\ast}\left(
z_{0}+\right)   &  =\left[  \partial_{z}\psi\left(  z_{0}-\right)  \right]
/m^{\ast}\left(  z_{0}-\right)  , \label{eq:match}%
\end{align}
where $z_{0}+$ and $z_{0}-$ denote the positions directly to the right and
left of the discontinuity \cite{1997plds.book.....D}. Using Eqs.
(\ref{eq:psi}) and (\ref{eq:match}), the amplitudes $A_{n+1}$ and $B_{n+1}$
can be related to $A_{n}$ and $B_{n}$ by%
\begin{equation}
\left(
\begin{array}
[c]{c}%
A_{n+1}\\
B_{n+1}%
\end{array}
\right)  =T_{n,n+1}\left(
\begin{array}
[c]{c}%
A_{n}\\
B_{n}%
\end{array}
\right)  , \label{eq:mat}%
\end{equation}
where the transfer matrix is with $\tilde{k}_{n}=k_{n}/m_{n}^{\ast}$ given by
\begin{align}
T_{n,n+1}  &  =T_{n\rightarrow n+1}T_{n}\left(  \Delta_{n}\right) \nonumber\\
&  =\left(
\begin{array}
[c]{cc}%
\frac{\tilde{k}_{n+1}+\tilde{k}_{n}}{2\tilde{k}_{n+1}}e^{\mathrm{i}k_{n}%
\Delta_{n}} & \frac{\tilde{k}_{n+1}-\tilde{k}_{n}}{2\tilde{k}_{n+1}%
}e^{-\mathrm{i}k_{n}\Delta_{n}}\\
\frac{\tilde{k}_{n+1}-\tilde{k}_{n}}{2\tilde{k}_{n+1}}e^{\mathrm{i}k_{n}%
\Delta_{n}} & \frac{\tilde{k}_{n+1}+\tilde{k}_{n}}{2\tilde{k}_{n+1}%
}e^{-\mathrm{i}k_{n}\Delta_{n}}%
\end{array}
\right)  . \label{eq:T1}%
\end{align}
Here,%
\begin{equation}
T_{n}\left(  \Delta_{n}\right)  =\left(
\begin{array}
[c]{cc}%
e^{\mathrm{i}k_{n}\Delta_{n}} & 0\\
0 & e^{-\mathrm{i}k_{n}\Delta_{n}}%
\end{array}
\right)  \label{eq:Th}%
\end{equation}
corresponds to the transfer matrix for a flat potential obtained from
Eq.\thinspace(\ref{eq:psi}), and%
\begin{equation}
T_{n\rightarrow n+1}=\frac{1}{2\tilde{k}_{n+1}}\left(
\begin{array}
[c]{cc}%
\tilde{k}_{n+1}+\tilde{k}_{n} & \tilde{k}_{n+1}-\tilde{k}_{n}\\
\tilde{k}_{n+1}-\tilde{k}_{n} & \tilde{k}_{n+1}+\tilde{k}_{n}%
\end{array}
\right)  \label{eq:Tst}%
\end{equation}
is the potential step matrix for the interface between the segments $n$ and
$n+1$ at $z_{0}=z_{n+1}$, derived from Eq.\thinspace(\ref{eq:match})
\cite{1997plds.book.....D}. For a structure divided into $N$ segments, the
relation between the amplitudes at the left and right boundaries of the
structure, $A_{0},B_{0}$ and $A_{N},B_{N}$, can be obtained by multiplying the
matrices for all segments%
\begin{align}
\left(
\begin{array}
[c]{c}%
A_{N}\\
B_{N}%
\end{array}
\right)   &  =T_{N-1,N}T_{N-2,N-1}\dots T_{0,1}\left(
\begin{array}
[c]{c}%
A_{0}\\
B_{0}%
\end{array}
\right) \nonumber\\
&  =\left(
\begin{array}
[c]{cc}%
T_{11} & T_{12}\\
T_{21} & T_{22}%
\end{array}
\right)  \left(
\begin{array}
[c]{c}%
A_{0}\\
B_{0}%
\end{array}
\right)  . \label{Eq:mat2}%
\end{align}

This equation must be complemented by appropriate boundary conditions. As
described in Section \ref{sec:BC}, for the QCL\ heterostructure we can
restrict our simulation to a limited number of periods (see Fig.
\ref{fig:window}) and assume $\psi=0$ at the boundaries of our simulation
window, corresponding to $A_{0}+B_{0}=0$ and $A_{N}+B_{N}=0$. The left
boundary condition can for example be enforced by setting $A_{0}=1$,
$B_{0}=-1$. The right boundary condition $A_{N}+B_{N}=0$ can then only be
satisfied if the energy dependent matrix elements $T_{11}(E)$, $T_{12}(E)$,
$T_{21}(E)$ and $T_{22}(E)$ in Eq.\thinspace(\ref{Eq:mat2}) assume certain
values. The corresponding energies $E$ are the eigenenergies of the bound
states \cite{1990IJQE...26.2025J}. Numerically, these eigenenergies are found
by the so-called shooting method. Here, the wave function at the right
boundary is computed from Eq.\thinspace(\ref{Eq:mat2}) as a function of
energy, $\psi_{N}(E)=A_{N}(E)+B_{N}(E)$, and the eigenergies are given by
$\psi_{N}(E)=0$. For the periodic QCL\ heterostructure, it is sufficient to
restrict the simulation to a single energy period (see Fig.\thinspace
\ref{fig:window}). In practice, Eq.\thinspace(\ref{Eq:mat2}) can only be
solved for a limited number of discrete energy points $E_{m}=E_{0}+m\Delta
_{E}$ with sufficiently close spacing $\Delta_{E}$. The eigenenergies are
located in intervals $E_{m}..E_{m+1}$ with $\psi_{N}(E_{m})\psi_{N}%
(E_{m+1})<0$, and can be determined more accurately by choosing a finer energy
grid in the corresponding intervals, or preferably by applying a root-finding
algorithm such as the bisection method \cite{1992nrca.book.....P}.

The accuracy of the transfer matrix scheme can generally be improved by
replacing Eq.\,(\ref{eq:T1}) with a symmetric transfer matrix,%

\begin{align}
T_{n,n+1}  &  =T_{n+1}\left(  \frac{\Delta_{n}}{2}\right)  T_{n\rightarrow
n+1}T_{n}\left(  \frac{\Delta_{n}}{2}\right) \nonumber\\
&  =\left(
\begin{array}
[c]{cc}%
\frac{\tilde{k}_{n+1}+\tilde{k}_{n}}{2\tilde{k}_{n+1}}e^{\mathrm{i}k_{n}%
^{+}\Delta_{n}} & \frac{\tilde{k}_{n+1}-\tilde{k}_{n}}{2\tilde{k}_{n+1}%
}e^{-\mathrm{i}k_{n}^{-}\Delta_{n}}\\
\frac{\tilde{k}_{n+1}-\tilde{k}_{n}}{2\tilde{k}_{n+1}}e^{\mathrm{i}k_{n}%
^{-}\Delta_{n}} & \frac{\tilde{k}_{n+1}+\tilde{k}_{n}}{2\tilde{k}_{n+1}%
}e^{-\mathrm{i}k_{n}^{+}\Delta_{n}}%
\end{array}
\right)  , \label{T2}%
\end{align}
where $k_{n}^{\pm}=\left(  k_{n}\pm k_{n+1}\right)  /2$ and $k_{n}%
=\sqrt{2m_{n}^{\ast}\left(  E-V_{n}\right)  }/\hbar$, $\tilde{k}_{n}%
=k_{n}/m_{n}^{\ast}$ \cite{2009IJQE...45..1059J}. Here, the band edge
discontinuities occuring between wells and barriers must be treated separately
by using the corresponding transfer matrix Eq.\,(\ref{eq:Tst}).

\subsubsection{\label{sec:fd}Finite difference method}

The finite difference method works by converting Eq.\,(\ref{eq:s1D}) to a
finite difference equation. As illustrated in Fig.\,\ref{fig:transfermat}(b),
a spatial grid\ with uniform spacing $\Delta_{z}$ is introduced, and the wave
function $\psi\left(  z\right)  $, potential $V\left(  z\right)  $ and
effective mass $m^{\ast}\left(  z\right)  $ are represented by the
corresponding values $\psi_{n}$, $V_{n}$ and $m_{n}^{\ast}$\ on the grid
points $z_{n}$. First order derivatives are approximated by $\partial_{z}%
\psi\left(  z_{n}+\Delta_{z}/2\right)  \approx\Delta\psi_{n+1/2}=\left(
\psi_{n+1}-\psi_{n}\right)  /\Delta_{z}$. Consequently, the term $\partial
_{z}\left(  m^{\ast}\right)  ^{-1}\partial_{z}\psi$ in Eq.\,(\ref{eq:s1D}) can
at $z=z_{n}$ be expressed as $\left(  \Delta\psi_{n+1/2}/m_{n+1/2}^{\ast
}-\Delta\psi_{n-1/2}/m_{n-1/2}^{\ast}\right)  /\Delta_{z}$. Using linear
interpolation $m_{n+1/2}^{\ast}=\left(  m_{n}^{\ast}+m_{n+1}^{\ast}\right)
/2$, we obtain the discretized form of Eq.\,(\ref{eq:s1D})
\cite{1998hqd.book.....F,1990PhRvB..4112047J},%
\begin{equation}
-s_{n}\psi_{n-1}+d_{n}\psi_{n}-s_{n+1}\psi_{n+1}=E\psi_{n} \label{eq:discr}%
\end{equation}
with
\begin{subequations}
\label{eq:sd}%
\begin{align}
s_{n}  &  =\frac{\hbar^{2}}{\Delta_{z}^{2}\left(  m_{n-1}^{\ast}+m_{n}^{\ast
}\right)  },\label{eq:sd1}\\
d_{n}  &  =\frac{\hbar^{2}}{\Delta_{z}^{2}}\left(  \frac{1}{m_{n-1}^{\ast
}+m_{n}^{\ast}}+\frac{1}{m_{n}^{\ast}+m_{n+1}^{\ast}}\right)  +V_{n}.
\label{eq:sd2}%
\end{align}
\end{subequations}
Here the grid should be chosen so that the band edge discontinuities occur
halfway between two adjacent grid points \cite{1998hqd.book.....F}, as
illustrated in Fig.\,\ref{fig:transfermat}(b). We again assume that the wave
function is zero at the boundaries of our simulation window, $\psi_{0}%
=\psi_{N\;}=0$. Writing Eq.\,(\ref{eq:discr}) in matrix form yields an
eigenvalue equation of the form $\left(  \mathbf{H}-E\mathbf{I}\right)
\mathbf{\psi}=\mathbf{0}$, where $\mathbf{\psi}$ is the wave function vector
with $\mathbf{\psi=}\left[  \psi_{1},\psi_{2},\dots,\psi_{N-1}\right]
^{\mathrm{T}}$, $\mathbf{I}$\ represents the identity matrix of size $N-1$,
and $\mathbf{H}$\ is the Hamiltonian matrix with the non-zero elements
$H_{n,n}=d_{n}$, $H_{n,n-1}=-s_{n}$, $H_{n,n+1}=-s_{n+1}$. This equation can
be solved by using a standard eigenvector solver for a tridiagonal matrix
problem \cite{1992nrca.book.....P}. However, if nonparabolicity is taken into
account, this is not directly possible since the effective masses $m_{n}%
^{\ast}$ depend on $E$, and a modified numerical scheme must be employed
\cite{2010JAP...108k3109C}. Alternatively, the shooting method discussed in
Section \ref{sec:transfermatrix} can be employed.

\subsection{Schr\"{o}dinger-Poisson equation system}

To lowest order, electron-electron interaction can be considered by
self-consistently solving Eq.\,(\ref{eq:s1D}) together with the Poisson
equation \cite{2000JAP....87.7931C,2008SeScT..23l5040L}
\begin{equation}
e^{-1}\partial_{z}\left[  \epsilon\left(  z\right)  \partial_{z}\tilde
{V}\left(  z\right)  \right]  =e\left[  n_{\mathrm{D}}\left(  z\right)
-\sum_{i}n_{i}^{\mathrm{s}}\left|  \psi_{i}\left(  z\right)  \right|
^{2}\right]  . \label{eq:poisson}%
\end{equation}
From a quantum mechanical point of view, this corresponds to a mean-field
treatment of the electron-electron interaction referred to as Hartree
approximation,\ representing the lowest order of a perturbation expansion in
the electron-electron interaction potential. In Eq.\,(\ref{eq:poisson}),
$\epsilon\left(  z\right)  $ is the permittivity which varies with
semiconductor composition and thus is also periodic, and $e$ denotes the
elementary charge. The right hand side of Eq.\,(\ref{eq:poisson}) corresponds
to the space charge $\rho$ in the QCL\ heterostructure due to the positively
charged donors with concentration $n_{\mathrm{D}}\left(  z\right)  $ and the
electrons, where $n_{i}^{\mathrm{s}}$ is the electron sheet density of level
$i$ with wave function $\psi_{i}\left(  z\right)  $. This charge distribution
in the structure gives rise to space charge effects, resulting in an
additional electrostatic potential energy $\tilde{V}\left(  z\right)  $\ which
causes conduction band bending \cite{2005ApPhL..86u1117J}.

The total potential $V$ in Eq.\thinspace(\ref{eq:s1D}) is then given by
$V=V_{0}+\tilde{V}$. Here, $V_{0}=V_{\mathrm{c}}-E_{\mathrm{p}}z/L_{\mathrm{p}%
}$ where $V_{\mathrm{c}}$ is the unbiased conduction band profile due to the
varying material composition, thus describing the wells and barriers, and the
term $-E_{\mathrm{p}}z/L_{\mathrm{p}}$ results from the applied bias. Since
the energy drop across a period is given by the external bias, $V\left(
z_{0}\right)  -V\left(  z_{0}+L_{\mathrm{p}}\right)  =E_{\mathrm{p}}$, we have
$\tilde{V}\left(  z_{0}\right)  =\tilde{V}\left(  z_{0}+L_{\mathrm{p}}\right)
$. Due to the charge neutrality in each period, $\int_{z_{0}}^{z_{0}%
+L_{\mathrm{p}}}\rho\mathrm{d}z=0$, we furthermore obtain $\partial_{z}%
\tilde{V}\left(  z_{0}\right)  =\partial_{z}\tilde{V}\left(  z_{0}%
+L_{\mathrm{p}}\right)  $, i.e., $\tilde{V}$\ has the periodicity of
$V_{\mathrm{c}}$. Thus we can restrict the solution of Eq.\thinspace
(\ref{eq:poisson}) to a single QCL period $z\in\left[  z_{0},z_{0}%
+L_{\mathrm{p}}\right]  $ and assume the boundary conditions $\tilde{V}\left(
z_{0}\right)  =\tilde{V}\left(  z_{0}+L_{\mathrm{p}}\right)  =0$.

Equation (\ref{eq:poisson}) can, for example, be solved by applying the finite
difference method. In analogy to Eqs.\thinspace(\ref{eq:discr}) and
(\ref{eq:sd}), we obtain%
\begin{equation}
\tilde{s}_{n}\tilde{V}_{n-1}-\tilde{d}_{n}\tilde{V}_{n}+\tilde{s}_{n+1}%
\tilde{V}_{n+1}=\rho_{n}, \label{eq:poisson_discretized}%
\end{equation}
with%
\begin{equation}
\rho_{n}=e\left[  n_{\mathrm{D},n}-\sum_{i}n_{i}^{\mathrm{s}}\left|
\psi_{i,n}\right|  ^{2}\right]  \label{eq:rho}%
\end{equation}
and
\begin{align}
\tilde{s}_{n}  &  =\frac{1}{2e\Delta_{z}^{2}}\left(  \epsilon_{n-1}%
+\epsilon_{n}\right)  ,\nonumber\\
\tilde{d}_{n}  &  =\frac{1}{2e\Delta_{z}^{2}}\left(  \epsilon_{n-1}%
+2\epsilon_{n}+\epsilon_{n+1}\right)  .
\end{align}
Equation (\ref{eq:poisson_discretized}) is then solved over a single QCL
period $z\in\left[  z_{0},z_{0}+L_{\mathrm{p}}\right]  $, with the grid points
$z_{n}$, $n=0\dots P$ which should coincide with the grid used for solving the
Schr\"{o}dinger equation Eq.\thinspace(\ref{eq:s1D}). Applying the boundary
conditions $\tilde{V}_{0}=\tilde{V}_{P}=0$, Eq.\thinspace
(\ref{eq:poisson_discretized}) can be written as a matrix equation
$\mathbf{M\tilde{V}}=\mathbf{\rho}$, where $\mathbf{\tilde{V}}$ and
$\mathbf{\rho}$ represent vectors with elements $\tilde{V}_{n}$ and $\rho_{n}%
$, respectively, with $n=1\dots\left(  P-1\right)  $. $\mathbf{M}$ is a matrix
with the non-zero elements $M_{n,n}=-\tilde{d}_{n}$, $M_{n,n-1}=\tilde{s}_{n}%
$, $M_{n,n+1}=\tilde{s}_{n+1}$. This equation can be efficiently solved using
an algorithm for tridiagonal equation systems \cite{1992nrca.book.....P}.

While $n_{i}^{\mathrm{s}}$ in Eq.\thinspace(\ref{eq:rho}) can in principle
only be determined by detailed carrier transport simulations, simpler and much
faster approaches are often adopted, e.g., for design optimizations of
experimental QCL structures over an extended parameter range. Frequently,
Fermi-Dirac statistics is applied
\cite{2000JAP....87.7931C,2008SeScT..23l5040L}, with%
\begin{equation}
n_{i}^{\mathrm{s}}=\frac{m_{i}^{\parallel}}{\pi\hbar^{2}}k_{\mathrm{B}%
}T_{\mathrm{L}}\ln\left\{  1+\exp\left[  \left(  \mu-\tilde{E}_{i}\right)
/\left(  k_{\mathrm{B}}T_{\mathrm{L}}\right)  \right]  \right\}  .
\label{eq:fermi}%
\end{equation}
Here we assume that the electron distribution is described by the lattice
temperature $T_{\mathrm{L}}$. Furthermore, $\mu$ is the chemical potential,
$k_{\mathrm{B}}$ denotes the Boltzmann constant, and $m_{i}^{\parallel}$ is
the effective mass associated with the $i$th subband, see Eq.\thinspace
(\ref{eq:mi}). If nonparabolicity effects are neglected, $m_{i}^{\ast}$ can
often be approximated by the value of the well material. In Eq.\thinspace
(\ref{eq:fermi}), the energy $\tilde{E}_{i}$ defined in Eq.\thinspace
(\ref{eq:Ei}) is used instead of the eigenenergy $E_{i}$, to correctly reflect
the invariance properties of the biased structure \cite{2009IJQE...45..1059J}.
Especially, this guarantees that the obtained results do not depend on the
choice of the elementary period in the heterostructure. The chemical potential
$\mu$ is found from the charge neutrality condition within one period, i.e.,%
\begin{equation}
n^{\mathrm{s}}=\int_{z_{0}}^{z_{0}+L_{\mathrm{p}}}n_{\mathrm{D}}%
\mathrm{d}z=\sum_{i}n_{i}^{\mathrm{s}}.
\end{equation}
This is done recursively by first determining a lower and an upper boundary
value for $\mu$ where $\sum_{i}n_{i}^{\mathrm{s}}<n^{\mathrm{s}}$ and
$\sum_{i}n_{i}^{\mathrm{s}}>n^{\mathrm{s}}$, respectively, and then finding
the exact $\mu$, e.g., by using the bisection method
\cite{1992nrca.book.....P}.

The total potential in the Schr\"{o}dinger equation (\ref{eq:s1D}) is given by
$V=V_{0}+\tilde{V}$, where $\tilde{V}$ has to be obtained by solving the
Poisson equation (\ref{eq:poisson}). On the other hand, the wave functions
$\psi_{i}$ in Eq.\,(\ref{eq:poisson}) must be determined from Eq.
(\ref{eq:s1D}). In practice, this is done by iteratively solving the
Schr\"{o}dinger and Poisson equations \cite{2000JAP....87.7931C}, initially
assuming $\tilde{V}=0$, until the results for $\tilde{V}$, $\psi_{i}$ and
$E_{i}$ converge. If the\ $n_{i}^{\mathrm{s}}$ are obtained from
self-consistent carrier transport simulations instead of using Eq.
(\ref{eq:fermi}), then the carrier transport simulation and the numerical
solution of the Schr\"{o}dinger-Poisson system have to be performed
iteratively until convergence is obtained. Such simulations are then referred
to as self-self-consistent approaches
\cite{2005ApPhL..86u1117J,2010JAP...107a3104J}.

\begin{figure}[ptb]
\includegraphics{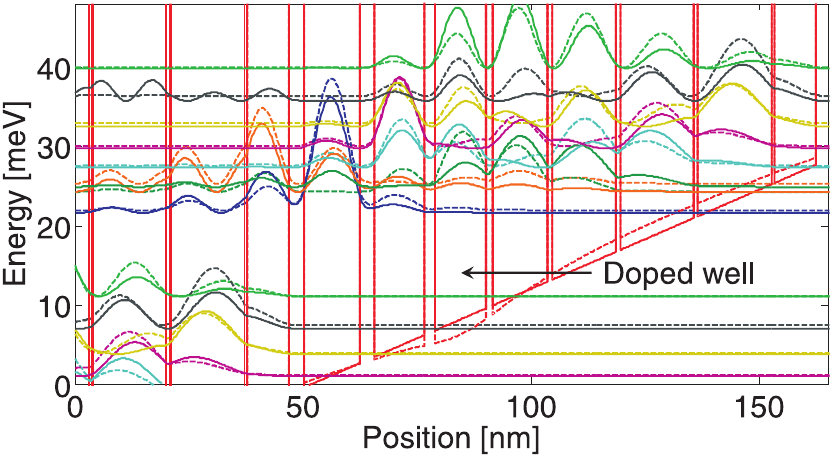}
\caption{{}(Color online) Conduction band profile and probability densities
for a bound-to-continuum terahertz QCL \cite{2003ApPhL..82.3165S}. Shown are
the results without considering space charge effects (solid lines) and for
space charge effects taken into accout, assuming thermally occupied subbands
(dashed lines).}%
\label{fig:spacecharge}%
\end{figure}

In Fig.\thinspace\ref{fig:spacecharge}, simulation results for a terahertz QCL
\cite{2003ApPhL..82.3165S} obtained by solving the Schr\"{o}dinger equation
(solid lines) and the Schr\"{o}dinger-Poisson system (dashed lines) are
compared. Although the populations obtained by Eq.\thinspace(\ref{eq:fermi})
usually deviate considerably from detailed carrier transport simulations, this
simplified approach can already give improved results, as compared to
completely neglecting space charge effects \cite{2010JAP...107a3104J}. The
reason is that the positively charged donors are usually localized in a
relatively small section, e.g., a single quantum well, while the electrons are
more or less distributed across the whole QCL period. This leads to
considerable conduction band bending due to space charge effects, as
illustrated in Fig.\thinspace\ref{fig:spacecharge}. While a more exact
determination of $n_{i}^{\mathrm{s}}$ yields a somewhat different electron
distribution, the overall result for $\tilde{V}$ will be similar as for the
simplified model based on Eq.\thinspace(\ref{eq:fermi}).

The method shown above is not the only one for solving the
Schr\"{o}dinger-Poisson system self-consistently under the condition of global
charge neutrality. A common alternative approach is to keep the chemical
potential fixed and solve the Poisson equation with Neumann boundary
conditions. These boundary conditions allow the Poisson potential to
self-adjust the density and maintain the global charge neutrality. Typically,
this requires a nonlinear realization of the Poisson equation to includes an
explicit potential dependence of the charge density. Such a nonlinear Poisson
equation can then effectively be solved with iterative methods such as the
predictor-corrector approach.~\cite{trellakis_predictor}

\section{\label{sec:Reso}Optical resonator modeling}

\begin{figure}[ptb]
\includegraphics{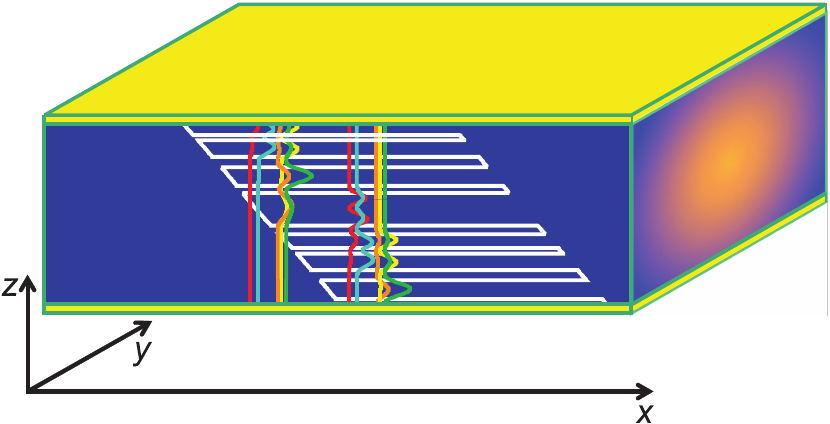}
\caption{{}(Color online) Typical waveguide resonator geometry of a QCL.}%
\label{fig:cavity}%
\end{figure}

The resonator spatially confines the radiation field, furthermore providing
optical outcoupling, beam shaping and frequency selection. While most
theoretical work has focused on the carrier transport in the gain medium,
there has also been progress in the modeling of the cavity. For example, the
resonator loss, which is crucial for the temperature performance of THz QCLs
\cite{fathololoumi2012terahertz,2008OExpr..16.3242B}, has been extracted from
finite element simulations of the resonator \cite{2005JAP....97e3106K}. In
Fig.\,\ref{fig:cavity}, a typical waveguide resonator geometry of a QCL is
sketched, and the used coordinate system is shown for reference. The
propagation direction of the optical field is denoted by $x$, $y$ refers to
the lateral direction, and $z$ indicates the growth direction of the
heterostructure. The resonator consists of materials with different
permittivities to obtain waveguiding and optical outcoupling. Thus, for
optical cavity simulations, the material permittivities must be known, which
in general depend on the frequency, doping level and temperature. In this
context, often the Drude model is employed with adequately chosen fitting
parameters \cite{2005JAP....97e3106K}. For intersubband optical transitions,
only the dipole matrix element in the $z$ direction where quantum confinement
occurs is nonzero, see Section \ref{sec:PDM}. Thus, only resonator modes with
an electric field component along the $z$ direction are amplified. For this
reason, surface emission, i.e., outcoupling through the $xy$-plane, can only
be obtained by using special outcoupling schemes
\cite{2003Sci...302.1374C,2006OExpr..1411672F}.

\subsection{Maxwell's equations}

The propagation of the electric and magnetic field vectors $\mathbf{E}\left(
\mathbf{x},t\right)  $ and $\mathbf{H}\left(  \mathbf{x},t\right)  $ is
generally described by Maxwell's equations. Here, $\mathbf{x}=\left[
x,y,z\right]  ^{\mathrm{T}}$ and $t$ are the position vector and time,
respectively. In the following, monochromatic fields are considered. We use
the physics convention, i.e., we assume a harmonic time dependence
$\propto\exp(-\mathrm{i}\omega t)$, where $\omega$ denotes the angular
frequency. The fields are then expressed as $\mathbf{H}\left(  \mathbf{x}%
,t\right)  =\Re\left\{  \mathbf{\hat{H}}\left(  \mathbf{x}\right)
\exp(-\mathrm{i}\omega t)\right\}  $, $\mathbf{E}\left(  \mathbf{x},t\right)
=\Re\left\{  \mathbf{\hat{E}}\left(  \mathbf{x}\right)  \exp(-\mathrm{i}\omega
t)\right\}  $, where $\mathbf{\hat{H}}$ and $\mathbf{\hat{E}}$ denote the
complex amplitude vectors. All equations can easily be converted to the
engineering convention [expressing the time dependence as $\propto
\exp(\mathrm{j}\omega t)$] by making the formal substitution $\mathrm{i}%
\rightarrow-\mathrm{j}$. For optical frequencies, typically the permeability
is given by its vacuum value $\mu_{0}$. Furthermore it is here assumed that
the dielectric constant can be described by a position dependent complex
scalar $\epsilon_{\mathrm{r}}\left(  \mathbf{x}\right)  $. Under these
conditions, Maxwell's equations simplify to
\begin{subequations}
\label{eq:Maxwell}%
\begin{align}
\mathbf{\nabla}\times\mathbf{\hat{H}}  &  =-\mathrm{i}\omega\epsilon
_{0}\epsilon_{\mathrm{r}}\mathbf{\hat{E}},\label{eq:Maxwell1}\\
\mathbf{\nabla}\times\mathbf{\hat{E}}  &  =\mathrm{i}\omega\mu_{0}%
\mathbf{\hat{H}}. \label{eq:Maxwell2}%
\end{align}
\end{subequations}
$\mathbf{\nabla B=}\mu_{0}\mathbf{\nabla}\left(  \mathbf{H}\right)  =0$ is
then automatically fulfilled, as can be seen by taking the divergence of Eq.
(\ref{eq:Maxwell2}). The equation $\mathbf{\nabla D}=\mathbf{\nabla}\left(
\epsilon_{0}\epsilon_{\mathrm{r}}\mathbf{E}\right)  =\rho$ delivers the charge
density $\rho$ and is not needed to compute $\mathbf{E}$ and $\mathbf{H}$.
Eqs.\thinspace(\ref{eq:Maxwell}) and the corresponding boundary conditions
define an eigenvalue problem, which yields the electromagnetic resonator
modes. General numerical approaches for solving Maxwell's equations include
the finite element \cite{FE} and finite difference time domain \cite{FDTD}
method, which also have been applied to the simulation of QCL cavities
\cite{2005JAP....97e3106K,2009OExpr..17..941B}.

\subsection{Two-dimensional waveguide model}

By eliminating $\mathbf{\hat{E}}$ from Eq.\thinspace(\ref{eq:Maxwell}), a wave
equation for $\mathbf{\hat{H}}$ can be derived \cite{1990ITMTT..38..722S}. In
many cases a waveguide geometry is used which does not depend on the
longitudinal $x$ direction, i.e., $\epsilon_{\mathrm{r}}=\epsilon_{\mathrm{r}%
}\left(  y,z\right)  $. Optical outcoupling is then obtained through the
cleaved semiconductor facets which serve as partly transparent mirrors. Since
the resonator length of up to a few mm is large as compared to the transverse
resonator dimensions, the computation of the transverse mode profile in the
$yz$-plane can be decoupled from the propagation coordinate and reduces to a
2D problem. The facet transmittance is then calculated based on the obtained
transverse field distribution. For a constant waveguide geometry in
propagation direction $x$, we can assume a field dependence $H_{y,z}\left(
\mathbf{x},t\right)  =\Re\left\{  \hat{H}_{y,z}\left(  y,z\right)  \exp\left(
\mathrm{i}\beta x-\mathrm{i}\omega t\right)  \right\}  $ with the complex
propagation constant $\beta$, and analogously for the electric field. The wave
equation then reduces to two coupled differential equations for the transverse
field components%
\begin{equation}
\left(  \partial_{p}^{2}+\partial_{q}^{2}\right)  \hat{H}_{p}+\frac
{\partial_{q}\epsilon_{\mathrm{r}}}{\epsilon_{\mathrm{r}}}\left(  \partial
_{p}\hat{H}_{q}-\partial_{q}\hat{H}_{p}\right)  =\left(  \beta^{2}-k_{0}%
^{2}\epsilon_{\mathrm{r}}\right)  \hat{H}_{p} \label{eq:wave2D}%
\end{equation}
with $k_{0}=\omega\sqrt{\mu_{0}\epsilon_{0}}=\omega/c$, $p=y,z$ and $q=z,y$
\cite{1990ITMTT..38..722S}. The longitudinal component $H_{x}$ is obtained
from $\mathbf{\nabla H}=0$. Furthermore, the electric field components can be
calculated from Eq.\thinspace(\ref{eq:Maxwell1}). The solution of
Eq.\thinspace(\ref{eq:wave2D}), along with the corresponding boundary
condition of vanishing fields for $y^{2}+z^{2}\rightarrow\infty$, constitutes
an eigenvalue problem, and the corresponding solutions $\hat{H}_{y,z}$ and
$\beta$ correspond to the waveguide modes. As mentioned above, only resonator
modes with an electric field component along the $z$ direction are amplified.
This is fulfilled for the transverse magnetic (TM) modes, which are
characterized by a vanishing magnetic field in the propagation direction,
i.e., $H_{x}=0$.

A waveguide mode is frequently characterized using three parameters, the
overlap (or field confinement) factor $\Gamma$, the waveguide loss coefficient
$a_{\mathrm{w}}$, and the mirror or outcoupling loss coefficient
$a_{\mathrm{m}}$, with the total power loss coefficient $a=a_{\mathrm{w}%
}+a_{\mathrm{m}}$. Based on these quantities, the threshold gain is given by
$g_{\mathrm{th}}=a/\Gamma$. Furthermore, for QCL simulations including the
optical cavity field, these parameters enter the simulation of the QCL\ gain
medium to describe the properties of the waveguide mode
\cite{2010ApPhL..96a1103J,2013SeScT..28j5008P}. $\Gamma$, $a_{\mathrm{w}}$ and
$a_{\mathrm{m}}$ can be obtained from the mode solutions of Eq.
(\ref{eq:wave2D}). The waveguide loss arises from the absorption in the
waveguide layers and is given by $a_{\mathrm{w}}=2\Im\left\{  \beta\right\}
$. The overlap factor corrects for the fact that the mode only partially
overlaps with the gain medium. To reflect the fact that the gain medium only
couples to the $E_{z}$ component, the overlap factor is defined as \cite{QCLs}%
\begin{equation}
\Gamma=\frac{\iint_{S_{\mathrm{g}}}\left|  \hat{E}_{z}\right|  ^{2}%
\mathrm{d}y\mathrm{d}z}{\iint_{-\infty}^{\infty}\left|  \mathbf{\hat{E}}%
\right|  ^{2}\mathrm{d}y\mathrm{d}z}, \label{eq:over}%
\end{equation}
where we integrate over the gain medium cross section area $S_{\mathrm{g}}$ in
the enumerator.

Strictly speaking, the calculation of the facet transmission constitutes a
full 3D problem, since the facets introduce an abrupt change in $x$
direction. However, since the QCL resonator length is large as compared to its
transverse dimensions, the computation of the transverse mode profile in the
$yz$-plane can be decoupled from the $x$ coordinate, as mentioned above. Only
for sufficently wide transverse waveguide dimensions, the facet reflectance
$R$ can be estimated from Fresnel's formula
\begin{equation}
R=\left|  n_{\mathrm{eff}}-1\right|  ^{2}/\left|  n_{\mathrm{eff}}+1\right|
^{2},
\end{equation}
with the effective refractive index defined as $n_{\mathrm{eff}}=\beta/k_{0}$.
In general, modal effects lead to an increased reflectance
\cite{2005JAP....97e3106K}. Various methods have been developed to extract $R$
from the mode solutions provided by Eq.\thinspace(\ref{eq:wave2D})
\cite{butler1974radiation,1972IJQE....8..470I,1993IPTL....5..148K,1971JAP....42.4466R}%
. The description of the outcoupling loss by a distributed coefficient
$a_{\mathrm{m}}$\ is obtained from $R=\exp\left(  -a_{\mathrm{m}}L\right)  $
with the resonator length $L$, yielding
\begin{equation}
a_{\mathrm{m}}=-\ln\left(  R\right)  /L. \label{eq:am}%
\end{equation}
If one facet is reflection coated and the light is outcoupled only at one
side, we obtain $a_{\mathrm{m}}=-\ln\left(  R\right)  /\left(  2L\right)  $.

\subsection{One-dimensional waveguide slab model}

If the waveguide width in lateral $y$ direction significantly exceeds its
thickness, the waveguide calculations can be reduced to a 1D problem with
$\epsilon_{\mathrm{r}}=\epsilon_{\mathrm{r}}\left(  z\right)  $, corresponding
to the simulation of a slab waveguide structure \cite{2010IJQE...46..618D}.
This applies for example to typical THz metal-metal waveguide resonators,
where the vertical $z$ dimension of around $10\,%
\mu
\mathrm{m}$ is often significantly smaller than the lateral $y$ dimension
($\approx25-200\,%
\mu
\mathrm{m}$) \cite{2005JAP....97e3106K}. For TM modes, in addition to
$H_{x}=0$ we have approximately $H_{z}\approx0$, and the $y$ component is
given by $H_{y}\left(  x,z\right)  =\Re\left\{  \hat{H}_{y}\left(  z\right)
\exp\left(  \mathrm{i}\beta x-\mathrm{i}\omega t\right)  \right\}  $, where
$\beta$ denotes the complex propagation constant. The one-dimensional wave
equation is then obtained from Eq.\thinspace(\ref{eq:Maxwell}) or
Eq.\thinspace(\ref{eq:wave2D}) as
\begin{equation}
\epsilon_{\mathrm{r}}\partial_{z}\left(  \epsilon_{\mathrm{r}}^{-1}%
\partial_{z}\hat{H}_{y}\right)  =\left(  \beta^{2}\mathbf{-}\epsilon
_{\mathrm{r}}k_{0}^{2}\right)  \hat{H}_{y}. \label{eq:wave1D}%
\end{equation}
The electric field amplitude is with Eq.\thinspace(\ref{eq:Maxwell1}) given
by
\begin{subequations}
\label{eq:Ezx}%
\begin{align}
\hat{E}_{z}  &  =-\frac{\beta}{\omega\epsilon_{0}\epsilon_{\mathrm{r}}%
}\hat{H}_{y},\label{eq:Ez}\\
\hat{E}_{x}  &  =-\frac{\mathrm{i}}{\omega\epsilon_{0}\epsilon_{\mathrm{r}}%
}\partial_{z}\hat{H}_{y}. \label{eq:Ex}%
\end{align}
\end{subequations}
From Eq.\thinspace(\ref{eq:wave1D}) we see that both $H_{y}$ and $\left(
1/\epsilon_{\mathrm{r}}\right)  \partial_{z}H_{y}$ must be continuous,
corresponding to the continuity of the field components $H_{y}$ and, with Eq.
(\ref{eq:Ex}), $E_{x}$ parallel to the layers.

Equation (\ref{eq:wave1D}) can be brought into the same form as the
one-dimensional Schr\"{o}dinger equation (\ref{eq:s1D}) by associating
$\hat{H}_{y}\left(  z\right)  $, $\epsilon_{\mathrm{r}}\left(  z\right)  $,
and $\epsilon_{\mathrm{r}}^{-1}\beta^{2}\mathbf{-}k_{0}^{2}$ with $\psi\left(
z\right)  $, $-2m^{\ast}\left(  z\right)  /\hbar^{2}$, and $E-V\left(
z\right)  $, respectively. Thus we can use the transfer matrix method
introduced in Section \ref{sec:transfermatrix}\ also for solving
Eq.\thinspace(\ref{eq:wave1D}). We start by dividing the waveguide in $z$
direction into layers $n$, $z_{n}\leq z<z_{n}+\Delta_{n}=z_{n+1}$, with
constant relative permittivites $\epsilon_{\mathrm{r}}^{\left(  n\right)  }$.
Although the gain medium itself consists of different layers, it can be
described by its total layer thickness and a single effective dielectric
constant since the individual layers are so thin that the electromagnetic wave
cannot resolve the structure. For TM modes, the inverse effective dielectric
constant is given by $\epsilon_{\mathrm{r},\mathrm{eff}}^{-1}=\left(
\Delta_{\mathrm{b}}\epsilon_{\mathrm{r},\mathrm{b}}^{-1}+\Delta_{\mathrm{w}%
}\epsilon_{\mathrm{r},\mathrm{w}}^{-1}\right)  /\left(  \Delta_{\mathrm{b}%
}+\Delta_{\mathrm{w}}\right)  $ where $\Delta_{\mathrm{b}}$ and $\Delta
_{\mathrm{w}}$ denote the total thickness of the barriers and wells in the
gain medium, respectively, and $\epsilon_{\mathrm{r},\mathrm{b}}$,
$\epsilon_{\mathrm{r},\mathrm{w}}$ are the corresponding dielectric constants
\cite{1999poet.book.....B,1994ApOpt..33.7875B}. In analogy to Eq.\thinspace
(\ref{eq:psi}) for solving the Schr\"{o}dinger equation, we can write the
solution of Eq.\thinspace(\ref{eq:wave1D}) in layer $n$ as
\cite{2010IJQE...46..618D}
\begin{equation}
\hat{H}_{y}^{\left(  n\right)  }=A_{n}\exp\left[  \mathrm{i}k_{n}\left(
z-z_{n}\right)  \right]  +B_{n}\exp\left[  -\mathrm{i}k_{n}\left(
z-z_{n}\right)  \right]  ,
\end{equation}
with $k_{n}=\left(  \epsilon_{\mathrm{r}}^{\left(  n\right)  }k_{0}%
^{2}\mathbf{-}\beta^{2}\right)  ^{1/2}$. The propagation through the segment
$n$ is then described by the matrix $T_{n}\left(  \Delta_{n}\right)  $ defined
in Eq.\thinspace(\ref{eq:Th}). As mentioned above, Eq.\thinspace
(\ref{eq:wave1D}) implies the continuity of $H_{y}$ and $\left(
1/\epsilon_{\mathrm{r}}\right)  \partial_{z}H_{y}$ across layer boundaries.
These matching conditions between two layers can be expressed in terms of
matrix Eq.\thinspace(\ref{eq:Tst}) by choosing $\tilde{k}_{n}=k_{n}%
/\epsilon_{\mathrm{r}}^{\left(  n\right)  }$. There is however one fundamental
difference between Eq.\thinspace(\ref{eq:s1D}) and Eq.\thinspace
(\ref{eq:wave1D}): Equation (\ref{eq:wave1D}) generally has complex
eigenvalues $\beta^{2}$ since $\epsilon_{\mathrm{r}}$ becomes complex for
materials with optical loss or gain. Consequently, the shooting method
described in Section \ref{sec:transfermatrix}\ is not applicable, and a
complex root finding algorithm has to be used.

The transmittance $T=1-R$ through the facet can be approximately computed from
the mode profile $\hat{H}_{y}\left(  z\right)  $. For TM polarization, we
obtain from the boundary value method \cite{butler1974radiation}%
\begin{align}
T  &  =\left[  \int_{-\infty}^{\infty}\frac{\Re\left\{  \beta\right\}
\Re\left\{  \epsilon_{\mathrm{r}}\left(  z\right)  \right\}  +\Im\left\{
\beta\right\}  \Im\left\{  \epsilon_{\mathrm{r}}\left(  z\right)  \right\}
}{\left|  \epsilon_{\mathrm{r}}\left(  z\right)  \right|  ^{2}}\left|
\hat{H}_{y}\left(  z\right)  \right|  ^{2}\mathrm{d}z\right]  ^{-1}\nonumber\\
&  \times8\pi\left|  \beta\right|  ^{2}k_{0}\int_{-k_{0}}^{k_{0}}\frac
{\sqrt{1-k_{z}^{2}/k_{0}^{2}}\left|  \Phi\left(  k_{z}\right)  \right|
^{2}\left|  \Phi^{\prime}\left(  k_{z}\right)  \right|  ^{2}}{\left|
k_{0}\sqrt{1-k_{z}^{2}/k_{0}^{2}}\Phi\left(  k_{z}\right)  +\beta\Phi^{\prime
}\left(  k_{z}\right)  \right|  ^{2}}\mathrm{d}k_{z}%
\end{align}
with the Fourier transforms%
\begin{align}
\Phi\left(  k_{z}\right)   &  =\frac{1}{2\pi}\int_{-\infty}^{\infty
}\hat{H}_{y}\left(  z\right)  \exp\left(  -\mathrm{i}k_{z}z\right)
\mathrm{d}z,\nonumber\\
\Phi^{\prime}\left(  k_{z}\right)   &  =\frac{1}{2\pi}\int_{-\infty}^{\infty
}\frac{\hat{H}_{y}\left(  z\right)  }{\epsilon_{\mathrm{r}}\left(  z\right)
}\exp\left(  -\mathrm{i}k_{z}z\right)  \mathrm{d}z.
\end{align}

Several extensions of the one-dimensional mode calculations presented above
are available. For high aspect ratios of the waveguide cross section, the 2D
mode calculation can be approximately reduced to 1D problems in $y$ and $z$
direction, respectively, by using the effective refractive index method
\cite{chiang1991performance}. A generalization of the transfer matrix scheme
for solving the two-dimensional wave equation, Eq.\,(\ref{eq:wave2D}), is the
film mode matching method, which is especially efficient for waveguides with a
rectangular geometry \cite{1994PApOp...3..381S}. Waveguides with a periodic
structure in propagation direction, such as a grating for surface emission,
can be treated with coupled mode theory \cite{2006IJQE...42..257S}.

\section{\label{sec:Over}Overview and classification of carrier transport models}

Depending on the intended goals, carrier transport models with varying degrees
of complexity have been used for the QCL gain medium simulation, ranging from
simple rate equation approaches to fully quantum mechanical descriptions. The
central task is to determine the optical gain, which is proportional to the
population inversion between the upper and the lower laser level, and the
current through the heterostructure as a function of the applied bias voltage.
Most simulation approaches require the eigenenergies and wave functions of the
energy levels in the QCL heterostructure as an input, which are determined by
solving the Schr\"{o}dinger equation or the Schr\"{o}dinger-Poisson equation
system (see Section \ref{sec:SP}).

\begin{figure}[ptb]
\includegraphics{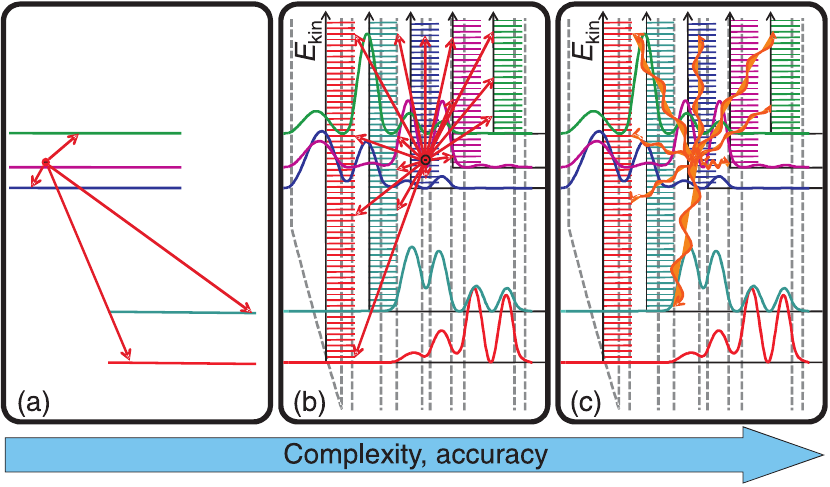}
\caption{{}(Color online) Illustration of various carrier transport models
with different levels of complexity and accuracy: (a) Rate equation approach;
(b) ensemble Monte Carlo method; (c) 3D density matrix/NEGF approach.}%
\label{fig:overv}%
\end{figure}

In Fig.\thinspace\ref{fig:overv}, various theoretical descriptions of the gain
medium with different levels of complexity are illustrated. The most basic
model of laser gain media in general is the rate equation approach
\cite{1986lase.book.....S}, which is also frequently applied to QCLs
\cite{2002IJQE...38..533F,2008IJQE...44...12Y,2002JAP....91.9019I,2001JAP....89.3084D,2012SeScT..27f5009S,2013SeScT..28j5008P,2005JAP....97h4506M}%
. The electron transport is simply described by transition rates between the
relevant energy levels in the gain medium. In the case of the QCL, these are
the electron subbands, and the corresponding transitions are referred to as
intersubband transitions. The application of the rate equation model to the
QCL is illustrated in Fig.\thinspace\ref{fig:overv}(a). In addition to the
transitions induced by the optical lasing field, nonradiative transitions
occur due to various scattering mechanisms, such as the interaction of the
electrons with phonons, other electrons or defects in the semiconductor
lattice. In the simplest case, these scattering rates are experimental or
empirical input parameters\ \cite{2002IJQE...38..533F,2008IJQE...44...12Y}. In
more advanced implementations, the transition rates are computed based on the
corresponding Hamiltonian \cite{2002JAP....91.9019I,2001JAP....89.3084D}, only
relying on well known material parameters such as the effective mass. Thus, no
specific experimental input is required, and no free ''fitting'' parameters
are available. A direct numerical solution is here not possible since the
transition rates depend on the a priori unknown electron populations. Rather,
the steady state solution must be found by simulating the temporal evolution
of the system until convergence is reached, or by using an iterative scheme.
Such advanced carrier transport methods are generally referred to as
self-consistent approaches. Apart from providing an intuitive description,
rate equation models are numerically very efficient, and thus are frequently
employed for the design and optimization of experimental QCL structures
\cite{2005JAP....97h4506M}. An extension are the Maxwell-Bloch equations,
which include the carrier-light interaction based on the density matrix
formalism \cite{2003noop.book.....B}. This approach is used to investigate
coherence effects between the laser field and the gain medium, relevant in
particular for the pulse formation\ in mode locked QCLs
\cite{2007PhRvA..75c1802W,2010OExpr..1813616G,2009PhRvL.102b3903M}.

In a quantum well heterostructure such as the QCL\ gain medium, the electrons
are only confined in the growth direction. Thus, energy quantization occurs in
one dimension, while the electrons can still freely move in two dimensions.
Due to this free in-plane motion in the quantum well, the electrons have
kinetic energy in addition to the eigenenergy of the corresponding quantized
level, as illustrated in Fig.\thinspace\ref{fig:overv}(b). For a more detailed
modeling and an improved understanding of the carrier transport in QCLs, the
in-plane motion of the electrons must be taken into account. Such
three-dimensional (3D) simulation approaches also consider intrasubband
transitions occuring between different kinetic energies within a subband, and
yield the electron distribution in each subband in addition to the level
populations. On the other hand, the complexity increases significantly as
compared to one-dimensional (1D) descriptions such as above discussed rate
equation approach, since transitions are now characterized by the initial and
final subbands as well as the corresponding kinetic energies. An example for a
self-consistent 3D approach is the ensemble Monte Carlo (EMC) method. Here,
the intrasubband processes are fully taken into account, and the scattering
rates are self-consistently evaluated based on the corresponding Hamiltonian.
Combining versatility and reliability with relative computational efficiency,
the EMC approach has been widely used for the analysis, design and
optimization of QCLs
\cite{2000ApPhL..76.2265I,2001ApPhL..79.3920K,2003ApPhL..83..207C,2004ApPhL..84..645C,2002ApPhL..80..920C,2006ApPhL..88f1119L,2007JAP...101f3101G,2007JAP...102k3107G,matyas2010temperature,matyas2011photon,2009JAP...105l3102J,2008pssc..5..221J,2009JPhD...42b5101L,2005JAP....97d3702B}%
.

In both the rate equation and EMC approach, the carrier transport is described
by scattering-induced transitions of discrete electrons between the quantized
energy levels, corresponding to a hopping transport model. Such methods are
referred to as semiclassical, since the energy quantization in the QCL
heterostructure is considered, but quantum coherence effects and quantum
mechanical dephasing are not included. Various quantum transport simulation
approaches have been developed which take into account quantum correlations
between the energy levels. One example is the density matrix method. Its 1D
version can be seen as a generalization of the rate equation approach, and is
frequently used for the analysis and optimization of THz QCLs
\cite{fathololoumi2012terahertz,2009PhRvB..80x5316K,2010PhRvB..81t5311D,2010NJPh...12c3045T}%
. In its 3D form \cite{2001PhRvL..87n6603I,2009PhRvB..79p5322W} illustrated in
Fig.\thinspace\ref{fig:overv}(c), it corresponds to a generalization of
semiclassical approaches based on the Boltzmann transport equation such as EMC
\cite{2001PhRvL..87n6603I}. Also the nonequilibrium Green's function (NEGF)
method, considered the most general quantum transport approach, has been
applied to the simulation of QCL structures
\cite{2002PhRvB..66h5326W,banit_wacker,2009PhRvB..79s5323K,2009ApPhL..95w1111S,Kubis_assess,Schmielau_ktyp,Kolek_openQCL,2012ApPhL.101u1113W,Wacker_JSTQE}%
. Quantum transport approaches are numerically much more demanding than their
semiclassical counterparts. On the other hand, quantum coherence effects can
play a pronounced role especially in THz QCLs where the energetic spacing
between the quantized levels is relative small
\cite{2005JAP....98j4505C,2009PhRvB..79p5322W}, while they are less relevant
in mid-infrared QCLs \cite{2001PhRvL..87n6603I}.

\begin{table}[ptb]
\caption{Classification of carrier transport modeling techniques. The
corresponding section number is given in brackets.}%
\label{tab:overv}%
\begin{ruledtabular}
\begin{tabular}
{lll}  &Semiclassical&Quantum transport\\ \hline\multicolumn{3}{l} {Empirical}
\\ \phantom{aa}1D&Rate equations (\ref{sec:PRE})&1D density matrix
(\ref{sec:DM})\\  &Maxwell-Bloch\footnote{Only the carrier-light interaction
is modeled using a density matrix formalism, while scattering is treated based
on rate equations.} (\ref{sec:PDM} )& \\ \multicolumn{3}{l}{Self-consistent}
\\ \phantom{aa}1D&Rate equations (\ref{sec:SCRE})&1D density matrix
(\ref{sec:DM})\\ \phantom{aa}3D&Monte Carlo (\ref{sec:EMC})&3D density matrix
(\ref{sec:DM})\\ & &NEGF (\ref{sec:NEGF})\\
\end{tabular}
\end{ruledtabular}
\end{table}

In Table \ref{tab:overv}, a classification of the different simulation
approaches covered in this review is given. Here we differentiate between
semiclassical schemes based on hopping transport between the quantized energy
levels, and quantum transport approaches taking into account quantum
correlations. We divide the methods into empirical approaches relying on
empirical or experimental input parameters, and self-consistent schemes which
evaluate the transition rates based on the corresponding Hamiltonian.
Furthermore, we distinguish between 1D modeling techniques only considering
the subband populations and intersubband transitions, and 3D approaches also
taking into account the electron distribution in the subbands and the
intrasubband dynamics.

\section{\label{sec:Phen}Empirical approaches}

The most basic approach for modeling the electron dynamics in a laser is to
use experimental or empirical transition rates between the relevant energy
levels in the laser gain medium \cite{1986lase.book.....S}. For QCLs, these
levels correspond to the quantized eigenstates of the heterostructure, which
have to be found by solving the Schr\"{o}dinger equation, Eq.\,(\ref{eq:s1D}).
The electron dynamics is then described by rate equations
\cite{1986lase.book.....S}. Often, only the nonradiative transitions are
considered which occur due to various scattering mechanisms, such as the
interaction of the electrons with phonons, other electrons or defects in the
semiconductor lattice. Simulations not including the optical cavity field can
be used to investigate under which conditions sufficient optical gain is
obtained, so that lasing operation can start at all. In this way, parameters
such as the threshold current density and maximum operating temperature can be
extracted. To investigate the lasing operation itself, including optical
output powers, saturation effects or the intrinsic linewidth
\cite{2008IJQE...44...12Y}, the optical field has to be included as well.
Modified empirical scattering-rate approaches have been developed for specific
types of QCLs \cite{2010SeScT..25d5025S}. Maxwell-Bloch equations which are a
generalization of the rate equation approach can be used to model the coherent
interaction between the laser field and the gain medium
\cite{2003noop.book.....B}, e.g., to investigate the formation of optical
instabilities in mode-locked QCLs \cite{2007PhRvA..75c1802W}.

\subsection{\label{sec:PRE}Empirical rate equations}

The rate equations for a laser are given by \cite{1986lase.book.....S}%
\begin{equation}
\mathrm{d}_{t}n_{i}^{\mathrm{s}}=\sum_{j\neq i}\tau_{ji}^{-1}n_{j}%
^{\mathrm{s}}-\tau_{i}^{-1}n_{i}^{\mathrm{s}}+\sum_{j\neq i}\left(
-W_{ij}^{\mathrm{opt}}n_{i}^{\mathrm{s}}+W_{ji}^{\mathrm{opt}}n_{j}%
^{\mathrm{s}}\right)  . \label{eq:rate}%
\end{equation}
The first two terms contain the relaxation transitions due to scattering,
e.g., the interaction of the electrons with phonons, other electrons or
defects in the semiconductor lattice. Also spontaneous emission can be
included here. The scattering rate from a level $j$ to $i$ is often expressed
in terms of an inverse scattering lifetime $\tau_{ji}^{-1}$, where $\tau
_{i}^{-1}=\sum_{j\neq i}\tau_{ij}^{-1}$ indicates the total inverse lifetime
of level $i$. The last sum contains the lasing transitions, where
$W_{ij}^{\mathrm{opt}}$ are the stimulated optical transition rates for those
transitions where an optical field at or near the corresponding frequency
$\left|  \omega_{ij}\right|  =\left|  E_{i}-E_{j}\right|  /\hbar$ is present
\cite{1986lase.book.....S}. The rates $W_{ij}^{\mathrm{opt}}$ are proportional
to the optical intensities in the corresponding lasing modes. Typically, only
one or a few transitions contribute to lasing. Furthermore, $n_{i}%
^{\mathrm{s}}$ is the electron sheet density of subband $i$, i.e., the
electron number divided by the in-plane cross section area $S$. This quantity
is often used in QCL heterostructures where energy quantization occurs in one
dimension and the electrons can still freely move in in-plane direction.

Commonly, the QCL\ heterostructure is designed strictly periodically, as
illustrated in Fig.\thinspace\ref{fig:qcl}. Apart from fabrication tolerances,
a periodic model is valid for the central QCL\ periods far away from the
contacts, if effects such as domain formation \cite{2006PhRvB..73c3311L} and
local variations of the optical field intensity can be neglected. The sum in
Eq.\thinspace(\ref{eq:rate}) can then be restricted to one representative
central period,%
\begin{equation}
\mathrm{d}_{t}n_{i}^{\mathrm{s}}=\sum_{j\neq i}\hat{\tau}_{ji}^{-1}%
n_{j}^{\mathrm{s}}-\tau_{i}^{-1}n_{i}^{\mathrm{s}}+\sum_{j\neq i}\left(
-\hat{W}_{ij}^{\mathrm{opt}}n_{i}^{\mathrm{s}}+\hat{W}_{ji}^{\mathrm{opt}%
}n_{j}^{\mathrm{s}}\right)  \label{eq:rate2}%
\end{equation}
with $i,j=1..N$, where $N$ is the number of subbands in each period. Here,
$\hat{\tau}_{ji}^{-1}=\sum_{n\in\mathbb{Z}}\tau_{j,i+nN}^{-1}$ includes the
transitions to all equivalent levels in the different periods, and analogously
for $\hat{W}_{ji}^{\mathrm{opt}}$. The total sheet density in each period is
determined by the doping sheet density,%
\begin{equation}
n^{\mathrm{s}}=\sum_{i=1}^{N}n_{i}^{\mathrm{s}}. \label{eq:ns}%
\end{equation}
The steady state solution is obtained by setting $\mathrm{d}_{t}%
n_{i}^{\mathrm{s}}=0$. If we are not interested in the lasing operation
itself, but only if sufficient inversion for lasing is obtained, stimulated
optical effects can be excluded, $\hat{W}_{ij}^{\mathrm{opt}}=\hat{W}_{ji}%
^{\mathrm{opt}}=0$. The subband populations $n_{i}^{\mathrm{s}}$ ($i=1..N$)
can then be found by solving the linear equation system Eq.\thinspace
(\ref{eq:rate2}) with $\mathrm{d}_{t}n_{i}^{\mathrm{s}}=0$ and $i=1..(N-1)$,
complemented by Eq.\thinspace(\ref{eq:ns}) to obtain a linearly independent
system and thus a unique solution. To include lasing, this system has to be
complemented by equations describing the optical intensities in the lasing
modes, since the $\hat{W}_{ij}^{\mathrm{opt}}$ are intensity dependent.

\begin{figure}[ptb]
\includegraphics{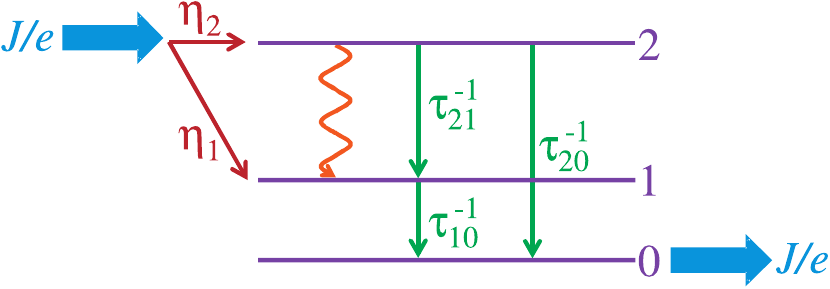}
\caption{{}(Color online) Three-level system model for a QCL.}%
\label{fig:three}%
\end{figure}

Often the empirical rate equation is restricted to the most important subbands
and transitions to obtain compact analytical results. A common model is the
three-level system \cite{2002IJQE...38..533F}, illustrated in Fig.
\ref{fig:three}. This description only includes the upper laser level $2$,
lower laser level $1$ and a reservoir level $0$ representing the extraction
and injector subbands. A fraction $\eta_{2}$ of the current density $J$ is
injected into the upper laser level, and $\eta_{1}$ into the lower laser
level. Furthermore, only the transition rates $\tau_{21}^{-1}$, $\tau
_{20}^{-1}$ and $\tau_{10}^{-1}$ are included, with the inverse upper and
lower laser level lifetimes $\tau_{2}^{-1}=\tau_{21}^{-1}+\tau_{20}^{-1}$,
$\tau_{1}^{-1}=\tau_{10}^{-1}$. With Eq.\thinspace(\ref{eq:rate}), the rate
equations for the upper and lower laser level are then obtained as%
\begin{align}
\mathrm{d}_{t}n_{2}^{\mathrm{s}}  &  =\eta_{2}J/e-\tau_{2}^{-1}n_{2}%
^{\mathrm{s}}-\sigma_{21}^{\mathrm{opt}}I\left(  n_{2}^{\mathrm{s}}%
-n_{1}^{\mathrm{s}}\right)  ,\nonumber\\
\mathrm{d}_{t}n_{1}^{\mathrm{s}}  &  =\eta_{1}J/e-\tau_{1}^{-1}n_{1}%
^{\mathrm{s}}+\tau_{21}^{-1}n_{2}^{\mathrm{s}}+\sigma_{21}^{\mathrm{opt}%
}I\left(  n_{2}^{\mathrm{s}}-n_{1}^{\mathrm{s}}\right)  , \label{eq:rate3}%
\end{align}
where $\sigma_{21}^{\mathrm{opt}}$ is the cross section for stimulated
emission from level $2$ to $1$ and $I$ is the optical intensity of the laser
radiaton at frequency $\omega_{21}=\left(  E_{2}-E_{1}\right)  /\hbar$. If
lasing is neglected, $I=0$, the steady state population inversion is with
$\mathrm{d}_{t}=0$ in Eq.\thinspace(\ref{eq:rate3}) obtained as
\cite{2002IJQE...38..533F}%
\begin{equation}
n_{2}^{\mathrm{s}}-n_{1}^{\mathrm{s}}=\frac{J}{e}\left[  \eta_{2}\tau
_{2}\left(  1-\frac{\tau_{1}}{\tau_{21}}\right)  -\eta_{1}\tau_{1}\right]  .
\end{equation}
The three-level model can be extended to take into account further effects
such as thermal backfilling of the lower laser level and backscattering from
level $1$ to level $2$ in terahertz QCLs, which can be modeled by introducing
additional lifetimes $\tau_{01}$ and $\tau_{12}$, respectively
\cite{2008IJQE...44...12Y}.

\subsection{\label{sec:PDM}Maxwell-Bloch equations}

Bloch equations are a generalization to the rate equation approach where the
interaction of the laser level electrons with the optical field is modeled
using a density matrix formalism rather than scattering rates
\cite{2003noop.book.....B}. In this way, optical nonlinearities and coherence
effects between the laser field and the gain medium can be considered, while
the carrier transport due to nonradiative mechanisms is included by the
corresponding lifetimes as for the rate equations. To describe the optical
propagation, the Bloch equations are complemented by Maxwell's or related
equations, such as the wave equation. This model has for example been used to
investigate the formation of optical instabilities in QCLs
\cite{2007PhRvA..75c1802W,2010OExpr..1813616G}, and to study the possibilty of
self-induced transparency modelocking \cite{2009PhRvL.102b3903M}. The
nonlinear optical dynamics in the gain medium also plays an important role for
the recently demonstrated QCL-based frequency combs \cite{2012Natur.492..229H}.

The interaction of a classical optical field with a two-level system is in the
density matrix formalism described by \cite{2003noop.book.....B}
\begin{align}
\partial_{t}\rho_{21}  &  =-\mathrm{i}\omega_{21}\rho_{21}-\mathrm{i}%
\hbar^{-1}d_{12}^{\ast}E_{z}\Delta_{21}-\gamma_{21}\rho_{21},\nonumber\\
\partial_{t}\Delta_{21}  &  =2\mathrm{i}\hbar^{-1}\left(  d_{12}^{\ast}%
\rho_{21}^{\ast}-d_{12}\rho_{21}\right)  E_{z}-\gamma_{E}\left(  \Delta
_{21}-\Delta_{21}^{\mathrm{eq}}\right)  , \label{eq:bloch}%
\end{align}
where the asterisk denotes the complex conjugate. $E_{z}\left(  x,t\right)  $
represents the field component in the growth direction $z$ of the
heterostructure since the other components do not interact with the gain
medium, and $d_{12}=-e\left\langle 1\right|  z\left|  2\right\rangle $ is the
corresponding dipole matrix element of the laser transition. Furthermore,
$\rho_{ij}\left(  x,t\right)  =\rho_{ji}^{\ast}\left(  x,t\right)
=\left\langle i\right|  \hat{\rho}\left(  x,t\right)  \left|  j\right\rangle $
are the density matrix elements, and $\Delta_{21}\left(  x,t\right)
=\rho_{22}\left(  x,t\right)  -\rho_{11}\left(  x,t\right)  $ is the
inversion. The density matrix can be normalized so that $\rho_{ii}%
=n_{i}^{\mathrm{s}}/n^{\mathrm{s}}$ gives the relative population of subband
$i$. Dissipative processes are phenomenologically included by adding decay
terms with relaxation rates $\gamma_{E}$ and $\gamma_{21}$, describing the
energy relaxation and dephasing, respectively. $\Delta_{21}^{\mathrm{eq}}$ is
the equilibrium inversion which the system approaches for $E_{z}=0$. If the
gain medium is not homogeneous along the propagation direction, as is the case
for self-induced transparency mode locking structures
\cite{2009PhRvL.102b3903M}, the parameters $d_{12}$, $\gamma_{E}$,
$\gamma_{21}$, $\omega_{21}$, and $\Delta_{21}^{\mathrm{eq}}$ depend on the
propagation coordinate $x$.

Assuming weak nonlinearity and inhomogeneity, the optical field propagation
can be described by the wave equation \cite{2003noop.book.....B}. For
propagation in $x$ direction it is given by
\begin{equation}
\left(  \partial_{x}^{2}-c^{-2}n_{0}^{2}\partial_{t}^{2}\right)
E_{z}=\epsilon_{0}^{-1}c^{-2}\partial_{t}^{2}P_{z}. \label{eq:wave}%
\end{equation}
$P_{z}\left(  x,t\right)  $ is the polarization component in $z$ direction due
to the lasing transition, given by $P_{z}=\left(  n^{\mathrm{s}}%
/L_{\mathrm{p}}\right)  \left(  d_{12}\rho_{21}+d_{12}^{\ast}\rho_{21}^{\ast
}\right)  $. Here, $L_{\mathrm{p}}$ is the length of a single QCL\ period and
$n^{\mathrm{s}}/L_{\mathrm{p}}$ thus corresponds to the average electron
concentration \cite{2010OExpr..1813616G}. Furthermore, $n_{0}$ denotes the
refractive index of the gain medium material.

$E_{z}$ and $\rho_{21}$ are typically expressed by their slowly varying
envelope functions,
\begin{align}
E_{z}\left(  x,t\right)   &  =\frac{1}{2}\hat{E}_{z}\left(  x,t\right)
\exp\left[  \mathrm{i}\left(  k_{21}x-\omega_{21}t\right)  \right]
+c.c.,\nonumber\\
\rho_{21}\left(  x,t\right)   &  =\eta_{21}\left(  x,t\right)  \exp\left[
\mathrm{i}\left(  k_{21}x-\omega_{21}t\right)  \right]  , \label{eq:ampl}%
\end{align}
where $k_{21}=\omega_{21}n_{0}/c$ and c.c. denotes the complex conjugate.
Inserting Eq.\thinspace(\ref{eq:ampl}) into Eq.\thinspace(\ref{eq:bloch}) and
neglecting the rapidly oscillating terms $\propto\exp\left(  \pm
2\mathrm{i}\omega_{21}t\right)  $, we obtain the density matrix equations in
the rotating wave approximation \cite{2003noop.book.....B}
\begin{subequations}
\label{eq:bloch2}%
\begin{align}
\partial_{t}\eta_{21}  &  =-\mathrm{i}\left(  2\hbar\right)  ^{-1}d_{12}%
^{\ast}\hat{E}_{z}\Delta_{21}-\gamma_{21}\eta_{21},\label{eq:bloch2a}\\
\partial_{t}\Delta_{21}  &  =\mathrm{i}\hbar^{-1}\left(  \eta_{21}^{\ast
}d_{12}^{\ast}\hat{E}_{z}-\eta_{21}d_{12}\hat{E}_{z}^{\ast}\right)
-\gamma_{E}\left(  \Delta_{21}-\Delta_{21}^{\mathrm{eq}}\right)  .
\label{eq:bloch2b}%
\end{align}
\end{subequations}
Inserting Eq.\thinspace(\ref{eq:ampl}) into Eq.\thinspace(\ref{eq:wave})
yields with the slowly varying amplitude approximation $\left|  \partial
_{x}^{2}\hat{E}_{z}\right|  \ll\left|  k_{21}\partial_{x}\hat{E}_{z}\right|
$, $\left|  \partial_{t}^{2}\hat{E}_{z}\right|  \ll\left|  \omega_{21}%
\partial_{t}\hat{E}_{z}\right|  $
\cite{2003noop.book.....B,2010OExpr..1813616G}
\begin{equation}
\partial_{x}\hat{E}_{z}+n_{0}c^{-1}\partial_{t}\hat{E}_{z}=\mathrm{i}%
\frac{\omega_{21}\Gamma}{\epsilon_{0}cn_{0}}\frac{n^{\mathrm{s}}%
}{L_{\mathrm{p}}}d_{12}\eta_{21}-\frac{1}{2}a\hat{E}_{z}, \label{eq:wave2}%
\end{equation}
furthermore assuming $\left|  \partial_{t}^{2}\eta_{21}\right|  ,\left|
\omega_{21}\partial_{t}\eta_{21}\right|  \ll\left|  \omega_{21}^{2}\eta
_{21}\right|  $. Here, the overlap factor $\Gamma$, Eq.\thinspace
(\ref{eq:over}), has been added to correct for the fact that the optical mode
only partially overlaps with the gain medium, and a loss term with the power
loss coefficient $a$ has been included. For applying the Maxwell-Bloch
equation model to QCLs, typically the simplifying assumption is made that the
lower laser level is depopulated very efficiently, $\tau_{1}\rightarrow0$,
resulting in $n_{1}^{\mathrm{s}}=0$ in Eq.\thinspace(\ref{eq:rate3})
\cite{2010OExpr..1813616G}. The inversion is then directly given by the upper
laser level population, $\Delta_{21}=n_{2}^{\mathrm{s}}$. By comparison of
Eq.\thinspace(\ref{eq:bloch2}) with Eq.\thinspace(\ref{eq:rate3}), we obtain
$\Delta_{21}^{\mathrm{eq}}=\tau_{2}\eta_{2}J/\left(  n^{\mathrm{s}}e\right)
$, $\gamma_{E}=\tau_{2}^{-1}$.

The Maxwell-Bloch equations Eqs.\,(\ref{eq:bloch2}), (\ref{eq:wave2}) are a
versatile approach to describe the optical dynamics in QCLs, and additional
effects can straightforwardly be implemented. For example, the influence of
spatial hole burning due to the standing wave modes in a linear cavity has
been extensively studied
\cite{2007PhRvA..75c1802W,2010OExpr..1813616G,2009ApPhL..95g1109T}, and
dispersion as well as saturable absorption have been added to the model
\cite{2010OExpr..18.5639T}.

\subsubsection{Optical gain and transition rates}

The Bloch equations Eq.\thinspace(\ref{eq:bloch2}) can be used to derive the
optical gain coefficient and transition rate associated with stimulated
emission and absorption \cite{2003noop.book.....B}. For a monochromatic
electromagnetic field, the stationary solution of Eq.\thinspace
(\ref{eq:bloch2a}) is obtained by setting $\partial_{t}\eta_{21}=0$,%
\begin{equation}
\eta_{21}=-\frac{\mathrm{i}}{2\hbar\gamma_{21}}d_{12}^{\ast}\hat{E}_{z}%
\Delta_{21}. \label{eq:eta12}%
\end{equation}
Multiplying Eq.\thinspace(\ref{eq:wave2}) from left with $\hat{E}_{z}^{\ast}$
and adding the complex conjugate, we obtain with Eq.\thinspace(\ref{eq:eta12}%
)
\begin{equation}
\partial_{x}I+n_{0}c^{-1}\partial_{t}I=\Gamma gI-aI, \label{eq:int}%
\end{equation}
with the power gain coefficient%
\begin{equation}
g=\frac{\omega_{21}}{\hbar\gamma_{21}\epsilon_{0}cn_{0}}\frac{n^{\mathrm{s}}%
}{L_{\mathrm{p}}}\left|  d_{12}\right|  ^{2}\left(  \rho_{22}-\rho
_{11}\right)  . \label{eq:gain0}%
\end{equation}
Here we have replaced the electric field by the optical intensity
$I=\epsilon_{0}cn_{0}\left|  \hat{E}_{z}\right|  ^{2}/2$. The optical power
inside the resonator is given by $P=IS_{\mathrm{g}}/\Gamma$, where
$S_{\mathrm{g}}$ is the cross section area of the gain medium and
$S_{\mathrm{g}}/\Gamma$ corresponds to the effective area of the waveguide
mode. Frequently, $I$ is assumed to change only slightly along the resonator,
i.e., the intensity is averaged over the $x$ coordinate. This assumption is
valid especially for the case of moderate output coupling at the facets where
the mirror loss can be described by a distributed coefficient $a_{\mathrm{m}}%
$, Eq.\thinspace(\ref{eq:am}). Equation\thinspace(\ref{eq:int}) then
simplifies to%
\begin{equation}
n_{0}c^{-1}\partial_{t}I=\Gamma gI-aI. \label{eq:int2}%
\end{equation}
The transition rate due to the optical field is with $\left.  \partial
_{t}\Delta_{21}\right|  _{\mathrm{opt}}=2\left.  \partial_{t}\rho_{22}\right|
_{\mathrm{opt}}=-2\left.  \partial_{t}\rho_{11}\right|  _{\mathrm{opt}}$
obtained from Eqs.\thinspace(\ref{eq:bloch2b}) and (\ref{eq:eta12}) as
\begin{align}
\left.  \partial_{t}\rho_{22}\right|  _{\mathrm{opt}}  &  =\frac{1}%
{2}\mathrm{i}\hbar^{-1}\left(  \eta_{21}^{\ast}d_{12}^{\ast}\hat{E}_{z}%
-\eta_{21}d_{12}\hat{E}_{z}^{\ast}\right) \nonumber\\
&  =\mathbf{-}\frac{1}{\hbar^{2}\gamma_{21}\epsilon_{0}cn_{0}}\left|
d_{12}\right|  ^{2}\left(  \rho_{22}-\rho_{11}\right)  I. \label{eq:Wopt}%
\end{align}
The contribution $\propto\rho_{22}I$ leading to a reduction of $\rho_{22}$ is
due to stimulated emission, while the contribution $\propto\rho_{11}I$
corresponds to absorption. For a slightly detuned optical field at a frequency
$\omega\neq\omega_{21}$, Eqs.\thinspace(\ref{eq:gain0}) and (\ref{eq:Wopt})
can be adapted by replacing $\gamma_{21}^{-1}$ with $\pi\mathcal{L}\left(
\omega\right)  $, where $\mathcal{L}\left(  \omega\right)  $ is the Lorentzian
lineshape function \cite{2003noop.book.....B}
\begin{equation}
\mathcal{L}\left(  \omega\right)  =\frac{1}{\pi}\frac{\gamma_{21}}{\gamma
_{21}^{2}+\left(  \omega-\left|  \omega_{21}\right|  \right)  ^{2}}.
\label{eq:Lor}%
\end{equation}
Thus we obtain with the sheet densities $n_{1}^{\mathrm{s}}=n^{\mathrm{s}}%
\rho_{11}$ and $n_{2}^{\mathrm{s}}=n^{\mathrm{s}}\rho_{22}$%
\begin{equation}
g=\frac{\pi\omega}{\hbar\epsilon_{0}cn_{0}L_{\mathrm{p}}}\left|
d_{12}\right|  ^{2}\left(  n_{2}^{\mathrm{s}}-n_{1}^{\mathrm{s}}\right)
\mathcal{L}\left(  \omega\right)  \label{eq:gain}%
\end{equation}
and%
\begin{equation}
\left.  \partial_{t}n_{2}^{\mathrm{s}}\right|  _{\mathrm{opt}}=\mathbf{-}%
\frac{\pi}{\epsilon_{0}cn_{0}\hbar^{2}}\left|  d_{12}\right|  ^{2}\left(
n_{2}^{\mathrm{s}}-n_{1}^{\mathrm{s}}\right)  I\mathcal{L}\left(
\omega\right)  . \label{eq:Wstim}%
\end{equation}
Comparison with Eq.\thinspace(\ref{eq:rate}) yields for the stimulated optical
transition rates%
\begin{equation}
W_{ij}^{\mathrm{opt}}=W_{ji}^{\mathrm{opt}}=\frac{\pi}{\epsilon_{0}cn_{0}%
\hbar^{2}}\left|  d_{12}\right|  ^{2}I\mathcal{L}\left(  \omega\right)  .
\label{eq:Wijopt}%
\end{equation}

\section{\label{sec:Self}Self-consistent semiclassical approaches}

Advanced self-consistent simulation approaches only rely on well known
material parameters such as the effective mass, and no further specific
experimental or empirical input is required. This also means that no
adjustable parameters are available to fit the simulation results to
experimental data. These approaches are based on the evaluation of the
transitions between the various states due to different scattering mechanisms,
including the interaction of the electrons with phonons, impurities and other
electrons. The associated scattering rates are computed based on the
corresponding Hamiltonian. A direct numerical solution of the resulting
equations is not possible since the transition rates depend on the initially
unknown electron populations. Rather, the steady state solution must be
self-consistently found by simulating the temporal evolution of the system
until convergence is reached, or by using iterative schemes. These methods
rely on the subband wave functions and eigenenergies found by solving the
Schr\"{o}dinger or Schr\"{o}dinger-Poisson equation, as described in Section
\ref{sec:SP}. The carrier transport is then modeled by transitions of discrete
electrons between the quantized energy levels, also referred to as hopping
transport. Thus, these methods are called semiclassical, since the energy
quantization in the QCL heterostructure is considered, but quantum coherence
effects and dephasing mechanisms are not included. Formally, semiclassical
carrier transport descriptions can be derived from the more general density
matrix formalism by neglecting the contribution of the off-diagonal matrix
elements, i.e., only considering the diagonal elements corresponding to the
occupations of the states \cite{2001PhRvL..87n6603I,2005PhRvB..72l5347I}. In
the following, the most relevant scattering mechanisms in the QCL will be
discussed. Furthermore, the self-consistent rate equation approach and the EMC
method will be described which are the two most widely used advanced
semiclassical QCL simulation schemes.

\subsection{\label{sec:scat}Scattering mechanisms and transition rates}

\begin{figure}[ptb]
\includegraphics{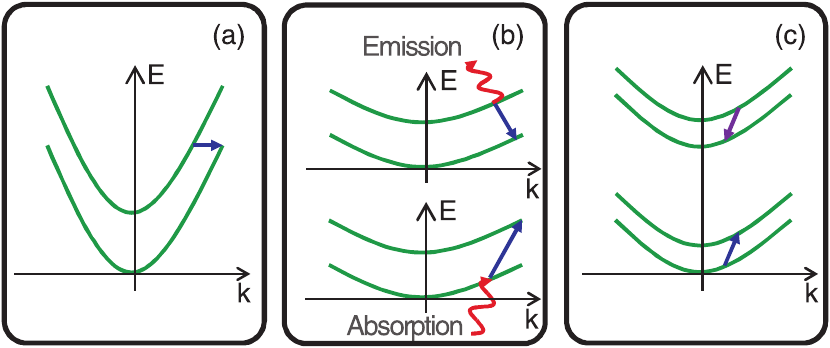}
\caption{(Color online) Different classes of scattering processes: (a) elastic
scattering; (b) inelastic scattering; (c) carrier-carrier scattering.}%
\label{fig:scattering}%
\end{figure}

\begin{figure}[ptb]
\includegraphics{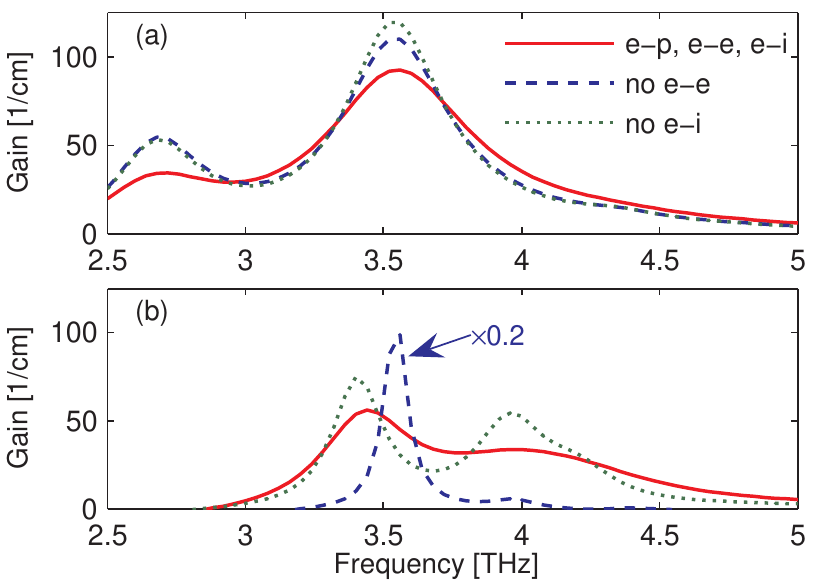}
\caption{(Color online) Simulation results for the spectral gain vs frequency,
as obtained by evaluating electron-phonon (e-p), electron-impurity (e-i) and
electron-electron (e-e) scattering (solid curves) and neglecting e-e (dashed
curves) or e-i (dotted curves) scattering. a) 3.4 THz resonant phonon
depopulation structure; b) 3.5 THz bound-to-continuum structure
\cite{2009JAP...105l3102J}. Reprinted with permission from J. Appl. Phys.
\textbf{105}, 123102 (2009). Copyright 2009 American Institute of Physics.}%
\label{fig:scattering2}%
\end{figure}

A transition of a carrier from one state to another due to a perturbation is
referred to as a scattering process. The perturbation is described by a
corresponding potential $V$, which can be static or time dependent
\cite{1997plds.book.....D}. This results in different classes of scattering
processes, illustrated in Fig.\,\ref{fig:scattering} for the case of
one-dimensional electron confinement as in QCL heterostructures. Elastic
scattering, shown in Fig.\,\ref{fig:scattering}(a), occurs for time constant
potentials. Here, the carrier energy is conserved. Relevant mechanisms in QCLs
include impurity, interface roughness, and alloy scattering. For potentials
with harmonic time dependence, $V\propto\cos\left(  \omega_{0}t\right)  $, the
carrier energy is changed by $\mp\hbar\omega_{0}$, corresponding to the case
of emission and absorption, respectively. This is referred to as inelastic
scattering [see Fig.\,\ref{fig:scattering}(b)]. An important inelastic
mechanism is optical phonon scattering, and also the interaction with photons
can be viewed as an inelastic scattering process. A special case are
intercarrier processes illustrated in Fig.\,\ref{fig:scattering}(c) such as
electron-electron scattering, where two electrons are involved in the
scattering event. Figure \ref{fig:scattering2} shows the influence of various
scattering mechanisms on the spectral gain obtained from an EMC simulation for
two different terahertz QCL designs \cite{2009JAP...105l3102J}.

QCLs are n-type devices which are typically based on direct bandgap
semiconductors. Thus, in the following we restrict our treatment of scattering
to the conduction band $\Gamma$ valley where also the lasing transitions
occur. However, we note that under certain conditions transitions to other
valleys can affect QCL\ operation
\cite{2006ApPhL..89s1119G,2007JAP...101f3101G,2007JAP...102k3107G,2008JAP...103g3101G}%
. Scattering causes an electron transition from an initial state $\left|
i\mathbf{k}\right\rangle $ to a final state $\left|  j\mathbf{k}^{\prime
}\right\rangle $ in the QCL\ heterostructure, where $\mathbf{k}=\left(
k_{x},k_{y}\right)  ^{\mathrm{T}}$ and $\mathbf{k}^{\prime}=\left(
k_{x}^{\prime},k_{y}^{\prime}\right)  ^{\mathrm{T}}$ are the corresponding
in-plane wave vectors. The states are described by their wave functions Eq.
(\ref{eq:psi3D}) and eigenenergies Eq.\thinspace(\ref{eq:Ekin}). For the
initial state, we have\ $\psi_{\mathrm{3D},i}=S^{-1/2}\psi_{i}\left(
z\right)  \exp\left(  \mathrm{i}\mathbf{kr}\right)  $ with $\mathbf{r}=\left(
x,y\right)  ^{\mathrm{T}}$ and $E_{i\mathbf{k}}=E_{i}+\hbar^{2}\mathbf{k}%
^{2}/\left(  2m_{i}^{\parallel}\right)  $ where $\psi_{i}\left(  z\right)  $
and $E_{i}$ are obtained from Eq.\thinspace(\ref{eq:s1D}) as described in
Section \ref{sec:SE}, and analogously for the final state.

For elastic scattering processes, $V$ is constant. The corresponding matrix
element is defined as%
\begin{align}
V_{j\mathbf{k}^{\prime},i\mathbf{k}} &  =\left\langle j\mathbf{k}^{\prime
}\right|  V\left|  i\mathbf{k}\right\rangle \nonumber\\
&  =S^{-1}\int_{S}\int_{-\infty}^{\infty}V\psi_{j}^{\ast}\psi_{i}\exp\left[
\mathrm{i}\left(  \mathbf{k}-\mathbf{k}^{\prime}\right)  \mathbf{r}\right]
\mathrm{d}^{2}r\mathrm{d}z.\label{eq:Vfi}%
\end{align}
For inelastic processes, we assume a harmonic potential of the form
$V=V_{0}\exp\left(  \mathrm{i}\mathbf{Qx}-\mathrm{i}\omega_{0}t\right)
+V_{0}^{\ast}\exp\left(  -\mathrm{i}\mathbf{Qx}+\mathrm{i}\omega_{0}t\right)
$, where $\mathbf{Q}$ is for example the phonon wave vector. The matrix
element can in analogy with Eq.\thinspace(\ref{eq:Vfi}) be obtained as%
\begin{align}
\left(
\begin{array}
[c]{c}%
V_{j\mathbf{k}^{\prime},i\mathbf{k}}^{+}\left(  Q\right)  \\
V_{j\mathbf{k}^{\prime},i\mathbf{k}}^{-}\left(  Q\right)
\end{array}
\right)   &  =\left\langle j\mathbf{k}^{\prime}\right|  \left(
\begin{array}
[c]{c}%
V_{0}\exp\left(  \mathrm{i}\mathbf{Qx}\right)  \\
V_{0}^{\ast}\exp\left(  -\mathrm{i}\mathbf{Qx}\right)
\end{array}
\right)  \left|  i\mathbf{k}\right\rangle \nonumber\\
&  =S^{-1}\int_{S}\int_{-\infty}^{\infty}\psi_{j}^{\ast}\psi_{i}\left(
\begin{array}
[c]{c}%
V_{0}\exp\left(  \mathrm{i}\mathbf{Qx}\right)  \\
V_{0}^{\ast}\exp\left(  -\mathrm{i}\mathbf{Qx}\right)
\end{array}
\right)  \nonumber\\
&  \times\exp\left[  \mathrm{i}\left(  \mathbf{k}-\mathbf{k}^{\prime}\right)
\mathbf{r}\right]  \mathrm{d}^{2}r\mathrm{d}z.\label{eq:Vfi2}%
\end{align}
Assuming bound wave functions $\psi_{i,j}\left(  z\right)  $\ in Eqs.
(\ref{eq:Vfi}) and (\ref{eq:Vfi2}), the integration over the $z$ coordinate
can be taken from $-\infty$ to $\infty$. This is also consistent with treating
the gain medium as an infinitely extended periodic heterostructure.
Furthermore, the in-plane cross section $S$ is assumed to be macrosopic, and
thus the integration can for scattering rate calculations be extended from
$-\infty$ to $\infty$ also in $x$ and $y$ direction.

The transition rate from an initial state $\left|  i\mathbf{k}\right\rangle $
to a final state $\left|  j\mathbf{k}^{\prime}\right\rangle $ is obtained from
Fermi's golden rule \cite{1997plds.book.....D}, given by
\begin{equation}
W_{i\mathbf{k},j\mathbf{k}^{\prime}}=\frac{2\pi}{\hbar}\left|  V_{j\mathbf{k}%
^{\prime},i\mathbf{k}}\right|  ^{2}\delta\left(  E_{j\mathbf{k}^{\prime}%
}-E_{i\mathbf{k}}\right)  \label{eq:golden}%
\end{equation}
for elastic scattering processes and%
\begin{equation}
W_{i\mathbf{k},j\mathbf{k}^{\prime}}^{\pm}\left(  Q\right)  =\frac{2\pi}%
{\hbar}\left|  V_{j\mathbf{k}^{\prime},i\mathbf{k}}^{\mp}\left(  Q\right)
\right|  ^{2}\delta\left(  E_{j\mathbf{k}^{\prime}}-E_{i\mathbf{k}}\pm
\hbar\omega_{0}\right)  \label{eq:golden2}%
\end{equation}
for inelastic processes. Here, $W_{i\mathbf{k},j\mathbf{k}^{\prime}}^{+}$
corresponds to the emission rate, caused by the component $\exp\left(
\mathrm{i}\omega_{0}t\right)  $, and $W_{i\mathbf{k},j\mathbf{k}^{\prime}}%
^{-}$ refers to absorption due to the component $\exp\left(  -\mathrm{i}%
\omega_{0}t\right)  $. The Dirac $\delta$ function ensures energy
conservation.$\ $For elastic processes, we obtain from $E_{j\mathbf{k}%
^{\prime}}=E_{i\mathbf{k}}$
\begin{equation}
\left|  \mathbf{k}^{\prime}\right|  =k^{\prime}=\left(  \frac{m_{j}%
^{\parallel}}{m_{i}^{\parallel}}k^{2}+2m_{j}^{\parallel}\frac{E_{i}-E_{j}%
}{\hbar^{2}}\right)  ^{1/2}. \label{eq:ks_el}%
\end{equation}
Analogously, energy conservation yields for inelastic scattering%
\begin{equation}
k^{\prime}=\left(  \frac{m_{j}^{\parallel}}{m_{i}^{\parallel}}k^{2}%
+2m_{j}^{\parallel}\frac{E_{i}-E_{j}\mp\hbar\omega_{0}}{\hbar^{2}}\right)
^{1/2}. \label{eq:ks_in}%
\end{equation}

The computation of the total transition rates from an initial state $\left|
i\mathbf{k}\right\rangle $ to a final subband $j$ involves the summation over
wave vectors. These sums can be converted to integrals, introducing a factor
of $L_{d}/\left(  2\pi\right)  $ per dimension where the device length $L_{d}$
in the corresponding direction is assumed to be large enough that quantization
effects can be neglected \cite{harrison}. An example is the summation over the
final in-plane wave vector $\mathbf{k}^{\prime}$ which is two-dimensional. It
can furthermore be advantagenous to express $\mathbf{k}^{\prime}\ $in polar
coordinates $\left|  \mathbf{k}^{\prime}\right|  \ $and $\phi$, and introduce
a kinetic energy variable $E_{j}^{\mathrm{kin}}=\hbar^{2}\left|
\mathbf{k}^{\prime}\right|  ^{2}/\left(  2m_{j}^{\parallel}\right)  $. Thus we
obtain
\begin{align}
\sum_{\mathbf{k}^{\prime}}\dots &  \rightarrow\frac{S}{\left(  2\pi\right)
^{2}}\iint_{-\infty}^{\infty}\dots\mathrm{d}^{2}k^{\prime}\nonumber\\
&  =\frac{Sm_{j}^{\parallel}}{\left(  2\pi\hbar\right)  ^{2}}\int_{0}^{\infty
}\int_{0}^{2\pi}\dots\mathrm{d}\phi\mathrm{d}E_{j}^{\mathrm{kin}}.
\label{eq:sum}%
\end{align}
Spin degeneracy is not considered here, since for single-electron scattering
processes the spin is conserved. For electron-electron scattering the spin
degeneracy must however be taken into account, as more closely discussed in
the corresponding section. Analogously, summation over a three-dimensional
wave vector, such as the phonon wave vector $\mathbf{Q}$, can in a crystal
lattice of volume $\Omega_{\mathrm{c}}$ be approximated by%
\begin{equation}
\sum_{\mathbf{Q}}\dots\rightarrow\frac{\Omega_{\mathrm{c}}}{\left(
2\pi\right)  ^{3}}\iiint_{-\infty}^{\infty}\dots\mathrm{d}^{3}Q.
\label{eq:sumQ}%
\end{equation}

\subsection{Phonon scattering}

\begin{figure}[ptb]
\includegraphics{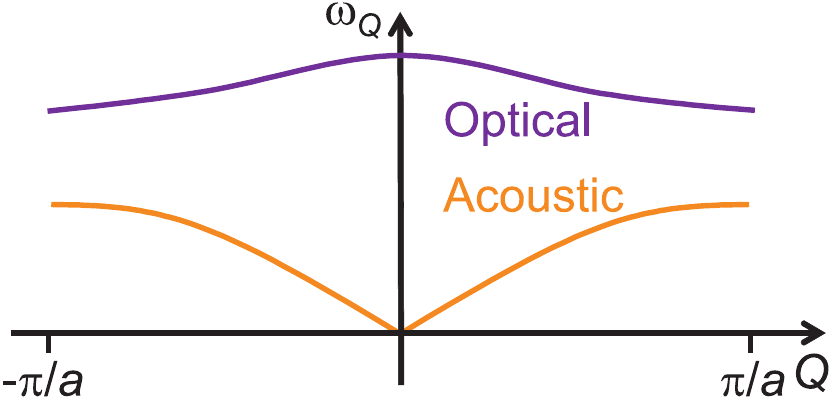}
\caption{{}(Color online) Schematic illustration of the dispersion relation
for an acoustic and optical phonon branch. This simplified example corresponds
to a one-dimensional lattice with $N_{\mathrm{u}}=2$ and lattice constant
$a$.}%
\label{fig:phonon}%
\end{figure}

A phonon is a quasiparticle associated with the lattice vibrations in a
crystal, representing an excited quantum mechanical state in the quantization
of the vibrational modes. Classically speaking, for an atom located at
position $\mathbf{x}$, lattice vibrations are described by a displacement
vector $\mathbf{u=U}\sin\left(  \mathbf{Qx}-\omega_{\mathbf{Q}}t\right)  $,
with the amplitude $\mathbf{U}$, wave vector $\mathbf{Q}$, and angular
frequency $\omega_{\mathbf{Q}}$. The normal modes are the solutions of
$\mathbf{u}$ for which the lattice uniformly oscillates at a single frequency
$\omega_{\mathbf{Q}}$, and a phonon corresponds to an elementary vibrational
motion. The associated relation between wave vector $\mathbf{Q}$ and frequency
$\omega_{\mathbf{Q}}$, the so-called dispersion relation, defines a phonon
branch, as illustrated in Fig.\,\ref{fig:phonon}. Acoustic modes are sound
waves where two consecutive atoms move in the same direction, and we have
$\omega_{\mathbf{Q}}=0$ for $\mathbf{Q=0}$. For optical modes, two consecutive
atoms in the same unit cell move in opposite direction, and $\omega
_{\mathbf{Q}}$ at $\mathbf{Q=0}$ typically corresponds to infrared optical
frequencies. Furthermore, the wave propagation can be predominantly
longitudinal ($\mathbf{Q}\parallel\mathbf{U}$) or transverse ($\mathbf{Q}%
\perp\mathbf{U}$); the corresponding branches are then referred to as
longitudinal or transverse branches. In three dimensions, there is a single
longitudinal acoustic (LA) and $2$ transverse acoustic (TA) branches as well
as $N_{\mathrm{u}}-1$ longitudinal optical (LO) and $2\left(  N_{\mathrm{u}%
}-1\right)  $ transverse optical (TO) branches, where $N_{\mathrm{u}}$ denotes
the number of atoms per unit cell. For example, GaAs has two atoms per unit
cell, i.e., $N_{\mathrm{u}}=2$.

The lattice vibrations lead to a perturbation of the carriers and thus carrier
scattering. Depending on the mechanisms, there are different types of phonon
scattering. Non-polar phonon scattering occurs in all crystals and is due to
acoustic and (for $N_{\mathrm{u}}\geq2$) TO\ phonons. Here the lattice
vibrations lead to a time dependent change of the conduction (and valence)
band energy. Polar scattering only occurs in polar semiconductors, e.g., III-V
semiconductors such as GaAs. It is due to LO phonons, where the out-of-phase
movement of the neighbouring atoms of different types causes a local dipole
moment, resulting in oscillating electric fields. Covalent semiconductors such
as group IV materials do not exhibit polar scattering.

The dominant phonon scattering mechanism in QCLs is due to LO phonons. Since
the optical phonon deformation potential approaches zero at the conduction
band $\Gamma$ point due to spherical symmetry, TO phonon scattering is
negligible for the $\Gamma$ valley; it can however play a significant role for
intervalley transitions and intravalley scattering in the $X$ and valence band
$\Gamma$ minimum \cite{1956PhRv..104.1281H,Hess,Chang}. Also acoustic phonons
tend to play a secondary role in QCLs \cite{2009PSSCR...6..579N}.

\subsubsection{\label{sec:nonpolar_scat}Non-polar phonon scattering}

Non-polar phonon scattering occurs in all crystals and is due to acoustic and
(for $N_{\mathrm{u}}\geq2$) TO\ phonons. The energy of the vibrating atom is
the sum of its potential and kinetic energy, which is equal to twice its
average kinetic energy. Thus, the total energy of the vibrational mode becomes
for $\mathbf{u=U}\sin\left(  \mathbf{Qx}-\omega_{\mathbf{Q}}t\right)  $
\begin{equation}
E=N_{\mathrm{a}}m\left\langle \left|  \partial_{t}\mathbf{u}\right|
^{2}\right\rangle _{t}\mathbf{=}\frac{1}{2}N_{\mathrm{a}}mU^{2}\omega
_{\mathbf{Q}}^{2},
\end{equation}
where the number of atoms with mass $m$ in a crystal lattice of volume
$\Omega_{\mathrm{c}}$ and density $\rho_{\mathrm{c}}$ is $N_{\mathrm{a}%
}=\Omega_{\mathrm{c}}\rho_{\mathrm{c}}/m$. Thus we obtain the amplitude for a
mode occupied by a single phonon of energy $E=\hbar\omega_{\mathbf{Q}}$%

\begin{equation}
U=\sqrt{\frac{2\hbar}{\Omega_{\mathrm{c}}\rho_{\mathrm{c}}\omega_{\mathbf{Q}}%
}}. \label{eq:U0}%
\end{equation}

\paragraph{Acoustic phonons}

Since phonon confinement does not play an important role for the cascade
structures (see discussion in Section~\ref{sec:LO}), scattering from bulk
phonons obeying Bose distribution is considered. Lawaetz has shown in
Ref.~\onlinecite{Lawaetz} that the deformation potential method of Bardeen and
Shockley~\cite{Bardeen_shockley} may be applied to the scattering of electrons
with acoustic phonons. This method is basically a Taylor expansion of the
scattering potential in the phonon momentum $\mathbf{Q}$. In the case of
vanishing screening, Kittel and Fong derive in Ref.~\onlinecite
{Kittel_theory} the change of the electronic energy in lowest order of the
lattice deformation. Since for acoustic phonons, adjacent atoms move in the
same direction, regions of compression or dilatation extend over several
lattice sites, and the crystal can be described as an elastic continuum. The
resulting strain gives rise to a time dependent change of the conduction band
energy $V=\delta V_{\mathrm{c}}$. The corresponding constant of
proportionality is the deformation potential $\Xi$ (for valleys without
rotational symmetry, $\Xi$ becomes a tensor), i.e., $V=\Xi\mathbf{\nabla u}$
for small displacements \cite{1997plds.book.....D}. We obtain%
\begin{equation}
V=\Xi\mathbf{QU}\cos\left(  \mathbf{Qx}-\omega_{\mathbf{Q}}t\right)  ,
\end{equation}
i.e., only LA phonons contribute since $\mathbf{QU}=0$ for transverse phonons.
For acoustic phonons, we have $\omega_{\mathbf{Q}}=0$ for $Q=\left|
\mathbf{Q}\right|  =0$, and the dispersion relation can be described by the
linear approximation $\omega_{\mathbf{Q}}=v_{\mathrm{s}}Q$ for small $Q$,
where $v_{\mathrm{s}}$ is the longitudinal sound velocity (see Fig.
\ref{fig:phonon}). With Eq.\thinspace(\ref{eq:U0}), the perturbation potential
is thus%
\begin{equation}
V=M(Q)\left\{  \exp\left[  \mathrm{i}\left(  \mathbf{Qx}-\omega_{\mathbf{Q}%
}t\right)  \right]  +\exp\left[  -\mathrm{i}\left(  \mathbf{Qx}-\omega
_{\mathbf{Q}}t\right)  \right]  \right\}  ,\label{eq:V_ac}%
\end{equation}
with the amplitude%
\begin{equation}
M(Q)=\Xi\sqrt{\frac{\hbar Q^{2}}{2\Omega_{\mathrm{c}}\rho_{\mathrm{c}}%
\omega_{\mathbf{Q}}}}=\Xi\sqrt{\frac{\hbar Q}{2\Omega_{\mathrm{c}}%
\rho_{\mathrm{c}}v_{\mathrm{s}}}}.\label{eq:M_ac}%
\end{equation}

Equation (\ref{eq:golden2}) gives the transition rate for an electron in an
initial state $\left|  i\mathbf{k}\right\rangle $, which is scattered to a
final state $j$ with a wave vector $\mathbf{k}^{\prime}$ by a phonon with wave
vector $\mathbf{Q}=\left[  \mathbf{q},q_{z}\right]  ^{\mathrm{T}}$, where
$\mathbf{q}$ corresponds to the in-plane component. Using the definition of
the three-dimensional wave functions in Eq.\,(\ref{eq:psi3D}), the
corresponding matrix element of the perturbation potential Eq.\,(\ref{eq:V_ac}%
) is with Eq.\,(\ref{eq:Vfi2}) given by%
\begin{equation}
V_{j\mathbf{k}^{\prime},i\mathbf{k}}^{\pm}=\frac{M(Q)}{S}\int_{S}\exp\left[
\mathrm{i}\left(  \mathbf{k}\pm\mathbf{q}-\mathbf{k}^{\prime}\right)
\mathbf{r}\right]  \mathrm{d}^{2}rF_{ji}^{\pm}\left(  q_{z}\right)
\label{matrix}%
\end{equation}
with the form factor
\begin{equation}
F_{ji}^{\pm}\left(  q_{z}\right)  =\int_{-\infty}^{\infty}\psi_{j}^{\ast
}\left(  z\right)  \exp\left(  \pm\mathrm{i}q_{z}z\right)  \psi_{i}\left(
z\right)  \mathrm{d}z. \label{eq:ff_ac}%
\end{equation}
For bound states, $\psi_{i,j}$ can be chosen to be real, and we have $\left|
F_{ji}^{+}\left(  q_{z}\right)  \right|  ^{2}=\left|  F_{ji}^{-}\left(
q_{z}\right)  \right|  ^{2}=\left|  F_{ji}\left(  q_{z}\right)  \right|  ^{2}%
$. The scattering rate obtained from Eq.\,(\ref{eq:golden2}) is thus%
\begin{align}
W_{i\mathbf{k},j\mathbf{k}^{\prime}}^{\pm}\left(  \mathbf{Q}\right)   &
=\frac{\left(  2\pi\right)  ^{3}}{\hbar S}\left(  N_{\mathbf{Q}}+\frac{1}%
{2}\pm\frac{1}{2}\right)  \left|  M(Q)\right|  ^{2}\left|  F_{ji}\left(
q_{z}\right)  \right|  ^{2}\nonumber\\
&  \times\delta\left(  \mathbf{k\mp q-k}^{\prime}\right)  \delta\left(
E_{j\mathbf{k}^{\prime}}-E_{i\mathbf{k}}\pm\hbar\omega_{\mathbf{Q}}\right)  ,
\label{eq:W_acs}%
\end{align}
where $W_{i\mathbf{k},j\mathbf{k}^{\prime}}^{+}$ and $W_{i\mathbf{k}%
,j\mathbf{k}^{\prime}}^{-}$ refer to emission and absorption, respectively.
Here we have used that $\left|  \int_{S}\exp\left[  \mathrm{i}\left(
\mathbf{k}\pm\mathbf{q}-\mathbf{k}^{\prime}\right)  \mathbf{r}\right]
\mathrm{d}^{2}r\right|  ^{2}$ can be approximated by $4\pi^{2}S\delta\left(
\mathbf{k\mp q-k}^{\prime}\right)  $ for sufficiently large in-plane cross
sections $S$. Since $M(Q)$ refers to a single phonon, the phonon occupation
number of a mode in thermal equilibrium has been added in Eq.\,(\ref{eq:W_acs}%
), given by the Bose-Einstein distribution%
\begin{equation}
N_{\mathbf{Q}}=\left[  \exp\left(  \frac{\hbar\omega_{\mathbf{Q}}%
}{k_{\mathrm{B}}T_{\mathrm{L}}}\right)  -1\right]  ^{-1}. \label{eq:NQ}%
\end{equation}
For emission the factor $N_{\mathbf{Q}}+1$ is used to also include spontaneous
emission processes.

The total transition rate from a given initial state $\left|  i\mathbf{k}%
\right\rangle $ to a subband $j$ is obtained by summing over all wave vectors
$\mathbf{k}^{\prime}$ and $\mathbf{Q}$. These sums can be converted to
integrals using Eqs.\,(\ref{eq:sum}) and (\ref{eq:sumQ}). With
Eqs.\,(\ref{eq:W_acs}) and (\ref{eq:M_ac}), we thus obtain the total
transition rate from a given initial state $\left|  i\mathbf{k}\right\rangle $
to a subband $j$ \cite{harrison}
\begin{align}
W_{i\mathbf{k},j}^{\pm}  &  =\frac{\Xi^{2}}{8\pi^{2}\rho_{\mathrm{c}%
}v_{\mathrm{s}}}\int Q\left(  N_{\mathbf{Q}}+\frac{1}{2}\pm\frac{1}{2}\right)
\left|  F_{ji}\left(  q_{z}\right)  \right|  ^{2}\nonumber\\
&  \times\delta\left[  E_{j,\mathbf{k\mp q}}-E_{i\mathbf{k}}\pm\hbar
\omega_{\mathbf{Q}}\right]  \mathrm{d}^{3}Q\mathbf{.} \label{eq:W_acs2}%
\end{align}

Since acoustic phonons do not carry much energy, often the quasi-elastic
approximation is applied, treating acoustic phonon scattering as elastic
process. This is achieved by neglecting the phonon energy term $\pm\hbar
\omega_{\mathbf{Q}}$ in Eqs.\thinspace(\ref{eq:W_acs}) and (\ref{eq:W_acs2}).
Besides, we can approximate Eq.\thinspace(\ref{eq:NQ}) as
\begin{equation}
N_{\mathbf{Q}}\approx N_{\mathbf{Q}}+1\approx\frac{k_{\mathrm{B}}%
T_{\mathrm{L}}}{\hbar\omega_{\mathbf{Q}}}\approx\frac{k_{\mathrm{B}%
}T_{\mathrm{L}}}{\hbar v_{\mathrm{s}}Q}\label{eq:equipart}%
\end{equation}
for all but the lowest temperatures, which is referred to as equipartition
approximation. Thus we obtain%
\begin{align}
W_{i\mathbf{k},j\mathbf{k}^{\prime}}^{\pm}\left(  \mathbf{Q}\right)   &
=\Xi^{2}\frac{4\pi^{3}k_{\mathrm{B}}T_{\mathrm{L}}}{\Omega_{\mathrm{c}}%
\rho_{\mathrm{c}}v_{\mathrm{s}}^{2}S\hbar}\left|  F_{ji}\left(  q_{z}\right)
\right|  ^{2}\nonumber\\
&  \times\delta\left(  \mathbf{k\mp q-k}^{\prime}\right)  \delta\left(
E_{j\mathbf{k}^{\prime}}-E_{i\mathbf{k}}\right)  .\label{eq:W_ac}%
\end{align}
With Eqs.\thinspace(\ref{eq:sum}) and (\ref{eq:sumQ}), we furthermore obtain
$W_{i\mathbf{k},j}^{\pm}\left(  \mathbf{Q}\right)  =0$ for $E_{j}%
>E_{i\mathbf{k}}$, and otherwise%
\begin{equation}
W_{i\mathbf{k},j}^{\pm}=\Xi^{2}\frac{k_{\mathrm{B}}T_{\mathrm{L}}%
m_{j}^{\parallel}}{4\pi\rho_{\mathrm{c}}v_{\mathrm{s}}^{2}\hbar^{3}}%
\int_{-\infty}^{\infty}\left|  F_{ji}\left(  q_{z}\right)  \right|
^{2}\mathrm{d}q_{z}.
\end{equation}
The final wave vector magnitude $k^{\prime}$ is then given by Eq.
(\ref{eq:ks_el}). With Eq.\thinspace(\ref{eq:ff_ac}), we obtain
\begin{equation}
\int_{-\infty}^{\infty}\left|  F_{ji}\left(  q_{z}\right)  \right|
^{2}\mathrm{d}q_{z}=2\pi\int_{-\infty}^{\infty}\left|  \psi_{j}\left(
z\right)  \right|  ^{2}\left|  \psi_{i}\left(  z\right)  \right|
^{2}\mathrm{d}z.\label{eq:Ffi}%
\end{equation}
Thus the total transition rate is given by%
\begin{equation}
W_{i\mathbf{k},j}=\Xi^{2}\frac{k_{\mathrm{B}}T_{\mathrm{L}}m_{j}^{\parallel}%
}{\rho_{\mathrm{c}}v_{\mathrm{s}}^{2}\hbar^{3}}\int_{-\infty}^{\infty}\left|
\psi_{j}\left(  z\right)  \right|  ^{2}\left|  \psi_{i}\left(  z\right)
\right|  ^{2}\mathrm{d}z,\label{eq:W_ac2}%
\end{equation}
which includes both emission and absorption and thus contains an additional
factor of $2$.

\paragraph{Transverse optical phonons}

Transverse optical phonons are treated in a similar manner as acoustic
phonons. Here, adjacent atoms move in opposite directions. Thus, the crystal
vibrations cannot be described as an elastic continuum as for acoustic
phonons, but correspond rather to oscillations of the two sublattices with
respect to each other. As a consequence, the potential change is directly a
function of the displacement vector $\mathbf{u}$, $\delta V_{\mathrm{c}%
}=\mathbf{D}_{\mathrm{TO}}\mathbf{u=}D_{\mathrm{TO}}u$. Here, $\mathbf{D}%
_{\mathrm{TO}}$ is the optical deformation potential field which assumes a
different value for different kinds of intra- and intervalley transitions, and
$D_{\mathrm{TO}}$ is the component parallel to $\mathbf{u}$. For optical
phonons, $\omega_{\mathbf{Q}}$ reaches an extremum $\omega_{\mathbf{Q}}%
=\omega_{\mathrm{TO}}$ for $Q=0$, and for small $Q$, the dispersion relation
can thus be approximated by $\omega_{\mathbf{Q}}=\omega_{\mathrm{TO}}$ (see
Fig.\thinspace\ref{fig:phonon}). The scattering rate for an electron from an
initial state $\left|  i\mathbf{k}\right\rangle $ to a final state $\left|
j\mathbf{k}^{\prime}\right\rangle $ due to phonons with wave vector
$\mathbf{Q}$ is again given by Eq.\thinspace(\ref{eq:W_acs}), with
\begin{equation}
M(Q)=D_{\mathrm{TO}}\sqrt{\frac{\hbar}{2\Omega_{\mathrm{c}}\rho_{\mathrm{c}%
}\omega_{\mathrm{TO}}}}.
\end{equation}
In analogy to Eq.\thinspace(\ref{eq:W_acs2}), we obtain the total transition
rate%
\begin{align}
W_{i\mathbf{k},j}^{\pm} &  =\frac{D_{\mathrm{TO}}^{2}}{8\pi^{2}\rho
_{\mathrm{c}}\omega_{\mathrm{TO}}}\int\left(  N_{\mathbf{Q}}+\frac{1}{2}%
\pm\frac{1}{2}\right)  \left|  F_{ji}\left(  q_{z}\right)  \right|
^{2}\nonumber\\
&  \times\delta\left[  E_{j,\mathbf{k\mp q}}-E_{i\mathbf{k}}\pm\hbar
\omega_{\mathrm{TO}}\right]  \mathrm{d}^{3}Q.\label{eq:W_TOs2}%
\end{align}
Approximating $\omega_{\mathbf{Q}}=\omega_{\mathrm{TO}}$, the phonon number
$N_{\mathbf{Q}}$ in Eq.\thinspace(\ref{eq:W_TOs2}) does\ not depend on
$\mathbf{Q}$, and thus we can set $N_{\mathbf{Q}}=N_{\mathrm{Ph}}$. With
Eq.\thinspace(\ref{eq:Ffi}), the result then simplifies to
\begin{align}
W_{i\mathbf{k},j}^{\pm} &  =\frac{D_{\mathrm{TO}}^{2}m_{j}^{\parallel}}%
{2\rho_{\mathrm{c}}\omega_{\mathrm{TO}}\hbar^{2}}\left(  N_{\mathrm{Ph}}%
+\frac{1}{2}\pm\frac{1}{2}\right)  \nonumber\\
&  \times\int_{-\infty}^{\infty}\left|  \psi_{j}\left(  z\right)  \right|
^{2}\left|  \psi_{i}\left(  z\right)  \right|  ^{2}\mathrm{d}z\label{W_TO2}%
\end{align}
for $E_{j}\leq E_{i\mathbf{k}}\mp\hbar\omega_{\mathrm{TO}}$, and becomes $0$ otherwise.

\subsubsection{\label{sec:LO}Longitudinal optical phonons}

As for TO\ phonons, adjacent atoms move in opposite directions for
LO\ phonons. Again the dispersion relation can be approximated with a constant
value $\omega_{\mathbf{Q}}=\omega_{\mathrm{LO}}$ for small $Q$ (see Fig.
\ref{fig:phonon}), where $\hbar\omega_{\mathrm{LO}}$ is the LO\ phonon energy
at $Q=0$. Polar semiconductors contain different types of atoms carrying
effective charges. For two different atoms with masses $m_{\mathrm{A}}$ and
$m_{\mathrm{B}}$ in a unit cell, these effective charges are given by $\pm
q_{\mathrm{eff}}$ with $q_{\mathrm{eff}}^{2}=\epsilon_{0}\omega_{\mathrm{LO}%
}^{2}\mu n_{\mathrm{c}}^{-1}\left(  \epsilon_{\mathrm{r},\infty}^{-1}%
-\epsilon_{\mathrm{r},0}^{-1}\right)  $\ \cite{1997plds.book.....D}. Here,
$\mu^{-1}=m_{\mathrm{A}}^{-1}+m_{\mathrm{B}}^{-1}$ is the inverse reduced
mass, and $n_{\mathrm{c}}$ is the number of cells per unit volume.
Furthermore, $\epsilon_{\mathrm{r},0}$ denotes the static dielectric constant,
and $\epsilon_{\mathrm{r},\infty}$ is the dielectric constant at very high
frequencies where the ions cannot respond to a field anymore and thus only the
electronic polarization remains. For LO waves, the out-of-phase movement of
the atoms causes a local dipole moment $\mathbf{p}\left(  t\right)
=q_{\mathrm{eff}}\mathbf{u}\left(  t\right)  $, where $\mathbf{u=U}\sin\left(
\mathbf{Qx}-\omega_{\mathbf{Q}}t\right)  $ now indicates the relative
displacement of the atoms in the unit cell; by contrast, transverse waves do
not generate dipoles. The amplitude for excitation by a single phonon is given
by Eq.\thinspace(\ref{eq:U0}) where $\rho_{\mathrm{c}}$ is replaced by $\mu
n_{\mathrm{c}}$. The resulting ionic polarization is $\mathbf{P}%
_{\mathrm{ion}}=\mathbf{p}n_{\mathrm{c}}$, and the electric field $\mathbf{E}$
can be obtained from the material equation $\mathbf{D}=\epsilon_{0}%
\mathbf{E}+\mathbf{P}_{\mathrm{ion}}$ with $\mathbf{D=0}$ due to the absence
of free carriers. With the potential $V=e\int\mathbf{E}\mathrm{d}\mathbf{x}$,
the resulting scattering rate from a state $\left|  i\mathbf{k}\right\rangle $
to $\left|  j\mathbf{k}^{\prime}\right\rangle $ is again given by Eq.
(\ref{eq:W_acs}), where%
\begin{equation}
M(Q)=\frac{1}{2Q}e\sqrt{\epsilon_{0}^{-1}\left(  \epsilon_{\mathrm{r},\infty
}^{-1}-\epsilon_{\mathrm{r},0}^{-1}\right)  }\sqrt{\frac{2\hbar\omega
_{\mathrm{LO}}}{\Omega_{\mathrm{c}}}}.\label{eq:MPO}%
\end{equation}

The total transition rate from a given initial state $\left|  i\mathbf{k}%
\right\rangle $ to a subband $j$ is again obtained by summing over all wave
vectors $\mathbf{k}^{\prime}$ and $\mathbf{Q}$, using Eqs.\thinspace
(\ref{eq:sum}) and (\ref{eq:sumQ}). For $E_{j}\leq E_{i\mathbf{k}}\mp
\hbar\omega_{\mathrm{LO}}$, we obtain%
\begin{align}
W_{i\mathbf{k},j}^{\pm}  &  =\frac{m_{j}^{\parallel}\omega_{\mathrm{LO}}e^{2}%
}{8\pi^{2}\hbar^{2}\epsilon_{0}}\left(  \epsilon_{\mathrm{r},\infty}%
^{-1}-\epsilon_{\mathrm{r},0}^{-1}\right)  \left(  N_{\mathrm{Ph}}+\frac{1}%
{2}\pm\frac{1}{2}\right) \nonumber\\
&  \times\int_{0}^{2\pi}J\left(  q\right)  \mathrm{d}\theta, \label{eq:W_LO2}%
\end{align}
and $W_{i\mathbf{k},j}^{\pm}=0$ otherwise. Here,%
\begin{align}
J\left(  q\right)   &  =\frac{\pi}{q}\iint_{-\infty}^{\infty}\psi_{j}^{\ast
}\left(  z\right)  \psi_{j}\left(  z^{\prime}\right)  \psi_{i}\left(
z\right)  \psi_{i}^{\ast}\left(  z^{\prime}\right) \nonumber\\
&  \times\exp\left(  -q\left|  z-z^{\prime}\right|  \right)  \mathrm{d}%
z\mathrm{d}z^{\prime} \label{eq:J}%
\end{align}
with the in-plane phonon wave vector magnitude%
\begin{equation}
q=\left(  k^{2}+k^{\prime2}-2kk^{\prime}\cos\theta\right)  ^{1/2},
\end{equation}
where $k^{\prime}$ is given by Eq.\thinspace(\ref{eq:ks_in}) with $\omega
_{0}=\omega_{\mathrm{LO}}$.

\paragraph{Corrections due to screening, hot phonons, phonon modes, and
lattice heating}

In a heterostructure, the assumption of bulk phonon modes is justified for
acoustic phonons, but not necessarily for LO\ phonons. The LO phonon modes
arising from different dielectric constants in the well and barrier material
can be described by microscopic and macroscopic approaches. The macroscopic
dielectric continuum model, which is widely used in this context, yields slab
modes which are confined in each layer, and interface modes corresponding to
exponential solutions around the interfaces \cite{2001JAP....90.5504W}.
However, it has been shown that the bulk mode approximation is valid for most
QCL\ designs \cite{2003ApPhL..83..207C,2008JAP...103g3101G}. Specifically, for
not too different compositions of the barrier and well material and thus
similar values of \ $\epsilon_{\mathrm{r},0}$, the use of bulk modes is a good
approximation \cite{2003ApPhL..83..207C}.

Screening can be included by changing Eq.\thinspace(\ref{eq:MPO}) to
\cite{1988PhRvB..37.2578G,2000JaJAP..39.6601P}%
\begin{equation}
M(Q)=\frac{Q}{2\left(  Q^{2}+q_{\mathrm{s}}^{2}\right)  }e\sqrt{\epsilon
_{0}^{-1}\left(  \epsilon_{\mathrm{r},\infty}^{-1}-\epsilon_{\mathrm{r}%
,0}^{-1}\right)  }\sqrt{\frac{2\hbar\omega_{\mathrm{LO}}}{\Omega_{\mathrm{c}}%
}}, \label{eq:screened_LO_M}%
\end{equation}
where $q_{\mathrm{s}}$ is the inverse screening length. For Eq.\thinspace
(\ref{eq:J}), we then obtain with $\tilde{q}=\sqrt{q^{2}+q_{\mathrm{s}}^{2}}$%
\begin{align}
J\left(  q\right)   &  =\frac{\pi}{\tilde{q}}\iint\limits_{-\infty}^{\infty
}\psi_{j}^{\ast}\left(  z\right)  \psi_{j}\left(  z^{\prime}\right)  \psi
_{i}\left(  z\right)  \psi_{i}^{\ast}\left(  z^{\prime}\right)  \exp\left(
-\tilde{q}\left|  z-z^{\prime}\right|  \right) \nonumber\\
&  \times\left(  1-\frac{\left|  z-z^{\prime}\right|  q_{\mathrm{s}}^{2}%
}{2\tilde{q}}-\frac{q_{\mathrm{s}}^{2}}{2\tilde{q}^{2}}\right)  \mathrm{d}%
z\mathrm{d}z^{\prime}. \label{eq:screened_LO_J}%
\end{align}
The simplest model is to use Debye screening with the bulk inverse Debye
screening length%
\begin{equation}
q_{\mathrm{s}}=\sqrt{\frac{n_{\mathrm{e}}e^{2}}{\epsilon_{0}\epsilon
_{\mathrm{r},0}k_{\mathrm{B}}T_{\mathrm{L}}}},
\end{equation}
where $n_{\mathrm{e}}$ is the averaged three-dimensional electron density.
However, Debye screening is generally only valid for high temperatures
\cite{2008ApPhL..92h1102N}, since $q_{\mathrm{s}}$ diverges for $T_{\mathrm{L}%
}\rightarrow0$.

Deviations from the equilibrium phonon distribution Eq.\thinspace(\ref{eq:NQ})
due to phonon emission and absorption in the heterostructure, also referred to
as hot phonons, are frequently considered in simulations
\cite{2002PhyB..314..336C,2006ApPhL..88f1119L}. The hot phonon effect is
commonly incorporated by using a relaxation time approach, describing the
decay of LO phonons into other phonon modes by a corresponding lifetime
\cite{1987ApPhL..50.1251L}.

Typically, in simulations it is assumed that the lattice temperature of the
gain medium is identical to a given heat sink temperature or to the room
temperature. However, this approximation is only legitimate for pulsed
operation at low duty cycle, and generally fails for continuous wave operation
\cite{2001ApPhL..78.2095S}. Lattice heating due to the dissipated electrical
power can be self-consistently modeled by coupled carrier transport and
thermal simulations \cite{2006IJQE...42..857E,ISI:000302238400015}, yielding
the actual temperature profile. The temperature distribution is computed by
solving the heat diffusion equation, where the heat generation rate is
obtained from the carrier transport simulation based on the dissipated
electrical power or the LO phonon scattering rate.

\subsection{\label{sec:ee}Electron-electron scattering}

Electron-electron scattering arises from the collision of electrons with other
electrons. More precisely, this scattering mechanism can be divided into two
main contributions, the short-range interaction between two electrons via the
screened Coulomb potential and the long-range electron-plasmon coupling
\cite{MC}. The latter is usually neglected in QCL simulations and is also not
considered here, since it only becomes relevant at elevated doping levels.

Many-electron effects are to lowest order already considered in the Poisson
equation, Eq.\thinspace(\ref{eq:poisson}), corresponding to a mean field
treatment of charges. In the semiclassical treatment, an inclusion of higher
order effects is obtained by corresponding scattering rates. Electron-electron
scattering is much more computationally demanding than the single-electron
processes, hampering its inclusion in quantum mechanical simulations of QCLs
beyond the mean-field approximation \cite{2009PhRvB..79s5323K}.

In semiclassical simulations, electron-electron scattering is typically
implemented as a two-electron process
\cite{1988PhRvB..37.2578G,1995PhRvB..5116860M}. An electron in an initial
state $\left|  i\mathbf{k}\right\rangle $ scatters to a final state $\left|
j\mathbf{k}^{\prime}\right\rangle $, accompanied by a transition of a second
electron from a state $\left|  i_{0}\mathbf{k}_{0}\right\rangle $ to $\left|
j_{0}\mathbf{k}_{0}^{\prime}\right\rangle $. Nonparabolicity effects have for
example been considered by Bonno et al. \cite{2008JAP...104e3719B} To
facilitate the treatment, we assume here an identical mass $m^{\ast}$ for all
subbands. The total scattering rate from $\left|  i\mathbf{k}\right\rangle $
to a subband $j$ is then obtained as
\cite{1988PhRvB..37.2578G,1995PhRvB..5116860M}%
\begin{equation}
W_{i\mathbf{k},j}=\frac{m^{\ast}}{4\pi\hbar^{3}S}\sum_{i_{0},j_{0}%
,\mathbf{k}_{0}}f_{i_{0}}\left(  \mathbf{k}_{0}\right)  \int_{0}^{2\pi
}\mathrm{d}\theta\left|  M_{ii_{0}jj_{0}}\left(  q\right)  \right|  ^{2},
\label{eq:R_ee}%
\end{equation}
with the in-plane cross section area $S$ and carrier distribution function
$f_{i_{0}}\left(  \mathbf{k}_{0}\right)  $ for the state\ $\left|
i_{0}\mathbf{k}_{0}\right\rangle $. $\theta$ is the angle between
$\mathbf{g}=\mathbf{k}_{0}-\mathbf{k}$ and $\mathbf{g}^{\prime}=\mathbf{k}%
_{0}^{\prime}-\mathbf{k}^{\prime}$, and $\mathbf{q=k-k}^{\prime}$ (with
$q=\left|  \mathbf{q}\right|  $) denotes the exchanged wave vector. For the
definition of the Coulomb matrix element, different conventions exist
\cite{1988PhRvB..37.2578G,2009JAP...106f3115N}. Here we use the widely
employed convention of Goodnick and Lugli
\cite{1988PhRvB..37.2578G,1995PhRvB..5116860M,2005JAP....97d3702B,2006ApPhL..89u1115L}%
, where the bare Coulomb matrix element is with $\epsilon=\epsilon_{0}%
\epsilon_{\mathrm{r},0}$ given by%
\begin{align}
V_{ii_{0}jj_{0}}^{\mathrm{b}}\left(  q\right)   &  =\frac{e^{2}}{2\epsilon
q}\int_{-\infty}^{\infty}\mathrm{d}z\int_{-\infty}^{\infty}\mathrm{d}%
z^{\prime}\left[  \psi_{i}\left(  z\right)  \psi_{i_{0}}\left(  z^{\prime
}\right)  \right. \nonumber\\
&  \times\left.  \psi_{j}^{\ast}\left(  z\right)  \psi_{j_{0}}^{\ast}\left(
z^{\prime}\right)  \exp\left(  -q\left|  z-z^{\prime}\right|  \right)
\right]  . \label{eq:Vbare}%
\end{align}
Electron-electron scattering is sensitive with respect to the spin.
Furthermore, screening has to be considered, caused by the response of the
other electrons to changes in the electrostatic field associated with the
scattering process. Often, only scattering of electron pairs with antiparallel
spin is considered, since the contribution of carriers with parallel spin is
smaller \cite{1988PhRvB..37.2578G,1994SeScT...9..478M}. Furthermore, basic
models consider screening by a constant screening wave number $q_{\mathrm{s}}%
$. For the transition matrix element $M_{ii_{0}jj_{0}}$, we then obtain
\begin{equation}
\left|  M_{ii_{0}jj_{0}}\left(  q\right)  \right|  ^{2}=\frac{1}{2}\left|
\frac{q}{q+q_{\mathrm{s}}}V_{ii_{0}jj_{0}}^{\mathrm{b}}\left(  q\right)
\right|  ^{2}, \label{eq:M2}%
\end{equation}
where the factor $q/\left(  q+q_{\mathrm{s}}\right)  $ corrects the bare
Coulomb matrix element for screening effects, and the factor $1/2$ arises from
neglecting parallel spin contributions. From energy conservation, we obtain
for the exchanged wave vector with $g=\left|  \mathbf{g}\right|  $ and
$g_{0}^{2}=4m^{\ast}\left(  E_{i}+E_{i_{0}}-E_{j}-E_{j_{0}}\right)  /\hbar
^{2}$
\begin{equation}
q=\frac{1}{2}\left[  2g^{2}+g_{0}^{2}-2g\left(  g^{2}+g_{0}^{2}\right)
^{1/2}\cos\theta\right]  ^{1/2}.
\end{equation}
A frequently used basic screening model is the single subband screening
approximation, assuming that most electrons in each period reside in a single
subband $i$, corresponding to the ground state. The obtained screening wave
number is then \cite{2005JAP....97d3702B,2006ApPhL..89u1115L}%
\begin{equation}
q_{\mathrm{s}}=\frac{e^{2}}{2\epsilon}\frac{m^{\ast}}{\pi\hbar^{2}}%
f_{i}\left(  \mathbf{k}=\mathbf{0}\right)  . \label{eq:qsss}%
\end{equation}

\subsubsection{Advanced spin and screening models}

Advanced implementations of spin and screening effects have been developed for
a more accurate description of electron-electron scattering
\cite{2010JAP...107a3104J}. Screening models with varying degrees of
sophistication are employed to obtain the screened Coulomb matrix elements
$V_{ii_{0}jj_{0}}^{\mathrm{s}}$ from $V_{ii_{0}jj_{0}}^{\mathrm{b}}$ in Eq.
(\ref{eq:Vbare}). In the random phase approximation (RPA), $V_{ii_{0}jj_{0}%
}^{\mathrm{s}}\left(  q\right)  $ is found by solving the equation system
\cite{1999PhRvB..5915796L}%
\begin{equation}
V_{ii_{0}jj_{0}}^{\mathrm{s}}=V_{ii_{0}jj_{0}}^{\mathrm{b}}+\sum_{mn}%
V_{imjn}^{\mathrm{b}}\Pi_{mn}V_{mi_{0}nj_{0}}^{\mathrm{s}}. \label{eq:Vscr}%
\end{equation}
Here, $\Pi_{mn}\left(  q\right)  $ is the polarizability tensor, given in the
long wavelength limit ($q\rightarrow0$) by%

\begin{equation}
\Pi_{mn}=\left\{
\begin{array}
[c]{cc}%
\frac{n_{m}^{\mathrm{s}}-n_{n}^{\mathrm{s}}}{E_{m}-E_{n}}, & m\neq n,\\
-\frac{m^{\ast}}{\pi\hbar^{2}}f_{n}\left(  0\right)  , & m=n.
\end{array}
\right.  \label{eq:Pi}%
\end{equation}
Commonly, simplified screening models using a constant screening wave number
$q_{\mathrm{s}}$ are employed to avoid the numerical load associated with
solving Eq.\,(\ref{eq:Vscr}) \cite{2005JAP....97d3702B,2006ApPhL..89u1115L}.
$V_{ii_{0}jj_{0}}^{\mathrm{s}}$ is then obtained by replacing the prefactor
$e^{2}/\left(  2\epsilon q\right)  $ in Eq.\,(\ref{eq:Vbare}) by
$e^{2}/\left[  2\epsilon\left(  q+q_{\mathrm{s}}\right)  \right]  $. This
implementation is also used in Eq.\,(\ref{eq:M2}). The single subband
screening model, Eq. (\ref{eq:qsss}), can be obtained from Eq.\,(\ref{eq:Vscr}%
) by assuming that most of the electrons reside in a single subband, and
screening is caused only by this subband
\cite{2005JAP....97d3702B,2006ApPhL..89u1115L}. A somewhat improved approach
which considers all subbands equally is the modified single subband model
\cite{2006ApPhL..89u1115L} with%
\begin{equation}
q_{\mathrm{s}}=\frac{e^{2}}{2\epsilon}\frac{m^{\ast}}{\pi\hbar^{2}}\sum
_{i}f_{i}\left(  \mathbf{k}=\mathbf{0}\right)  , \label{eq:qsmss}%
\end{equation}
where $i$ sums over the subbands in one period. An alternative approach
consists in treating intersubband scattering as unscreened, thus applying Eq.
(\ref{eq:qsmss}) only to the intrasubband matrix elements
\cite{2007JAP...101f3101G}.

For collisions of electrons with parallel spin,\ interference\ occurs between
$V_{ii_{0}jj_{0}}^{\mathrm{s}}\ $and the 'exchange' matrix element
$V_{ii_{0}j_{0}j}^{\mathrm{s}}$ \cite{1994SeScT...9..478M}. Accounting for
this exchange effect, we obtain \cite{1995PhRvB..5116860M,1994SeScT...9..478M}%
\begin{align}
\left|  M_{ii_{0}jj_{0}}\right|  ^{2} &  =\frac{p_{\mathrm{a}}}{2}\left[
\left|  V_{ii_{0}jj_{0}}^{\mathrm{s}}\left(  q^{-}\right)  \right|
^{2}+\left|  V_{ii_{0}j_{0}j}^{\mathrm{s}}\left(  q^{+}\right)  \right|
^{2}\right]  \nonumber\\
&  +\frac{p_{\mathrm{p}}}{2}\left|  V_{ii_{0}jj_{0}}^{\mathrm{s}}\left(
q^{-}\right)  -V_{ii_{0}j_{0}j}^{\mathrm{s}}\left(  q^{+}\right)  \right|
^{2},\label{eq:M2ex}%
\end{align}
where%
\begin{equation}
q^{\pm}=\frac{1}{2}\left[  2g^{2}+g_{0}^{2}\pm2g\left(  g^{2}+g_{0}%
^{2}\right)  ^{1/2}\cos\theta\right]  ^{1/2},
\end{equation}
and $p_{\mathrm{a}}=p_{\mathrm{p}}=1/2$ are the probabilities for antiparallel
and parallel spin collisions, respectively. There are two common approaches to
implement electron-electron scattering without explicitly considering the spin
dependence. One method which has been used in Eq.\thinspace(\ref{eq:M2}) is to
completely neglect the parallel spin collisions
\cite{1988PhRvB..37.2578G,1994SeScT...9..478M}, implying $p_{\mathrm{a}}=1/2$,
$p_{\mathrm{p}}=0$ in Eq.\thinspace(\ref{eq:M2ex}). This approach tends to
overestimate the exchange effect. Another common method is to ignore the
exchange effect, i.e., parallel spin collisions are treated the same way as
antiparallel spin contributions \cite{1994SeScT...9..478M}. This corresponds
to $p_{\mathrm{a}}=1$, $p_{\mathrm{p}}=0$ in Eq.\thinspace(\ref{eq:M2ex}).

\begin{figure}[ptb]
\includegraphics{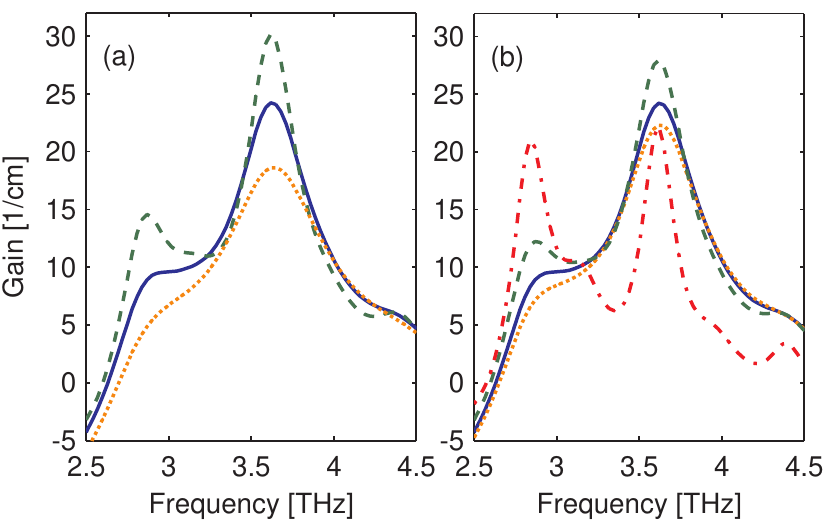}
\caption{(Color online) EMC simulation results for the spectral gain vs
frequency of a bound-to-continuum terahertz QCL \cite{2003ApPhL..82.3165S} at
a lattice temperature $T_{\mathrm{L}}=10\,\mathrm{K}$
\cite{2010JAP...107a3104J}. (a) Comparison of different screening models: RPA
(solid curve), modified single subband model for all matrix elements (dashed
curve) or for the intrasubband elements only (dotted curve), thus treating
intersubband elements as unscreened. (b) Comparison of different spin
implementations. Shown are results for fully taking into account (solid curve)
and ignoring the exchange effect (dotted curve), and ignoring parallel spin
collisions (dashed curve). For comparison, the result obtained with no
electron-electron scattering is also displayed (dash-dotted curve). Reprinted
with permission from J. Appl. Phys. \textbf{107}, 013104 (2010). Copyright
2010 American Institute of Physics.}%
\label{fig:ee}%
\end{figure}

Figure \ref{fig:ee} contains EMC simulations of the gain spectra for different
implementations of screening and spin effects \cite{2010JAP...107a3104J}. In
Fig.\,\ref{fig:ee}(a), the exchange effect is fully included, but different
screening models are employed. The reference curve is based on the exact
evaluation of the RPA (solid curve). Applying the simplified screening model
Eq.\,(\ref{eq:qsmss}) to all matrix elements overestimates the screening of
the intersubband elements, thus leading to an underestimation of scattering.
The resulting spectral gain profile (dashed curve) features an excessively
narrow gain bandwidth and enhanced peak gain. On the other hand, completely
ignoring the screening effect for the intersubband matrix elements
overestimates the intersubband scattering, thus resulting in a lowered and
\ broadened gain profile (dotted curve). In Fig.\,\ref{fig:ee}(b), screening
is included in the RPA. Results are shown for $p_{\mathrm{a}}=p_{\mathrm{p}%
}=1/2$ (solid curve), $p_{\mathrm{a}}=1$, $p_{\mathrm{p}}=0$ (dotted curve),
$p_{\mathrm{a}}=1/2$, $p_{\mathrm{p}}=0$ (dashed curve), and $p_{\mathrm{a}%
}=p_{\mathrm{p}}=0$ (dash-dotted curve). The last case, which corresponds to
completely neglecting electron-electron scattering in the simulation, yields
two narrow gain spikes at around $2.8$ and $3.6\,\mathrm{THz}$, largely
deviating from the experimental electroluminescence
measurements.\cite{2003ApPhL..82.3165S} Ignoring the exchange effect (dotted
curve) leads to an overestimation of the scattering, and thus a reduced peak
gain and increased gain bandwidth. On the other hand, completely neglecting
parallel spin collisions (dashed curve) leads to an underestimation of
scattering and thus an enhanced peak gain.

\subsection{\label{sec:elast}Elastic scattering processes}

Impurity scattering has been shown to play an important role in QCLs
\cite{2004ApPhL..84..645C,2008ApPhL..92h1102N,2009JAP...106f3115N}, where it
is often the dominant elastic scattering process. Also interface roughness can
have a considerable impact on QCL operation
\cite{2008PSSCR...5..232K,2009JAP...105l3102J,Kubis_Klimeck_APL,Deutsch}.
Alloy scattering occurs in semiconductor alloys such as AlGaAs and other
ternary materials, and usually only has to be considered for the well material
since the wave functions are largely localized in the wells \cite{Bastard:88}.
For elastic intrasubband scattering in the conduction band $\Gamma$
valley,\ we have $k=k^{\prime}$ due to energy conservation, i.e., elastic
intrasubband scattering does not affect the carrier distribution due to
in-plane isotropy.

\subsubsection{\label{sec:elast_impur}Impurity scattering}

Impurity scattering in QCLs arises from the doping, e.g., the ionized donors
in the heterostructure. Charged impurities are to lowest order already
considered in the Poisson equation, Eq.\thinspace(\ref{eq:poisson}),
corresponding to a mean field treatment of charges. An inclusion of higher
order effects is in the semiclassical treatment obtained by corresponding
scattering rates. The perturbation potential due to a charged impurity at
position $\left(  x^{\prime},y^{\prime},z^{\prime}\right)  ^{\mathrm{T}}$ is
with $\mathbf{r}^{\prime}=\left(  x^{\prime},y^{\prime}\right)  ^{\mathrm{T}}$
and $\epsilon=\epsilon_{0}\epsilon_{\mathrm{r},0}$ given by the Coulomb
potential%
\begin{equation}
V=-\left(  4\pi\epsilon\right)  ^{-1}e^{2}\left(  \left|  \mathbf{r}%
-\mathbf{r}^{\prime}\right|  ^{2}+\left|  z-z^{\prime}\right|  ^{2}\right)
^{-1/2}.
\end{equation}
For the definition of the corresponding matrix element, various conventions
exist \cite{2006ApPhL..89u1115L,2009JAP...106f3115N}. In analogy to Eq.
(\ref{eq:Vbare}) for electron-electron scattering, we define the bare matrix
element as \cite{2006ApPhL..89u1115L}
\begin{equation}
V_{ij}^{\mathrm{b}}\left(  z^{\prime},q\right)  =-\frac{e^{2}}{2\epsilon
q}\int_{-\infty}^{\infty}\psi_{i}\left(  z\right)  \psi_{j}^{\ast}\left(
z\right)  \exp\left(  -q\left|  z-z^{\prime}\right|  \right)  \mathrm{d}%
z.\label{eq:V_imp}%
\end{equation}
The scattering rate is obtained from Eq.\thinspace(\ref{eq:golden}) as%
\begin{align}
W_{i\mathbf{k},j\mathbf{k}^{\prime}} &  =\frac{2\pi}{S\hbar}\int\left|
V_{ij}^{\mathrm{b}}\left(  z^{\prime},q\right)  \right|  ^{2}n_{\mathrm{D}%
}\left(  z^{\prime}\right)  \mathrm{d}z^{\prime}\delta\left(  E_{j\mathbf{k}%
^{\prime}}-E_{i\mathbf{k}}\right)  \nonumber\\
&  =\frac{\pi e^{4}}{2S\epsilon^{2}\hbar q^{2}}F_{ij}\left(  q\right)
\delta\left(  E_{j\mathbf{k}^{\prime}}-E_{i\mathbf{k}}\right)
\label{eq:W_imp}%
\end{align}
with the form factor%
\begin{align}
F_{ij}\left(  q\right)    & =\int\mathrm{d}z^{\prime}n_{\mathrm{D}}\left(
z^{\prime}\right)  \nonumber\\
& \times\left[  \int_{-\infty}^{\infty}\psi_{i}\left(  z\right)  \psi
_{j}^{\ast}\left(  z\right)  \exp(-q\left|  z-z^{\prime}\right|
)\mathrm{d}z\right]  ^{2}.
\end{align}
Here we have summed over the donors to include the effect of all ionized
impurities, corresponding to an integration $S\int\dots n_{\mathrm{D}}\left(
z\right)  \mathrm{d}z$, with the doping concentration $n_{\mathrm{D}}\left(
z\right)  $. Furthermore, an additional factor $S^{-2}$ has been added since
the prefactor $S^{-1}$ is omitted in Eq.\thinspace(\ref{eq:V_imp}). Due to
energy conservation $E_{i\mathbf{k}}=E_{j\mathbf{k}^{\prime}}$, the final wave
vector magnitude $k^{\prime}$ is given by Eq.\thinspace(\ref{eq:ks_el}).

The total transition rate from a given initial state $\left|  i\mathbf{k}%
\right\rangle $ to a subband $j$ is found by summing over all final wave
vectors $\mathbf{k}^{\prime}$ using Eq.\thinspace(\ref{eq:sum}). We obtain
\cite{2006ApPhL..89u1115L,1988JaJAP..27..563T}%
\begin{equation}
W_{i\mathbf{k},j}=\frac{m_{j}^{\parallel}e^{4}}{4\pi\epsilon^{2}\hbar^{3}}%
\int_{0}^{\pi}\frac{F_{ij}\left(  q\right)  }{q^{2}}\mathrm{d}\theta,
\label{eq:W_imp2}%
\end{equation}
with $q\left(  \theta\right)  =\left|  \mathbf{k-k^{\prime}}\right|  =\left(
k^{2}+k^{\prime2}-2kk^{\prime}\cos\theta\right)  ^{1/2}$. With Eq.
(\ref{eq:ks_el}), this yields
\begin{equation}
q\left(  \theta\right)  =\left[  \left(  1+\frac{m_{j}^{\parallel}}%
{m_{i}^{\parallel}}\right)  k^{2}+q_{0}^{2}-2k\left(  \frac{m_{j}^{\parallel}%
}{m_{i}^{\parallel}}k^{2}+q_{0}^{2}\right)  ^{1/2}\cos\theta\right]  ^{1/2}
\label{eq:q_theta}%
\end{equation}
where $q_{0}^{2}=2m_{j}^{\parallel}\left(  E_{i}-E_{j}\right)  /\hbar^{2}$.
Values of $\theta$ resulting in complex $q\left(  \theta\right)  $ should be
excluded from the integration in Eq.\thinspace(\ref{eq:W_imp2}) since
scattering is then impossible.

\paragraph{Screening}

As for electron-electron scattering discussed in Section \ref{sec:ee}, the
random phase approximation can also be used to obtain the screened matrix
elements $V_{ij}^{\mathrm{s}}\left(  z^{\prime},q\right)  $ for impurity
scattering from the bare elements, $V_{ij}^{\mathrm{b}}\left(  z^{\prime
},q\right)  $ in Eq.\thinspace(\ref{eq:V_imp}). We obtain
\cite{2006ApPhL..89u1115L,2009JAP...106f3115N}%
\begin{equation}
V_{ij}^{\mathrm{s}}=V_{ij}^{\mathrm{b}}+\sum_{mn}V_{imjn}^{\mathrm{b}}\Pi
_{mn}V_{mn}^{\mathrm{s}}\label{eq:Vscr2}%
\end{equation}
with $V_{imjn}^{\mathrm{b}}$ and $\Pi_{mn}$ defined in Eqs.\thinspace
(\ref{eq:Vbare}) and (\ref{eq:Pi}), respectively. In Eq.\thinspace
(\ref{eq:W_imp}), $V_{ij}^{\mathrm{b}}\left(  z^{\prime},q\right)  $ has then
to be replaced by $V_{ij}^{\mathrm{s}}\left(  z^{\prime},q\right)  $, and also
Eq.\thinspace(\ref{eq:W_imp2}) has to be changed correspondingly. As for
electron-electron scattering, also here often simplified screening models
using a constant screening wave number $q_{\mathrm{s}}$ are employed to avoid
the numerical load associated with solving Eq.\thinspace(\ref{eq:Vscr2})
\cite{1988JaJAP..27..563T,2009JAP...106f3115N}. Common models to handle
screening are the model of Brooks and Herring~\cite{Brooks} and the model of
Conwell and Weisskopf.~\cite{Conwell_Weisskopf} While the model of Conwell and
Weisskopf is based on a Rutherford-like scattering of electrons on bare
Coulomb potentials with a cut-off radius, Brooks and Herring describe the
impurity potential as being screened by free carriers. This latter approach is
valid, if the constant screening length is much larger than the electronic
wave length, which requires high temperatures and low carrier densities. For
higher densities and low temperatures, the incorporation of more realistic
screening has been shown to be essential,~\cite{Krieger_impur} whereas the
limit of negligible free carrier screening is better described by the approach
of Conwell and Weisskopf.~\cite{MC} A detailed overview of the various
refinements to the approach of Brooks and Herring has been given in
Ref.~\onlinecite{chatto}. Electron scattering from screened impurities due to
Brooks and Herring will be discussed in more detail in
Sec.~\ref{sec:NEGF_Brooks_imp}.

\subsubsection{\label{sec:elast_interface}Interface roughness scattering}

\begin{figure}[ptb]
\includegraphics{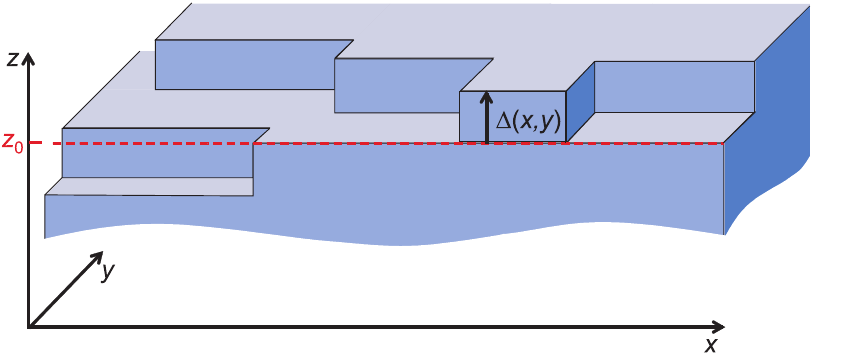}
\caption{{}(Color online) Interface between two heterostructure layers
consisting of different materials. The local deviation from a perfect
interface due to roughness is expressed by $\Delta\left(  x,y\right)  $.}%
\label{fig:Delta}%
\end{figure}

It has been shown that scattering from rough interfaces can change the QCL
performance
significantly~\cite{2008PSSCR...5..232K,2009JAP...105l3102J,Kubis_Klimeck_APL,Deutsch}%
. There are two fundamentally different models for the interface roughness
scattering in literature. In the first model, rough interfaces of quantum
wells cause fluctuating well widths and fluctuating confinement energies of
resonant states. In this model, the scattering potential is identified with
the change of the well eigenenergies.~\cite{1987ApPhL..51.1934S, Schnabel_IR}
In the second model, fluctuations of the band offset yields the scattering
potential.~\cite{Roblin_IR, nemo1d} The latter model may be favored for
generality, since it does not require the existence of confined states, nor
the distinguishability of level broadening by rough interfaces from other
mechanisms. B. R. Nag has shown in Ref.~\onlinecite{nag} that both models
agree well for quantum wells of various dimensions and materials. Thus, we
consider interface roughness scattering due to the imperfections in the
interface between the barrier and well material in the heterostructure,
causing a local deviation of the interface $\Delta\left(  x,y\right)
=\Delta\left(  \mathbf{r}\right)  $ from its average position $z_{0}$ as
illustrated in Fig.\thinspace\ref{fig:Delta}. Since it is not feasible to
measure $\Delta\left(  \mathbf{r}\right)  $ directly, the interface roughness
is characterized by its the standard deviation $\Delta$ and correlation length
$\Lambda$. Typically, a Gaussian autocorrelation function is assumed
\cite{1987ApPhL..51.1934S},%
\begin{align}
\left\langle \Delta\left(  \mathbf{r}\right)  \Delta\left(  \mathbf{r}%
^{\prime}\right)  \right\rangle  & =\frac{1}{S}\int\Delta\left(
\mathbf{r}\right)  \Delta\left(  \mathbf{r+d}\right)  \mathrm{d}%
^{2}r\nonumber\\
& =\Delta^{2}\exp\left(  -\frac{\mathbf{d}^{2}}{\Lambda^{2}}\right)
\label{eq:corr}%
\end{align}
with $\mathbf{d=r}^{\prime}-\mathbf{r}$. A change $\Delta\left(
\mathbf{r}\right)  $ of the interface position corresponds to a perturbing
potential \
\begin{equation}
V=\pm V_{\mathrm{o}}\left[  H\left(  z-z_{0}\right)  -H\left(  z-z_{0}%
-\Delta\left(  \mathbf{r}\right)  \right)  \right]  ,
\end{equation}
where $z_{0}$ and $V_{\mathrm{o}}$ are the average interface position and the
band offset, respectively. The ''$+$'' (''$-$'') sign is obtained if the
barrier (well) is located at $z<z_{0}$. Furthermore, $H$ denotes the Heaviside
step function. Since $\Delta\left(  \mathbf{r}\right)  $ is small, i.e., on
the order of a monolayer, we can approximate $\psi_{i,j}\left(  z_{0}%
+\Delta\left(  \mathbf{r}\right)  \right)  \approx\psi_{i,j}\left(
z_{0}\right)  $. Thus we obtain with Eq.\thinspace(\ref{eq:Vfi})
\begin{equation}
V_{j\mathbf{k}^{\prime},i\mathbf{k}}=\pm\frac{V_{\mathrm{o}}}{S}\psi
_{i}\left(  z_{0}\right)  \psi_{j}^{\ast}\left(  z_{0}\right)  \int
\mathrm{d}^{2}r\Delta\left(  \mathbf{r}\right)  \exp\left(  \mathrm{i}%
\mathbf{qr}\right)
\end{equation}
and%
\begin{align}
\left|  V_{j\mathbf{k}^{\prime},i\mathbf{k}}\right|  ^{2} &  =\frac
{V_{\mathrm{o}}^{2}}{S^{2}}\left|  \psi_{i}\left(  z_{0}\right)  \psi
_{j}^{\ast}\left(  z_{0}\right)  \right|  ^{2}\nonumber\\
&  \times\int\mathrm{d}^{2}d\left[  \exp\left(  \mathrm{i}\mathbf{qd}\right)
\int\mathrm{d}^{2}r\Delta\left(  \mathbf{r}\right)  \Delta\left(
\mathbf{r}+\mathbf{d}\right)  \right]
\end{align}
with $\mathbf{q}=\mathbf{k}-\mathbf{k}^{\prime}$. Assuming a Gaussian
autocorrelation Eq.\thinspace(\ref{eq:corr}), we obtain from Eq.\thinspace
(\ref{eq:golden}) the scattering rate%
\begin{align}
W_{i\mathbf{k},j\mathbf{k}^{\prime}} &  =\frac{2\pi^{2}}{\hbar S}%
V_{\mathrm{o}}^{2}\Delta^{2}\Lambda^{2}\sum_{n}\left|  \psi_{i}\left(
z_{n}\right)  \psi_{j}^{\ast}\left(  z_{n}\right)  \right|  ^{2}\nonumber\\
&  \times\exp\left(  -\frac{1}{4}\Lambda^{2}q^{2}\right)  \delta\left(
E_{j\mathbf{k}^{\prime}}-E_{i\mathbf{k}}\right)  ,
\end{align}
where we additionally sum over all interfaces, located at positions $z_{n}$.
Due to the the energy conservation $E_{i\mathbf{k}}=E_{j\mathbf{k}^{\prime}}$,
the final wave vector magnitude $k^{\prime}$ is given by Eq.\thinspace
(\ref{eq:ks_el}).

The total transition rate from a given initial state $\left|  i\mathbf{k}%
\right\rangle $ to a subband $j$ is found by summing over all final wave
vectors $\mathbf{k}^{\prime}$ using Eq.\thinspace(\ref{eq:sum}). We obtain%
\begin{align}
W_{i\mathbf{k},j}  &  =\frac{m_{j}^{\parallel}}{\hbar^{3}}V_{\mathrm{o}}%
^{2}\Delta^{2}\Lambda^{2}\sum_{n}\left|  \psi_{i}\left(  z_{n}\right)
\psi_{j}^{\ast}\left(  z_{n}\right)  \right|  ^{2}\nonumber\\
&  \times\int_{0}^{\pi}d\theta\exp\left(  -\frac{1}{4}\Lambda^{2}q^{2}\right)
H\left(  q^{2}\right)  ,
\end{align}
where $q\left(  \theta\right)  $ is again given by Eq.\thinspace
(\ref{eq:q_theta}), and values of $\theta$ with $q^{2}<0$ should be excluded
from the integration since then scattering cannot occur.

\begin{figure}[ptb]
\includegraphics{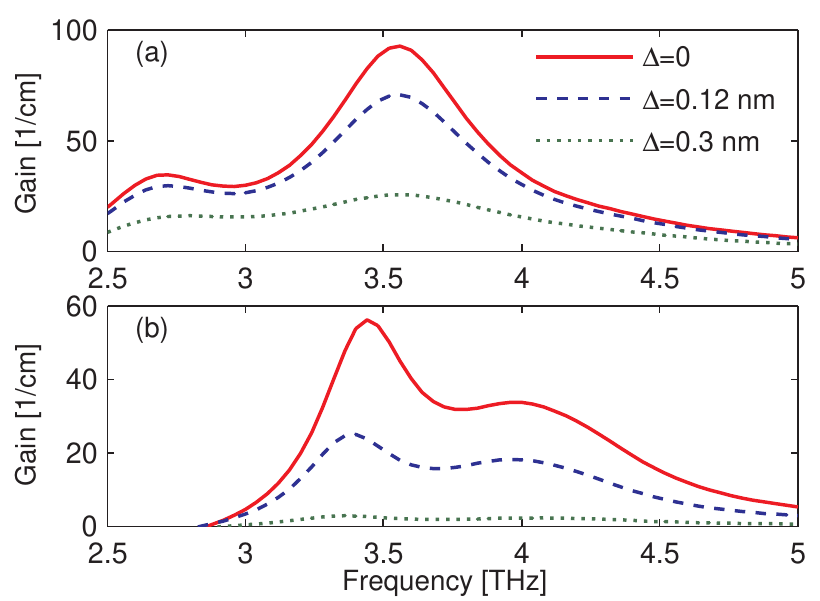}
\caption{(Color online) Simulation results for the spectral gain vs frequency,
as obtained for a correlation length $\Lambda=10\,\mathrm{nm}$ and different
values of the interface roughness mean height $\Delta$
\cite{2009JAP...105l3102J}. a) 3.4 THz resonant phonon depopulation structure;
b) 3.5 THz bound-to-continuum structure. Reprinted with permission from J.
Appl. Phys. \textbf{105}, 123102 (2009). Copyright 2009 American Institute of
Physics.}%
\label{fig:IR}%
\end{figure}

The characterization of interface roughness by the parameters $\Lambda$ and
$\Delta$ is somewhat phenomenological, since it is difficult to directly
extract them from experimental measurements of the interface profile. Rather,
they are typically chosen to reproduce the measured optical transition
linewidths \cite{2005ApPhL..86f2113T}. For GaAs-based terahertz QCLs,
frequently used values are $\Lambda\approx10\,\mathrm{nm}$ and $\Delta
\approx0.1\,\mathrm{nm}$ \cite{2008ApPhL..92h1102N,2008PSSCR...5..232K}. In
Fig.\ \ref{fig:IR}, EMC simulation results for the gain profile are shown for
two different types of terahertz QCLs \cite{2009JAP...105l3102J}. In addition
to scattering of electrons with phonons, impurities and other electrons,
interface roughness scattering with different values of the interface
roughness mean height has been considered. Results are shown for $\Delta=0$
(no interface roughness scattering), $0.12\,\mathrm{nm}$ and $0.3\,\mathrm{nm}%
$ (corresponding to one monolayer of GaAs). Depending on the assumed value of
$\Delta$, interface roughness can lead to considerable broadening of the gain
profile and reduction in the peak gain. The phonon depopulation structure in
Fig. \ref{fig:IR}(a) is less sensitive to interface roughness than the
bound-to-continuum structure in Fig.\thinspace\ref{fig:IR}(b), since the wave
functions in the minibands of the latter structure are less localized and thus
extend across more interfaces.

\subsubsection{\label{sec:elas_alloy}Alloy scattering}

In a ternary semiconductor alloy A$_{x}$B$_{1-x}$C such as Al$_{x}$Ga$_{1-x}%
$As, the A and B atoms are randomly distributed on the corresponding lattice
sites. Thus, for a zinc-blende structure, each unit cell contains an atom of
type A or B and an atom of type C. The potential in the semiconductor is then
given by \cite{1963PhRv..132.1047A,Bastard:88}%
\begin{equation}
V_{\mathrm{c}}\left(  \mathbf{x}\right)  =\sum_{n}V_{n}\left(  \mathbf{x}%
-\mathbf{x}_{n}\right)  ,
\end{equation}
where $\mathbf{x}_{n}$ denotes the position of unit cell $n$. Furthermore,
$V_{n}=V_{\mathrm{AC}}$ ($V_{n}=V_{\mathrm{BC}}$) if the unit cell contains an
atom of type A (B), where $V_{\mathrm{AC}}$ and $V_{\mathrm{BC}}$ are the
potentials of the binary materials AC and BC. The conduction band profile of
alloys is typically considered in virtual crystal approximation, obtained by
averaging over the contributions of the two alloy materials AC and BC,%
\begin{equation}
V_{\mathrm{c,VCA}}=\sum_{n}\left[  xV_{\mathrm{AC}}\left(  \mathbf{x}%
-\mathbf{x}_{n}\right)  +\left(  1-x\right)  V_{\mathrm{BC}}\left(
\mathbf{x}-\mathbf{x}_{n}\right)  \right]  .
\end{equation}
The perturbing potential corresponds then to the difference between the real
conduction band energy $V_{\mathrm{c}}\left(  \mathbf{x}\right)  $ and
$V_{\mathrm{c,VCA}}$ obtained by the virtual crystal approximation,%
\begin{equation}
V=V_{\mathrm{c}}-V_{\mathrm{c,VCA}}=\sum_{n}c_{n}\delta V\left(
\mathbf{x-x}_{n}\right)  ,\label{eq:V_al}%
\end{equation}
with $\delta V\left(  \mathbf{x}\right)  =V_{\mathrm{AC}}\left(
\mathbf{x}\right)  -V_{\mathrm{BC}}\left(  \mathbf{x}\right)  $ and
$c_{n}=\left(  1-x\right)  $ ($c_{n}=-x$) if the unit cell contains an atom of
type A (B).

The scattering potential $\delta V$ is effective within the unit cell, and
also screening effects are usually neglected due to the short-range nature of
the scattering potential \cite{Bastard:88}. Thus, on the scale of the envelope
functions $\psi\left(  \mathbf{x}\right)  $ which are slowly varying on the
atomic scale, $\delta V$ can be modeled as a $\delta$ function potential
\cite{Bastard:88}%
\begin{equation}
\delta V\left(  \mathbf{x}\right)  =\delta V\Omega_{0}\delta\left(
\mathbf{x}\right)  ,\label{eq:dV}%
\end{equation}
where $\Omega_{0}$ denotes the unit cell volume, given by $\Omega_{0}=a^{3}/4$
for a zinc-blende structure with lattice constant $a$. Furthermore, $\delta V$
is the alloy scattering potential, which approximately corresponds to the
conduction band offset of the binary materials involved, if they are not
strongly lattice-mismatched \cite{Bastard:88}. Generally, however, $\delta V$
deviates from the conduction band differences \cite{Kubis_Klimeck_APL}, and
should be estimated from empirical tightbinding
parameters~\cite{Mehrotra_alloy,BoykinIR,HedgeIR,KrishnamurthyIR}. Using Eqs.
(\ref{eq:V_al}) and (\ref{eq:dV}), the matrix element Eq.\thinspace
(\ref{eq:Vfi}) becomes with $\mathbf{q=k}-\mathbf{k}^{\prime}$%
\begin{equation}
V_{j\mathbf{k}^{\prime},i\mathbf{k}}=S^{-1}\delta V\Omega_{0}\sum_{n}c_{n}%
\psi_{j}^{\ast}\left(  z_{n}\right)  \psi_{i}\left(  z_{n}\right)  \exp\left(
\mathrm{i}\mathbf{qr}_{n}\right)  ,
\end{equation}
and the square modulus is given by%
\begin{align}
\left|  V_{j\mathbf{k}^{\prime},i\mathbf{k}}\right|  ^{2} &  =S^{-2}\left(
\delta V\Omega_{0}\right)  ^{2}\sum_{n}\sum_{m}\big\{c_{n}c_{m}\exp\left[
\mathrm{i}\mathbf{q}\left(  \mathbf{r}_{n}-\mathbf{r}_{m}\right)  \right]
\nonumber\\
&  \times\psi_{j}^{\ast}\left(  z_{n}\right)  \psi_{j}\left(  z_{m}\right)
\psi_{i}\left(  z_{n}\right)  \psi_{i}^{\ast}\left(  z_{m}\right)  \big\}.
\end{align}
For calculating the averge of $\left|  V_{j\mathbf{k}^{\prime},i\mathbf{k}%
}\right|  ^{2}$, the expectation value $\left\langle c_{n}c_{m}\right\rangle $
must be determined. Each unit cell has the probability $x$ of containing an
atom of type A, and $1-x$ of containing an atom of type B. Thus we obtain
$\left\langle c_{n}c_{m}\right\rangle =0$ for $m\neq n$ and $\left\langle
c_{n}c_{m}\right\rangle =x\left(  1-x\right)  $ for $m=n$. Furthermore
replacing the sum over the unit cell positions $\sum_{\mathbf{x}_{n}}f\left(
z_{n}\right)  $ by an integral $\Omega_{0}^{-1}\int f\left(  z\right)
\mathrm{d}^{3}x=\Omega_{0}^{-1}S\int f\left(  z\right)  \mathrm{d}z$, we
obtain
\begin{align}
\left\langle \left|  V_{j\mathbf{k}^{\prime},i\mathbf{k}}\right|
^{2}\right\rangle  &  =S^{-1}\Omega_{0}\left(  \delta V\right)  ^{2}x\left(
1-x\right)  \nonumber\\
&  \times\int_{-\infty}^{\infty}\left|  \psi_{j}\left(  z\right)  \psi
_{i}\left(  z\right)  \right|  ^{2}\mathrm{d}z.
\end{align}
From Eq.\thinspace(\ref{eq:golden}), we obtain the scattering rate%
\begin{align}
W_{i\mathbf{k},j\mathbf{k}^{\prime}} &  =\frac{2\pi}{\hbar S}\delta\left(
E_{j\mathbf{k}^{\prime}}-E_{i\mathbf{k}}\right)  \nonumber\\
&  \times\int_{-\infty}^{\infty}\Omega_{0}\left(  \delta V\right)
^{2}x\left(  1-x\right)  \left|  \psi_{j}\psi_{i}\right|  ^{2}\mathrm{d}z,
\end{align}
where $\Omega_{0}\left(  z\right)  $, $\delta V\left(  z\right)  $ and
$x\left(  z\right)  $ have been taken into the integral to account for varying
alloy compositions along the growth direction $z$. Due to the energy
conservation $E_{i\mathbf{k}}=E_{j\mathbf{k}^{\prime}}$, the final wave vector
magnitude $k^{\prime}$ is given by Eq.\thinspace(\ref{eq:ks_el}). The total
transition rate from a given initial state $\left|  i\mathbf{k}\right\rangle $
to a subband $j$ is with Eq.\thinspace(\ref{eq:sum})%
\begin{equation}
W_{i\mathbf{k},j}=m_{j}^{\parallel}\hbar^{-3}\int_{-\infty}^{\infty}\Omega
_{0}\left(  \delta V\right)  ^{2}x\left(  1-x\right)  \left|  \psi_{j}\psi
_{i}\right|  ^{2}\mathrm{d}z
\end{equation}
for $E_{j}<E_{i\mathbf{k}}$, and $0$ otherwise.

\subsection{\label{sec:SCRE}Self-consistent rate equation approach}

In the self-consistent rate equation approach, the intersubband scattering
rates in the rate equation Eq.\thinspace(\ref{eq:rate}) or (\ref{eq:rate2})
are self-consistently determined based on the corresponding Hamiltonian. Thus,
this approach only relies on well known material parameters such as the
effective mass, and not on experimental or empirical lifetimes as in Section
\ref{sec:PRE}. Offering a compromise between accuracy and predictive power on
the one hand and relative numerical efficiency on the other hand, the
self-consistent rate equation approach is widely used for the simulation of
QCLs
\cite{2002JAP....91.9019I,2002JAP....92.6921H,2001JAP....89.3084D,2005ApPhL..86u1117J,2012SeScT..27f5009S,2013SeScT..28j5008P}%
. Although extensions of the self-consistent rate equation approach to include
the light field have been presented \cite{2013SeScT..28j5008P}, the optical
cavity field is neglected in most cases. Such simulations yield the
unsaturated population inversion or gain, indicating if lasing can start at
all, but giving no information about the actual lasing operation.

\subsubsection{Intersubband scattering rates}

The number of electrons per unit energy and area in a thermalized subband $i$
of a 2D system is given by $n_{i}^{E}f_{i}^{\mathrm{FD}}$ with the Fermi-Dirac
distribution%
\begin{equation}
f_{i}^{\mathrm{FD}}\left(  E_{i\mathbf{k}}\right)  =\left\{  \exp\left[
\left(  E_{i\mathbf{k}}-E_{i}^{\mathrm{F}}\right)  /\left(  k_{\mathrm{B}%
}T_{i}\right)  \right]  +1\right\}  ^{-1},\label{eq:fermi2}%
\end{equation}
where $E_{i}^{\mathrm{F}}$ is here a ''quasi'' Fermi energy describing the
kinetic energy distribution of the electrons within the subband $i$
\cite{harrison}, and $T_{i}$ is the associated electron temperature.
Furthermore, $n_{i}^{E}$ is the density of states per unit area and energy in
a 2D system \cite{1997plds.book.....D}, and $E_{i\mathbf{k}}=E_{i}+\hbar
^{2}k^{2}/\left(  2m_{i}^{\parallel}\right)  $ is the energy of the electrons
in subband $i$ with an in-plane wave vector $\mathbf{k}$. The scattering rate
from an initial subband $i$ to a final subband $j$ is then obtained by
averaging over the carrier distribution. Assuming a Fermi-Dirac distribution
in the initial and final state, we obtain \cite{harrison}%
\begin{equation}
\left(  \tau_{ij}^{\left(  m\right)  }\right)  ^{-1}=\frac{\int_{E_{i}%
}^{\infty}f_{i}^{\mathrm{FD}}\left(  E\right)  W_{i\mathbf{k},j}^{\left(
m\right)  }\mathrm{d}E}{\int_{E_{i}}^{\infty}f_{i}^{\mathrm{FD}}\left(
E\right)  \mathrm{d}E}.
\end{equation}
Here $n_{i}^{E}$ has been omitted in the enumerator and denominator because it
is constant, $n_{i}^{E}=m_{i}^{\parallel}/\left(  \pi\hbar^{2}\right)  $ for
$E\geq E_{i}$. The index $m$ denotes the corresponding scattering mechanism.
$W_{i\mathbf{k},j}^{\left(  m\right)  }$ is for example given by
Eqs.\thinspace(\ref{eq:W_LO2}) and (\ref{eq:R_ee}) for LO phonon and
electron-electron scattering, respectively. The total rate is obtained by
summing over the individual contributions of the different scattering
mechanisms, i.e., $\tau_{ij}^{-1}=\sum_{m}\left(  \tau_{ij}^{\left(  m\right)
}\right)  ^{-1}$.

The sheet density in a subband $i$, i.e., electron number per unit area, is
given by%
\begin{align}
n_{i}^{\mathrm{s}} &  =\int_{E_{i}}^{\infty}f_{i}^{\mathrm{FD}}\left(
E\right)  n_{i}^{E}\mathrm{d}E\nonumber\\
&  =\frac{m_{i}^{\parallel}k_{\mathrm{B}}T_{i}}{\pi\hbar^{2}}\left\{
\frac{E_{i}^{\mathrm{F}}-E_{i}}{k_{\mathrm{B}}T_{i}}+\ln\left[  1+\exp\left(
-\frac{E_{i}^{\mathrm{F}}-E_{i}}{k_{\mathrm{B}}T_{i}}\right)  \right]
\right\}  .\label{eq:n2D_rate}%
\end{align}
Furthermore expressing the carrier energy in terms of the in-plane wave
vector, we obtain with $\mathrm{d}E=\hbar^{2}\left(  m_{i}^{\parallel}\right)
^{-1}k\mathrm{d}k$ \cite{harrison}%
\begin{equation}
\left(  \tau_{ij}^{\left(  m\right)  }\right)  ^{-1}=\frac{\int_{0}^{\infty
}f_{i}^{\mathrm{FD}}\left[  E_{i}+\hbar^{2}k^{2}/\left(  2m_{i}^{\parallel
}\right)  \right]  W_{i\mathbf{k},j}^{\left(  m\right)  }k\mathrm{d}k}{\pi
n_{i}^{\mathrm{s}}}.\label{eq:Wif}%
\end{equation}
An additional factor $\left\{  1-f_{j}^{\mathrm{FD}}\left[  E_{i}+\hbar
^{2}k^{2}/\left(  2m_{i}^{\parallel}\right)  \pm E_{0}\right]  \right\}  $ can
be included in the integral of Eq.\thinspace(\ref{eq:Wif}), accounting for the
final state blocking due to Pauli's exclusion principle \cite{harrison}. This
correction is only relevant for high doping levels. Here $E_{0}$ is $0$ for
elastic scattering mechanisms, and corresponds to the TO or LO phonon energy
for optical phonon absorption (''$+$'' sign) and emission (''$-$'' sign),
respectively. The Fermi energy can be calculated from Eq.\thinspace
(\ref{eq:n2D_rate}),%
\begin{equation}
E_{i}^{\mathrm{F}}-E_{i}=k_{\mathrm{B}}T_{i}\ln\left[  \exp\left(  \frac
{n_{i}^{\mathrm{s}}\pi\hbar^{2}}{m_{i}^{\parallel}k_{\mathrm{B}}T_{i}}\right)
-1\right]  .
\end{equation}

For lightly doped semiconductors, as is often the case in QCLs, we have
$\left(  E_{i\mathbf{k}}-E_{i}^{\mathrm{F}}\right)  \gg k_{\mathrm{B}}T_{i}$
in Eq.\thinspace(\ref{eq:fermi2}), which then approaches a classical
Maxwell-Boltzmann distribution%
\begin{equation}
f_{i}^{\mathrm{MB}}\left(  E_{i\mathbf{k}}\right)  =\exp\left[  -\left(
E_{i\mathbf{k}}-E_{i}^{\mathrm{F}}\right)  /\left(  k_{\mathrm{B}}%
T_{i}\right)  \right]  .\label{eq:MB}%
\end{equation}
Under this condition, Eq.\thinspace(\ref{eq:Wif}) simplifies to%
\begin{equation}
\left(  \tau_{ij}^{\left(  m\right)  }\right)  ^{-1}=\frac{\hbar^{2}}%
{m_{i}^{\parallel}k_{\mathrm{B}}T_{i}}\int_{0}^{\infty}\exp\left(
-\frac{\hbar^{2}k^{2}}{2m_{i}^{\parallel}k_{\mathrm{B}}T_{i}}\right)
W_{i\mathbf{k},j}^{\left(  m\right)  }k\mathrm{d}k.\label{eq:Wif2}%
\end{equation}

Numerically, the integration of Eq.\thinspace(\ref{eq:Wif}) or Eq.\thinspace
(\ref{eq:Wif2}) is performed from $E_{i}$ up to a sufficiently large maximum
value, e.g., the highest value of the simulated potential profile $V\left(
z\right)  $ \cite{harrison}.

\subsubsection{Rate equations}

In the self-consistent rate equation approach, typically the QCL is modeled as
a biased periodic heterostructure, excluding effects such as domain formation
\cite{2006PhRvB..73c3311L}. Then the simulation can be restricted to a single
representative period far away from the contacts, additionally applying
periodic boundary conditions \cite{2001JAP....89.3084D}. This corresponds to
solving Eq.\thinspace(\ref{eq:rate}) with the self-consistently calculated
scattering rates, Eq.\thinspace(\ref{eq:Wif}). While Eq.\thinspace
(\ref{eq:rate}) in principle includes the transitions to all equivalent levels
in the different periods, in practice only scattering between the central
period and adjacent periods has to be considered. Frequently the $1\frac{1}%
{2}$ period model is used, which applies to QCLs where a period consists of an
injector region and an active region
\cite{2002JAP....91.9019I,2001JAP....89.3084D}. Here, for the active region
states of the central period, only scattering transitions involving other
states of the central period and the right-neighboring injector region are
considered in Eq.\thinspace(\ref{eq:rate}). Analogously, for the states of the
injector region, only scattering involving states of the central period and
the left-neighboring active region are taken into account. Since the
scattering rates Eq.\thinspace(\ref{eq:Wif}) in general depend on the electron
densities $n_{i}^{\mathrm{s}}$, a direct solution of Eq.\thinspace
(\ref{eq:rate}) is not possible, and an iterative scheme is commonly used. For
simulations without lasing included, i.e., $W_{ij}^{\mathrm{opt}}=0$ in Eq.
(\ref{eq:rate}), setting $\mathrm{d}_{t}n_{i}^{\mathrm{s}}=0$ yields the
steady state solution \cite{2001JAP....89.3084D}
\begin{equation}
n_{i}^{\mathrm{s}}=\left.  \sum_{j\neq i}n_{j}^{\mathrm{s}}\tau_{ji}%
^{-1}\right/  \sum_{j\neq i}\tau_{ij}^{-1},\label{eq:iteration}%
\end{equation}
where $i=1..N$ refers to the central period containing $N$ subbands. The
summation index $j$ now only includes subbands in the central period and
adjacent periods, as discussed above. For subbands outside the central period,
the sheet density $n_{j}^{\mathrm{s}}$ of the equivalent level in the central
subband is used. For the numerical computation of the subband populations,
initially identical electron densities $n_{i}^{\mathrm{s}}=n^{\mathrm{s}}/N$
are assumed. Each iteration involves computing the scattering rates with Eq.
(\ref{eq:Wif}) or Eq.\thinspace(\ref{eq:Wif2}), calculating the new values for
$n_{i}^{\mathrm{s}}$ using Eq.\thinspace(\ref{eq:iteration}), and
renormalizing the sheet densities so that Eq.\thinspace(\ref{eq:ns}) is
fulfilled. Convergence can be accelerated by combining the sheet densities of
the previous two iterations, $\xi n_{i}^{\mathrm{s,new}}+\left(  1-\xi\right)
n_{i}^{\mathrm{s,old}}$, as input for the next iteration
\cite{2002JAP....91.9019I}, where the relaxation parameter is typically chosen
as $\xi=0.5$. The simulation has converged when the obtained sheet densities
do not significantly change anymore between iterations.

\subsubsection{Kinetic energy balance method}

Since $T_{i}$ in Eq.\thinspace(\ref{eq:fermi2}) is unknown, the scattering
rates Eq. (\ref{eq:Wif}) have to be evaluated assuming that the electron
temperature in each subband is equivalent to the lattice temperature,
$T_{i}=T_{\mathrm{L}}$. However, the electron temperature can significantly
exceed $T_{\mathrm{L}}$ in quantum cascade lasers \cite{2004ApPhL..84.3690S}.
Thus, the rate equation model has been extended to account for electron
heating, typically assuming an identical electron temperature $T_{i}%
=T_{\mathrm{e}}$ for all subbands \cite{2002JAP....92.6921H}. The kinetic
electron energy generation rate per period and unit device in-plane cross
section is given by \cite{2002JAP....92.6921H}%
\begin{equation}
R^{E}=\sum_{i=1}^{N}\sum_{j}\sum_{m}n_{i}^{\mathrm{s}}\left(  \tau
_{ij}^{\left(  m\right)  }\right)  ^{-1}\left(  E_{i}-E_{j}+E_{0}^{\left(
m\right)  }\right)  ,\label{eq:RE}%
\end{equation}
where $i$ sums over the subbands of the central period and $j$ also includes
states from adjacent periods, as discussed above. Furthermore, $m$ sums over
the different scattering contributions, such as LO phonon emission
($E_{0}^{\left(  m\right)  }=-E_{\mathrm{LO}}$) and absorption ($E_{0}%
^{\left(  m\right)  }=E_{\mathrm{LO}}$), elastic scattering mechanisms
($E_{0}^{\left(  m\right)  }=0$) and electron-electron scattering. For
inelastic scattering mechanisms such as phonon-induced transitions, also
intrasubband contributions $i=j$ have to be included
\cite{2002JAP....92.6921H}. Electron-electron scattering is often neglected in
Eq.\thinspace(\ref{eq:RE}) since the net kinetic energy does not change for
the important cases where both electrons stay within their respective subbands
or swap the subbands. Also photon transitions do not change the kinetic energy
for parabolic subbands due to $\mathbf{k}$ conservation
\cite{2013SeScT..28j5008P}. The numerical evaluation proceeds as follows. The
average electron temperature corresponds to the value of $T_{\mathrm{e}}$
where $R^{E}=0$ in Eq.\thinspace(\ref{eq:RE}). The rate equation is
self-consistently solved as described above for an initial guess of
$T_{\mathrm{e}}$, e.g., $T_{\mathrm{e}}=T_{\mathrm{L}}$. Based on the obtained
$n_{i}^{\mathrm{s}}$ and $\tau_{ij}^{\left(  m\right)  }$, $R^{E}$ in
Eq.\thinspace(\ref{eq:RE}) is calculated. This procedure is repeated and the
guess for $T_{\mathrm{e}}$ is iteratively improved until $R_{E}\approx0$ is
obtained. Since the assumption of a single effective electron temperature is
not always adequate in QCLs \cite{2005ApPhL..86k1115V}, extended approaches
have been developed allowing for different effective temperatures $T_{i}$ in
the individual subbands \cite{2012SeScT..27f5009S}.

\subsection{\label{sec:EMC}Ensemble Monte Carlo method}

Semiclassically, the carrier transport between states $\left|  i\mathbf{k}%
\right\rangle $ in a quantum well system is given by the Boltzmann equation
\cite{2005RPPh...68.2533I}%
\begin{equation}
\mathrm{d}_{t}f_{i\mathbf{k}}=\sum_{j}\sum_{\mathbf{k}^{\prime}}\left(
W_{j\mathbf{k}^{\prime},i\mathbf{k}}f_{j\mathbf{k}^{\prime}}-W_{i\mathbf{k}%
,j\mathbf{k}^{\prime}}f_{i\mathbf{k}}\right)  \label{eq:boltz}%
\end{equation}
with the scattering rates $W_{i\mathbf{k},j\mathbf{k}^{\prime}}=\sum
_{m}W_{i\mathbf{k},j\mathbf{k}^{\prime}}^{\left(  m\right)  }$, where $m$ sums
over the individual contributions of the different scattering mechanisms. The
distribution function $f_{i\mathbf{k}}\left(  t\right)  $ represents the
probability of the state $\left|  i\mathbf{k}\right\rangle $ being occupied at
a given time $t$. Eq.\thinspace(\ref{eq:boltz}) corresponds to an extended
version of the rate equations Eq.\thinspace(\ref{eq:rate}), where $i$ and
$j$\ have been replaced by $i\mathbf{k}$ and $j\mathbf{k}^{\prime}$,
respectively.\ The physical quantites of interest, such as the sheet densities
$n_{i}^{\mathrm{s}}$ and current density can be extracted from $f_{i\mathbf{k}%
}\left(  t\right)  $. For the numerical evaluation of Eq.\thinspace
(\ref{eq:boltz}), typically the ensemble Monte Carlo (EMC) method is used,
which is based on statistical sampling of the scattering events for a large
ensemble of carriers \cite{MC}, here $N_{\mathrm{e}}\approx10^{4}..10^{5}$
electrons. The large number of carriers considered allows for an extraction of
the physical quantities as function of $t$ by statistical averaging over the
carrier ensemble. Thus, this method is also applicable to time dependent
processes where we cannot use temporal averaging as for ergodic systems
\cite{MC}. Furthermore, electron-electron scattering can be implemented as a
two-electron process where a second electron is randomly chosen as scattering
partner \cite{1988PhRvB..37.2578G}.

\subsubsection{Simulation technique}

\begin{figure}[ptb]
\includegraphics{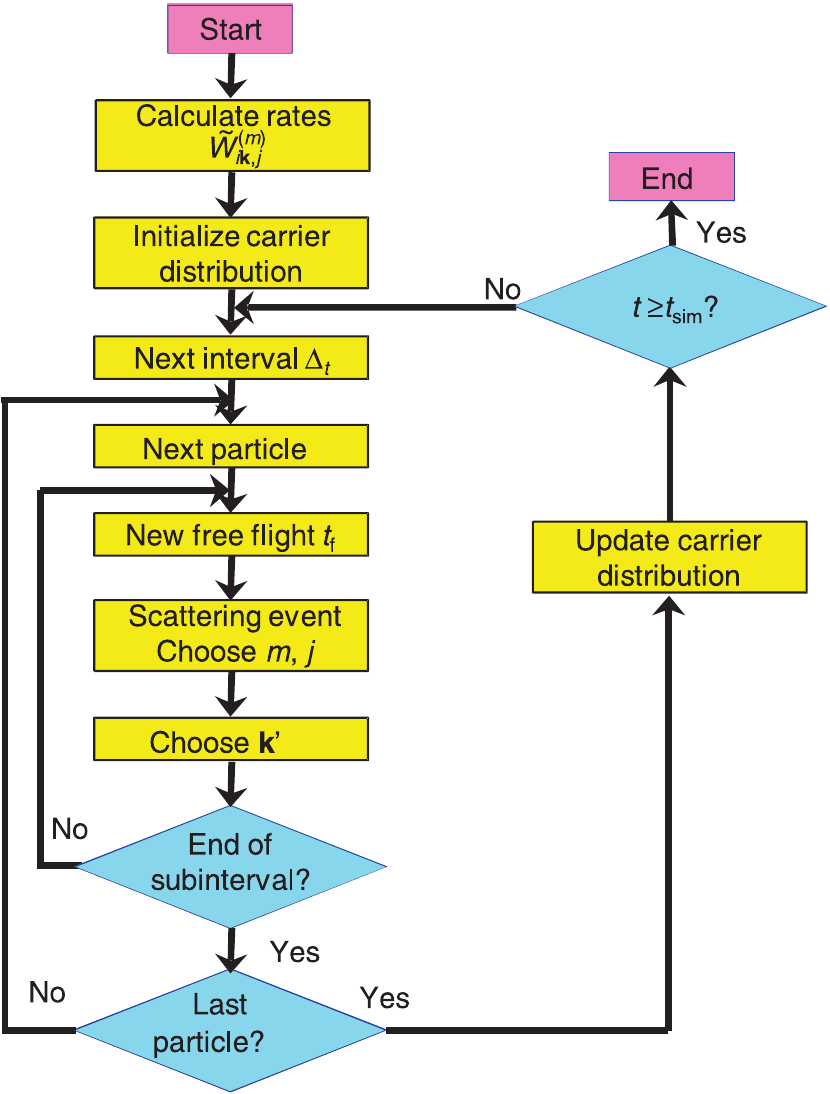}
\caption{{}(Color online) Schematic diagram of the EMC algorithm}%
\label{fig:EMC}%
\end{figure}

Figure \ref{fig:EMC} contains a schematic diagram of the EMC algorithm. The
system dynamics is evaluated up to a time $t_{\mathrm{sim}}$, which must be
chosen long enough to ensure convergence to the stationary solution. The
simulation is divided into subintervals $\Delta_{t}$, where the scattering
dynamics is subsequently evaluated. $\Delta_{t}$ should be chosen so small
that the average electron distribution does not change significantly over the
time interval, but big enough to include several scattering events per
electron on average. Assuming periodic boundary conditions, we can restrict
our simulation to a few periods. Each electron is characterized by its subband
$i$ and in-plane wave vector $\mathbf{k}$. All rates $W_{i\mathbf{k}%
,j}^{\left(  m\right)  }$ for the various scattering mechanisms $m$ are
computed and tabulated at the beginning of the simulation to save
computational resources. Here, we have to introduce a discrete grid for the
wave vector. Since in the conduction band $\Gamma$ valley, all states $\left|
i\mathbf{k}\right\rangle $ in subband $i$ with the same value $k=\left|
\mathbf{k}\right|  $ are equivalent due to in-plane isotropy, it is practical
to use the kinetic energy $E_{\mathrm{kin}}=$ $\hbar^{2}\mathbf{k}^{2}/\left(
2m_{i}^{\parallel}\right)  $ instead. The kinetic energy grid then divides the
energy axis into segments $n$ of widths $\Delta_{E}^{\left(  n\right)  }$
centered around discrete energies $E_{\mathrm{kin}}^{\left(  n\right)  }$ with
$k^{\left(  n\right)  }=\left(  2m_{i}^{\parallel}E_{\mathrm{kin}}^{\left(
n\right)  }\right)  ^{1/2}/\hbar$. However, some rates $W_{i\mathbf{k}%
,j}^{\left(  m\right)  }$ such as Eq.\thinspace(\ref{eq:R_ee}) depend on the
initially unknown carrier distribution itself, i.e., $W_{i\mathbf{k}%
,j}^{\left(  m\right)  }=W_{i\mathbf{k},j}^{\left(  m\right)  }\left(
t\right)  $. This problem can be overcome by tabulating an upper estimate
$\tilde{W}_{i\mathbf{k},j}^{\left(  m\right)  }$ for time dependent scattering
rates and compensating for the too high value by introducing artificial
''self-scattering'', as described further below.

An important quantity is the carrier distribution function, which can directly
be obtained from $N_{i}^{\left(  n\right)  }$, denoting the number of
simulated electrons in the $n$th energy cell of subband $i$ at a time $t$.
With the density of states per unit area and energy in a 2D system $n_{i}%
^{E}=m_{i}^{\parallel}/\left(  \pi\hbar^{2}\right)  $
\cite{1997plds.book.....D}, the number of available states in the $n$th energy
cell is $n_{i}^{E}\Delta_{E}^{\left(  n\right)  }S$. The simulated device
in-plane cross section is with the sheet doping density per period
$n^{\mathrm{s}}$ given by $S=N_{\mathrm{e}}/\left(  n^{\mathrm{s}%
}N_{\mathrm{p}}\right)  $, where $N_{\mathrm{p}}$\ corresponds to the number
of periods over which the $N_{\mathrm{e}}$ simulated electrons are
distributed. The carrier distribution function is then approximately given as%
\begin{equation}
f_{i}\left(  E_{\mathrm{kin}},t\right)  =\frac{n^{\mathrm{s}}N_{\mathrm{p}%
}N_{i}^{\left(  n\right)  }\left(  t\right)  }{n_{i}^{E}\Delta_{E}^{\left(
n\right)  }N_{\mathrm{e}}}, \label{eq:fiE}%
\end{equation}
where $n$ indicates the energy cell containing the value $E_{\mathrm{kin}}$.
Eq.\thinspace(\ref{eq:fiE}) can also be expressed as a function of $k=\left(
2m_{i}^{\parallel}E_{\mathrm{kin}}\right)  ^{1/2}/\hbar$. Additional temporal
averaging of the carrier distribution function Eq.\thinspace(\ref{eq:fiE}),
for example over a simulation subinterval $\Delta_{t}$, further reduces the
fluctuations resulting from the treatment of the carriers as discrete particles.

The simulation is initialized by assigning values $i$ and $\mathbf{k}$ to each
electron in the ensemble. Here, in principle arbitrarily chosen values can be
used since Eq.\thinspace(\ref{eq:boltz}) will converge to its stationary
solution after a sufficiently long simulation time $t_{\mathrm{sim}}$
independently of the chosen initial conditions. However, the required
$t_{\mathrm{sim}}$ to obtain convergence can be reduced by assuming a suitable
initial electron distribution which is not too far from the converged
solution. For example, the carriers can be distributed equally within the
subbands, and a thermalized distribution Eq.\thinspace(\ref{eq:MB}) in each
subband can be assumed. The Monte Carlo method is based on a stochastic
evaluation of the scattering events, assuming that the electron undergoes
scattering after a randomly selected free flight time $t_{\mathrm{f}}$. For a
time independent scattering rate $\tau_{0}^{-1}$, the free flight time is
given by $t_{\mathrm{f}}=-\tau_{0}\ln\left(  r\right)  $, where $r$ is a
random number evenly distributed between $0$ and $1$ \cite{MC}. Here we choose
$\tau_{0}^{-1}\geq\sum_{m}\sum_{j}\tilde{W}_{i\mathbf{k},j}^{\left(  m\right)
}$, corresponding to an upper bound for the total outscattering rate
$W_{i\mathbf{k}}\left(  t\right)  =\sum_{m}\sum_{j}W_{i\mathbf{k},j}^{\left(
m\right)  }\left(  t\right)  $, where $j=i$ is included in the summations to
account for intrasubband scattering. For each ensemble electron, the
scattering dynamics is evaluated over a simulation subinterval $\Delta_{t}$,
where the last free flight is continued in the next subinterval. After each
scattering event, the final subband $j$ of the scattered electron and
scattering mechanism $m$ are randomly selected based on their associated
probability $\tau_{0}W_{i\mathbf{k},j}^{\left(  m\right)  }\left(  t\right)
$. The too high amount of scattering assumed is corrected by introducing
artificial ''self-scattering'', which occurs with a probability $1-\tau
_{0}\sum_{m}\sum_{j}W_{i\mathbf{k},j}^{\left(  m\right)  }\left(  t\right)
$\ and does not change the carrier state at all \cite{MC}. Pauli's exclusion
principle can be considered by subsequently rejecting the occured scattering
event with a probability $f_{j\mathbf{k}^{\prime}}$. The final wave vector
magnitude $k^{\prime}$ is obtained from Eq.\thinspace(\ref{eq:ks_el})\ for
elastic scattering, and from Eq.\thinspace(\ref{eq:ks_in}) for inelastic
processes. Electron-electron scattering is in EMC typically implemented as a
two-electron process with a randomly selected partner electron and scattering
angle. To conserve energy for each scattering event individually, also the
partner electron has to undergo scattering, i.e., the final wave vector has to
be determined for both participating electrons \cite{1988PhRvB..37.2578G}.

The periodic boundary conditions can for example be implemented by simulating
three periods, where the $N_{\mathrm{e}}$ electrons are located in the central
period, i.e., $N_{\mathrm{p}}=1$ in Eq.\thinspace(\ref{eq:fiE}). Electrons
scattered to the first or third period are automatically injected into the
equivalent subband of the central period \cite{2000ApPhL..76.2265I}. By
counting the difference $\Delta_{N}$ of electrons scattered from the central
period to the left- and right-neighbouring period, respectively, over a
simulation subinterval $\Delta_{t}$, the current density can be computed as%
\begin{equation}
J=\frac{\Delta_{N}e}{\Delta_{t}S}=\frac{\Delta_{N}en^{\mathrm{s}}%
N_{\mathrm{p}}}{\Delta_{t}N_{\mathrm{e}}}.
\end{equation}

\subsection{Inclusion of the optical cavity field}

The optical gain and photon-induced transition rate are given by Eqs.
(\ref{eq:gain}) and (\ref{eq:Wijopt}) for a transition from a level $i$ to a
level $j$, which here correspond to states $\left|  i\mathbf{k}\right\rangle $
and $\left|  j\mathbf{k}^{\prime}\right\rangle $, respectively. The dipole
matrix element then becomes%
\begin{equation}
d_{i\mathbf{k},j\mathbf{k}^{\prime}}=-e\left\langle i\mathbf{k}\right|
z\left|  j\mathbf{k}^{\prime}\right\rangle =d_{ij}S^{-1}\int_{S}\exp\left[
\mathrm{i}\left(  \mathbf{k}^{\prime}-\mathbf{k}\right)  \mathbf{r}\right]
\mathrm{d}^{2}r
\end{equation}
with
\begin{equation}
d_{ij}=-e\left\langle i\right|  z\left|  j\right\rangle =-e\int_{-\infty
}^{\infty}\psi_{i}^{\ast}\left(  z\right)  z\psi_{j}\left(  z\right)
\mathrm{d}z,
\end{equation}
and for $\left| d_{i\mathbf{k},j\mathbf{k}^{\prime}}\right| ^{2}$, we obtain%
\begin{equation}
\left| d_{i\mathbf{k},j\mathbf{k}^{\prime}}\right| ^{2}=4\pi^{2}\left|  d_{ij}\right|
^{2}S^{-1}\delta\left(  \mathbf{k}^{\prime}\mathbf{-k}\right)  .\label{eq:d2}%
\end{equation}
Here, we have used that $\left|  \int_{S}\exp\left[  \mathrm{i}\left(
\mathbf{k}^{\prime}\mathbf{-k}\right)  \mathbf{r}\right]  \mathrm{d}%
^{2}r\right|  ^{2}$ can be approximated by $4\pi^{2}S\delta\left(
\mathbf{k}^{\prime}\mathbf{-k}\right)  $ for sufficiently large in-plane cross
sections $S$. The photon-induced transition rate given in Eq.\thinspace
(\ref{eq:Wijopt}) now becomes%
\begin{equation}
W_{i\mathbf{k},j\mathbf{k}^{\prime}}^{\mathrm{opt}}=\frac{4\pi^{3}}%
{\epsilon_{0}cn_{0}\hbar^{2}S}\left|  d_{ij}\right|  ^{2}\delta\left(
\mathbf{k}^{\prime}\mathbf{-k}\right)  \sum_{m}I_{m}\mathcal{L}_{ij}\left(
\omega_{m},k\right)  ,\label{eq:W_opt}%
\end{equation}
where we sum over all relevant cavity modes with frequencies $\omega_{m}$ and
intensities $I_{m}$ to account for multimode lasing. Adapting Eq.\thinspace
(\ref{eq:Lor}) to the present case, the definition of the Lorentzian lineshape
function becomes
\begin{equation}
\mathcal{L}_{ij}\left(  \omega,k\right)  =\frac{1}{\pi}\frac{\gamma
_{ij}\left(  k\right)  }{\gamma_{ij}^{2}\left(  k\right)  +\left[
\omega-\left|  \omega_{ij}\left(  k\right)  \right|  \right]  ^{2}%
},\label{eq:Lor2}%
\end{equation}
where $\omega_{ij}\left(  k\right)  =\left(  E_{i\mathbf{k}}-E_{j\mathbf{k}%
}\right)  /\hbar$ denotes the resonance frequency and $\gamma_{ij}\left(
k\right)  $ is the optical linewidth of the transition, which is given by
\begin{equation}
\gamma_{ij}\left(  k\right)  =\frac{1}{2}\left(  \sum_{\ell\neq i}%
W_{i\mathbf{k},\ell}+\sum_{\ell\neq j}W_{j\mathbf{k},\ell}\right)
\label{eq:gij}%
\end{equation}
when only the lifetime broadening contributions are considered
\cite{2009JAP...105l3102J}. The total photon-induced transition rate from a
given initial state $\left|  i\mathbf{k}\right\rangle $ to a subband $j$ is
found from Eq.\thinspace(\ref{eq:W_opt}) by summation over all final wave
vectors $\mathbf{k}^{\prime}$ using Eq.\thinspace(\ref{eq:sum}),%
\begin{equation}
W_{i\mathbf{k},j}^{\mathrm{opt}}=\frac{\pi}{\epsilon_{0}cn_{0}\hbar^{2}%
}\left|  d_{ij}\right|  ^{2}\sum_{m}I_{m}\mathcal{L}_{ij}\left(  \omega
_{m},k\right)  .\label{eq:W_opt2}%
\end{equation}
The transition rate due to spontaneous photon emission can also be directly
calculated from Eq.\thinspace(\ref{eq:W_opt2}) \cite{jirauschek2010monte}, but
is usually negligible compared to other scattering mechanisms in QCLs. The
gain contribution at frequency $\omega$ of a single electron in state $\left|
i\mathbf{k}\right\rangle $, i.e., $n_{i}^{\mathrm{s}}=1/S$, is obtained from
Eq.\thinspace(\ref{eq:gain}) by summing over the transitions to all available
final states $\left|  j\mathbf{k}^{\prime}\right\rangle $. With Eqs.\thinspace
(\ref{eq:d2}) and (\ref{eq:sum}), we obtain%
\begin{equation}
g\left(  \omega\right)  =\frac{\pi\omega}{\hbar\epsilon_{0}cn_{0}%
SL_{\mathrm{p}}}\sum_{j}\frac{\omega_{ij}\left(  k\right)  }{\left|
\omega_{ij}\left(  k\right)  \right|  }\left|  d_{ij}\right|  ^{2}%
\mathcal{L}_{ij}\left(  \omega,k\right)  .\label{eq:gain1}%
\end{equation}
For $E_{i\mathbf{k}}<E_{j\mathbf{k}}$, we have $\omega_{ij}\left(  k\right)
/\left|  \omega_{ij}\left(  k\right)  \right|  =-1$, indicating absorption and
thus resulting in a negative contribution to the optical gain
\cite{2010ApPhL..96a1103J}. For EMC simulations, the gain can be evaluated by
summing Eq.\thinspace(\ref{eq:gain1}) over all simulated electrons $n$ in the
corresponding states $\left|  i_{n}\mathbf{k}\right\rangle $ of the central
simulation period(s) \cite{2010ApPhL..96a1103J},%
\begin{equation}
g\left(  \omega\right)  =\frac{\pi\omega}{\hbar\epsilon_{0}cn_{0}%
SL_{\mathrm{p}}}\sum_{n}\sum_{j}\frac{\omega_{i_{n}j}\left(  k\right)
}{\left|  \omega_{i_{n}j}\left(  k\right)  \right|  }\left|  d_{i_{n}%
j}\right|  ^{2}\mathcal{L}_{i_{n}j}\left(  \omega,k\right)  ,\label{eq:gain2}%
\end{equation}
where $j$ also includes states in adjacent periods. $S$ then corresponds to
the simulated device in-plane cross section, which is for a sheet doping
density per period $n^{\mathrm{s}}$ given by $S=N_{\mathrm{e}}/\left(
n^{\mathrm{s}}N_{\mathrm{p}}\right)  $, where $N_{\mathrm{P}}$\ corresponds to
the number of periods over which the $N_{\mathrm{e}}$ simulated electrons are distributed.

For rate equations, the electron wave vector dependence is not considered, and
thus a $k$ independent averaged value $\gamma_{ij}$ has to be taken in Eq.
(\ref{eq:Lor2}). Furthermore, $\omega_{ij}$ does not depend on the wave vector
if nonparabolicity effects are neglected. Then the gain simplifies to Eq.
(\ref{eq:gain}), thus becoming%
\begin{equation}
g=\frac{\pi\omega}{\hbar\epsilon_{0}cn_{0}L_{\mathrm{p}}}\sum_{i,j}\left|
d_{ij}\right|  ^{2}\left(  n_{i}^{\mathrm{s}}-n_{j}^{\mathrm{s}}\right)
\mathcal{L}_{ij}\left(  \omega\right)  .\label{eq:gain3}%
\end{equation}

The intensity evolution for a mode $m$ is again given by Eq.\thinspace
(\ref{eq:int2}),%
\begin{equation}
n_{0}c^{-1}\partial_{t}I_{m}=\Gamma_{m}g\left(  \omega_{m}\right)  I_{m}%
-a_{m}I_{m}.\label{eq:Im}%
\end{equation}

\begin{figure}[ptb]
\includegraphics{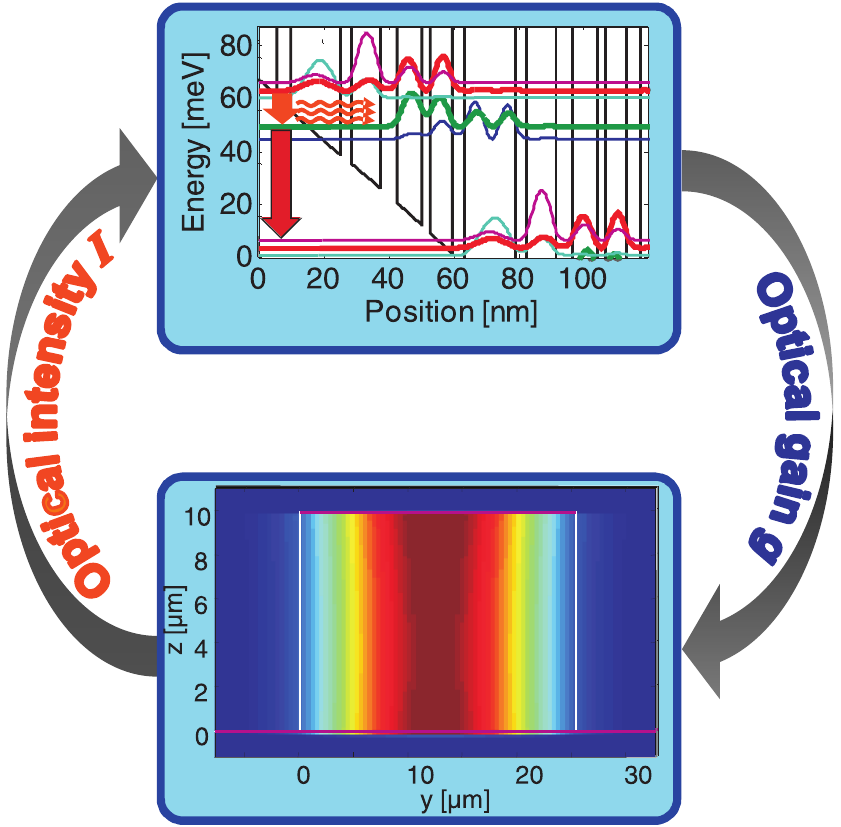}
\caption{{}(Color online) Illustration of coupled simulations including both
the carrier transport and optical cavity field.}%
\label{fig:coupled}%
\end{figure}

The carrier transport and intensity evolution have to be simulated by a
coupled approach, as illustrated in Fig.\,\ref{fig:coupled}. The carrier
transport simulation based on rate equations or EMC now includes the
photon-induced transition rates Eq.\,(\ref{eq:W_opt2}), which depend on the
intensities given by Eq.\,(\ref{eq:Im}). On the other hand, Eq.\,(\ref{eq:Im})
depends on the gain which is extracted from the carrier transport simulations
by using Eq.\,(\ref{eq:gain2}) or Eq.\,(\ref{eq:gain3}).

\subsection{Selected EMC simulation results}

In the following, some EMC simulation\ results are presented for both
mid-infrared and terahertz QCLs, illustrating the versatility of this
approach. Examples without inclusion of the optical cavity field are shown in
Sections \ref{sec:scat}--\ref{sec:elast}, see Figs.\thinspace
\ref{fig:scattering2}, \ref{fig:ee} and \ref{fig:IR} where results for the
unsaturated optical gain are displayed. Here we focus on coupled simulations
of the carrier transport and the optical cavity field.

\begin{figure}[ptb]
\includegraphics{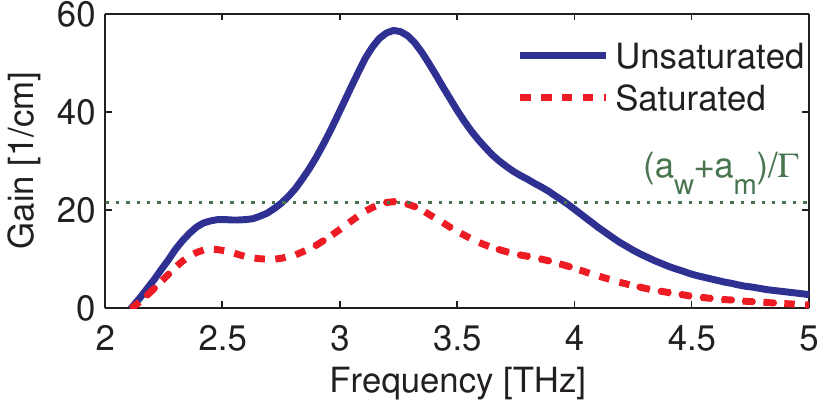}
\caption{{}(Color online) Simulated unsaturated and saturated power gain
coefficient vs frequency. The dashed line indicates the threshold gain
\cite{2010ApPhL..96a1103J}. Reprinted with permission from Appl. Phys. Lett.
\textbf{96}, 011103 (2010). Copyright 2010 American Institute of Physics.}%
\label{fig:gsat}%
\end{figure}

\begin{figure}[ptb]
\includegraphics{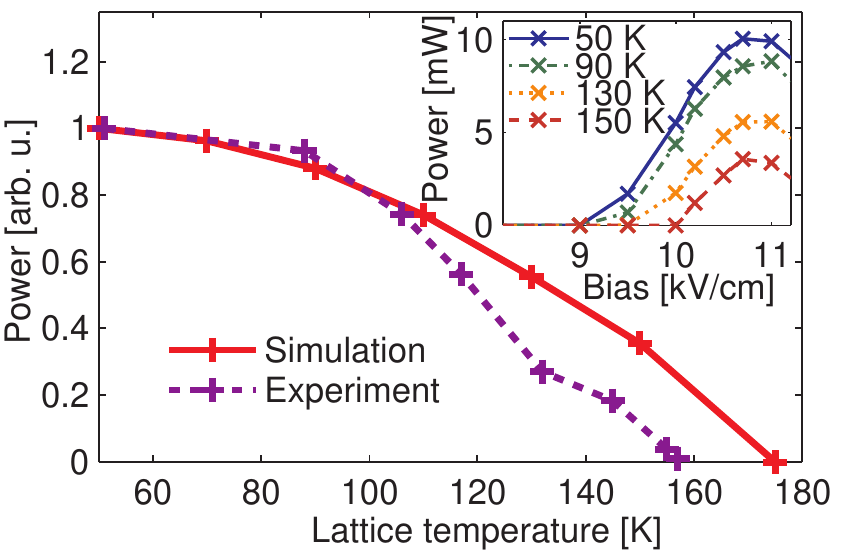}
\caption{{}(Color online) Comparison of the measured optical power vs lattice
temperature \cite{2005OExpr..13.3331W} to EMC simulation results; the inset
contains the simulated power vs applied bias for various lattice temperatures
\cite{2010ApPhL..96a1103J}. Reprinted with permission from Appl. Phys. Lett.
\textbf{96}, 011103 (2010). Copyright 2010 American Institute of Physics.}%
\label{fig:lase}%
\end{figure}

The EMC method has been applied to study the optical power and gain saturation
in terahertz QCLs, yielding good agreement with experimental data
\cite{2010ApPhL..96a1103J}. Here, examples are presented for a high
temperature QCL lasing up to $164\,\mathrm{K}$ \cite{2005OExpr..13.3331W}. In
Fig.\,\ref{fig:gsat}, the simulated unsaturated spectral gain curve is
compared to the result obtained for carrier-light coupling included,
demonstrating gain saturation and clamping at the threshold gain value due to
the lasing field. In Fig.\,\ref{fig:lase}, the obtained optical output power
as a function of lattice temperature is compared to experimental data,
demonstrating good agreement with experiment.

\begin{figure}[ptb]
\includegraphics{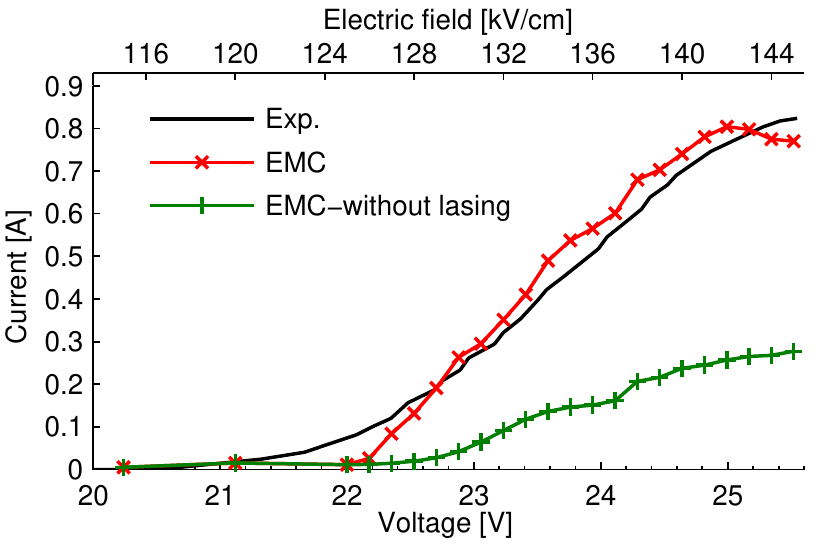}
\caption{(Color online) The measured current-voltage characteristics for a
high-efficiency mid-infrared QCL \cite{2010NaPho...4...99B} is compared to
results obtained from EMC simulations with and without lasing included
\cite{matyas2011photon}. Reprinted with permission from J. Appl. Phys.
\textbf{110}, 013108 (2011). Copyright 2011 American Institute of Physics.}%
\label{fig:mir}%
\end{figure}

Due to the stimulated optical transitions, the lasing field significantly
influences the carrier transport, affecting not only the subband populations
and thus the optical gain, but also the electric current. This is especially
the case for high efficiency mid-infrared QCLs. In Fig.\,\ref{fig:mir}, the
measured current density of a mid-infrared QCL with $\approx50\%$ wall-plug
efficiency \cite{2010NaPho...4...99B} is compared to EMC results with and
without carrier-light coupling included \cite{matyas2011photon}. Good
agreement with experiment is only obtained in the former case, while the
simulation neglecting photon-induced processes significantly underestimates
the current density, demonstrating the significant influence of these
processes on the carrier transport.

\section{\label{sec:DM}Density matrix approaches}

\begin{figure}[ptb]
\includegraphics{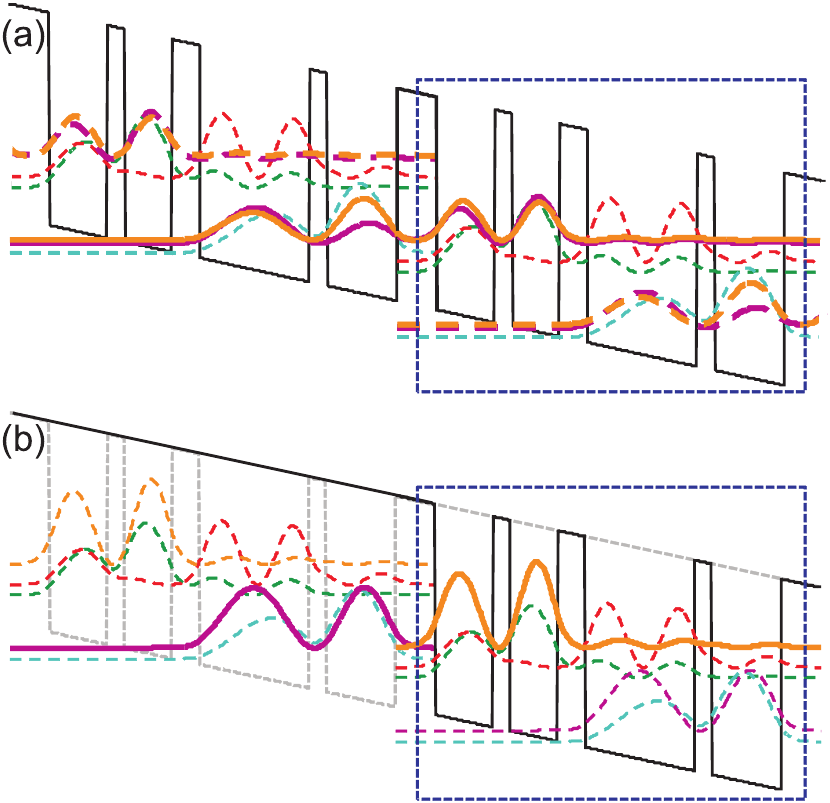}
\caption{{}(Color online) Conduction band profile and probability densities
for a terahertz QCL, computed based on (a) the actual potential $V$ modeled as
a periodic sequence of stages, and (b) the tight-binding conduction band
profile $V_{\mathrm{tb}}$ obtained by extending the barriers at the stage
boundaries to confine the wavefunctions within the stage. The rectangles
denote a single stage. The extended wavefunctions spanning the thick barrier
and the corresponding localized wavefunctions are marked by bold lines.}%
\label{fig:dm}%
\end{figure}

The one-dimensional density matrix approach is frequently used for the
analysis and optimization of QCLs
\cite{fathololoumi2012terahertz,2009PhRvB..80x5316K,2010PhRvB..81t5311D,2010NJPh...12c3045T}%
. Also three-dimensional versions have been developed
\cite{2001PhRvL..87n6603I,2009PhRvB..79p5322W}. One-dimensional density matrix
approaches can be seen as a quantum mechanical generalization of rate
equations, Eq.\thinspace(\ref{eq:rate}), to include effects such as resonant
tunneling and dephasing. In the semiclassical description, transport through a
barrier occurs instantaneously due to electrons being scattered into wave
functions which are spatially extended across the barrier
\cite{2005JAP....98j4505C}, see Fig.\thinspace\ref{fig:dm}(a). In the quantum
mechanical picture, the electron transport across the barrier is desribed by a
coherent superposition of the extended states with a narrow anticrossing
energy gap $\Delta_{E}$, resulting in a localized electron wavepacket. Due to
the coherent time evolution of these states, the wavepacket oscillates between
the left and right well with the so-called Rabi oscillation frequency
$\Omega=\Delta_{E}/\hbar$ \cite{2005JAP....98j4505C}, and the corresponding
tunneling time corresponds to half the duration of an oscillation cycle,
$\tau_{\mathrm{tun}}=\pi/\Omega$ \cite{2009PhRvB..80x5316K}. The semiclassical
picture is adequate as long as the Rabi oscillations are not significantly
dampened by dephasing, since then the wavepacket oscillates uniformly between
the wells and the averaged population distribution corresponds to the
semiclassical description. However, for strong dephasing with relaxation times
$\tau\leq$ $\tau_{\mathrm{tun}}$, the wave packet no longer oscillates.
Rather, the electrons accumulate in the left well, and the semiclassical
picture of a uniform electron distribution across the extended states is no
longer valid \cite{2005JAP....98j4505C}. Now the current transport is not
limited by the scattering-induced electron transport, but by tunneling through
the barrier. This effect is commonly referred to as wave function localization
due to dephasing, and occurs for thick barriers with narrow anticrossing gaps
$\Delta_{E}$ and thus long tunneling times $\tau_{\mathrm{tun}}$
\cite{2005JAP....98j4505C}. Especially in terahertz QCLs, thick injection
barriers are used with $\Delta_{E}\approx1~\mathrm{meV}$
\cite{2005JAP....98j4505C}, corresponding to $\tau_{\mathrm{tun}}$ of around
$2~\mathrm{ps}$, while the dephasing time can be estimated from the
spontaneous emission linewidth of approx $5~\mathrm{meV}$
\cite{2005JAP....98j4505C} to be $\approx0.3~\mathrm{ps}$.

Rather than describing localized wavepackets by a coherent superposition of
eigenstates, frequently localized wave functions are used
\cite{2009PhRvB..80x5316K,2010PhRvB..81t5311D,2010NJPh...12c3045T}, see
Fig.\thinspace\ref{fig:dm}(b). Here, the wave functions are computed for each
QCL period separately, by assuming that the thick injection barriers at the
left and right side of the period are infinitely thick
\cite{2009PhRvB..80x5316K}. The Schr\"{o}dinger equation, Eq.\thinspace
(\ref{eq:s1D}), is then solved for the corresponding tight-binding conduction
band profile $V_{\mathrm{tb}}$ rather than the actual potential $V$. The
corresponding Hamiltonian is obtained in the framework of tight-binding theory
\cite{Bastard:88}. Specifically, the Rabi frequency for a doublet of states
spanning the coupling barrier is given by $\Omega_{ij}=\hbar^{-1}\left\langle
i\right|  V-V_{\mathrm{tb}}\left|  j\right\rangle $ \cite{2009PhRvB..80x5316K}%
. Depending on the investigated QCL structure, resonant tunneling may have to
be considered at more than a single coupling barrier in each period
\cite{2010PhRvB..81t5311D,2010NJPh...12c3045T}. Then the period is subdivided
into smaller regions, and the tight binding formalism has to be applied accordingly.

Wave functions $\psi_{i}$ represent pure states, and thus are not adequate for
describing decoherence effects due to the interaction with the environment.
This can be accomplished by means of a density matrix $\rho_{ij}$ (see also
Sec.\thinspace\ref{sec:PDM}), which can represent both pure and mixed states
and thus is suitable to model dephasing. The diagonal terms $\rho_{ii}$
correspond to the occupation of level $i$, and the off-diagonal elements
$\rho_{ij}$ are the coherence or polarization terms for the doublet $i,j$
\cite{2005JAP....98j4505C}. The density matrix can be normalized so that the
diagonal terms correspond to the sheet density, $\rho_{ii}=n_{i}^{\mathrm{s}}%
$. The time evolution of the density matrix $\rho_{ij}\left(  t\right)  $\ is
described by the von Neumann equation%
\begin{equation}
\mathrm{i}\hbar\mathrm{d}_{t}\rho_{ij}=\sum_{\ell}\left(  H_{i\ell}\rho_{\ell
j}-\rho_{i\ell}H_{\ell j}\right)  ,\label{eq:dm}%
\end{equation}
where the Hamiltonian matrix elements are defined as $H_{ij}=\left\langle
i\right|  \hat{H}\left|  j\right\rangle $. For describing the carrier
transport in QCLs, the Hamiltonian is divided into two parts,
$\hat{H}=\hat{H}_{0}+\hat{H}^{\prime}$. Here $\hat{H}_{0}$ describes the
coherent evolution of the quantum system due to the conduction band potential
while $\hat{H}^{\prime}$ contains the perturbation potentials corresponding to
the various scattering mechanisms and introduces dissipation to the system.
$\hat{H}^{\prime}$ is commonly implemented in a somewhat phenomenological
manner by using transition and dephasing rates, similar as for the rate
equation approach discussed in Sections \ref{sec:PRE}\ and \ref{sec:SCRE}.
Equation (\ref{eq:dm}) can then be cast into the form
\cite{2010NJPh...12c3045T,2009PhRvB..80x5316K}%
\begin{align}
\mathrm{d}_{t}n_{i}^{\mathrm{s}} &  =\sum_{j\neq i}\tau_{ji}^{-1}%
n_{j}^{\mathrm{s}}-\tau_{i}^{-1}n_{i}^{\mathrm{s}}+\sum_{j}\mathrm{i}%
\Omega_{ij}\left(  \rho_{ij}-\rho_{ji}\right)  ,\nonumber\\
\mathrm{d}_{t}\rho_{ij} &  =\mathrm{i}\Omega_{ij}\left(  n_{i}^{\mathrm{s}%
}-n_{j}^{\mathrm{s}}\right)  -\mathrm{i}\rho_{ij}\omega_{ij}-\gamma_{ij}%
\rho_{ij},\label{eq:dm2}%
\end{align}
where $\omega_{ij}=\left(  E_{i}-E_{j}\right)  /\hbar$, and $\tau_{i}%
^{-1}=\sum_{j\neq i}\tau_{ij}^{-1}$ indicates the total inverse lifetime of
level $i$. The scattering rates $\tau_{ij}^{-1}$ have already been discussed
in Sections \ref{sec:PRE}\ and \ref{sec:SCRE}. The dephasing rate is given by
\cite{2005JAP....98j4505C}%
\begin{equation}
\gamma_{ij}=\left(  \tau_{i}^{-1}+\tau_{j}^{-1}\right)  /2+\tau_{\mathrm{pure}%
,ij}^{-1}.\label{eq:tau_pure}%
\end{equation}
Here, $\left(  \tau_{i}^{-1}+\tau_{j}^{-1}\right)  /2$ is the lifetime
broadening contribution [see also Eq.\thinspace(\ref{eq:gij})], and
$\tau_{\mathrm{pure},ij}^{-1}$ contains the pure dephasing. The Rabi frequency
$\Omega_{ij}$ is nonzero only for doublets spanning a coupling barrier;
specifically, $\Omega_{ii}=0$. In Eq.\thinspace(\ref{eq:dm2}), resonant
tunneling is assumed to be independent of the in-plane wave vector, and the
rates $\tau_{ij}^{-1},\gamma_{ij}$ are averaged over the kinetic electron
distribution within the subbands \cite{2009PhRvB..80x5316K}. The stationary
solution of Eq.\thinspace(\ref{eq:dm2}) is obtained by setting $\mathrm{d}%
_{t}=0$, yielding \cite{2010NJPh...12c3045T}%
\begin{equation}
\sum_{j\neq i}\left[  \tau_{ji}^{-1}n_{j}^{\mathrm{s}}+R_{ij}\left(
n_{j}^{\mathrm{s}}-n_{i}^{\mathrm{s}}\right)  \right]  -\tau_{i}^{-1}%
n_{i}^{\mathrm{s}}=0\label{eq:dmrate}%
\end{equation}
with%
\begin{equation}
R_{ij}=\frac{2\Omega_{ij}^{2}\gamma_{ij}^{-1}}{1+\omega_{ij}^{2}\gamma
_{ij}^{-2}}.\label{eq:Rij}%
\end{equation}
Furthermore, Eq.\thinspace(\ref{eq:ns}) has to be fulfilled, i.e., the total
sheet density in each period is determined by the doping sheet density
$n^{\mathrm{s}}$.

Equation (\ref{eq:dm2}) can be solved using empirical rates, or implemented in
a self-consistent manner by calculating the scattering rates $\tau_{ij}^{-1}$
as discussed in Section \ref{sec:SCRE}. The pure dephasing rate $\tau
_{\mathrm{pure},ij}^{-1}$ in Eq.\thinspace(\ref{eq:tau_pure}) can be computed
based on intrasubband scattering transitions \cite{2010NJPh...12c3045T,unuma},
but is often treated by assuming an empirical value
\cite{2005JAP....98j4505C,2010PhRvB..81t5311D}. In a simplified model, the
transport across the barrier can be restricted to the coupling between two
states $1$ and $2$. For this case, Eq.\thinspace(\ref{eq:dmrate}) simplifies
with $\tau_{12}^{-1}=\tau_{21}^{-1}=0$ to $\tau_{2}^{-1}n_{2}^{\mathrm{s}%
}+R_{12}\left(  n_{2}^{\mathrm{s}}-n_{1}^{\mathrm{s}}\right)  =0$. The current
density through the barrier is then given by $J=eR_{12}\left(  n_{1}%
^{\mathrm{s}}-n_{2}^{\mathrm{s}}\right)  =e\tau_{2}^{-1}n_{2}^{\mathrm{s}}$.
With Eq.\thinspace(\ref{eq:Rij}) and $n_{1}^{\mathrm{s}}+n_{2}^{\mathrm{s}%
}=n^{\mathrm{s}}$, we obtain \cite{Kaz}%
\begin{equation}
J=\frac{2en^{\mathrm{s}}\Omega_{12}^{2}\gamma_{12}^{-1}}{1+\omega_{12}%
^{2}\gamma_{12}^{-2}+4\Omega_{12}^{2}\gamma_{12}^{-1}\tau_{2}}.
\end{equation}

In the three-dimensional density matrix method, additionally the in-plane wave
vector $\mathbf{k}$ is taken into account, i.e., the electrons are described
by states $\left|  i\mathbf{k}\right\rangle $ rather than $\left|
i\right\rangle $. Consequently, the density matrix is given by $\rho
_{i\mathbf{k},j\mathbf{k}^{\prime}}$, and $i,j,\ell$\ in Eq.\thinspace
(\ref{eq:dm}) have to be replaced by $i\mathbf{k}$, $j\mathbf{k}^{\prime}$ and
$\ell\mathbf{k}^{\prime\prime}$, respectively. Three-dimensional density
matrix approaches can be seen as a quantum mechanical generalization of the
Boltzmann equation given in Eq. (\ref{eq:boltz}). Various approaches based on
the three-dimensional density matrix have been developed for QCL simulation,
where the scattering mechanisms are self-consistently implemented based on the
corresponding Hamiltonians \cite{2001PhRvL..87n6603I,2009PhRvB..79p5322W}.
Specifically, also a hybrid density matrix-Monte Carlo approach has been
introduced, where the tunneling transport through the coupling barrier is
treated based on the density matrix formalism, while scattering inside each
period is treated semiclassically using an EMC approach
\cite{2005JAP....98j4505C}.

\section{\label{sec:NEGF}Non-equilibrium Green's function formalism}

\subsection{\label{sec:NEGF_intro}General scope of the non-equilibrium Green's
function method}

Quantum cascade devices utilize charge transport in structures of the
nanometer length scale. In this regime, quantum effects such as coherent
tunneling, interference and confinement play a very important role for the
transport physics. At the same time, however, the devices are run at finite
temperatures, which support a significant amount of scattering with phonons.
Since most devices are doped, alloyed and/or based on heterojunctions of
different materials, impurity and alloy disorder as well as surface and
interface roughness influence the transport, too. All these scattering effects
share the fact that the full quantum information (i.e. the full phonon phase,
the precise position of the impurities, etc.) is either unknown or lost in the
statistics of large numbers of scattering events. Consequently, these effects
contribute to incoherent scattering and dephasing.

It is well established that the nonequilibrium Green's function theory (NEGF)
is the most general scheme for the prediction of coherent and incoherent
quantum transport. Since its introduction in the
1960s,~\cite{Schwinger,kadanoff_book, Keldysh_1965} this formalism has been
successfully applied on a great variety of transport problems. These problems
include, but are not limited to spin,~\cite{Niko_PRL05,suzuki_spin,spin_Kubis}
phonon~\cite{Xu_phonon,Yamamoto_CNT_phonon,Luisier_phonon_phonon} and electron
transport~\cite{Luisier_transistor,Do_2006}, covering different materials such
as metals~\cite{Chen_andreev,Ke_metal}, semiconductors~\cite{Bulusu_2007},
topological insulators,~\cite{2012PhysRevB..86.085131H} and even various
dimensionalities such as layered structures,~\cite{wacker}
nanotubes,~\cite{Yamamoto_CNT_phonon,Lazzeri_hot_phonon_CNT}
fullerenes,~\cite{Li_C60} and molecules.~\cite{Sato_organic,Damle_2002} For
electronic transport, the NEGF method has been implemented in a large variety
of representations, ranging from envelope function approximations such as the
effective mass~\cite{2009LaserPhysics..19.762K,2009JCompElec..8.267B} and
\textbf{k.p} method,~\cite{2013IEEETransonElecDev..60.2111H} to atomistic
representations such as tight
binding~\cite{2013IEEETransonElecDev..60.2171M,2013ApplPhysLett..102.193501J}
and even density functional theory models.~\cite{Thygesen} In most cases, the
NEGF method is used for transport in open systems, i.e. devices connected to
spin, charge or heat reservoirs via semi-infinite leads. Nevertheless, two
different approaches have been successfully applied to mimic the
field-periodic conditions cascade structures are facing.~\cite{lee,IEEE_Harrison,2009PhRvB..79s5323K,Kolek_openQCL}

It is this high flexibility of NEGF that offers choosing specialized basis
functions such as e.g. Wannier-Stark states for cascade structures,~\cite{lee}
wire modes for homogeneous wire systems~\cite{Luisier_transistor} or generic
basis functions for general cases within the low rank
approximation.~\cite{ZengLRA} It is true for all these cases that the closer
the basis functions match the actual quasi particles of the system, the fewer
basis functions are required to reliably predict the device performance and
the more efficient the numerical implementation of NEGF will be. It is worth
to mention here that the NEGF method natively contains more information than
semiclassical methods such as the Boltzmann equation. In fact, it has already
been shown in the 1960s how to approximate the NEGF equations to yield the
Boltzmann equation.~\cite{kadanoff_book}

The NEGF method, however, faces one major drawback compared to most other
methods: the numerical solution of the NEGF equations is expensive both in
terms of memory and CPU time. The numerical costs are particularly high if
incoherent scattering in the self-consistent Born approximation is considered.
For this reason, most numerical implementations of NEGF on real devices do not
include electron-electron scattering beyond the first order (Hartree)
approximation. Higher orders of electron-electron scattering (i.e., exchange
terms) are required to model energy transfer during inelastic
electron-electron scattering, but their numerical load typically prohibits the
implementation on concrete transport problems.

\subsection{\label{sec:overview_NEGF}Overview of the non-equilibrium Green's
function method}

\subsubsection{Fundamental equations and observables}

As discussed in Section~\ref{sec:SE}, electrons in quantum cascade lasers can
be successfully described within the effective mass approximation. Hereby, the
devices are typically considered as laterally homogeneous quantum well
heterostructures. The electronic structure is then represented with the
Hamiltonian%
\begin{align}
\hat{H}_{0}  &  =\frac{-\hbar^{2}}{2}\partial_{z}\frac{1}{m^{\ast}\left(
z,E\right)  }\partial_{z}+\frac{\hbar^{2}k^{2}}{2m^{\parallel}\left(
z,E\right)  }+V\left(  z\right)  ,\nonumber\\
V\left(  z\right)   &  =V_{\mathrm{c}}\left(  z\right)  -e\Phi\left(
z\right)  ,
\end{align}
where $k$ is the lateral electron momentum, $\Phi\left(  z\right)  $ the
electrostatic potential, and $V_{\mathrm{c}}\left(  z\right)  $ denotes the
material and position dependent conduction band edge, including the band
offsets. Note that the effective mass is energy-dependent to include the
nonparabolocity as described in detail in Sec.~\ref{sec:nonparabolicity}. In
the stationary limit, the NEGF method describes transport with the electronic
retarded and lesser Green's function ${G}^{R}$ and ${G}^{<}$, respectively.
These functions solve four coupled partial differential equations that read
in operator form for a given energy $E$ and in-plane 
momentum $\mathbf{k}$~\cite{Haug}
\begin{subequations}
\label{set1}%
\begin{align}
G^{R}  &  =\left(  E\hat{1}-\hat{H}_{0}-\Sigma^{R}\right)  ^{-1}%
,\label{set1a}\\
G^{<}  &  =G^{R}\Sigma^{<}G^{R\dag},\label{set1b}\\
\Sigma^{<}  &  =G^{<}D^{<},\label{set1c}\\
\Sigma^{R}  &  =G^{R}D^{R}+G^{R}D^{<}+G^{<}D^{R}. \label{set1d}%
\end{align}
\end{subequations}
Here, $\Sigma^{R}$ and $\Sigma^{<}$ denote the retarded and lesser
self-energies and $D^{R}$ and $D^{<}$ are the sum of retarded and lesser
Green's function of the environment. Equations~(\ref{set1a}) and (\ref{set1b})
are also referred to as Dyson and Keldysh equation, respectively. The
expressions for the self-energies, Eqs.~(\ref{set1c}) and (\ref{set1d}), are
discussed in Sec.~\ref{sec:NEGF_scattering} for various scattering mechanisms.
We note that in that section and in the rest of this paper all Green's functions
and self-energies are given in real space representation which requires a 
transformation of Eqs.~(\ref{set1}) from operator space into real space. 
Thus, the Green's functions and self-energies given in the following have the dimensions 
$\mathrm{eV}^{-1} \mathrm{m}^{-1}$ and 
$\mathrm{eV} \mathrm{m}^{-1}$, respectively, unless stated otherwise.
Green's functions and self-energies are solved in the self-consistent Born
approximation, i.e., they are solved iteratively until convergence is
achieved. Most of the known current conserving simplifications of the
self-energies do not ease the convergence of the self-consistent Born
iterations, but reduce the numerical burden of each individual iteration only.
In contrast to methods that require the solution of the Schr\"{o}dinger
equation, the solutions of Eqs.~(\ref{set1}) do not require to solve an
eigenvalue problem. Consequently, energy dependent effective masses in the
Hamiltonian $\hat{H}_{0}$ do not increase the numerical complexity of NEGF.
The Green's functions and self-energies are functions of two spatial
coordinates $z,z^{\prime}$, the absolute value of the lateral momentum $k$ and
the energy $E$. The energy and spatially resolved spectral function $A\left(
z,E\right)  =\mathrm{i}\left[  G^{R}\left(  z,z,0,E\right)  -G^{R\dag}\left(
z,z,0,E\right)  \right]  $ shows width and location of resonant states in the
system. This indicates that $G^{R}$ contains the information of resonant
states and the density of states. The $G^{<}$ function contains in addition
information of the occupancy of the electronic states. Consequently, occupancy
related observables such as density, current and optical gain are dependent on
$G^{<}$: the spatially and energy-resolved density $n\left(  z,E\right)  $ and
current density $J\left(  z,E\right)  $ are defined in relation to the density
$n\left(  z\right)  $ and current density $J\left(  z\right)  $, respectively%
\begin{align}
&  n\left(  z\right)  =\int\mathrm{d}En\left(  z,E\right) \nonumber\\
&  =\frac{2}{\left(  2\pi\right)  ^{3}}\text{$\Im$}\int\mathrm{d}%
E\int\mathrm{d}^{2}kG^{<}\left(  z,z,k,E\right)  , \label{n_of_z_E}%
\end{align}%
\begin{align}
&  J\left(  z\right)  =\int\mathrm{d}EJ\left(  z,E\right) \nonumber\\
&  =-\frac{\hbar e}{\left(  2\pi\right)  ^{3}}\lim_{z^{\prime}\rightarrow
z}\int\mathrm{d}E\int\mathrm{d}^{2}k\frac{1}{m^{\ast}\left(  z,E\right)
}\nonumber\\
&  \times\text{$\Re$}\left(  \partial_{z}-\partial_{z^{\prime}}\right)
G^{<}\left(  z,z^{\prime},k,E\right)  . \label{j_of_z_E}%
\end{align}
The current density in Eq.~(\ref{j_of_z_E}) is only correct as long as the
kinetic energy operator is the only term in the Hamiltonian that does not
commute with the position operator. This can be seen from the fundamental
probability current density operator in real space $\mathbf{x}$ and time $t$
representation (see e.g. Refs.~\onlinecite{ Fetter,Mahan})%
\begin{equation}
\left\langle \hat{J}\left(  \mathbf{x}_{1},t_{1}\right)  \right\rangle
=\lim_{t_{2}\rightarrow t_{1}}\lim_{\mathbf{x}_{2}\rightarrow\mathbf{x}_{1}%
}\mathrm{i}\hbar\hat{v}G^{<}\left(  \mathbf{x}_{1},t_{1};\mathbf{x}_{2}%
,t_{2}\right)  \label{def_general_j}%
\end{equation}
with the velocity operator%
\begin{equation}
\hat{v}=\frac{\mathrm{i}}{\hbar}\left[  \hat{H},\hat{x}\right]  _{-}.
\label{def_velo}%
\end{equation}
Please note that the Green's function in Eq.~(\ref{def_general_j}) is given 
in real space $\mathbf{x}$ and time $t$ representation and has the 
dimension $\mathrm{eV}^{-1}\mathrm{m}^{-3} \mathrm{s}^{-1}$.
The optical field amplitude absorption coefficient $\alpha\left(
z,\omega\right)  $ for a photon of frequency $\omega$ is a function of the
permittivity $\epsilon\left(  z,\omega\right)  $~\cite{jackson},
\begin{align}
&  \alpha\left(  z,\omega\right) \nonumber\\
&  =\Im\left[  \frac{\epsilon\left(  z,\omega\right)  }{\epsilon_{0}}\right]
\frac{\sqrt{2}\omega}{c}\left\{  \Re\left[  \frac{\epsilon\left(
z,\omega\right)  }{\epsilon_{0}}\right]  +\left|  \frac{\epsilon\left(
z,\omega\right)  }{\epsilon_{0}}\right|  \right\}  ^{-1/2}.
\end{align}
Here, $\epsilon_{0}$ is the vaccum permittivity and $c$ denotes the speed of
light. The optical absorption coefficient is usually defined with respect to
the field intensity ($a$) rather than with respect to the field
amplitude~\cite{1986lase.book.....S}, see Sec.~(\ref{sec:Reso}). In
particular, in the gain regime one often refers to the power gain $-a$. These
quantities are related by $a=2\alpha$. The permittivity depends on the complex
conductance $\sigma\left(  z,\omega\right)  $ and the materials dielectric
constant $\epsilon_{r}\left(  z\right)  $,%

\begin{equation}
\epsilon\left(  z,\omega\right)  =\epsilon_{0}\epsilon_{r}\left(  z\right)
+\mathrm{i}\sigma\left(  z,\omega\right)  /\omega.
\end{equation}
Before lasing starts and the perturbation $\delta\hat{V}\left(  \omega\right)
$ due to the optical field is still small, the optical absorption can be
extracted from the linear response of the Green's functions. To first order,
the change of the lesser Green's function is given by (for a given
$\mathbf{k}$)~\cite{2002PhRvB..66h5326W}
\begin{align}
&  \delta G^{<}\left(  \omega,E\right) \nonumber\\
&  =G^{R}\left(  E+\hbar\omega\right)  \delta\hat{V}\left(  \omega\right)
G^{<}\left(  E\right) \nonumber\\
&  +G^{<}\left(  E+\hbar\omega\right)  \delta\hat{V}\left(  \omega\right)
G^{R\dag}\left(  E\right) \nonumber\\
&  +G^{R}\left(  E+\hbar\omega\right)  \delta\Sigma^{R}\left(  E+\hbar
\omega,-E\right)  G^{<}\left(  E\right) \nonumber\\
&  +G^{<}\left(  E+\hbar\omega\right)  \delta\Sigma^{R\dag}\left(
E+\hbar\omega,-E\right)  G^{R\dag}\left(  E\right) \nonumber\\
&  +G^{R}\left(  E+\hbar\omega\right)  \delta\Sigma^{<}\left(  E+\hbar
\omega,-E\right)  G^{R\dag}\left(  E\right)  . \label{delta_Glesser}%
\end{align}
Changes to the scattering self-energies $\delta\Sigma$ due to the photonic
perturbation $\delta\hat{V}$ are first order vertex corrections.

The wavelengths of terahertz lasers are much larger than the typical
dimensions of the QCL periods. Thus, for the electron transport calculation,
the optical electric field can be considered constant in the active device.
Since the QCL laser light is usually linearly polarized in transport direction
$z$, the perturbing potential reads in Coulomb gauge%
\begin{equation}
\delta\hat{V}\left(  \omega\right)  =\frac{\hbar e}{m^{\ast}\left(  z\right)
\omega}E_{z}\left(  \omega\right)  \partial_{z} \label{delta_V_of_omega}%
\end{equation}
with the photon electric field component in $z$ direction $E_{z}$. This
results in the change of the current density $\delta J\left(  z,\omega\right)
$ in linear order of $\delta\hat{V}\left(  \omega\right)  $%
\begin{align}
&  \delta J\left(  z_{1},\omega\right) \nonumber\\
&  =-\lim_{z_{2}\rightarrow z_{1}}\frac{\hbar^{2}e}{m^{\ast}\left(
z_{1}\right)  }\left(  \partial_{z_{1}}-\partial_{z_{2}}\right) \nonumber\\
&  \times\int\frac{\mathrm{d}E}{2\pi\hbar}\int\frac{\mathrm{d}^{2}k}{\left(
2\pi\right)  ^{2}}\delta G^{<}\left(  z_{1},z_{2},k,\omega,E\right)
\nonumber\\
&  -\frac{2\hbar e^{2}E_{z}\left(  \omega\right)  }{m^{\ast}\left(
z_{1}\right)  \omega}\int\frac{\mathrm{d}E}{2\pi\hbar}\int\frac{\mathrm{d}%
^{2}k}{\left(  2\pi\right)  ^{2}}G^{<}\left(  z_{1},z_{1},k,E\right)  .
\label{delta_j}%
\end{align}
The first term of the last equation results from the current operator in
Eq.~(\ref{j_of_z_E}) applied on the change of the lesser Green's functions
$\delta G^{<}$. The second term results from the change of the velocity and of
the current operator due to the fact that $\delta\hat{V}\left(  \omega\right)
$ does not commute with the position operator [see Eq.~(\ref{def_velo})]. The
quotient of the perturbation of the current density $\delta J\left(
z,\omega\right)  $ and the electric field $E_{z}\left(  \omega\right)  $ of
the photon gives us the optical conductance%
\begin{equation}
\sigma\left(  z,\omega\right)  =\frac{\delta J\left(  z,\omega\right)  }%
{E_{z}\left(  \omega\right)  }.
\end{equation}
If vertex corrections $\delta\Sigma$ are ignored for simplicity,
Eqs.~(\ref{delta_V_of_omega}), (\ref{delta_Glesser}), and (\ref{delta_j}) can
be combined into an equation for the optical conductance%
\begin{align}
&  \sigma\left(  z_{1},\omega\right) \nonumber\\
&  =\lim_{z_{2}\rightarrow z_{1}}\frac{\hbar^{2}e^{2}}{m^{\ast}\left(
z_{1}\right)  ^{2}\left(  2\pi\right)  ^{3}\omega}\left(  \partial_{z_{1}%
}-\partial_{z_{2}}\right)  \int\mathrm{d}E\mathrm{d}^{2}k\mathrm{d}%
z_{3}\nonumber\\
&  \times\left[  G^{R}\left(  z_{1},z_{3},k,E+\hbar\omega\right)  \left.
\partial_{z^{\prime}}G^{<}\left(  z^{\prime},z_{2},k,E\right)  \right|
_{z^{\prime}=z_{3}}\right. \nonumber\\
&  \left.  +G^{<}\left(  z_{1},z_{3},k,E+\hbar\omega\right)  \left.
\partial_{z^{\prime}}G^{A}\left(  z^{\prime},z_{2},k,E\right)  \right|
_{z^{\prime}=z_{3}}\right] \nonumber\\
&  -2e^{2}\int\frac{\mathrm{d}E\mathrm{d}^{2}k}{\left(  2\pi\right)
^{3}m^{\ast}\left(  z_{1}\right)  \omega}G^{<}\left(  z_{1},z_{1},k,E\right)
. \label{opt_sigma}%
\end{align}
When the self-consistently solved Green's functions are used for the optical
conductance, Eq.~(\ref{opt_sigma}) fully accounts for the self-consistently
calculated electron states and their non-equilibrium state occupations. The
vertex corrections $\delta\Sigma$ in Eq.~(\ref{delta_Glesser}) increase the
numerical load significantly, since they require self-consistent iterations of
$\delta G$ with $\delta\Sigma$. Both $\delta G$ and $\delta\Sigma$ depend on
the optical frequency $\omega$ in addition to the ''standard'' dependence of
the Green's function [e.g. $\delta\Sigma=\delta\Sigma\left(  z,z^{\prime
},k,E,\omega\right)  $]. Thus, such self-consistency is expensive in terms of
memory and time. It has been shown that vertex corrections narrow the optical
linewidths and increase the peak height of the absorption coefficient in
terahertz QCLs.~\cite{banit_wacker} Calculations without $\delta\Sigma$ may
not predict quantitative values of $\alpha\left(  z,\omega\right)  $, but can
serve for qualitative predictions only. If the interaction with the photonic
field should be considered beyond linear response - such as in the case of
electron transport during lasing, time dependent NEGF\ with a periodically
oscillating electric field is required. Details of this approach can be found
in Ref.~\onlinecite{Wacker_JSTQE}.

The balance between the benefits of the NEGF method and its numberical load is
strongly device physics dependent. If the electron transport is clearly
dominated by incoherent scattering and tunneling across multiple barriers is
negligible, semiclassical models are clearly more efficient than NEGF.
Otherwise, if scattering is neglible, the NEGF method has to compete with
numerically more efficient methods such as the Schr\"{o}dinger equation or the
quantum transmitting boundary method.~\cite{1992PhRvB..45.3583T} If scattering
in low order captures the physics and energy resolved information is not
desired, the density matrix method is more efficient than NEGF as well. The
strength of NEGF is that it allows to predict transport in all these regimes
and gives deep insight to the ongoing processes in any of the before mentioned
situations. It will be discussed in Sec.~(\ref{sec:NEGF_result}) in detail
that a single terahertz QCL can move from the ballistic to the scattering
dominated regime with the applied electric field. Therefore, it is appropriate
to use NEGF on these devices.

\subsubsection{Different basis representations and low rank approximations}

The solution of the NEGF equations Eqs.~(\ref{set1}) requires many
matrix-matrix products and matrix inversions. Both scale cubically with $N$,
the number of degrees of freedom the equations are discretized in. As a
result, highly resolved NEGF\ calculations can easily require modern
supercomputers.~\cite{2011IEEETransonNano..10.1464S} It is obvious that any
reduction of $N$ will reduce the number of floating point operations
signficantly.~\cite{ZengLRA} Real space discretization of Eqs.~(\ref{set1})
faces the challenge that any device feature such as differing widths of
semiconductor layers has to match to the chosen resolution. Particularly small
device features require an inhomogeneous grid in real space, increasing the
numerical complexity.

The most common technique to reduce $N$ is to transform the NEGF equations
into a system of basis functions that are equal or close to the actual
propagating eigenstates of the system's Hamiltonian. The more the basis
functions agree with the eigenfunctions of the system, the less the basis
functions couple with each other and the fewer functions need to be included
in the actual transport calculation. If the target basis has less states than
the rank $N$ of the original (real space) representation, the transformation
matrices are rectangular. In the ideal case, when the basis functions are
isolated from each other, their individual contribution to transport can be
solved and summed. Prominent examples of such rank reductions are the mode
space approaches for transport in nanowire structures. If the basis functions
are coupled, it is common to still limit the number of considered functions.
In this case, however, this approach gives only approximate results. In
general cases, transformations with rectangular matrices that reduce the rank
of the equations are called low rank approximations.~\cite{ZengLRA}

In the framework of cascade devices, a small number $N$ of Wannier or
Wannier-Stark basis functions have proven to represent quantum transport well
enough for reliable predictions. For instance, Lee et al. were able to solve
the IV characteristics and optical output performance of terahertz QCLs with
only $5$ states per period.~\cite{wacker_nondiagscatt} Although it is a common
and numerically very efficient technique, caution is required when results of
the low rank space are transformed back into the original real space
representation if the rank of the two differ a lot. Numerically, the back
transformation produces results with the high resolution of the original
space, but the reliability of the data in the real space resolution strongly
depends on how well the basis functions of the low rank space match the
physics of the device. Unphysical oscillations of the current density with
real space can indicate such reliability issues. Increasing the number of
considered basis functions in the low rank space can improve the reliability
of the approximate results.~\cite{wacker_nondiagscatt,ZengLRA}

\subsection{\label{sec:NEGF_boundaries}Boundary conditions for cascade systems}

Boundary conditions in stationary NEGF always fall into two
categories:\ boundary conditions for $G^{R}$ that are related to the density
of states and boundary conditions for $G^{<}$ that are related to state
occupancy and particle distributions. Since quantum cascade lasers consist of
many repetitions of the same building block -- a sequence of quantum wells and
barriers -- NEGF is often applied on cascade structures with field periodic
boundary conditions.~\cite{lee, Schmielau_ktyp, Harrison_qdot} Field periodic
boundary conditions for the retarded Green's function can be mimicked by
solving the NEGF equations in a system of $3$ or more periods: the innermost
period is considered as the actual active device while the boundary periods
are considered as charge suppliers only. The Poisson equation is solved with
periodic boundary condition and an applied homogeneous electric field is added
to the resulting potential. Since the source sided periods are on average on a
higher potential (and contrary for the drain sided periods), the system
effectively faces field periodic charge boundary conditions. The initial guess
for the lesser Green's function $G^{<}$ is also field periodic. Since the
system is closed, self-consistent Born iterations of the higher order
scattering terms may change the number of electrons. To maintain global charge
neutrality, the $G^{<}$ function needs rescaling between
iterations.~\cite{Harrison_qdot} Although this type of boundary treatment is
appealing for cascade systems, care on the numerical details has to be taken:
In the first iteration of the self-consistent Born approximation the system is
purely coherent. Then, the equation for $G^{R}$ is van-Hove singular whenever
the energy $E$ agrees with a resonance of the system. Otherwise, the resulting
retarded Green's function is real and the spectral function $A$ vanishes
(i.e., no electronic states are found). To solve this issue, a small
artificial retarded scattering self-energy is typically included in
Eqs.~(\ref{set1}). This self-energy maintains a minimal broadening of
resonances. This artificial broadening has to be negligible compared to the
actual scattering self-energies. Note that purely coherent and field periodic
calculations would clearly yield divergent electron energies in nonequilibrium.

Most applications of NEGF tackle transport in open systems. In open systems,
the active device is connected via semi-infinite leads with charge reservoirs
that supply or collect electrons with a constant, typically equilibrium
distribution. Technically, the boundary conditions of $G^{R}$ and $G^{<}$ are
then introduced by including contact self-energies $\Sigma_{\mathrm{con}}^{R}$
and $\Sigma_{\mathrm{con}}^{<}$ in the equations for $G^{R}$ and $G^{<}$ in
Eqs.~(\ref{set1}), respectively. In this case, the large number of cascade
periods can be mimicked by considering one or a few periods as the actual
active device and including the potential landscape and material sequences of
the adjacent periods in the calculation of $\Sigma^{R}$%
.~\cite{2009PhRvB..79s5323K,Kolek_openQCL} In this way, the presence of the
leads broadens resonant states in the active device and allows for the
solution of purely ballistic transport without adding artifical retarded
self-energies. It also allows for the assessment of individual scattering
mechanisms, since scattering and its broadening of the states is not essential
for numerical convergence and can be turned off selectively. However, the
boundary condition for $G^{<}$ requires a model for the electron distribution
in the leads. It is common in NEGF for open systems to assume equilibrium
Fermi distributions. In the case of cascade devices, however, such equilibrium
distributions do not include possible heating effects within previous cascade
periods and therefore allow results of nonperiodic electron distributions:
When heating effects exist, the electrons do not dissipate all energy they
gain while traversing the potential drop of a single period. Then, the
electron distributions at source and drain side of each period differ.~\cite{2009PhRvB..79s5323K}

Alternatively, an electron distribution can be read out of $G^{<}$ at device
positions that are equivalent to the lead/device interfaces. In this case, the
nonequilibrium distribution is defined as
\begin{equation}
f\left(  z,k,E\right)  \equiv-\mathrm{i}G^{<}\left(  z,z,k,E\right)  /A\left(
z,z,k,E\right)  , \label{def_f_of_z}%
\end{equation}
with the spectral function%
\begin{equation}
A\left(  z,z^{\prime},k,E\right)  =\mathrm{i}\left[  G^{R}\left(  z,z^{\prime
},k,E\right)  -G^{R\dag}\left(  z,z^{\prime},k,E\right)  \right]  .
\label{spectral_function}%
\end{equation}
In equilibrium, the function $f\left(  z,k,E\right)  $ equals the Fermi
distribution $f\left(  E,\mu\right)  $.~\cite{Haug} When the distribution
function of source ($z=0$) and drain ($z=L$) are assumed to be equivalent to
the distribution functions at positions in the device that are a single
QCL\ period apart
\begin{align*}
&  f\left(  0,k,E\right)  \overset{!}{=}f\left(  nL_{\mathrm{p}}%
,k,E+ne\phi\right) \\
&  f\left(  L,k,E\right)  \overset{!}{=}f\left(  L-nL_{\mathrm{p}}%
,k,E-ne\phi\right)
\end{align*}
field periodic boundary conditions for $G^{<}$ are
mimicked.~\cite{Kolek_openQCL} Here, $L_{\mathrm{p}}$ is the length of a
QCL\ period, $\phi$ is the potential drop per period and $n$ is the number of
explicitly considered QCL periods. To ensure global charge neutrality, the
Poisson equation is solved with the applied electric field at both boundaries.

\subsection{\label{sec:NEGF_scattering}Scattering self-energies}

In the following, the resulting expressions for the self-energies, given in
Eqs.~(\ref{set1c}) and (\ref{set1d}), are summarized for the most important
scattering mechanisms in QCLs. If multiple scattering mechanisms are
considered, the individual self-energies of the respective scattering
mechanisms can be summed up and the summed self-energy is then used in
Eq.~(\ref{set1}). Some of the scattering self-energies are numerically so
expensive to implement that approximations are inevitable. Some approximations
are shown that ease the numerical burden significantly but still faithfully
reproduce the scattering rates expected from Fermi's golden rule. It is worth
mentioning that all presented approximations have been carefully assessed.
Approximations that violate Pauli blocking, underestimate effects of
nonlocalility of quantum mechanics or miss important characteristics of the
scattering potentials have been avoided. Coherent effects or the balance
between incoherent and coherent QCL physics are to the best of the authors'
knowledge not affected by the presented approximations. A detailed discussion
of the validity of approximations in NEGF can be found in Ref.~\onlinecite
{Kubis_assess}. This section ends with a brief discussion on general
approximations of scattering self-energies and some numerical details for
efficient implementations.

\subsubsection{Scattering from longitudinal acoustic phonons}

To avoid many-particle Green's functions, the phonon gas is assumed to remain
unchanged by the electron propagation. This requires applying the perturbation
potential of Eq.~(\ref{eq:V_ac}) in second order and to group the phonon
creation with annihilation operators.~\cite{Landau} The resulting products of
phonon and electron operators are then translated into the electron and phonon
Green's functions in Eqs.~(\ref{set1}) for the lesser and retarded
self-energies~\cite{Haug,Mahan,wacker}%
\begin{align}
&  \Sigma^{<}\left(  z_{3},z_{4},k,E\right) \nonumber\\
&  =\frac{1}{\left(  2\pi\right)  ^{3}}\int\mathrm{d}^{2}q\mathrm{d}%
q_{z}\left|  C_{\mathbf{Q}}\right|  ^{2}\mathrm{e}^{\mathrm{i}q_{z}\left(
z_{3}-z_{4}\right)  }\nonumber\\
&  \times\left[  N_{\mathbf{Q}}G^{<}\left(  z_{3},z_{4},\left|  \mathbf{k}%
-\mathbf{q}\right|  ,E-\hbar\omega_{\mathbf{Q}}\right)  \right. \nonumber\\
&  \left.  +\left(  1+N_{\mathbf{Q}}\right)  G^{<}\left(  z_{3},z_{4},\left|
\mathbf{k}-\mathbf{q}\right|  ,E+\hbar\omega_{\mathbf{Q}}\right)  \right]  ,
\label{general_phonon_<}%
\end{align}%
\begin{align}
&  \Sigma^{R}\left(  z_{3},z_{4},k,E\right) \nonumber\\
&  =\frac{1}{\left(  2\pi\right)  ^{3}}\int\mathrm{d}^{2}q\mathrm{d}%
q_{z}\left|  C_{\mathbf{Q}}\right|  ^{2}\mathrm{e}^{\mathrm{i}q_{z}\left(
z_{3}-z_{4}\right)  }\nonumber\\
&  \times\left[  N_{\mathbf{Q}}G^{R}\left(  z_{3},z_{4},\left|  \mathbf{k}%
-\mathbf{q}\right|  ,E+\hbar\omega_{\mathbf{Q}}\right)  \right. \nonumber\\
&  \left.  +\left(  1+N_{\mathbf{Q}}\right)  G_{0}^{R}\left(  z_{3}%
,z_{4},\left|  \mathbf{k}-\mathbf{q}\right|  ,E-\hbar\omega_{\mathbf{Q}%
}\right)  \right] \nonumber\\
&  +\frac{1}{\left(  2\pi\right)  ^{3}}\int\mathrm{d}^{2}q\mathrm{d}%
q_{z}\left|  C_{\mathbf{Q}}\right|  ^{2}\mathrm{e}^{\mathrm{i}q_{z}\left(
z_{3}-z_{4}\right)  }\nonumber\\
&  \times\left[  \frac{1}{2}G_{0}^{<}\left(  z_{3},z_{4},\left|
\mathbf{k}-\mathbf{q}\right|  ,E-\hbar\omega_{\mathbf{Q}}\right)  \right.
\nonumber\\
&  \left.  -\frac{1}{2}G_{0}^{<}\left(  z_{3},z_{4},\left|  \mathbf{k}%
-\mathbf{q}\right|  ,E+\hbar\omega_{\mathbf{Q}}\right)  \right] \nonumber\\
&  -\frac{\mathrm{i}}{\left(  2\pi\right)  ^{4}}\int\mathrm{d}E^{\prime}%
\int\mathrm{d}^{2}q\mathrm{d}q_{z}\left|  C_{\mathbf{Q}}\right|
^{2}\mathrm{e}^{\mathrm{i}q_{z}\left(  z_{3}-z_{4}\right)  }\nonumber\\
&  \times G_{0}^{<}\left(  z_{3},z_{4},\left|  \mathbf{k}-\mathbf{q}\right|
,E-E^{\prime}\right) \nonumber\\
&  \times\Pr\left(  \frac{1}{E^{\prime}+\hbar\omega_{\mathbf{Q}}}-\frac
{1}{E^{\prime}-\hbar\omega_{\mathbf{Q}}}\right)  . \label{general_phonon_R}%
\end{align}
Here, $\mathbf{Q}=\left[  \mathbf{q},q_{z}\right]  ^{\mathrm{T}}$ is the
phonon wave vector with the in-plane component $\mathbf{q}$, and $Q=\left|
\mathbf{Q}\right|  =\sqrt{q^{2}+q_{z}^{2}}$. The coupling constant for a
phonon of wave vector $\mathbf{Q}$ and phonon frequency $\omega_{\mathbf{Q}}$
is denoted with $C_{\mathbf{Q}}$, and the the analytical phonon Green's
functions are already inserted.~\cite{Landau} It is worth noting that
Eqs.~(\ref{general_phonon_<}) and (\ref{general_phonon_R}) are valid for
electrons scattering from any type of bulk equilibrium phonons. In the case of
the deformation potential perturbation and a linearized phonon dispersion
relation,~\cite{MC} the coupling constant reads [similar to Eq.~(\ref{eq:M_ac}%
)]
\begin{equation}
\left|  C_{\mathbf{Q}}\right|  ^{2}=\frac{\hbar\Xi^{2}}{2\rho_{\mathrm{c}%
}\omega_{\mathbf{Q}}}Q^{2}. \label{unscreened_LApot}%
\end{equation}
Even in this approximate shape, Eqs.(\ref{general_phonon_<}) and
(\ref{general_phonon_R}) require numerical solution of the three-dimensional
$\mathbf{Q}$ integrals. Since that is typically beyond numerical feasibility,
further approximations are commonly applied.

\paragraph{\label{sec:NEGF_el_ac_ph}Elastic acoustic phonon scattering}

A rather common approach \cite{nemo1d, Reggiani} is to neglect energy changes
of the electrons in the scattering process with the phonons,%
\begin{equation}
E\pm\hbar\omega_{\mathbf{Q}}\approx E.
\end{equation}
In addition, high temperatures are assumed~\cite{Ridley}%
\begin{equation}
k_{\mathrm{B}}T\gg\hbar\omega_{\mathbf{Q}},
\end{equation}
which allows to apply the equipartition approximation to the Bose distribution
(see also Sec.~\ref{sec:nonpolar_scat})%
\begin{equation}
N_{\mathbf{Q}}+1\approx N_{\mathbf{Q}}\approx\frac{k_{\mathrm{B}}T}%
{\hbar\omega_{\mathbf{Q}}}=\frac{k_{\mathrm{B}}T}{\hbar v_{s}Q}.
\label{high_temp}%
\end{equation}

\begin{figure}[ptb]
\includegraphics[width=8.6cm]{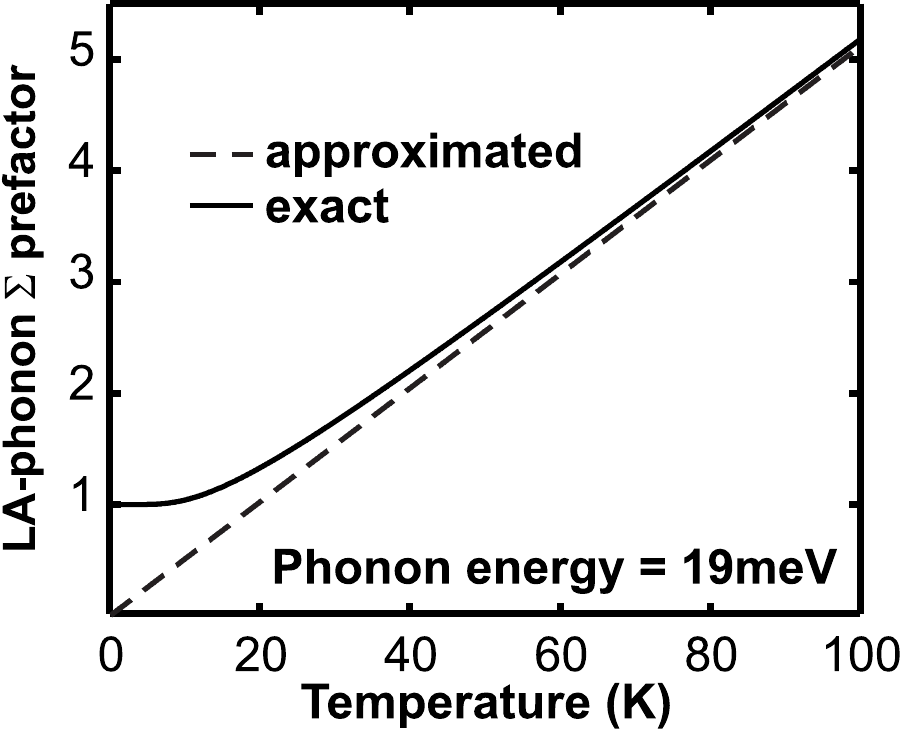}
\caption{Phonon-number-dependent prefactor in the integral of
Eqs.~(\ref{general_phonon_<}) and (\ref{general_phonon_R}) at $19$~meV phonon
energy as a function of the bath temperature. The high temperature
approximation described in the text (dashed) converges to the exact solution
(solid) at approximately $30$~K. The lower the phonon energy, the faster both
curves converge.}%
\label{hot_phonon}%
\end{figure}

The lower the phonon energy, the better justified the equipartition
approximation. In the case of GaAs the acoustic phonon energy extends up to
approximately $19$~meV. At this phonon energy, the equipartition approximation
is well justified for phonon bath temperatures above approximately $30$~K, as
illustrated in Fig.~\ref{hot_phonon}. This figure shows the exact (solid) as
well as the approximated [Eq.~(\ref{high_temp}), dashed] result of $\left(
1+2N_{\mathbf{Q}}\right)  $ at $19$~meV.

With all these approximations the formula for the lesser self-energy finally
reads%
\begin{align}
&  \Sigma^{<}\left(  z_{3},z_{4},k,E\right) \nonumber\\
&  =\frac{1}{\left(  2\pi\right)  ^{3}}\frac{k_{\mathrm{B}}T\Xi^{2}}%
{\rho_{\mathrm{c}}v_{s}^{2}}\nonumber\\
&  \times\int\mathrm{d}^{2}q\mathrm{d}q_{z}\mathrm{e}^{\mathrm{i}q_{z}\left(
z_{3}-z_{4}\right)  }G^{<}\left(  z_{3},z_{4},\left|  \mathbf{k}%
-\mathbf{q}\right|  ,E\right)  .
\end{align}
When the integral over $q_{z}$ is approximated~\cite{1988PhRvB..37.2578G} to
run from $-\infty$ to $\infty$ it can be solved analytically which results in
a local scattering self-energy%
\begin{align}
&  \Sigma^{<}\left(  z_{3},z_{4},k,E\right) \nonumber\\
&  =\frac{1}{\left(  2\pi\right)  ^{2}}\frac{k_{\mathrm{B}}T\Xi^{2}}%
{\rho_{\mathrm{c}}v_{s}^{2}}\delta\left(  z_{3}-z_{4}\right)  \int
\mathrm{d}^{2}lG^{<}\left(  z_{3},z_{4},l,E\right)  . \label{self_ac_ela}%
\end{align}
This scattering self-energy describes elastic scattering processes. Therefore,
terms involving the lesser Green's function in the formula of the retarded
self-energy Eq.~(\ref{general_phonon_R}) vanish exactly and the retarded
self-energy can be obtained from Eq.~(\ref{self_ac_ela}) by replacing the
''$<$''\ with ''$R$''.

\paragraph{Inelastic acoustic phonon scattering}

An elastic approximation of the scattering with acoustic phonons misses an
important physical effect: Inelastic scattering with acoustic phonons allows
dissipation of arbitrarily small amounts of energy. In particular, when the
electronic energy is insufficient to emit polar optical phonons, such an
elastic treatment of acoustic phonons may underestimate electron
thermalization. Setting the energy of acoustic phonons to a constant, but
finite value still limits the minimum amount of dissipated energy.~\cite{lee}
Such small amounts of dissipated energy become important when the electrons do
not carry enough energy to dissipate polar optical phonons. A full
thermalization of such electrons requires the possibility to dissipate
arbitrarily small amounts of energy.

\begin{figure}[ptb]
\includegraphics[width=8.6cm]{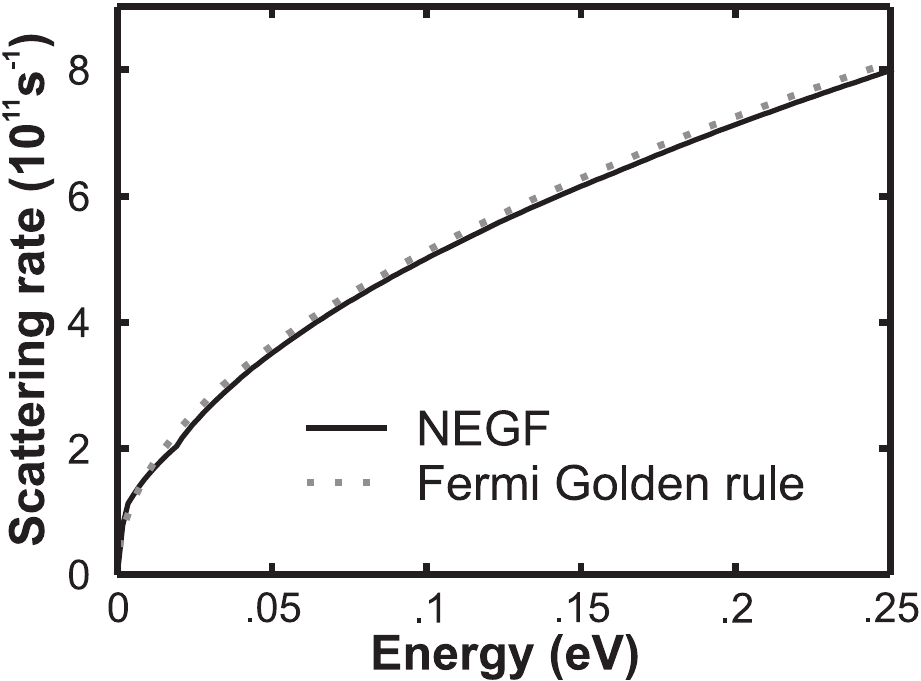}
\caption{On-shell scattering rate of electrons by longitudinal acoustic
phonons in bulk $n$-doped GaAs at $300$\thinspace K and $n=2\times
10^{18}\,\mathrm{cm}^{-3}$. The Fermi golden rule (dotted) and the NEGF
calculation (full line) match perfectly.}%
\label{lac_scrate}%
\end{figure}

A simple approximate inclusion of inelasticity is to replace the Green's
function in Eq.~(\ref{self_ac_ela}) with an approximate Green's function
$\tilde{G}$, averaged over the energy range of $\pm$ one acoustic phonon%
\begin{align}
&  \tilde{G}^{<,R}\left(  z_{3},z_{4},l,\tilde{E}\right) \nonumber\\
&  =\frac{1}{2\hbar\omega_{\mathrm{D}}}\int_{E-\hbar\omega_{\mathrm{D}}%
}^{E+\hbar\omega_{\mathrm{D}}}\mathrm{d}E^{\prime}G^{<,R}\left(  z_{3}%
,z_{4},l,E^{\prime}\right)  ,\nonumber\\
\forall\tilde{E}  &  \in\left[  E-\hbar\omega_{\mathrm{D}},E+\hbar
\omega_{\mathrm{D}}\right]  . \label{aver_Green}%
\end{align}
Hereby, the Debye frequency
\begin{equation}
\omega_{\mathrm{D}}=\left(  \frac{3\rho_{N}}{4\pi}\right)  ^{1/3}v_{s}%
\end{equation}
with the number density $\rho_{N}$ limits the width of that average. In
connection with the equipartition approximation of Sec.~\ref{sec:NEGF_el_ac_ph},
the numerical benefit from this approximation is threefold. First, the $q_{z}$
integral in Eq.~(\ref{general_phonon_<}) can be solved analytically and yields
a local scattering self-energy similar to Eq.~(\ref{self_ac_ela}),%
\begin{align}
&  \Sigma^{<}\left(  z_{3},z_{4},k,E\right) \nonumber\\
&  =\frac{1}{\left(  2\pi\right)  ^{2}}\frac{\Xi^{2}k_{\mathrm{B}}T}%
{\rho_{\mathrm{c}}v_{s}^{2}2\hbar\omega_{\mathrm{D}}}\delta\left(  z_{3}%
-z_{4}\right) \nonumber\\
&  \times\int_{E-\hbar\omega_{\mathrm{D}}}^{E+\hbar\omega_{\mathrm{D}}%
}\mathrm{d}E^{\prime}\int\mathrm{d}^{2}lG^{<}\left(  z_{3},z_{4},l,E^{\prime
}\right)  . \label{inel_approx_acL}%
\end{align}
Second, scattering with acoustic phonons is inelastically implemented and
allows dissipatation of arbitrarily small energies. Third, all terms
containing $G^{<}$ in Eq.~(\ref{general_phonon_R}) vanish exactly and the
retarded self-energy is independent of the lesser Green's functions,%
\begin{align}
&  \Sigma^{R}\left(  z_{3},z_{4},k,E\right) \nonumber\\
&  =\frac{1}{\left(  2\pi\right)  ^{2}}\frac{\Xi^{2}k_{\mathrm{B}}T}%
{\rho_{\mathrm{c}}v_{s}^{2}2\hbar\omega_{\mathrm{D}}}\delta\left(  z_{3}%
-z_{4}\right) \nonumber\\
&  \times\int_{E-\hbar\omega_{\mathrm{D}}}^{E+\hbar\omega_{\mathrm{D}}%
}\mathrm{d}E^{\prime}\int\mathrm{d}^{2}lG^{R}\left(  z_{3},z_{4},l,E^{\prime
}\right)  . \label{inel_approx_acR}%
\end{align}
If all $\Sigma^{R}$ are independent of $G^{<}$, the retarded functions can be
solved in advance of the lesser Green's functions and self-energies. Once the
retarded self-energy of a homogeneous system in equilibrium is known, the
bulk, on-shell scattering rate can be extracted: In homogeneous systems, the
self-energies depend only on the difference of the propagation coordinates
($r=z_{3}-z_{4}$). A Fourier transform of the imaginary part of the retarded
self-energy with respect to $r$ agrees then with the scattering
rate~\cite{wacker}
\begin{equation}
\tilde{\Gamma}\left(  \mathbf{k},k_{z},E\right)  =-\frac{2}{\hbar}\Im
\int\mathrm{d}r\exp\left(  \mathrm{i}k_{z}r\right)  \Sigma^{R}\left(
r,k,E\right)  . \label{scat_rate}%
\end{equation}
If $\tilde{\Gamma}$ is evaluated at
\begin{equation}
k_{z}=\sqrt{2m^{\ast}\left(  E\right)  E/\hbar^{2}-k^{2}m^{\ast}\left(
E\right)  /m^{\parallel}\left(  E\right)  },
\end{equation}
it agrees with the on-shell scattering rate of bulk electrons with energy
\begin{equation}
E=\frac{\hbar^{2}}{2}\left[  \frac{k^{2}}{m^{\parallel}\left(  E\right)
}+\frac{k_{z}^{2}}{m^{\ast}\left(  E\right)  }\right]  .
\end{equation}
Figure~\ref{lac_scrate} shows that scattering rate for the self-energy
Eq.~(\ref{inel_approx_acR}). It nicely reproduces the on-shell scattering rate
of Fermi's golden rule.

\begin{figure}[ptb]
\includegraphics[width=8.6cm]{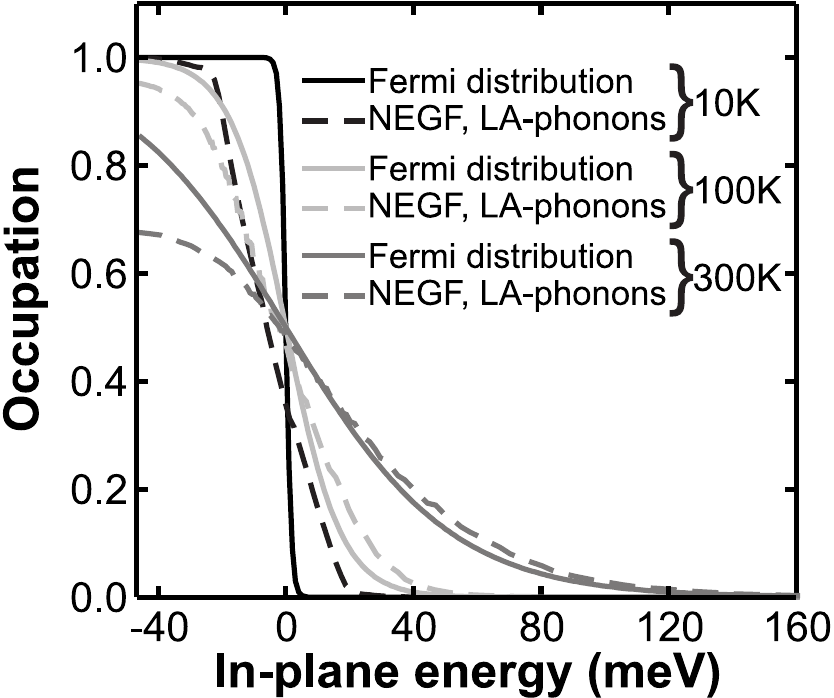}\caption{In-plane electron
distribution in the first state of a $10$~nm wide In$_{.0165}$Ga$_{.9835}$As
quantum well at equilibrium and various temperatures. The dashed lines show
occupations resulting from NEGF calculations when the approximate inelastic
scattering with LA-phonons of Eqs.~(\ref{inel_approx_acR}) and
(\ref{inel_approx_acL}) is the only included incoherent scattering mechanism.
The Fermi distributions are depicted with solid lines. The zero in energy
marks the chemical potential.}%
\label{LA_check}%
\end{figure}

In spite of the scattering rate, the presented approximations of inelastic
longitudinal acoustic phonons lead to an incorrect electron distribution: It
can be shown that any phonon distribution that deviates from the Bose
distribution [such as the approximate one of Eq.~(\ref{high_temp})] will cause
the equilibrium electron distribution to deviate from the Fermi
distribution.~\cite{Kubis_phd} This is illustrated in\ Fig.~\ref{LA_check} as
it shows the electronic occupation of the first state of an unbiased $10$~nm
wide In$_{.0165}$Ga$_{.9835}$As quantum well that is surrounded by $10$~nm
thick GaAs layers at various temperatures. In this calculation, only
scattering on acoustic phonons given by Eqs.~(\ref{inel_approx_acL}) and
(\ref{inel_approx_acR}) is included. The zero of energy is set to the chemical
potential of the device and energies below the conduction band edge are
neglected. The higher the device temperature, the better is the agreement
between the high-energy tail and the Fermi distribution. However, the
occupations at energies below the chemical potential are underestimated at any temperature.

Apart from these deviations from the Fermi distribution,
Eqs.~(\ref{inel_approx_acL}) and (\ref{inel_approx_acR}) are numerically
efficient and describe the probability for inelastic scattering with acoustic
phonons well enough. As soon as another inelastic scattering mechanism is
included that is significantly stronger than the acoustic phonon scattering,
such deviations from the Fermi distribution become negligible.

\subsubsection{Scattering from polar longitudinal optical phonons}

\begin{figure}[ptb]
\includegraphics[width=8.6cm]{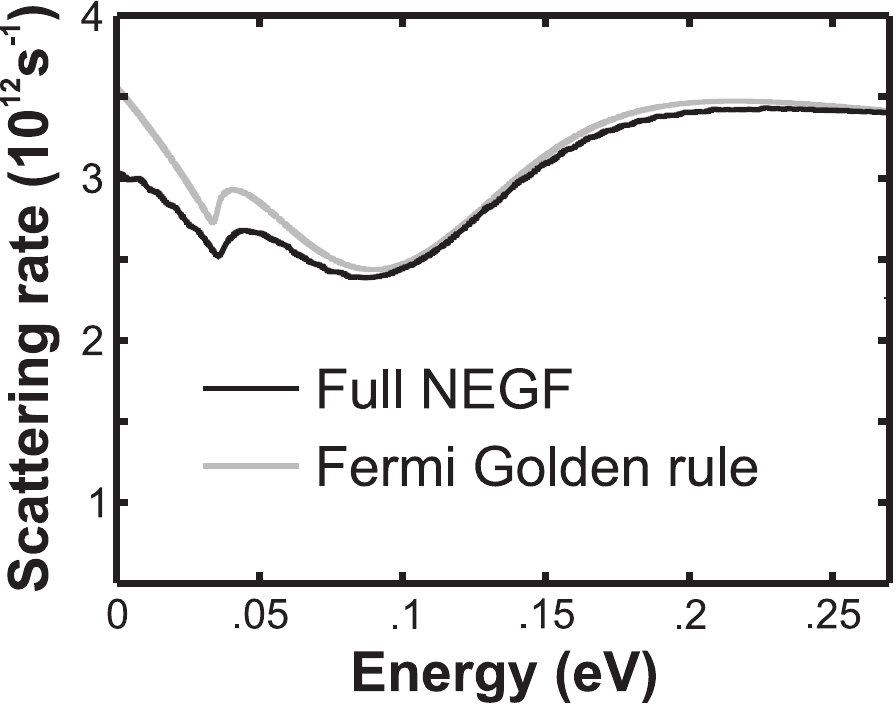}\caption{On-shell scattering
rate of electrons with screened polar optical phonons in homogeneous n-doped
GaAs at $300$\thinspace K. The doping concentration is $2\times10^{18}%
\,\mathrm{cm}^{-3}$ and the screening length is set to $5\,\mathrm{nm}$ for
comparison. Results of NEGF calculations (black) agree nicely with Fermi's
golden rule (grey). The scattering rate below the chemical potential at
approximately $90$\thinspace meV originates mainly from the scattering of
conduction band holes.}%
\label{first_pop_rate}%
\end{figure}

The scattering of electrons from polar optical phonons is discussed in detail
in Sec.~\ref{sec:LO}. With the screened polar optical scattering potential of
Eq.~(\ref{eq:screened_LO_M}) the lesser self-energy reads similar to
Eq.~(\ref{eq:screened_LO_J})%
\begin{align}
&  \Sigma^{<}\left(  z_{3},z_{4},k,E\right) \nonumber\\
&  =\frac{\gamma\pi}{\left(  2\pi\right)  ^{3}}\int\mathrm{d}^{2}%
l\frac{\mathrm{e}^{-\sqrt{\left|  \mathbf{k}-\mathbf{l}\right|  ^{2}%
+q_{\mathrm{s}}^{2}}\left|  z_{3}-z_{4}\right|  }}{\sqrt{\left|
\mathbf{k}-\mathbf{l}\right|  ^{2}+q_{\mathrm{s}}^{2}}}\nonumber\\
&  \times\left[  1-\frac{q_{\mathrm{s}}^{2}\left|  z_{3}-z_{4}\right|
}{2\sqrt{\left|  \mathbf{k}-\mathbf{l}\right|  ^{2}+q_{\mathrm{s}}^{2}}}%
-\frac{q_{\mathrm{s}}^{2}}{2\left(  \left|  \mathbf{k}-\mathbf{l}\right|
^{2}+q_{\mathrm{s}}^{2}\right)  }\right] \nonumber\\
&  \times\left[  N_{\mathrm{Ph}}G^{<}\left(  z_{3},z_{4},l,E-E_{\mathrm{LO}%
}\right)  \right. \nonumber\\
&  \left.  +\left(  1+N_{\mathrm{Ph}}\right)  G^{<}\left(  z_{3}%
,z_{4},l,E+E_{\mathrm{LO}}\right)  \right]  \label{LO_selfL}%
\end{align}
with the LO phonon energy $E_{\mathrm{LO}}=\hbar\omega_{\mathrm{LO}}$, phonon
occupation number $N_{\mathrm{Ph}}$ given in Eq.~(\ref{eq:NQ}), and%
\begin{equation}
\gamma=e^{2}\frac{E_{\mathrm{LO}}}{2\epsilon_{0}}\left(  \frac{1}%
{\epsilon_{\mathrm{r},\infty}}-\frac{1}{\epsilon_{\mathrm{r},0}}\right)  .
\end{equation}
Analogously, the formula for the retarded self-energy can be derived to
\begin{align}
&  \Sigma^{R}\left(  z_{3},z_{4},k,E\right) \nonumber\\
&  =\frac{\gamma\pi}{\left(  2\pi\right)  ^{3}}\int\mathrm{d}^{2}%
l\frac{\mathrm{e}^{-\sqrt{\left|  \mathbf{k}-\mathbf{l}\right|  ^{2}%
+q_{\mathrm{s}}^{2}}\left|  z_{3}-z_{4}\right|  }}{\sqrt{\left|
\mathbf{k}-\mathbf{l}\right|  ^{2}+q_{\mathrm{s}}^{2}}}\nonumber\\
&  \times\left[  1-\frac{q_{\mathrm{s}}^{2}\left|  z_{3}-z_{4}\right|
}{2\sqrt{\left|  \mathbf{k}-\mathbf{l}\right|  ^{2}+q_{\mathrm{s}}^{2}}}%
-\frac{q_{\mathrm{s}}^{2}}{2\left(  \left|  \mathbf{k}-\mathbf{l}\right|
^{2}+q_{\mathrm{s}}^{2}\right)  }\right] \nonumber\\
&  \times\left[  \left(  1+N_{\mathrm{Ph}}\right)  G^{R}\left(  z_{3}%
,z_{4},l,E-E_{\mathrm{LO}}\right)  \right. \nonumber\\
&  +N_{\mathrm{Ph}}G^{R}\left(  z_{3},z_{4},l,E+E_{\mathrm{LO}}\right)
+\frac{1}{2}G^{<}\left(  z_{3},z_{4},l,E-E_{\mathrm{LO}}\right) \nonumber\\
&  -\frac{1}{2}G^{<}\left(  z_{3},z_{4},l,E+E_{\mathrm{LO}}\right)
+\mathrm{i}\int\frac{\mathrm{d}\tilde{E}}{2\pi}G^{<}\left(  z_{3}%
,z_{4},l,\tilde{E}\right) \nonumber\\
&  \left.  \times\left(  \mathrm{\Pr}\frac{1}{E-\tilde{E}-E_{\mathrm{LO}}%
}-\mathrm{\Pr}\frac{1}{E-\tilde{E}+E_{\mathrm{LO}}}\right)  \right]  .
\label{LO_selfR}%
\end{align}
Svizhenko et al. have shown in one-dimensional systems that the principal
value integrals of the last line in Eq.~(\ref{LO_selfR}) shift the energies of
resonant states~\cite{Svizhenko}. When that shift is not differing
significantly between different confined states, it leads only to a rigorous
shift of the current-voltage characteristics. Therefore and since the
principal value integrals are numerically expensive to solve, these principal
value integrals are often neglected.~\cite{lee, Harrison_qdot, nemo1d} The
on-shell scattering rate that corresponds to Eq.~(\ref{LO_selfR}) is
illustrated in Fig.~\ref{first_pop_rate}. It shows results for a homogeneously
n-doped GaAs device with $n=2\times10^{18}~$cm$^{-3}$. It is worth emphasizing
that the self-energies of the polar optical phonon scattering are finite when
the two propagation coordinates $z_{3}$ and $z_{4}$ differ. This nonlocality
of the scattering originates from the long-range nature of the Coulomb
potential. Screening can efficiently limit this effect as can be seen from the
exponents in Eqs.~(\ref{LO_selfL}) and (\ref{LO_selfR}).

In Fig.~\ref{first_pop_rate}, NEGF is solved in equilibrium at room
temperature while only scattering with polar optical phonons is included. The
black line in Fig.~\ref{first_pop_rate} shows the scattering rate $\Gamma$
resulting from Eq.~(\ref{scat_rate}) while the gray line shows results of
Fermi's golden rule (see e.g. Ref.~\onlinecite{Ridley}). Due to the high
electron density, the chemical potential lies at approximately $90$~meV above
the conduction band edge (at $0~$meV). In this case, holes in the conduction
band at energies lower than the chemical potential contribute to the total
scattering rate.~\cite{Kubis_assess} Therefore, the Fermi golden rule result
is a sum of scattering probabilities of four kinds of scattering events:
emission and absorption of LO-phonons by electrons and holes in the conduction
band. When present, all these scattering mechanisms are automatically included
in the self-energy of Eq.~(\ref{LO_selfR}).

\subsubsection{\label{sec:NEGF_Brooks_imp}Brooks-Herring scattering on charged impurities}

It has been discussed already in Sec.~\ref{sec:elast_impur} that for electrons
scattering from charged impurities, the perturbing potential $V_{\text{imp}%
}\left(  \mathbf{x};\mathbf{s}_{1},\mathbf{s}_{2},\ldots,\mathbf{s}%
_{N}\right)  $ at position $\mathbf{x}$ is created by the $N$ impurities at
the position $\mathbf{s}_{1},\ldots,\mathbf{s}_{N}$. In this section, more
details on the perturbation of Green's functions due to impurity scattering
are presented. It is particularly relevant since the NEGF formulation of
impurity scattering differs from the semiclassical picture in
Sec.~\ref{sec:elast_impur}: the scattering on charged impurities may influence
the electronic correlation function between distinct positions. This is also
true for scattering on polar optical phonons in Eqs.~(\ref{LO_selfL}) and
(\ref{LO_selfR}), but the following derivation illustrates nicely this effect.
If the impurities are randomly distributed, the probability density to find an
impurity at position $\mathbf{s}_{i}$ is given by%
\begin{equation}
P\left(  \mathbf{s}_{i}\right)  =\frac{n_{\mathrm{D}}\left(  \mathbf{s}%
_{i}\right)  }{N}. \label{inhom_imp_prob}%
\end{equation}
Here, the total number of impurities $N$ is given by the integral of the
impurity density over the total volume $V$%
\begin{equation}
\int_{V}n_{\mathrm{D}}\left(  \mathbf{s}\right)  \mathrm{d}^{3}s=N.
\end{equation}
As discussed in Sec.~\ref{sec:elast_impur}, the first order of the impurity
scattering is covered by the Poisson equation. For the second order, the
squared scattering potential is required. The actual distribution of the
impurities is unknown, thus the product of the perturbing potential has to be
averaged over the impurity positions which introduces the impurity potential
autocorrelation function%
\begin{align}
&  \left\langle V_{\text{imp}}\left(  \mathbf{x}_{3};\mathbf{s}_{1}%
,\mathbf{s}_{2},\ldots,\mathbf{s}_{N}\right)  V_{\text{imp}}\left(
\mathbf{x}_{4};\mathbf{s}_{1},\mathbf{s}_{2},\ldots,\mathbf{s}_{N}\right)
\right\rangle _{\text{imp}}\nonumber\\
&  =\frac{1}{N^{N}}\int\prod_{i=1}^{N}\mathrm{d}^{3}s_{i}n_{\mathrm{D}}\left(
\mathbf{s}_{i}\right)  \frac{1}{\left(  2\pi\right)  ^{6}}\nonumber\\
&  \times\int\mathrm{d}^{3}Q\mathrm{d}^{3}Q^{\prime}V_{\text{imp}}\left(
Q\right)  V_{\text{imp}}\left(  Q^{\prime}\right)  \sum_{j=1}^{N}%
\mathrm{e}^{\mathrm{i}\mathbf{Q}\cdot\left(  \mathbf{x}_{3}-\mathbf{s}%
_{j}\right)  }\mathrm{e}^{\mathrm{i}\mathbf{Q}^{\prime}\cdot\left(
\mathbf{x}_{4}-\mathbf{s}_{j}\right)  }.
\end{align}
With some algebra, this simplifies to
\begin{align}
&  \left\langle V_{\text{imp}}\left(  \mathbf{x}_{3};\mathbf{s}_{1}%
,\mathbf{s}_{2},\ldots,\mathbf{s}_{N}\right)  V_{\text{imp}}\left(
\mathbf{x}_{4};\mathbf{s}_{1},\mathbf{s}_{2},\ldots,\mathbf{s}_{N}\right)
\right\rangle _{\text{imp}}\nonumber\\
&  =\frac{1}{\left(  2\pi\right)  ^{6}}\int\mathrm{d}^{3}s\mathrm{d}%
^{3}Q\mathrm{d}^{3}Q^{\prime}n_{\mathrm{D}}\left(  \mathbf{s}\right)
V_{\text{imp}}\left(  Q\right) \nonumber\\
&  \times V_{\text{imp}}\left(  Q^{\prime}\right)  \mathrm{e}^{-\mathrm{i}%
\left(  \mathbf{Q}+\mathbf{Q}^{\prime}\right)  \cdot\mathbf{s}}\mathrm{e}%
^{\mathrm{i}\mathbf{Q}\cdot\mathbf{x}_{3}}\mathrm{e}^{\mathrm{i}%
\mathbf{Q}^{\prime}\cdot\mathbf{x}_{4}}. \label{impur_corr_exact_inhom}%
\end{align}

Typical cascade lasers are set in the regime of the approach of Brooks and
Herring (see discussion in Sec.~\ref{sec:elast_impur})~\cite{Brooks},
\begin{equation}
V_{\text{imp}}\left(  Q\right)  =\frac{\text{e}^{2}}{\epsilon}\frac{1}%
{Q^{2}+q_{\mathrm{s}}^{2}} \label{Vimp3dq}%
\end{equation}
with the inverse Debye screening length~$q_{\mathrm{s}}$. In these systems,
the doping profile is independent of in-plane positions\ [$n_{\mathrm{D}%
}\left(  \mathbf{s}\right)  =n_{\mathrm{D}}\left(  z\right)  $] and the
impurity potential correlation Eq.~(\ref{impur_corr_exact_inhom}) can be
simplified further,%
\begin{align}
&  \left\langle V_{\text{imp}}\left(  \mathbf{x}_{3};\mathbf{s}_{1}%
,\mathbf{s}_{2},\ldots,\mathbf{s}_{N}\right)  V_{\text{imp}}\left(
\mathbf{x}_{4};\mathbf{s}_{1},\mathbf{s}_{2},\ldots,\mathbf{s}_{N}\right)
\right\rangle _{\text{imp}}\nonumber\\
&  =\frac{1}{\left(  2\pi\right)  ^{2}}\int\mathrm{d}z\mathrm{d}%
^{3}Qn_{\mathrm{D}}\left(  z\right)  \mathrm{e}^{\mathrm{i}\mathbf{Q}%
\cdot\left(  \mathbf{x}_{3}-\mathbf{x}_{4}\right)  }\nonumber\\
&  \times V_{\text{imp}}\left(  Q,z_{3}-z\right)  V_{\text{imp}}\left(
-Q,z_{4}-z\right)  . \label{impur_corr_1D}%
\end{align}
The two-dimensional Fourier transform of the Debye-H\"{u}ckel potential with
respect to the in-plane coordinates reads~\cite{nemo1d}%
\begin{equation}
V_{\text{imp}}\left(  q,r_{z}\right)  =\frac{e^{2}}{2\epsilon}\frac
{\exp\left(  -\sqrt{q_{\mathrm{s}}^{2}+q^{2}}\left|  r_{z}\right|  \right)
}{\sqrt{q_{\mathrm{s}}^{2}+q^{2}}}. \label{Debye_Hueckel_potential1D}%
\end{equation}
In 3D real space representation, the elastic charged impurity scattering
self-energy is a product of the electronic Green's function with the
scattering potential correlation. After the Fourier transform with respect to
the in-plane momentum the scattering self-energies read~\cite{nemo1d}%
\begin{align}
&  \Sigma_{\text{imp}}^{<,R}\left(  z_{3},z_{4},k,E\right)  =\frac{e^{4}%
}{16\pi^{2}\epsilon^{2}}\nonumber\\
&  \times\int\mathrm{d}^{2}qF\left(  z_{3},z_{4},\left|  \mathbf{k}%
-\mathbf{q}\right|  \right)  G^{<,R}\left(  z_{3},z_{4},q,E\right)
,\nonumber\\
&  F\left(  z_{3},z_{4},\sqrt{q_{\mathrm{s}}^{2}+p^{2}}\right) \nonumber\\
&  =\int\mathrm{d}zn_{\mathrm{D}}\left(  z\right)  \frac{\mathrm{e}%
^{-\sqrt{q_{\mathrm{s}}^{2}+p^{2}}\left(  \left|  z_{3}-z\right|  +\left|
z_{4}-z\right|  \right)  }}{q_{\mathrm{s}}^{2}+p^{2}}. \label{ex_impur}%
\end{align}
Unfortunately, this result requires a three-dimensional integral for each
value of $\left(  z_{3},z_{4},k,E\right)  $. It turns out that a numerical
implementation of such a self-energy is very time consuming and an
approximation of the self-energy is necessary.

\paragraph{Averaged remote scattering}

The scattering self-energy of the last paragraph is only in so far ''exact'',
that the correlation function of impurities is not further approximated.
However, the assumption of a constant inverse screening length $q_{\mathrm{s}%
}$ is already a significant approximation and in reality, screening in
inhomogeneous devices is neither homogeneous, nor constant with respect to
momentum and frequency. Instead, it would be a more realistic ansatz to
describe the screening of charges with a polarization that depends on both
propagation coordinates and is capable to describe the many particle effects
correctly (see e.g. Ref.~\onlinecite{Mahan}). However, such a dielectric
function is numerically too demanding.

Given this fact, it appears questionable to put great efforts in an
''exact''\ numerical implementation of Eq.~(\ref{ex_impur}). Instead, the
position dependent number of charged impurities $n_{\mathrm{D}}$ in
Eq.~(\ref{ex_impur}) can be approximated with its average $\left\langle
n_{\mathrm{D}}\right\rangle _{\mathbf{x}_{3},\mathbf{x}_{4}}$ along the
shortest propagation path between both propagation coordinates $\mathbf{x}%
_{3}$ and $\mathbf{x}_{4}$. In this way, the influence of scattering at
inhomogeneously distributed charged impurities is approximated with an
effective scattering at homogeneously distributed impurities. This
approximation effectively affects the function $F$ of Eq.~(\ref{ex_impur}),
\begin{align}
&  F\left(  z_{3},z_{4},\sqrt{q_{\mathrm{s}}^{2}+p^{2}}\right) \nonumber\\
&  \approx n_{\mathrm{D}}\left(  z_{3},z_{4}\right)  \int\mathrm{d}%
z\frac{\mathrm{e}^{-\sqrt{q_{\mathrm{s}}^{2}+p^{2}}\left(  \left|
z_{3}-z\right|  +\left|  z_{4}-z\right|  \right)  }}{q_{\mathrm{s}}^{2}+p^{2}%
},\nonumber\\
&  \left\langle n_{\mathrm{D}}\right\rangle _{z_{3},z_{4}}\nonumber\\
&  =\left\{
\begin{array}
[c]{c}%
n_{\mathrm{D}}\left(  z_{3}\right)  \text{, for }z_{3}=z_{4}\\
\frac{1}{z_{4}-z_{3}}\int_{z_{3}}^{z_{4}}\mathrm{d}\zeta n_{\mathrm{D}}\left(
\zeta\right)  \text{, elsewhere}%
\end{array}
\right.  . \label{approximation_F_impur}%
\end{align}
With this approximation, the scattering self-energy simplifies to
\begin{align}
&  \Sigma^{<,R}\left(  z_{3},z_{4},k,E\right)  =\frac{\left\langle
n_{\mathrm{D}}\right\rangle _{z_{3},z_{4}}e^{4}}{4\epsilon^{2}\left(
2\pi\right)  ^{2}}\nonumber\\
&  \times\int\mathrm{d}^{2}q\left[  \frac{\left|  z_{3}-z_{4}\right|
+1/\sqrt{q_{\mathrm{s}}^{2}+q^{2}}}{q_{\mathrm{s}}^{2}+q^{2}}\mathrm{e}%
^{-\sqrt{q_{\mathrm{s}}^{2}+q^{2}}\left|  z_{3}-z_{4}\right|  }\right.
\nonumber\\
&  \left.  \times G^{<,R}\left(  z_{3},z_{4},\left|  \mathbf{k}-\mathbf{q}%
\right|  ,E\right)  \right]  . \label{self_impur}%
\end{align}

\begin{figure}[tb]
\includegraphics[width=8.6cm]{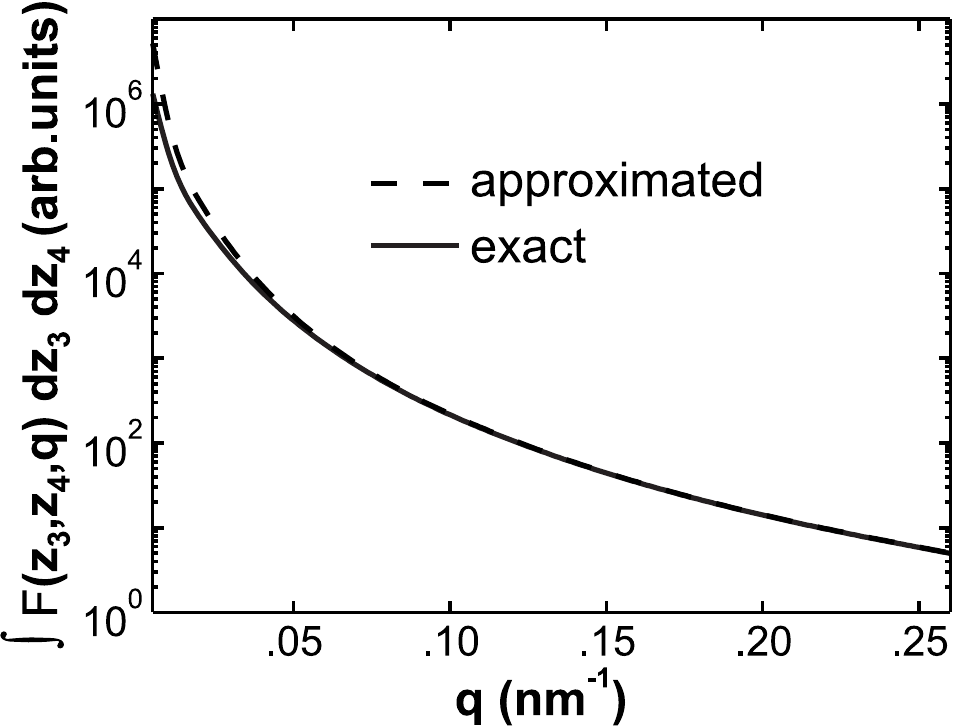}\caption{The function $F$ of
Eqs.~(\ref{ex_impur}) and (\ref{approximation_F_impur}) integrated over both
propagation coordinates. The approximation (dashed) of $F$ deviates from the
exact one (solid) only for small $q$.}%
\label{impur_approx_q}%
\end{figure}

To assess the validity of Eq.~(\ref{approximation_F_impur}), we compare the
exact solution of $F$ of Eq.~(\ref{ex_impur}) with the approximate $F$ of
Eq.~(\ref{approximation_F_impur}) in the case of a step like impurity density%
\begin{equation}
n_{\mathrm{D}}\left(  z\right)  \propto%
\genfrac{\{}{.}{0pt}{}{1,\forall\left|  z\right|  \leq7~\text{nm}%
}{0,~\text{else}}%
. \label{doping_example}%
\end{equation}
Figure~\ref{impur_approx_q} shows the approximated
[Eq.~(\ref{approximation_F_impur})] as well as the exact integral
[Eq.~(\ref{ex_impur})], integrated over $z_{3}$ and $z_{4}$ for various values
of $q$. The smaller the transferred momentum in devices with large screening
lengths is, the more important is the actual shape of the impurity density
$n_{\mathrm{D}}\left(  z\right)  $ and the more is the function $F$ affected
by the approximation in Eq.~(\ref{approximation_F_impur}).
Figure~\ref{impur_approx_q} shows that the approximation of
Eq.~(\ref{approximation_F_impur}) effectively overestimates scattering only
for rather small transferred in-plane momenta. If the screening length is
shorter than $20$~nm (which corresponds to $q_{\text{s}}>0.05~$nm$^{-1}$), the
discrepancy of the effective scattering strength is negligible.

\begin{figure}[tb]
\includegraphics[width=8.6cm]{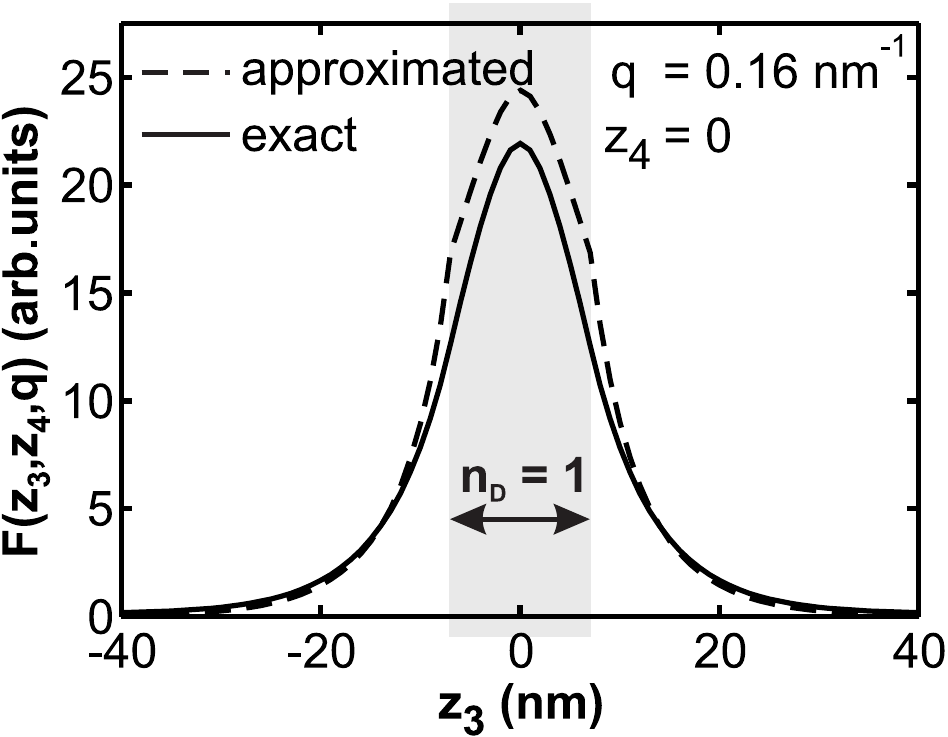}\caption{The
function $F\left(  z_{3},z_{4}=0,q\right)  $ of Eqs.~(\ref{ex_impur}) and
(\ref{approximation_F_impur}) at $q=0.16~$nm$^{-1}$. The impurity density
$n_{\mathrm{D}}\left(  z\right)  $ is given in Eq.~(\ref{doping_example}) and
is nonzero only in the gray shaded region. Note that the propagation
coordinate $z_{4}$ is centered in the compact support of $n_{\mathrm{D}%
}\left(  z\right)  $.}%
\label{impur_approx_z4_0}%
\end{figure}

\begin{figure}[tb]
\includegraphics[width=8.6cm]{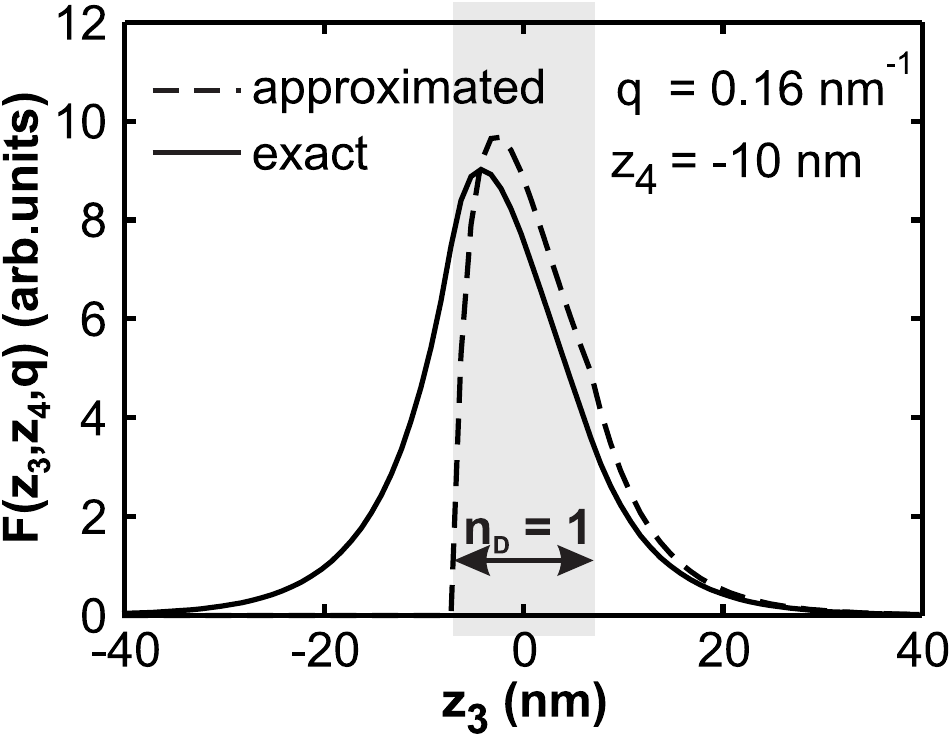}\caption{Same as
in Fig.~\ref{impur_approx_z4_0}, but with $z_{4}=-10$~nm, i.e., outside of the
compact support of $n_{\mathrm{D}}\left(  z\right)  $ (grey shaded region).}%
\label{impur_approx_z4_m10}%
\end{figure}

\begin{figure}[tb]
\includegraphics[width=8.6cm]{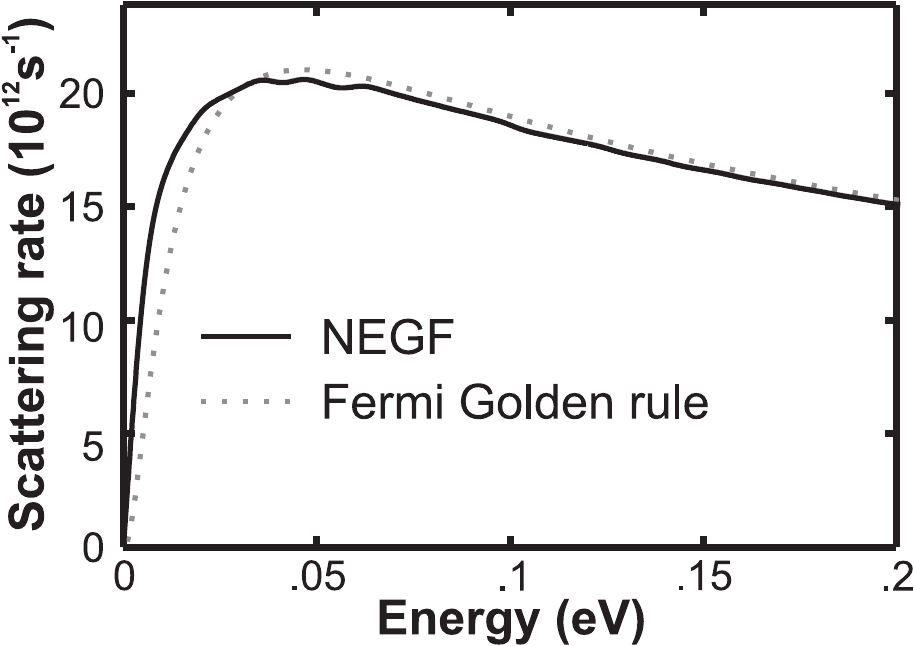}\caption{On-shell scattering rate
of electrons by charged impurities at $300$\thinspace K in homogeneous n-doped
GaAs. The doping concentration is $2\times10^{18}\,\mathrm{cm}^{-3}$. The
screening length was set constant to $5$\thinspace nm for comparison (instead
of the realistic $3$\thinspace nm). The dotted line (Fermi golden rule) and
the full line (NEGF calculation) agree nicely.}%
\label{imp_scrate}%
\end{figure}

Figure \ref{impur_approx_z4_0} shows $F$ in the case of the impurity density
in Eq.~(\ref{doping_example}) with $z_{4}$ centered in the compact support of
$n_{\mathrm{D}}\left(  z\right)  $. The direct path between $z_{3}$ and
$z_{4}$ may cross an area with nonzero $n_{\mathrm{D}}\left(  z\right)  $.
Then, the approximation of $F$ (dashed) overestimates the exact solution
(solid) for values of $z_{3}$ close to and within the compact support of
$n_{\mathrm{D}}\left(  z\right)  $. However, when the direct path between
$z_{3}$ and $z_{4}$ does not touch the compact support of $n_{\mathrm{D}}$,
i.e.,%
\begin{equation}
z_{3},z_{4}>7~\text{nm }\vee\text{\ }z_{3},z_{4}<-7~\text{nm,}
\label{remote_propagation}%
\end{equation}
the approximation of Eq.~(\ref{approximation_F_impur}) underestimates the
scattering (see Fig.~\ref{impur_approx_z4_m10}). Consequently, the
approximation in Eq.~(\ref{approximation_F_impur}) partly neglects remote
scattering at charged impurities. Nevertheless, nonlocality of quantum
mechanics as well as the self-consistent Born approximation ensure that a
propagation between points $z_{3}$ and $z_{4}$ that fulfill
Eq.~(\ref{remote_propagation}) is still affected by scattering at the charged
impurities. Figure~\ref{imp_scrate} illustrates that Eq.~(\ref{self_impur})
reproduces the on-shell scattering rate of Fermi's golden rule in a
homogeneous GaAs system.

\subsubsection{Scattering from rough interfaces}

The scattering potential has been discussed already in
Sec.~\ref{sec:elast_interface}. Since a detailed knowledge of the $\left(
x,y\right)  $-dependence of the realistic interface is futile, the interface
roughness potential $V_{\mathrm{IR}}\left(  \mathbf{x}\right)  $ has to be
averaged in the $\left(  x,y\right)  $-direction. When the mid-point between
the two materials is assumed to be randomly distributed within a roughness
interval $2\Delta$ the two possible values of $V_{\mathrm{IR}}\left(
x,y,z\right)  $, i.e., $\pm V_{\mathrm{o}}/2$ (where $V_{\mathrm{o}}$ is the
conduction band offset) are equally likely and the $\left(  x,y\right)
$-average of the roughness potential $\left\langle V_{\mathrm{IR}}\left(
z\right)  \right\rangle _{\left(  x,y\right)  }$ vanishes exactly.
Consequently, the lowest order scattering at rough interfaces is of second
order in $V_{\mathrm{o}}$. Its derivation is analog to the scattering at
charged impurities ($\forall z_{3},z_{4}$ within the roughness interval),
\begin{align}
&  \Sigma^{<,R}\left(  z_{3},z_{4},\mathbf{r}_{3},\mathbf{r}_{4},E\right)
\nonumber\\
&  =G^{<,R}\left(  z_{3},z_{4},\mathbf{r}_{3}\mathbf{,r}_{4},E\right)
\nonumber\\
&  \times\left\langle V_{\mathrm{IR}}\left(  z_{3},\mathbf{r}_{3}\right)
V_{\mathrm{IR}}\left(  z_{4},\mathbf{r}_{4}\right)  \right\rangle _{\left(
x,y\right)  }. \label{gen_self_IR}%
\end{align}
In general, the product of the perturbing potentials averaged with respect to
the actual interface configuration $\left\langle V_{\mathrm{IR}}\left(
z_{3},\mathbf{r}_{3}\right)  V_{\mathrm{IR}}\left(  z_{4},\mathbf{r}%
_{4}\right)  \right\rangle _{\left(  x,y\right)  }$ depends on the growth
coordinates $z_{3}$ and $z_{4}$. However, modern molecular beam epitaxy
techniques allow for (depending on the growth direction) almost monoatomic or
monomolecular resolved growth of layers, so that the typical roughness
interval extends only over a few Angstrom (see Sec.~\ref{sec:elast_interface}%
). If this is much smaller than the typical numerical resolution in envelope
function approximations (such as the effective mass representations), $\Delta$
is usually represented by a single discretization interval ($\forall
z_{3},z_{4}$ within the roughness interval)%
\begin{align}
&  \left\langle V_{\mathrm{IR}}\left(  x_{3},y_{3},z_{3}\right)
V_{\mathrm{IR}}\left(  x_{4},y_{4},z_{4}\right)  \right\rangle _{\left(
x,y\right)  }\nonumber\\
&  =\left\langle V_{\mathrm{IR}}\left(  x_{3},y_{3},C\right)  V_{\mathrm{IR}%
}\left(  x_{4},y_{4},C\right)  \right\rangle _{\left(  x,y\right)  }.
\label{IR_correlation}%
\end{align}
If $\Delta$ is smaller than the numerical grid spacing $s$, the self-energies
are finite only when its both propagation coordinates lie on the ideal
midpoint $C$. Deviations of the step height from the grid spacing can be
linearly interpolated
\begin{align}
&  \Sigma^{<,R}\left(  C,C,\mathbf{r}_{3},\mathbf{r}_{4},E\right) \nonumber\\
&  =\frac{2\Delta}{s}G^{<,R}\left(  C,C,\mathbf{r}_{3},\mathbf{r}_{4},E\right)
\nonumber\\
&  \times\left\langle V_{\mathrm{IR}}\left(  \mathbf{r}_{3},C\right)
V_{\mathrm{IR}}\left(  \mathbf{r}_{4},C\right)  \right\rangle _{\left(
x,y\right)  }. \label{self_IR_num_appr}%
\end{align}
If no discretization point lies directly on the steep interface (as in most
finite differences discretizations), the interface is at the center between
two adjacent grid points. Then, one half of this self-energy can be
distributed on each point adjacent to the interface. In this way, the Green's
function on the interface is approximated with the Green's function of the
respective adjacent grid point. It is worth noting that the prefactor of
$\Delta^{2}$ in formulas of Sec.~\ref{sec:elast_interface} originates from the
transformation of $z_{3}$ and $z_{4}$ to delocalized basis states (see
e.g.~Ref.~\onlinecite{lee}).

\paragraph{Gaussian roughness correlation}

As pointed out in Sec.~\ref{sec:elast_interface}, it is a common approximation
to assume a Gaussian-shaped in-plane interface roughness autocorrelation
$\left\langle V_{\mathrm{IR}}\left(  \mathbf{r}_{3},C\right)  V_{\mathrm{IR}%
}\left(  \mathbf{r}_{4},C\right)  \right\rangle _{\left(  x,y\right)  }$ (see
e.g. Refs.~\onlinecite{nag, unuma, Yamakawa_IR})%
\begin{align}
&  \left\langle V_{\mathrm{IR}}\left(  \mathbf{r}_{3},C\right)  V_{\mathrm{IR}%
}\left(  \mathbf{r}_{4},C\right)  \right\rangle _{\left(  x,y\right)
}\nonumber\\
&  =\frac{V_{\mathrm{o}}^{2}}{4}\mathrm{\exp}\left(  -\frac{\left|
\mathbf{r}_{3}-\mathbf{r}_{4}\right|  ^{2}}{\Lambda^{2}}\right)  ,
\end{align}
with the correlation length $\Lambda$ of the roughness in $\left(  x,y\right)
$-direction. When this function is inserted into Eq.~(\ref{self_IR_num_appr}),
Fourier transformed into the in-plane momentum space and integrated over the
in-plane scattering angle, the scattering self-energy reads~\cite{nemo1d}%
\begin{align}
&  \Sigma^{<,R}\left(  C,C,k,E\right)  =\frac{\Lambda^{2}}{4\pi}%
\frac{V_{\mathrm{o}}^{2}}{4}\exp\left(  -k^{2}\Lambda^{2}/4\right) \nonumber\\
&  \times\int\mathrm{d}^{2}q\mathrm{I}_{0}\left(  \frac{kq\Lambda^{2}}%
{2}\right) \nonumber\\
&  \times\exp\left(  -q^{2}\Lambda^{2}/4\right)  G^{<,R}\left(
C,C,q,E\right)  ,
\end{align}
with the modified Bessel function%
\begin{equation}
\mathrm{I}_{0}\left(  x\right)  =\frac{1}{\pi}\int_{0}^{\pi}\mathrm{d}\phi
\exp\left(  x\cos\phi\right)  .
\end{equation}

\paragraph{Exponential shaped roughness correlation}

Studies of different material systems have shown,~\cite{Yamakawa_IR,Kruithof}
that an exponential shape of the roughness autocorrelation may better
reproduce experimental data. Then, the autocorrelation of the scattering
potential reads%
\begin{align}
&  \left\langle V_{\mathrm{IR}}\left(  \mathbf{r}_{3},C\right)  V_{\mathrm{IR}%
}\left(  \mathbf{r}_{4},C\right)  \right\rangle _{\left(  x,y\right)
}\nonumber\\
&  =\frac{V_{\mathrm{o}}^{2}}{4}\mathrm{\exp}\left(  -\frac{\left|
\mathbf{r}_{3}-\mathbf{r}_{4}\right|  }{\lambda}\right)  ,
\end{align}
where $\lambda$ is the exponential correlation length. When the Fourier
transform of this function is inserted in Eq.~(\ref{self_IR_num_appr}) and the
angle integral of the convolution is performed, the self-energy reads%
\begin{align}
&  \Sigma^{<,R}\left(  C,C,k,E\right) \nonumber\\
&  =\frac{\lambda^{2}}{\pi}\frac{V_{\mathrm{o}}^{2}}{2}\int_{0}^{\infty
}\mathrm{d}q\mathrm{E}\left(  2\sqrt{\frac{\lambda^{2}kq}{1+\lambda^{2}\left(
k+q\right)  ^{2}}}\right) \nonumber\\
&  \times\frac{qG^{<,R}\left(  C,C,q,E\right)  }{\sqrt{1+\lambda^{2}\left(
k+q\right)  ^{2}}\left[  1+\lambda^{2}\left(  k-q\right)  ^{2}\right]  }.
\end{align}
Here, $\mathrm{E}\left(  x\right)  $ denotes the complete elliptical integral
of the second kind,%
\begin{equation}
\mathrm{E}\left(  x\right)  =\int_{0}^{x}\frac{\sqrt{1-k^{2}t^{2}}}{1-t^{2}%
}\mathrm{d}t.
\end{equation}

\subsubsection{Scattering from alloy disorder}

As discussed in Sec.~\ref{sec:elas_alloy}, the perturbing potential of
electrons scattering on alloy disorder is given by the difference of the real
conduction band offset of the material at position $z$ to the idealized one
[see Eq.~(\ref{eq:V_al})]. The derivation of the self-energies for the
scattering on alloy disorder is very analog to the scattering on charged
impurities: In a random alloy, the first order contribution of $\delta V$
vanishes and the second order in $\delta V$ gives the first nonvanishing
scattering self-energy. Band edge fluctuations
\begin{align}
&  \left\langle \delta V\left(  x_{3},y_{3},z_{3}\right)  \delta V\left(
x_{4},y_{4},z_{4}\right)  \right\rangle _{\left(  x,y\right)  }\nonumber\\
&  =\delta V^{2}\left\langle \delta x\left(  x_{3},y_{3},z_{3}\right)  \delta
x\left(  x_{4},y_{4},z_{4}\right)  \right\rangle _{\left(  x,y\right)  }%
\end{align}
are caused by concentration fluctuations $\delta x$. With the assumption of
local randomness within the volume $\Omega_{0}$ [see also Eq.~(\ref{eq:dV})]
\begin{equation}
\left\langle \delta x\left(  \mathbf{x}\right)  \delta x\left(  \mathbf{x}%
^{\prime}\right)  \right\rangle =\Omega_{0}x\left(  1-x\right)  \delta\left(
\mathbf{x}-\mathbf{x}^{\prime}\right)  ,
\end{equation}
the alloy scattering self-energy reads
\begin{align}
\Sigma^{<,R}\left(  z_{3},z_{4},k,E\right)   &  =\Omega_{0}x\left(
1-x\right)  \delta V^{2}\delta\left(  z_{3}-z_{4}\right) \nonumber\\
&  \times\int\mathrm{d}^{2}qG^{<,R}\left(  z_{3},z_{4},q,E\right)  .
\end{align}

\subsubsection{Inelastic electron-electron scattering}

In most of the NEGF\ implementations, the electron-electron interaction is
only included in the Hartree approximation. This mean field approach requires
solving the Poisson equation self-consistently with the electron distribution
within the active device. The inclusion of inelastic scattering, however,
requires solving the NEGF\ equations beyond the Hartree-Fock approximation, as
the Fock term gives only elastic scattering contributions. Higher order
correlation terms increase the numerical effort dramatically: Correlation
terms are proportional to convolutions of at least three single electron
Green's functions. Approximations of these convolution integrals typically
combine most of the electron Green's functions into the interaction term $W$
and assume a product ansatz for the self-energy, i.e., $GW$
approximations.~\cite{Hedin1965} Both terms of this approximation are subject
to further approximations. Assessments of typical approximations have shown
that lower order approximations may even yield more realistic results than
inclusions of higher order terms.~\cite{vBarth1996,vBarth1998} Examples for
NEGF\ implementations of inelastic electron-electron scattering on cascade
devices so far only include the plasmon pole approximation for $W$ and a
low-order $GW_{0}\,$%
approximation.~\cite{2009PhRvB..79s5323K,2009ApPhL..95w1111S} The challenge of
a numerically efficient implementation of electron-electron scattering in
NEGF\ is not conclusively solved in literature. The assessment of possible
approximations of electron-electron scattering in NEGF\ is beyond the scope of
this article. We refer the reader to the literature for more thorough discussions.~\cite{Thygesen,2009PhRvB..79s5323K,2009ApPhL..95w1111S}

\subsubsection{General remarks on scattering self-energies}

It is worth making a few remarks on the implementation of scattering in NEGF
in general. When NEGF is solved self-consistently with the Poisson equation,
convergence essentially requires meshes in energy and momentum that resolve
resonant states and van Hove singularities well. Most often that is only
possible with inhomogeneous and adaptive meshes in energy and
momentum.~\cite{nemo1d,Kubis_phd,Samarth} All scattering self-energies of the
previous sections are given in their analytical form. When they are
discretized in energy and momentum, they represent effective scattering
between energy and momentum intervals of different sizes. For numerical
current conservation, the discretization of the energy and momentum integrals
in the scattering self-energies has to carefully acount for these different
sizes.~\cite{Kubis_phd} In most NEGF implementations, current conservation is
ensured by solving the scattering self-energies in the self-consistent Born
approximation. Alternative approaches that solve the self-energies in current
conserving non-self-consistent approximations are the B\"{u}ttiker probe
approach~\cite{first_buettiker} and the lowest order approximation of
Ref.~\onlinecite{2012PhRvB..86.161404M}. These approaches are either limited
to close-to equilibrium situations or to weak scattering
only.~\cite{2012PhRvB..86.161404M,Kubis_phd} In other words, the Green's
functions that appear in the scattering self-energies of the previous
paragraphs are the full scattered Green's functions. Solving the Green's
functions and self-energies in the full self-consistent Born approximation
requires iterative solutions of the involved NEGF\ equations. Since this is
numerically challenging, it is tempting to either truncate the iterations
before convergence, or to apply lower order approximations. While some of the
low order approximations may still conserve the current, they can easily face
artifacts. For instance, the neglect of $G^{<}$ contributions to the inelastic
$\Sigma^{R}$ self-energy has been shown to violate the Pauli blocking.~\cite{nemo1d,2009PhRvB..79s5323K}

The full energy and momentum integrals of self-energies such as the polar
optical phonon scattering represent another high numerical burden. This has
motivated several authors to simplify these integrals with representative
transferred momenta. While this approximation eases the numerical load a lot,
it has to be treated with great care, since the integrand functions vary
significantly with the transferred momentum.~\cite{Schmielau_ktyp,Kubis_assess}

Nonlocal scattering mechanisms such as the charged impurity scattering cause a
much higher numerical load than local scattering mechanisms: Nonlocal
scattering increases the number of nonvanishing elements in the inverse of the
retarded Green's function in Eqs.~(\ref{set1}) and thereby increases the
number of floating point operations to solve $G^{R}$. Even the solutions of
the scattering self-energies obviously require more operations and more
memory, too. When nonlocal scattering is approximated with local scattering,
at least an appropriate compensation factor has to be introduced to avoid
underestimation of the scattering rate.~\cite{Klimeck,Kubis_assess,Luisier_elph}

\subsection{\label{sec:NEGF_result}Selected results of NEGF\ on terahertz QCLs}

In the following, some NEGF\ results for terahertz QCLs are presented that
illustrate typical and important features of NEGF. All results represent the
electron propagation in the QCLs in terms of stationary vertical transport in
laterally homogeneous quantum well heterostructures. The QCLs are considered
to be in contact with two charge reservoirs at $z=R$ and $z=L$, respectively.
Thereby, the charge transport is treated as a scattering problem from source
to drain, with the open device forming the scattering center. Cascade periods
that surround the active device are included within the contact self-energies
that supply electrons in equilibrium Fermi distributions. Incoherent
scattering of electrons on optical and acoustic phonons, charged impurities,
and rough interfaces is included. The electron-electron interaction in the
Hartree approximation is taken into account. Material parameters are taken
from Ref.~\onlinecite{texbook}. All results shown in this section are for
terahertz QCLs consisting of periodically repeated GaAs and Al$_{.15}%
$Ga$_{.85}$As layers. Each period consists of layers of the widths $\left(
30\right)  ~92~\left(  55\right)  ~80~\left(  27\right)  ~66~\left(
41\right)  ~155~\text{\AA}$, where the values in parentheses indicate the
Al$_{.15}$Ga$_{.85}$As barriers.~\cite{wien_doping_last} Only the widest well
is doped with a sheet doping density of $1.9\times10^{10}$~cm$^{-2}$. Only a
single period is explicitly calculated, whereas the remaining periods are
covered within the lead model.

\subsubsection{QCL\ work principle - energy resolved spectral function}

One important difference between NEGF\ and the density matrix method is that
the resulting Green's function and self-energies are energy resolved. This
additional information increases the numerical load, but it can unveil
important insight into the device physics. One of the energy resolved
quantities is the spectral function of Eq.~(\ref{spectral_function}) as it can
illustrate the mechanisms that are responsible for gain in the presently
studied QCL structures.

\begin{figure}[tb]
\includegraphics[width=8.6cm]{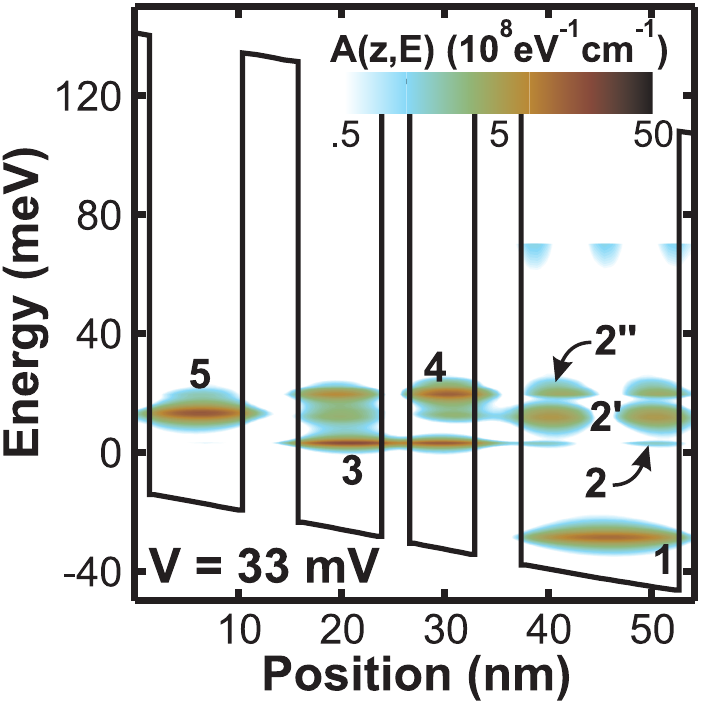}\caption{(Color online)
Contour plot of the spectral function $A\left(  z,E\right)  $ of the QCL, in
units of $10^{8}\,\text{eV}^{-1}\,\text{cm}^{-1}$, as a function of position
$z$ in nm and energy $E$ in meV. The zero in energy marks the chemical
potential of the source. The applied bias voltage is $33$~mV per period. The
solid line indicates the self-consistent potential profile. The spectral
function is only shown within the energy interval from $-50$ to $70$~meV. The
labels number the relevant resonances that are discussed in the main text.}%
\label{qcl_ldos_33mv}%
\end{figure}

Figure~\ref{qcl_ldos_33mv} depicts a contour plot of the energy and spatially
resolved spectral function of the terahertz QCL for vanishing lateral momentum
$k=0$ at a bias voltage of $33$~mV per period which is below but close to the
lasing threshold (of about $50$~mV per period). The maxima of the spectral
function represent resonant states. All states show a finite width and a fine
structure that results from the coherent and incoherent coupling of all well
states with one another. In other words, the width of the broadened levels
correspond to the total lifetime of the electrons in the respective device
state. The upper laser level (labeled by $\#4$) which is predominantly an
antibonding state is aligned with the confined state $\#5$ in the leftmost
source-sided quantum well and therefore gets filled by resonant tunneling. The
lower laser level $\#3$ gets efficiently emptied by two mechanisms. First, the
bonding state $\#3$ is aligned with the states $\#2$ and $\#2^{\prime}$ of the
rightmost well which allows its coherent depletion by tunneling. Second, the
energy difference between this state and the lowest resonance state ($\#1$)
matches approximately the energy of an LO phonon ($36$~meV) which leads to an
additional depletion by the resonant emission of LO phonons. At the shown
voltage the alignment is visible, but not fully established. The detuning of
the alignment leads to a strong coherent leakage which is in more detail
discussed below.

\subsubsection{Effect of incoherent scattering}

Transport calculations of mid-infrared QCLs suggest that the contribution of
coherent propagation to the charge transport in QCLs is insignificant compared
to the efficient incoherent scattering.~\cite{2001PhRvL..87n6603I}
Furthermore, Monte Carlo solutions of the semiclassical Boltzmann equation
that neglect (coherent) correlation effects between laser states have
successfully predicted charge transport in terahertz QCLs near
threshold.~\cite{2009JAP...105l3102J} Nevertheless, it can be shown that a
general answer to the question whether the transport in terahertz QCLs is
mainly coherent or incoherent cannot be given, since the balance between both
is sensitive to details of the device structure. It actually turns out that
the four well resonant phonon terahertz QCL of Fig.~\ref{qcl_ldos_33mv} is a
very instructive example for the interplay of coherent and incoherent transport.

\begin{figure}[tb]
\includegraphics[width=8.6cm]{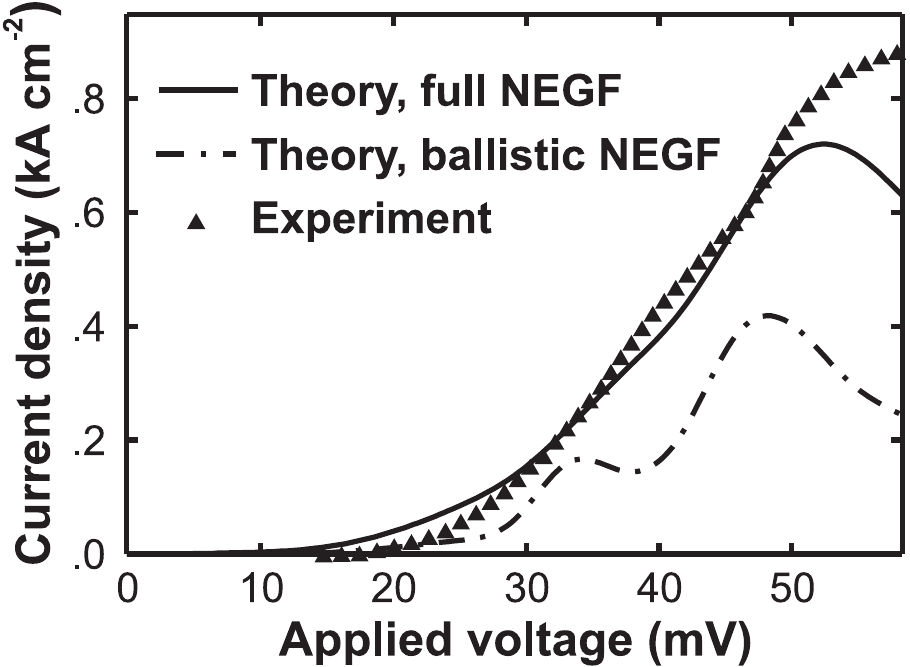}%
\caption{Experimental (triangles) and theoretically predicted (solid lines)
\textit{I-V} characteristics of the QCL of Fig.~\ref{qcl_ldos_33mv}. Ballistic
calculations (dash-dotted), i.e. NEGF\ calculations ignoring any incoherent
scattering mechanism, underestimate the current density. At the two ballistic
resonances, however, a large portion of the realistic current is coherent.}%
\label{qcl_iv_ballistic_compare}%
\end{figure}

A rather large portion of the current in the\ resonant phonon terahertz QCLs
of Ref.~\onlinecite{wien_doping_last} stems from coherent transport. This can
be deduced from Fig.~\ref{qcl_iv_ballistic_compare}, comparing experimental
and various theoretical results for the current density of this terahertz QCL.

The solid curve shows the calculation where phonon, impurity, interface
roughness and electron-electron scattering in the Hartree approximation have
been fully included. The dashed curve in Fig.~\ref{qcl_iv_ballistic_compare}
shows the limiting case where all scattering self-energies have been
artificially turned off. Obviously, incoherent scattering enhances the current
density in this QCL. In contrast, incoherent scattering reduces the current
density when no confining barriers are present (such as in transistor
devices). In fact, it has been shown that incoherent scattering decreases the
current in low resistive devices, whereas the current in high resistive
devices is increased by incoherent scattering.~\cite{first_buettiker}

\subsubsection{Coherent regime}

The two maxima of the ballistic current in Fig.~\ref{qcl_iv_ballistic_compare}
near $33$ and $48$~mV correspond to aligned laser states. At these voltages,
electronic states are generated that extend across the entire QCL period. Such
a delocalized state can be found in Fig.~\ref{qcl_ldos_33mv}. At a bias
voltage of $33$~mV per period, the lowest state in the injector well (labeled
with $\#5$) is aligned with the second state ($\#2^{\prime}$) of the rightmost
quantum well, i.e., the collector well. This alignment generates a finite
spectral function in the gap between the upper ($\#4$) and the lower ($\#3$)
laser level. In this way, the spectral function at the energy $E=12~$meV
remains finite in every quantum well of the active period. Thus, electrons of
this energy can coherently tunnel throughout the QCL\ period and maintain a
maximum in the ballistic \textit{I-V} characteristic.

Resonances in the current density caused by states that extend across the
total QCL\ period are already known in literature. In the area of Monte Carlo
simulations of QCLs, anticrossing of laser states also lead to such highly
delocalized states which eases the coherent multibarrier tunneling. In the
Monte Carlo formalism, however, the laser states are determined with a
Hermitian Schr\"{o}dinger equation which yields infinite state lifetimes.
Thus, when the alignment conditions of the above delocalized states are met,
the corresponding resonance in the \textit{I-V} characteristics predicted in
the Monte Carlo formalism is very large. It has been shown in Ref.~\onlinecite
{2005JAP....98j4505C} that a finite lifetime of the laser states reduces the
height of the artificial current peaks significantly. The finite lifetime of
the electrons in the ballistic calculations of
Fig.~\ref{qcl_iv_ballistic_compare} originates from the finite probability for
electrons to leave the device and thereby to ''decay'' into lead states. In
fact, the Schr\"{o}dinger equation that corresponds to the solution of the
Dyson equation Eq.~(\ref{set1a}) is non-Hermitian, irrespective whether
incoherent scattering is implemented or not. Thus, artificial spikes of the
\textit{I-V} characteristics cannot be seen in NEGF\ for open devices.

\subsubsection{Incoherent regime}

\begin{figure}[ptb]
\includegraphics[width=8.6cm]{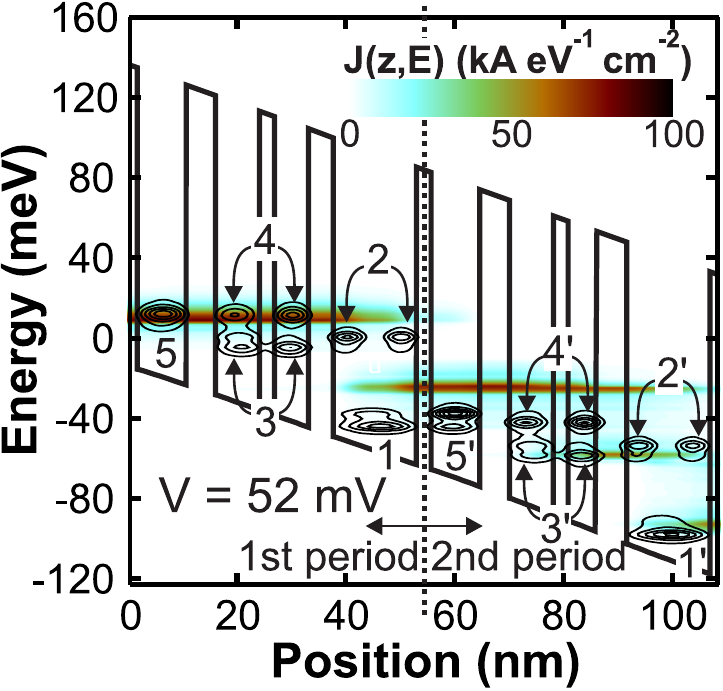}\caption{(Color online) Calculated
conduction band profile (thick line), contour lines of the energy and
spatially resolved spectral function $A(z,E)$ at vanishing in-plane momentum
in two adjacent periods of the QCL at a bias voltage of 52 mV per period in
the relevant energy range between -121 meV and 78 meV. The contour plot shows
the spatially and energy resolved current density $J(z,E)$ in the same energy
range. The zero in energy marks the chemical potential of the source. The
dotted line marks the boundary between first and second period. }%
\label{ecur_2periods}%
\end{figure}

When the applied voltage of the QCL in Fig.~\ref{qcl_ldos_33mv} exceeds
$33$~mV the completely delocalized state breaks apart and the ballistic
\textit{I-V} characteristic shows a negative differential resistivity. Such a
situation is depicted by the contour lines of Fig.~\ref{ecur_2periods} as they
show the spectral function of the same QCL as in Fig.~\ref{qcl_ldos_33mv}, but
at an applied bias voltage of $52$~mV per period. Here, the most prominent
maxima of the spectral function separate into two groups of partly delocalized
states: One group consists of the aligned injector ($\#5$ in
Fig.~\ref{qcl_ldos_33mv}) and upper laser level ($\#4$) and allows for the
coherent propagation from the source sided device boundary to the center of
each QCL period. The second group of delocalized states is generated by the
alignment of the lower laser level ($\#3$) and the second state of the
collector well ($\#2$) and eases the electronic propagation from the center of
each QCL period to its drain sided limit. Since the states of both groups are
energetically separated, electronic transitions between them require
dissipation of energy. Thus, at this bias voltage, the coherent propagation
throughout the total QCL period is suppressed, and the ballistic current is
significantly smaller than the corresponding result for incoherent scattering
included (see Fig.~\ref{qcl_iv_ballistic_compare}).

\subsubsection{Incommensurate transport periodicity - energy resolved current density}

Typically, electrons in quantum cascade devices are expected to follow the
periodicity of the structure. In theory, this is not necessarily the case, as
can be seen from the contour plot of the spectral function at vanishing
in-plane momentum shown in Fig.~\ref{ecur_2periods} for two adjacent QCL
periods. The states associated with the first and second QCL period are
labeled by numbers and primed numbers, respectively. In both periods, the
alignment of the states follows the scheme described in detail for
Fig.~(\ref{qcl_ldos_33mv}). In so far, the spectral function is periodic. In
particular, the energy difference between the lower laser levels (3 and 3')
and the lowest collector states (1 and 1') match the energy of an LO phonon in
both periods. Nevertheless, the carrier distribution deviates from the
geometric QCL periodicity. This can be seen by a contour graph of the local
energy resolved current density $J\left(  z,E\right)  $ [as defined in
Eq.~(\ref{j_of_z_E})] in Fig.~\ref{ecur_2periods}. The function $J\left(
z,E\right)  $ shows spatially constant (i.e. horizontal) stripes in regions
where the electrons propagate without dissipating energy. Disruptions of these
horizontal stripes mark positions where LO-phonons get emitted. The figure
shows that the number of emitted LO phonons is not equal for adjacent QCL
periods. When the electrons have passed the first period and traversed a
potential drop of $52~$meV, they have emitted only one LO-phonon of energy
$36$~meV. The included elastic and inelastic scattering mechanisms are not
able to dissipate the remaining $16$~meV within this QCL period. This is a
consequence of the good state alignment that supports efficient coherent
multi-barrier tunneling. Consequently, the electrons enter the second period
with an in-plane kinetic energy of $16$~meV. As can be seen in
Fig.~\ref{ecur_2periods}, the energy of the leftmost current stripe coincides
with states 5 and 4, whereas the following current stripe lies above the
corresponding states 5' and 4'. Thus, these propagating electrons are now able
to emit an LO-phonon, ending up in and occupying the lower laser level 3'.
This occupancy leads to the build-up of the current stripe near $z=80~$nm in
Fig.~\ref{ecur_2periods} that is absent in the first period. The electrons can
now tunnel resonantly from state 3' into 2' and scatter into the lowest
collector state 1' by the emission of an additional LO-phonon. Thus, the
electrons have emitted a total of $3$ LO phonons ($3\times36~\mathrm{meV}%
=108$~$\mathrm{meV}$) across $2$ QCL periods (voltage drop of $2\times
52~\mathrm{meV}=104$~$\mathrm{meV}$) and are finally fully thermalized. The
remaining small energy discrepancy can be gained from absorbing or emitting
acoustic phonons. This process is repeated in the subsequent QCL periods such
that a commensurable charge distribution with period two is established. Since
the detailed energy balance depends on the applied bias voltage, the carrier
density and current distribution may even become incommensurable with the
geometric periods. A consequence of this incomplete carrier thermalization is
a significant reduction in the occupation inversion and the optical gain in
every other period. The calculated gain shows a drop of approximately $65\%$
in the second period in Fig.~\ref{ecur_2periods}. It has been estimated that
electron-electron scattering cannot relax the electrons and restore the
periodicity of the carrier distribution to a single QCL
period.~\cite{2009PhRvB..79s5323K}. This is mainly due to the efficient
coherent tunneling of the electrons which supports resonant LO phonon emission
instead.~\cite{Kubis_Klimeck_APL} Recent experimental findings indicate the
heating of the electron gas described here.~\cite{Fujita}

\subsubsection{Temperature degradation - energy resolved density}

\begin{figure}[tb]
\includegraphics[width=8.6cm]{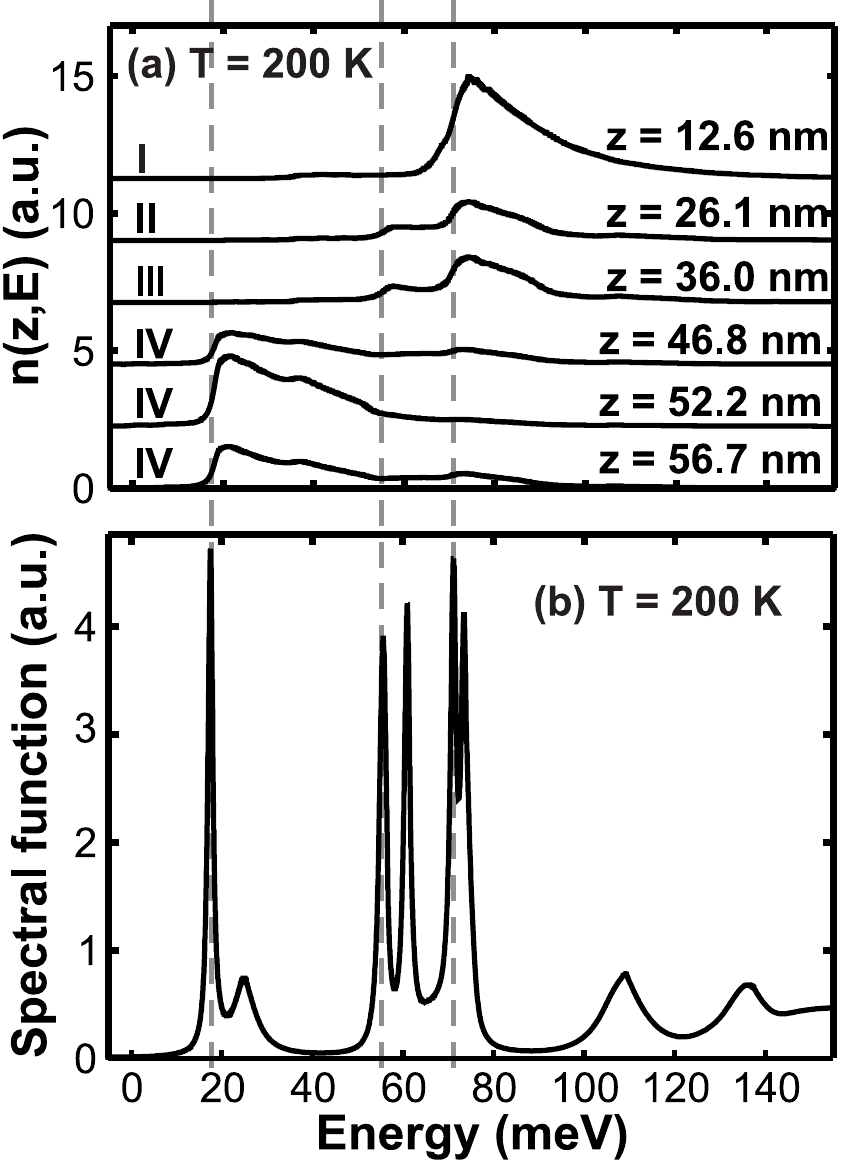}\caption{(a) Energy
resolved density at a bias of $50~$mV per period at a lattice temperature of
$200~$K for various spatial positions in the QCL of Fig.~\ref{qcl_ldos_33mv}.
The Roman numbers denote the injector ($\mathrm{I}$), the two active
($\mathrm{II}$ and $\mathrm{III}$) and the collector well ($\mathrm{IV}$). (b)
Spatially integrated spectral function at $200~$K; the peaks mark the energy
of resonant states.}%
\label{qcl_eden_temperature200}%
\end{figure}

\begin{figure}[tb]
\includegraphics[width=8.6cm]{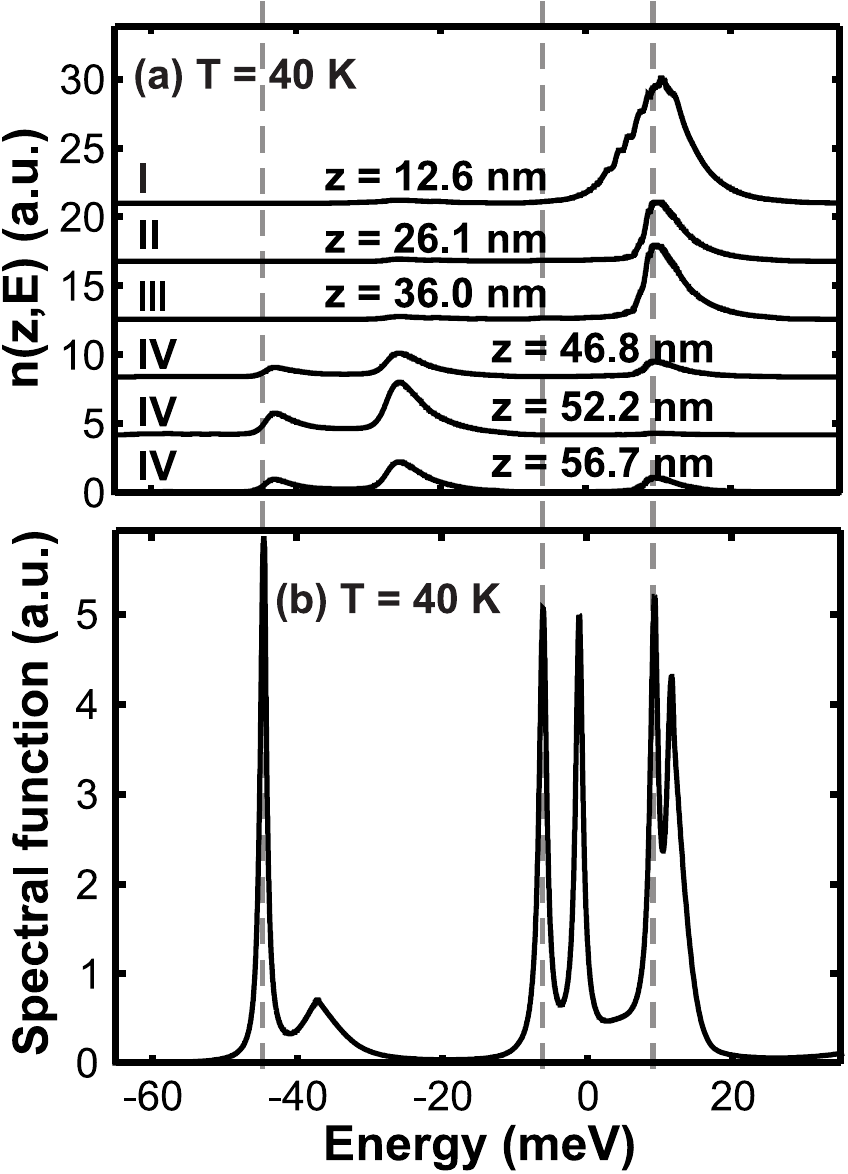}\caption{(a) Energy
resolved density and (b) spatially integrated spectral function for the same
QCL as in Fig.~\ref{qcl_eden_temperature200}, but at a lattice temperature of
$40~$K.}%
\label{qcl_eden_temperature40}%
\end{figure}

With increasing temperature, the number of electrons with high in-plane
kinetic energies becomes significant. This can be seen in
Fig.~\ref{qcl_eden_temperature200}(a) that shows the energy resolved electron
density $n\left(  z,E\right)  $ in the injector well (labeled with
$\mathrm{I}$), the two active quantum wells ($\mathrm{II}$ and $\mathrm{III}$)
and the collector well ($\mathrm{IV}$) of the four well terahertz QCL of
Fig.~\ref{qcl_ldos_33mv} at a temperature of $200$~K. The subbands are
depicted in Fig.~\ref{qcl_eden_temperature200}(b) which shows the spatially
integrated spectral function at vanishing in-plane momentum. The dashed lines
mark the subband energies of the lowest state in the QCL period ($E=17.4~$meV)
as well as the lower ($E=55.5~$meV and $E=60.9~$meV) and the upper laser
levels ($E=71.1~$meV and $E=73.4~$meV). Electrons with more than $20$~meV of
kinetic energy above the subband minima can be found in the upper laser level.
This kinetic energy suffices for electrons in the upper laser level to scatter
into the lower laser level by emitting an optical phonon (thermally activated
phonon emission). In addition, some electrons in the collector well
($\mathrm{IV}$) can reach sufficiently high kinetic energies to fill the lower
laser level. These mechanisms reduce the occupation inversion and the optical gain.

For comparison, the energy resolved density and the spectral function of this
QCL at $40~$K lattice temperature are shown in
Figs.~\ref{qcl_eden_temperature40}(a) and (b), respectively. It is easy to see
that both mechanisms, the thermally activated phonon emission and the thermal
backfilling are absent at this lower temperature.~\cite{APL_Yasuda}

\section{\label{sec:Concl}Conclusion and outlook}

The advancement of QCL simulation approaches is driven by an intrinsic
motivation to further improve the description and numerical implementation of
the underlying physical processes, and by experimental progress requiring
improved and extended simulations. A current research topic is the
consideration of the relevant quantum effects at an acceptable numerical
efficiency. Furthermore, the self-consistent inclusion of the laser field in
the simulation is of great interest, expecially in the context of frequency
conversion structures for the terahertz and near infrared regime. Besides,
alternative material systems extending the spectral range covered by QCLs or
enabling high temperature terahertz operation might necessitate a theoretical
treatment beyond the conduction band $\Gamma$ valley.

\subsection{Development of hybrid quantum-semiclassical and approximate
quantum simulation approaches}

Quantum coherence effects can play a pronounced role especially in terahertz
QCLs where the energetic spacing between the quantized levels is relatively
small \cite{2005JAP....98j4505C,2009PhRvB..79p5322W}. On the other hand,
quantum transport approaches such as the density matrix formalism or NEGF are
numerically much more demanding than their semiclassical counterparts.
Especially for full 3D simulations, this impedes their application to QCL
design and optimization. Thus, there have been various efforts to implement
certain aspects of the quantum transport theory into the semiclassical
description to increase its accuracy and range of validity, but preserving its
relative numerical efficiency. Examples include phenomenologically extending
the standard rate-equation models to include coherence
\cite{2009PhRvB..80s5317G}, incorporating energetic broadening of the
quantized states into EMC simulations
\cite{2009JAP...105h3722A,2013ApPhL.102a1101M}, and combining the
semiclassical Monte Carlo method with a density matrix approach
\cite{2005JAP....98j4505C,2012ApPhL.100a1108B}. Hybrid density matrix-Monte
Carlo methods exploit the fact that resonant tunneling dominates the transport
only in some regions, e.g., through a thick barrier such as an injection
barrier, while a semiclassical transport description is adequate in the rest
of the QCL structure \cite{2005JAP....98j4505C}. A special challenge in hybrid
quantum-semiclassical simulations will be a self-consistent description of
dephasing going beyond phenomenological dephasing time models
\cite{2005JAP....98j4505C}. Furthermore, free carrier absorption
\cite{1992JAP....72.4966B,2011PhRvB..84t5319W} and electron leakage into the
continuum of states \cite{2010PhyE...42.2628M,2012PhRvB..85s5326F} can have an
effect and thus should be adequately implemented into the simulations.

\subsection{Modeling of innovative QCL designs based on alternative material systems}

Innovative QCL designs based on alternative material systems hold the
potential of extending the spectral range covered by QCLs or enabling high
temperature terahertz operation. The exploration of alternative QCL designs
and subsequent systematic optimization is greatly facilitated by careful
modeling. Up to now, lasing has only been obtained for n-type QCLs, using
InGaAs/InAlAs on InP substrate or GaAs/AlGaAs on GaAs
\cite{2007NaPho...1..517W,2012NaPho...6..432Y}, and to a smaller extent
antimonides such as InAs/AlSb \cite{2003ApPhL..82.1003O,2010ApPhL..96n1110C}.
Since all these materials (apart from AlSb) have direct bandgaps, QCL
simulations have up to now focused on the conduction band $\Gamma$ valley.
Thus, for alternative material systems with an indirect bandgap or using
valence band transitions, the simulation methods will have to be
correspondingly extended and adapted. Furthermore, additional effects such as
strong polarization fields might have to be included.

\begin{figure}[ptb]
\includegraphics{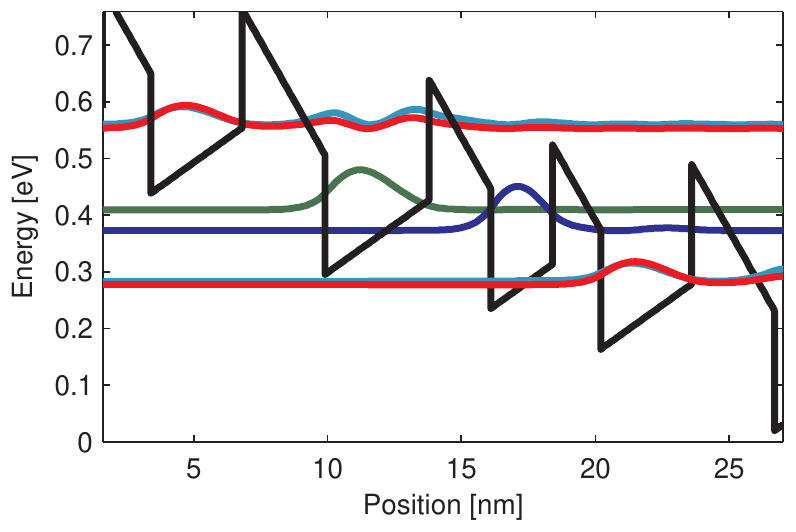}
\caption{{}(Color online) Conduction band profile and probability densities
for an experimental GaN/AlGaN structure.}%
\label{fig:gan}%
\end{figure}

One example for an alternative material system which could extend the
application range of QCLs is GaN/AlGaN. Due to the large conduction band
discontinuity, nitride-based QCL structures are promising candidates for short
wavelength applications \cite{2010JAP...108j3704S,2000ApPhL..77..334G}.
Furthermore, the GaN/AlGaN material system is also interesting for the
development of high temperature terahertz QCLs because of the large optical
phonon energies ($E_{\mathrm{LO}}\approx90\,\mathrm{meV}$)
\cite{2010JAP...108j3704S,2004ApPhL..84.2995J,2009JAP...105k3103B,2005SuMi...37..107S}%
. However, no working nitride-based QCL has been demonstrated to date, only
absorption and electroluminescence has been observed
\cite{2010JAP...108j3704S,2000ApPhL..77..334G}. First NEGF\ results on a
GaN-based QCL indicated a too large level broadening to maintain
lasing~\cite{2012JAP..111.083105Y}. The feasibility of terahertz GaN/AlGaN
QCLs has also been studied using self-consistent rate equation models
\cite{2004ApPhL..84.2995J,2005SuMi...37..107S} and Monte Carlo simulations
\cite{2009JAP...105k3103B}. For simulating such structures, the band bending
effects due to the strong intrinsic polarization fields have to be considered.
These are clearly visible in Fig.\thinspace\ref{fig:gan}, showing the
conduction band profile and probability densities for an experimental
GaN/AlGaN structure \cite{2010JAP...108j3704S}.

An alternative to the conventional n-type QCLs are designs employing
hole-based intersubband transitions in the valence band. Si/SiGe-based p-type
structures are interesting since this material system offers a high
integration potential and low process costs. For terahertz applications, a
major advantage versus GaAs-based terahertz emitters is that potentially
higher operating temperatures can be obtained due to the absence of LO phonon
scattering in type IV semiconductors \cite{2002ApPhL..81.1543L}. A working
Si/SiGe QCL has not yet been demonstrated; however, electroluminescence could
be achieved in the mid-infrared regime due to transitions between quantized
heavy hole states \cite{2000Sci...290.2227D}, and in the terahertz regime due
to light hole-heavy hole transitions \cite{2002ApPhL..81.1543L}. The
simulation of hole-based devices is far more elaborate than for comparable
n-type structures, since mixing effects of the heavy hole, light hole and
split-off band have to be considered in the carrier transport simulations and
the Schr\"{o}dinger solver. Thus, a more complex description than the
effective mass approximation typically used for n-type devices has to be
employed, such as \textbf{k.p} theory
\cite{2004JAP....96.6803I,2004PhRvB..69w5308I}. Due to the absence of working
p-type QCLs, only few theoretical studies based on rate equations
\cite{2001ApPhL..78..401F,2004JAP....96.6803I} and the Monte Carlo method
\cite{2004PhRvB..69w5308I} have been published. If p-type QCLs based on
Si/SiGe or another material system become technically feasible, the adaption
and efficient numerical implementation of the different simulation approaches
discussed in this review paper for the valence band will become a demanding task.

Alternatively, n-type terahertz QCLs based on Ge/SiGe have been considered
\cite{2007JAP...102i3103D}. Here, the lasing transition takes place in the
conduction band $L$ valleys, which are in contrast to the $\Gamma$ valley both
anisotropic and degenerate. Rate equation simulations have been performed to
investigate the feasibility of such designs \cite{2009ApPhL..95m1103L}.
Intersubband absorption has been experimentally observed in the conduction
band of Ge/SiGe quantum well structures \cite{2009ApPhL..95e1918D}, but up to
now no operating QCL has been demonstrated.

\subsection{Inclusion of the optical cavity field}

Simulations of the coupled electron and optical dynamics have mainly been
performed using one-dimensional approaches such as rate
\cite{2010SeScT..25d5025S,2013SeScT..28j5008P} and Maxwell-Bloch
\cite{2007PhRvA..75c1802W,2010OExpr..1813616G,2009PhRvL.102b3903M} equations
or the density matrix method \cite{2010NJPh...12c3045T}. On the other hand,
with very few exceptions
\cite{2010ApPhL..96a1103J,2012ApPhL.100a1108B,2012ApPhL.101u1113W} advanced
three-dimensional carrier transport simulations have focused on the electron
dynamics, completely ignoring the light field. However, the inclusion of the
lasing field is not only required to study the actual lasing operation, but
also to model terahertz and infrared frequency conversion QCL sources. Here,
the nonlinear optical properties of the QCL heterostructure must be adequately
implemented, and the modeling of the optical cavity field is crucial to
evaluate the nonlinear frequency conversion process. An example is terahertz
difference frequency generation in QCL structures, enabling room temperature
terahertz generation \cite{belkin2008room,2013OExpr..21..968L,a15} with
broadband frequency tunability \cite{a15,2012ApPhL.101y1121L}. A major goal is
here to push the available room temperature output power from currently
$120~\mathrm{\mu W}$ \cite{a15} to a few mW, as required for most technical
applications. Recently, such a QCL\ device has been modeled using an EMC
approach \cite{2013OExpr..21.6180J}. Artificial optical nonlinearities are
also attractive for extending QCL operation towards shorter wavelengths, e.g.,
by using frequency doubling \cite{2011IJQE...47..691V}.

The self-consistent inclusion of the optical cavity field requires adequate
electromagnetic modeling of the resonator to determine the mode solutions and
the corresponding overlap factor $\Gamma$ and losses $a_{\mathrm{w}}$,
$a_{\mathrm{m}}$. Increasingly, special resonator designs based on plasmonic
effects or exhibiting subwavelength structuring are used for beam shaping or
to enhance the efficiency and spectral purity. For example, the recent
performance improvement of terahertz difference frequency sources has largely
benefited from special cavity designs employing distributed feedback
structures \cite{2011ApPhL..99m1106L} or the Cherenkov effect
\cite{2012ApPhL.100y1104V,a15}. In addition to general electromagnetic
modeling approaches such as the finite element method, the development of
adapted methods for specific cavity types, which are numerically efficient and
provide more intuitive insight, is helpful for device simulation and optimization.%

\begin{acknowledgments}
We thank Prof. Irena Knezevic for valuable comments and suggestions. C.J. acknowledges support from P. Lugli at the TUM and funding by the Emmy Noether program of the Deutsche Forschungsgemeinschaft under Grant No. DFG JI115/1-1. T.K. acknowledges support by Hamamatsu Photonics K.K and the National Science Foundation (NSF) under Grants No. OCI-0749140, EEC-0228390, and ECCS-0701612. Computational resources of nanoHUB.org are gratefully acknowledged.
\end{acknowledgments}%

\end{document}